\documentclass[aps,prl,twocolumn,superscriptaddress,nofootinbib,preprintnumbers]{revtex4-1}

\usepackage{graphicx,color,amsmath,amssymb,bm}
\usepackage{slashed}
\usepackage{eqnarray}
\usepackage{comment}
\usepackage{amsmath}

\newcommand{\nocontentsline}[3]{}
\newcommand{\toclesslab}[3]{\bgroup\let\addcontentsline=\nocontentsline#1{#2\label{#3}}\egroup}
\newcommand{\tocless}[2]{\bgroup\let\addcontentsline=\nocontentsline#1{#2}\egroup}
\newcommand{\epar}{E_\parallel}

\newcommand{\bfE}{{\bf E}}
\newcommand{\bfB}{{\bf B}}
\newcommand{\edb}{{\bf E} \cdot {\bf B}}

\newcommand{\gagg}{g_{a\gamma\gamma}}

\newcommand{\nablabold}{\boldsymbol{\nabla}}

\newcommand{\ie}{\emph{i.e.~}}
\newcommand{\boldnabla}{\boldsymbol{\nabla}}

\newcommand{\ff}{\mathcal{F}}

\newcommand{\gammasc}{\gamma_{\rm sc}}
\newcommand{\deltaNS}

\newcommand{\apjl}{Astrophys. J. Lett.} 
\newcommand{\apjs}{Astrophys. J. Suppl. Ser.} 
\newcommand{\apss}{Astrophys. Space Sci.} 
\newcommand{\aap}{Astron. Astrophys.} 
\newcommand{\jcap}{J. Cosmol. Astropart. Phys.} 
\newcommand{\mnras}{MNRAS} 

\usepackage[export]{adjustbox}
\usepackage{tikz}
\usepackage[normalem]{ulem}
\usepackage{tkz-euclide}
\usetikzlibrary{decorations.pathmorphing}	
\tikzset{
    v/.style={decorate, decoration={snake, segment length=3mm, amplitude=0.75mm}, draw},
    f/.style={draw=black, postaction={decorate},
        decoration={markings,mark=at position .6 with {\arrow[very thick]{latex}}}},
    fb/.style={draw=black, postaction={decorate},
        decoration={markings,mark=at position .4 with {\arrowreversed[very thick]{latex}}}},
    fnar/.style={draw=black},
    g/.style={decorate, draw=black,
        decoration={coil,amplitude=3pt, segment length=3.5pt}},
    s/.style={dashed,draw=black, postaction={decorate},
        decoration={markings,mark=at position .55 with {\arrow[very thick]{latex}}}},
    sb/.style={dashed,draw=black, postaction={decorate},
        decoration={markings,mark=at position .55 with {\arrowreversed[draw=black,very thick]{latex}}}},
    snar/.style={dashed,draw=black,line width =1.25pt},
}

\usepackage{hyperref} 
\usepackage{amsmath}
\hypersetup{
    colorlinks=true,
    citecolor=blue,
    linkcolor=magenta,
    filecolor=magenta,
    urlcolor=blue,
}

\newcommand{\km}{{\, {\rm km}}}

\newcommand{\pc}{{\, {\rm pc}}}
\newcommand{\kpc}{{\, {\rm kpc}}}

\newcommand{\cm}{{\, {\rm cm}}}

\newcommand{\eV}{{\, {\rm eV}}}

\newcommand{\GeV}{{\, {\rm GeV}}}

\newcommand{\GHz}{{\, {\rm GHz}}}

\newcommand{\kms}{{\, {\rm km \, s^{-1}}}}

\newcommand{\ergpersecond}{{\, {\rm erg \, s^{-1}}}}

\newcommand{\rpc}{r_{\scriptscriptstyle{\rm PC}}}
\newcommand{\rns}{R_{\rm NS}}

\newcommand{\gcmag}{PSR J1745--2900}

\definecolor{mypurple}{RGB}{164,64,214}

\newcommand{\sr}[1]{\textcolor{orange}{[SR: #1]}}

\begin{document}

\preprint{DESY-25-050}
\title{Searching for Axion Dark Matter Near Relaxing Magnetars}

\author{Sandip Roy}
\email{sandiproy@princeton.edu}
\affiliation{Department of Physics, Princeton University, Princeton, NJ 08544, USA}

\author{Anirudh Prabhu}
\email{prabhu@princeton.edu}
\affiliation{Princeton Center for Theoretical Science, Princeton University, Princeton, NJ 08544, USA}

\author{Christopher Thompson}
\email{thompson@cita.utoronto.ca}
\affiliation{Canadian Institute for Theoretical Astrophysics, 60 St. George St., Toronto, ON M5S 3H8, Canada.}

\author{Samuel J. Witte}
\email{samuel.witte@physics.ox.ac.uk}
\affiliation{Rudolf Peierls Centre for Theoretical Physics, University of Oxford, Parks Road, Oxford OX1 3PU, UK}
\affiliation{Deutsches Elektronen-Synchrotron DESY, Notkestraße 85, 22607 Hamburg, Germany}
\affiliation{
II. Institute of Theoretical Physics, Universität Hamburg, 22761, Hamburg, Germany}

\author{Carlos Blanco}
\email{carlosblanco2718@princeton.edu}
\affiliation{Department of Physics, The Pennsylvania State University, University Park, PA 16802, USA}
\affiliation{Department of Physics, Princeton University, Princeton, NJ 08544, USA}
\affiliation{Stockholm University and The Oskar Klein Centre for Cosmoparticle Physics, Alba Nova, 10691 Stockholm, Sweden}

\author{Jonathan Zhang}
\email{jzhang@physics.utoronto.ca}
\affiliation{Department of Physics, McLennan Physical Labs, Toronto, ON M5S 1A7, Canada}

\date{\today}

\begin{abstract} 
Axion dark matter passing through the magnetospheres of magnetars can undergo hyper-efficient resonant mixing with low-energy photons, leading to the production of narrow spectral lines that could be detectable on Earth. Since this is a resonant process triggered by the spatial variation in the photon dispersion relation, the luminosity and spectral properties of the emission are highly sensitive to the charge and current densities permeating the magnetosphere. To date, a majority of the studies investigating this phenomenon have assumed a perfectly dipolar magnetic field structure with a near-field plasma distribution fixed to the minimal charge-separated force-free configuration. While this {may} be a reasonable treatment for the closed field lines of conventional radio pulsars, the strong magnetic fields around magnetars are believed to host processes that drive strong deviations from this minimal configuration. In this work, we study how realistic magnetar magnetospheres impact the electromagnetic emission produced from axion dark matter. Specifically, we construct charge and current distributions that are consistent with magnetar observations, and use these to recompute the prospective sensitivity of radio and sub-mm telescopes to axion dark matter. We demonstrate that the two leading models yield vastly different predictions for the frequency and amplitude of the spectral line, indicating systematic uncertainties in the plasma structure are significant. Finally, we discuss various observational signatures that can be used to differentiate the local plasma loading mechanism of an individual magnetar, which will be necessary if there is hope of using such objects to search for axions. 
\end{abstract}

\maketitle

{\hypersetup{linkcolor=blue}
\noindent 

\section{Introduction}

The quantum chromodynamics (QCD) axion, a pseudo-Nambu Goldstone boson arising from the spontaneous breaking of a global $U(1)$ symmetry, is among the most well-motivated candidates for new fundamental physics beyond the Standard Model~\cite{PQ1, PQ2, WeinbergAxion, WilczekAxion}. This is in part because this particle offers the most promising and long-standing solution to explain the absence of CP violation in QCD ({\emph{i.e.}}, the so-called `Strong CP problem'), and in part because it can naturally explain the existence of dark matter (it is cosmologically stable, feebly interacting, and can easily be produced non-thermally in the early Universe with the correct abundance)~\cite{PRESKILL1983127, Abbott1982, Fischler1982}. 

Today, enormous experimental efforts are underway to detect both the QCD axion, as well as the broader class of axion-like particles, which is often extended to include low-energy pseudoscalars that possess discrete shift symmetries that serve to protect the smallness of their mass. A majority of these efforts attempt to measure the coupling of the axion to electromagnetism, which at low energies arises via the dimension-five coupling

\begin{align}
\mathcal{L} \supset -\frac{\gagg}{4}  \, a \, F_{\mu \nu} \tilde{F}^{\mu \nu} = \gagg \, a \, \edb, 
\end{align}
where $a$ is the axion field, $\gagg$ is the axion-photon coupling constant, $F$ and $\tilde{F}$ are the electromagnetic field strength tensor and its dual, respectively, and $\bfE$ and $\bfB$ are the electric and magnetic field, respectively. For the QCD axion, $\gagg \equiv (E/N - 1.92) \times \alpha/(2\pi f_a)$, where $\alpha$ is the fine structure constant, $E/N$ are the electromagnetic/color anomalies of the ultraviolet theory, and $f_a$ is known as the axion decay constant (and is related to the symmetry breaking scale), while for the more general class of axions this coupling $\gagg$ is typically treated as a free parameter of the theory. 

This interaction allows axions to mix with photons in the presence of an external magnetic field, with the efficiency of this mixing being roughly set by a combination of the magnetic field strength, 
{\color{black} plasma frequency $\omega_p$}, the spatial extent of the {\color{black} magnetized plasma}, and the relative momentum difference of the axion and the photon, $\delta k \equiv |k_a - k_\gamma|$.   {\color{black} The interaction
is limited by an inherent mismatch in the axion and photon dispersion relations, which arises from the fact that the axion has a finite mass that is unconnected with the effective mass of the photon in the plasma -- see e.g.~\cite{Witte:2021arp,millar2021axionphotonUPDATED,McDonald:2023ohd,Gines:2024ekm,McDonald:2024uuh}.}




It has been known for some time that neutron stars (NSs) offer extremely promising conditions in which to search for axion-photon mixing~\cite{Pshirkov:2007st,Huang:2018lxq,Hook:2018iia,Safdi:2018oeu,Battye:2019aco,Leroy:2019ghm,Foster:2020pgt,Prabhu:2020yif,Buckley:2020fmh,Witte:2021arp,Battye:2021xvt,battye2021robust,Nurmi:2021xds,Foster:2022fxn,Witte:2022cjj,Battye:2023oac,McDonald:2023shx,noordhuis2023novel,Xue:2023ejt,Tjemsland:2023vvc,Noordhuis:2023wid,Caputo:2023cpv,Witte:2024akb,Kouvaris:2022guf,Maseizik:2024qly,Song:2024rru}; this is because they host the strongest magnetic fields in the observable Universe, and because they are surrounded by a dense plasma which can drive resonant axion-photon transitions, which occur when $\delta k \rightarrow 0$ (corresponding to the point where the axion and photon dispersion relation are degenerate). 

The general problem of interest is the following. Assuming axions contribute non-negligibly to the dark matter abundance in the Universe, axions will be continuously falling into, and out of, the gravitational wells produced by individual NSs in the galaxy. During the traversal of the magnetosphere, axion-photon mixing will be naturally enhanced by the strong background magnetic field sourced by the NS itself; while this enhancement alone is not typically sufficient to generate observable signatures, photon production can be further stimulated via local resonances which are triggered by the spatially varying photon dispersion relation. These resonances occur when the four momentum of the axion matches the four momentum of the photon, $k_a^\mu \simeq k_\gamma^\mu$, which in the limit of non-relativistic (or semi-relativistic) axions traversing a non-relativistic plasma (as is relevant, for example, in the closed field lines of pulsars) reduces to $m_a \simeq \omega_p$. 

For standard radio pulsars, these resonances are naturally encountered for axion masses $m_a \lesssim 5 \times 10^{-5}$ eV, corresponding to photons with frequencies $f \lesssim 10$ GHz. Photons sourced at these resonances tend to escape the magnetosphere, producing a radio signal that may be detected on Earth.  Since the initial axion velocity distribution in the galaxy prior to in-fall is extremely narrow, {\color{black} $\delta \omega / \omega \sim v^2 / 2 \sim 5 \times 10^{-7}$ (where $v \approx 300$ km/s is the typical root-mean-square (RMS) velocity of dark matter in the Milky Way),} 
one naively expects the observable signal to manifest as an extremely narrow radio line roughly situated at the axion mass (up to an apparent Doppler shift due to the relative motion of the NS) \footnote{Although axions pick up kinetic energy as they fall into the gravitational well of the NS, this effect is exactly canceled by the gravitational redshift of the produced photons climbing out of the well~\cite{Hook:2018iia}.}. Several factors modify this overly simplified picture; for example, low-energy photons can (in some systems) be absorbed, the refraction of the newly produced photons induces a sizable broadening of the spectral line (although the line typically still remains at the level of $\delta \omega / \omega \lesssim \mathcal{O}(10^{-4})$), the resonance itself contains a mild phase space dependence which broadens the resonance region, {\color{black} and so on} (see e.g. ~\cite{Witte2021,Battye:2021xvt,McDonald:2023shx,Tjemsland:2023vvc} for further details). Nevertheless, the description above paints the general picture which, at least at the qualitative level, holds true.

Existing searches have attempted to detect either the presence of such a line in isolated NSs~\cite{Foster:2020pgt,Battye:2021yue,Battye:2023oac}, or by searching for the collective signature of all radio lines produced from the population of NSs in dense stellar environments such as the Galactic Center (GC)~\cite{Foster:2022fxn}. {\color{black} In the latter case,} the signal appears as a forest of narrow lines~\cite{Safdi:2018oeu}, centered about the axion mass but Doppler broadened by the characteristic velocity dispersion of the NSs in the population (being of the order of $\delta \omega / \omega \sim \mathcal{O}(10^{-3})$). These searches have yielded leading constraints on axions in the mass range 15--35 $\mu$eV~\cite{Foster:2022fxn} and competitive constraints around $\sim 4  \, \mu$eV~\cite{Battye:2023oac}.\footnote{Axions can also be directly sourced in NS magnetospheres from small-scale oscillations in the electromagnetic fields~\cite{Prabhu:2020yif,noordhuis2023novel}, giving rise to anomalous radio signatures~\cite{Prabhu:2020yif,noordhuis2023novel,Noordhuis:2023wid,Caputo:2023cpv}, the formation of bound axion clouds~\cite{Noordhuis:2023wid,Witte:2024akb}, producing fast radio bursts~\cite {Prabhu:2023cgb}, and inducing a short-scale nulling of the coherent radio emission~\cite{Caputo:2023cpv}. Consequently, these searches do not rely on the assumption that axions comprise the dark matter.} 

To date, most studies have assumed that the magnetospheres of NSs can be described by the Goldreich-Julian (GJ) model \cite{GoldreichJulian1969}.  This assumes the NS is surrounded by the minimal charge density required to screen the rotationally-induced parallel electric field $E_{\parallel} \equiv \edb/|\bfB|$, corresponding to a charge number density  $n_{\rm GJ} \simeq -2 \boldsymbol{\Omega} \cdot \bfB/e$, where $e$ is the electron charge and $\boldsymbol{\Omega}$ is the angular velocity of the NS.
{\color{black} An underlying assumption in this description is that the magnetic field of the pulsar is frozen into its solid crust; while this is thought to be a reasonable assumption for typical pulsars, this assumption is expected to break down for NSs with the strongest magnetic fields.}

Magnetars are a class of magnetically-powered NSs, with surface magnetic fields potentially reaching and even exceeding values as high as $10^{15}$ G (see~\cite{WoodsThompson2006} for reviews). For such strong magnetic fields, the Lorentz force, ${\bf j} \times \bfB$ (${\bf j}$ being the current density) can shear the crust of the NS, launching powerful currents into the otherwise dormant closed magnetosphere~\cite{Thompson2000}. These currents power non-thermal emission in the form of hard X-ray, gamma-ray, infrared, optical, and radio emission~\cite{Kaspi_2017}. Intriguingly, coherent radio emission has been observed from some magnetars (e.g., PSR J1745--2900) at frequencies as high as $291$ GHz ($\approx 1.2$ meV/$2\pi$).  {\color {black} This suggests} that the magnetosphere hosts regions where the plasma frequency can reach the meV range.
In the context of axion searches, this would imply that 
{\color{black} magnetars are sites for resonant conversion} across a much wider range of {\color{black} $m_a$} than previously thought possible; should this be the case, magnetars may offer a unique avenue for probing one of the 
{\color{black} best motivated} and experimentally challenging {\color{black} parts of $m_a$ space}.  

The goal of this paper is to investigate how realistic magnetar magnetosphere models impact the detectability of spectral lines {\color{black} produced by} the resonant conversion of axion dark matter. We begin by reconstructing the spatially inhomogeneous charge and current densities in the two leading 
{\color{black} models of magnetar electrodynamics.}
{\color{black} This includes a self-consistent description of the
non-potential magnetic field outside the magnetar,} non-local pair production, 
and the distribution of relativistic plasma.
We fit these models to hard X-ray observations of nearby magnetars, and then apply a state-of-the-art ray tracing algorithm which computes the differential power and spectral properties produced by each plasma configuration. For each model, we determine the potential detectability of the spectral line using current and near-future radio and sub-mm telescopes. 

The outline of this paper is as follows. We begin in Section \ref{sec:magnetar_astrophysics} by discussing the 
observational evidence for {\color{black} strong} deviations in magnetar magnetospheres from the baseline GJ model, as well as early attempts 
{\color {black} at a self-consistent description of magnetar
electrodynamics.} In Section \ref{sec:jbundle} and Section \ref{sec:collisionalModel}, we present two self-consistent solutions to the magnetar circuit that have been proposed in the literature, review the 
{\color{black} relevant QED interactions,} and present analytic and numerical models of the magnetospheric plasma distributions. In Section \ref{sec:modelComp} we discuss the consistency of each model 
with observations of Galactic magnetars {\color{black} and
suggest some diagnostics.} Section \ref{sec:mixing} 
{\color{black} summarizes how,} given a fixed magnetosphere model, we compute the axion-induced spectral line; Section~\ref{sec:sense} illustrates the projected sensitivity for a variety of telescopes {\color{black} and a representative sample of magnetosphere models.} We conclude in Section~\ref{sec:conclude}. {In the Appendices, we provide additional details that support and extend the analysis in the main text. Appendix~\ref{sec:supercritical_conversion} examines how strong-field Euler–Heisenberg corrections to the photon dispersion relation affect resonant axion-photon mixing in ultra-strong magnetic fields. In Appendix~\ref{sec:raytracingappendix}, we present further results from our ray-tracing simulations of axion-to-photon conversion. Appendix~\ref{sec:telescope_appendices} details the telescope parameters used to estimate experimental sensitivity. In Appendix~\ref{sec:1D_approx}, we compare the results of the full 3D ray-tracing simulations with those obtained from the commonly used 1D approximation. Finally, Appendix~\ref{sec:conversion_turbulent} explores how turbulence in the magnetic field suppresses resonant axion-photon conversion.} 

\vspace{0.1in}

\vspace{0.1in}

\section{Magnetar Astrophysics} \label{sec:magnetar_astrophysics}

A growing body of observations indicates that magnetar electrodynamics are poorly described by the standard pulsar model. For instance, even in relatively quiescent states, magnetars emit spectrally hard ($E \gtrsim 10$ keV) X-ray continua with a luminosity far exceeding their spin-down luminosities, $\dot{E} = 4\pi^2 I_{\rm NS} \dot{P}/P^3$, where $I_{\rm NS} \approx 10^{45}$ g cm$^2$ is the NS moment of inertia, $P$ is the rotation period, and $\dot{P}$ is the rate of change of the rotation period~\cite{Kuiper_2006, Gotz2006, Enoto_2017}.

The hard X-ray emission is believed to be powered by large currents flowing in the closed magnetosphere.  
Independent support for this picture comes from observations of
order-of-magnitude variations in spin-down torque that can persist
for months (see, e.g.,~\cite{DibKaspi2014}), and optical/IR emission some $10^4$ times brighter than
expected from the low-energy (Rayleigh-Jeans) tail of a surface X-ray blackbody~\cite{Hulleman2001, Durant:2006qd}.

The required currents are carried by plasma whose density must exceed the minimum co-rotational charge density predicted by the GJ model by several orders of magnitude. The force-free condition, $\edb = 0$, combined with Gauss' law, provides an estimate of the minimum charge density in the magnetosphere,
\begin{align}\label{eq:rhoco}
    \rho_{\rm co} = - 2 \boldsymbol{\Omega} \cdot \bfB + \boldsymbol{\Omega} \cdot \left[ {\bf r} \times \boldsymbol{\nabla}\times \bfB  \right],
\end{align}
where ${\bm \Omega}$ is the spin angular velocity of the magnetar, and ${\bm B}$ is the local magnetic field. 
The first term on the right-hand side of Eq. \eqref{eq:rhoco} is the minimal co-rotational charge density, $\rho_{\rm GJ}$, in a potential magnetic field ($\boldsymbol{\nabla} \times \bfB = 0$);  the second term is a correction arising from
the magnetic twist. When the magnetosphere is weakly twisted, as expected {in} weaker-field radio pulsars, the second term is subdominant\footnote{An exception is if the magnetosphere contains twist on small scales. We explore this scenario in Appendix ~\ref{sec:conversion_turbulent}.}.


The net electron (positron) density, $n_{-(+)}$, is derived from the charge and current densities,
\begin{align}
    \rho &= e (n_+ - n_-), \\
    j_B &= \hat B\cdot(\boldnabla\times{\bf B})/4\pi = e(n_+ v_+ - n_- v_-), \label{eqn:jB}
\end{align}
where $e$ is the positron charge and $v_{-(+)}$ is the electron (positron) velocity along the field line. 
The bright X-ray emission of a magnetar implies the presence of strong currents with density
$j_B \gg |\rho_{\rm co}|c$, or equivalently with $|n_+-n_-| \ll |j_B|/ec$, except near the magnetic poles.

In the models considered in this paper, currents flowing in the closed magnetosphere are assumed to flow over only a small part ($\sim 1-10\%$) of the magnetar surface.  
This is motivated, in significant part, by detections of hot spots following bright X-ray outbursts (see~\cite{Beloborodov:2016mmx} for a list of hot spot sizes observed in transient magnetars). The persistence of the measured X-ray and IR emission, and torque changes, suggest the presence of a long-lived non-potential component of the magnetic field outside the star.  These localized currents
have been called a `j-bundle' or `magnetospheric arcades', depending on the geometry of the current-carrying regions. The term `j-bundle' typically refers to magnetic flux tubes whose footpoints form a roughly circular region near the magnetic pole of the magnetar surface. In contrast, an `arcade' describes magnetic field lines anchored in sheet-like regions characterized by a small extent in one angular direction (either polar, $\Delta \theta$, or azimuthal, $\Delta \phi$) and a significantly larger extent in the other (see Fig.~\ref{fig:collisional_model_geometries} for examples). These regions can, in principle, be located anywhere on the magnetar surface, not just near the poles. 

The collection of currents flowing in the magnetosphere forms a circuit with an enormous self-inductance. The large bolometric output following an outburst, $L_X \sim 10^{35}$  erg s$^{-1}$, assumed to be powered by magnetospheric particles bombarding the surface, combined with the particle residency time in the magnetosphere $t_{\rm res} \sim \rns \sim 3 \times 10^{-5}$ s, where $\rns \approx 10\km$ is the NS radius, implies a total energy $E_{\rm part} \sim 3\times 10^{30}$ erg. This energy is negligible compared to the total energy stored in the non-potential magnetic field, $E_{\rm non-pot} \sim B_\varphi^2 \rns^3 \sim 10^{42}-10^{45}$ erg, where $B_\varphi$ is the toroidal (non-potential) component of the magnetic field. As a result, the magnetosphere is well-described by force-free electrodynamics, with ${\bf j} \parallel {\bf B}$.

In the absence of a sufficient number of charged particles, the large self-inductance implies the presence of an enormous voltage drop along twisted field lines. The strength of this voltage is regulated by microphysical processes which supply the magnetosphere with electron-positron pairs. Two particular channels for $e^\pm$ pair production have been subject to particular scrutiny: (i) the current forms a double layer structure with voltage $\sim 100$ MeV--GeV along ${\bf B}$, in which relativistic $e^\pm$
spawn new pairs by scattering keV photons emitted from the magnetar surface at the first Landau resonance~\cite{BeloborodovThompson2007}; and (ii) collisional plasma containing trans-relativistic $e^\pm$, in which
pairs are regenerated self-consistently as annihilation photons gain energy by
repeated electron scattering~\cite{ThompsonKostenko2020}.

To some extent, these mechanisms of pair creation are not mutually consistent. In order for a double layer to form, the current must become slightly charge-starved, with density falling slightly below $|\boldsymbol{\nabla} \times \bfB|/e$~\cite{BeloborodovThompson2007}. The collisional plasma, on the other hand, is a powerful source of $\sim$ MeV
photons that bathe large parts of the magnetosphere outside the current-carrying zones~\cite{ThompsonKostenko2020, ZhangThompson2024}. These photons may produce pairs through $\gamma + \gamma \rightarrow e^+ + e^-$, the cross section of which is enhanced in super-Schwinger magnetic fields, $B > B_{\rm Q} = 4.41\times 10^{13}$ G~\cite{KostenkoThompson2018}. Pairs can be created in sufficient quantity to screen the field-aligned
voltage in places where the magnetic field is more weakly sheared, especially near the magnetic poles, where double layers would otherwise be expected to form.



Of additional importance is the fact that magnetar magnetospheres are dynamical over a range of scales. Twist may be transferred from the interior to the exterior impulsively due to a magnetohydrodynamic instability in the core or a sudden fracture of the crust in response to a build-up of magnetic stresses~\cite{ThompsonDuncan1995, ThompsonDuncan2001}, or gradually due to slow elastic, plastic, and thermal evolution of the magnetar crust~\cite{Thompson_2017}. Once twist has been injected into the magnetosphere it is dissipated either rapidly, through magnetospheric instabilities~\cite{Parfrey2012, Parfrey2013, Mahlmann:2019arj, Mahlmann_2023}, or gradually via Ohmic dissipation~\cite{BeloborodovThompson2007, Beloborodov_2009}. The evolution of twist in the magnetosphere may be associated with radiative events. Rapid injection and dissipation of twist in the magnetosphere can explain X-ray bursts~\cite{Gogus1999, Gogus2000, Rea2009, Rea_2010, Esposito2020} and fast radio bursts~\cite{Bochenek2020, CHIME1, CHIME2}. Following these outbursts are extended phases of `afterglow' emission that can endure on timescales from $\sim$ months--years~\cite{Kaspi_2017}. The ratio of energy radiated in outbursts to that in the afterglow emission can range from $\sim 10^{-2}-10^{2}$~\cite{Woods:2003si}. The long timescale associated with afterglow emission can be explained by gradual lengthening of crustal faults~\cite{Thompson_2017}, sudden deposition of heat into the crust~\cite{LyubarskyEichlerThompson2002, Li:2015epa}, or slow untwisting of the magnetosphere through Ohmic decay of currents~\cite{TLK02, Beloborodov_2009}. If the average injection timescale exceeds the average dissipation timescale, the endpoint of evolution is a quiescent magnetar, with only potential fields, and lower plasma density.


\subsection{Intensity of Magnetospheric Currents}

Given the importance of the location of magnetospheric twist and the plasma state surrounding a magnetar on the resonant conversion of axions, we pause to review a concrete model.  We consider a differential
axisymmetric twisting of an axisymmetric magnetic field, with poloidal component 
$B_r = (r^2\sin^2\theta)^{-1}\partial_\theta{\cal F}$, $B_\theta = -(r\sin\theta)^{-1}\partial_r{\cal F}$.
The function ${\cal F}(r,\theta)$ is defined as the total magnetic flux through a circular loop around the axis, from the axis out to angle $\theta$ at radius $r$. In the case of a dipole
with magnetic moment $\mu$, this is ${\cal F} = \mu\sin^2\theta/r$, which may be expressed in terms
of the maximum radius ${\cal R} = r/\sin^2\theta$ to which a field line extends, as ${\cal F} = \mu/{\cal R}$.

Consider a weak twisting of this field, with one hemisphere rotating by an angle $\Delta\phi({\cal F}) 
\ll 1$ with respect to the opposing hemisphere.  This creates a toroidal field $B_\phi \propto \Delta\phi$.   The twist may be smooth, with $\Delta\phi$ approximately constant
over some range of polar angle;  or alternatively, the crustal shearing may have significant power
on small scales, with $d\Delta\phi/d\ln{\cal F} \gg 1$.   We further neglect 
the rotation of the star, which is valid so long as $\Delta \phi \gg \Omega r/c$. Then, the force-free condition (${\bf j} \times {\bf B} = 0$) corresponds to
$\boldnabla\times{\bf B} = \lambda({\cal F}) {\bf B}$, with the coefficient $\lambda({\cal F})$ constant on each
magnetic field line, since ${\bf B}\cdot\boldnabla\lambda = 0$.  Expressing $B_\theta$ in terms of
${\cal F}$ gives 
\begin{equation}\label{eq:lambda}
    \lambda = {d(B_\phi r\sin\theta)\over d{\cal F}} = {2\over c}{dI\over d{\cal F}},
\end{equation}
where $I$ is the poloidal current closed by the flux surface. The twist is related to $\lambda$ by
integrating along a flux surface,
\begin{equation}
\Delta\phi = \int_{\theta_0}^{\theta_1} d\theta {B_\phi\over B_\theta\sin\theta} = 
(B_\phi r\sin\theta)_{\cal F}\int_{\theta_0}^{\theta_1} {d\theta\over B_\theta r\sin^2\theta},
\end{equation}
where $\theta_0$ and $\theta_1$ represent two foot points of the flux surface. Since the integrand on the right-hand side of the equation above only depends on the background poloidal field, we can use the expression for a magnetic dipole to obtain
\begin{equation}
   (B_\phi r\sin\theta)_{\cal F} = {\Delta\phi\over 2\mu}{\cal F}^2 \cos\theta_{\cal F}
\end{equation}
in a dipolar geometry where flux surface ${\cal F}$ intersects the magnetar surface at polar angle $\theta_{\cal F}$. Combining this with Eq. (\ref{eq:lambda}) gives 
\begin{eqnarray}
\lambda &=& {\Delta\phi\over{\cal R}} f(\Delta\phi), \quad
f(\Delta\phi) \equiv  1 + {1\over 4}{d\ln\Delta\phi\over d\ln\theta_{\cal F}},
\end{eqnarray}
when $\theta_{\cal F} \rightarrow 0$.
One observes that strong local gradients in the foot point motions translate into higher
current densities outside the star. The emergence of local gradients in foot point motions is well-motivated in some models of NS magnetic field evolution. For example, Hall drift in NS crusts can transfer magnetic energy from large to small scales (analogous to a turbulent cascade), amplifying the current density at small scales~\cite{GoldreichReisenegger1992, Cumming:2004mf, GourgouliatosCumming2013}.  Small-scale non-potential magnetic fields
$\delta B_\perp$ possess current fluctuations with intensity $J_\parallel \propto k_\perp \delta B_\perp$, where $k_\perp \sim \nabla_\perp$ is the wave number associated with gradients perpendicular to the background (non-potential) magnetic field. Depending on the spectrum of magnetic field fluctuations, $J_\parallel$ can grow with $k_\perp$, leading to the formation of intense, small-scale currents.

The minimal number density to conduct the current, achieved when electrons and positrons are counter-propagating with speed close to the speed of light, is 
\begin{eqnarray} \label{eqn:nmin_twist}
    n_{\rm min} &=& {J\over ec} = {B\over 4\pi e {\cal R}}\,\Delta\phi f(\Delta\phi) \\ \nonumber
    &=& 1.6\times 10^{16} B_{14} {\cal R}_6^{-1}\Delta\phi\, f(\Delta\phi)\quad{\rm cm}^{-3},
\end{eqnarray}
where $B_{14} = B/(10^{14} \ {\rm G})$, and ${\cal R}_6 = {\cal R}/(10^6 \ {\rm cm})$.

On a fixed flux surface, $n_{\rm min}$ and $B$ are proportional, scaling as $r^{-3}$ in a dipole geometry.
This density exceeds the co-rotation number density $n_{\rm GJ} \sim -2 \,\boldsymbol{\Omega}\cdot{\bf B}/e$ by 
a factor of the ratio of the light cylinder radius to the maximum radius of the field line, 
\begin{equation}
{n_{\rm min}\over n_{\rm GJ}} \sim 10^4 \,\Delta\phi\, f(\Delta\phi) {R_{\rm NS}\over {\cal R}}\left({P\over {\rm sec}}\right).
\end{equation}
If we assume momentarily that the plasma is, at most, semi-relativistic (we revisit the general case in a self-consistent manner in the following sections), then the plasma frequency
\begin{equation} \label{eq:wp}
    \omega_p \approx 1 \times 10^{-2} \, {\rm eV} \,[\Delta\phi\,f(\Delta\phi)]^{1/2} \left({R_{\rm NS}\over 
    {\cal R}}\right)^{1/2} B_{14}^{1/2} \, ,
\end{equation}
sets the characteristic axion mass for which resonant axion-photon transitions occur. The plasma frequency scales as $\sim r^{-3/2}$ along a flux surface and, for a uniform twist, as $\sim r^{-2}$
at fixed polar angle. Here, we see that the plasma frequency near the surface of the star can be around one hundred times larger than that predicted in the GJ model, suggesting such objects may be a powerful way to probe heavier axions.

\subsection{Implications for Axion Conversion}

While the current density is determined by the distribution of magnetospheric twist, this current does not uniquely define the full phase space of the plasma, and therefore the location and efficiency of resonant axion-photon mixing.  Resonant axion-photon mixing occurs when the four-momentum of the axion matches that of the associated electromagnetic mode, \ie $k_a^\mu= k_\gamma^\mu$. For a semi-relativistic axion passing through a magnetized plasma, this reduces to the condition that $m_a \simeq \omega_{p, {\rm eff}}$, where one can define an `effective plasma frequency' $\omega_{p, {\rm eff}}^2 \equiv \sum_s \left\langle \omega_{p,s}^2/\gamma_s^3 \right\rangle$, where the sum runs over the species $s$, $\omega_{p,s}^2 = e^2 n_s/m_e$ is the plasma frequency of that species, $n_s$ is the number density, $\gamma_s$ is the gamma factor of the flow of species $s$, and $\left\langle ... \right\rangle$ represents the average over the plasma distribution function (see Sec. \ref{sec:mixing} for additional details). Here, one can see that the location and efficiency of this resonant mixing process therefore depend on the full 6D phase space distribution of the magnetospheric plasma (although in a very strong magnetic field, particle motion is effectively confined to flow along magnetic field lines, reducing the dimensionality of the phase space), which is not uniquely constrained in the discussion above.

In Sections \ref{sec:jbundle} and \ref{sec:collisionalModel}, we describe concrete models of the double layer and collisional plasma states, which are then used in Section \ref{sec:mixing} to compute the intensity of
the photon line emerging from resonant conversion of accreted dark matter axions.  The properties of the magnetar circuit described above have the following implications for conversion of dark matter axions to photons (as demonstrated in the remainder of this paper):

\vspace{0.1in}

\noindent 1. Most of the detectable resonant conversion is concentrated outside the j-bundles or magnetospheric
arcades. In situations where collisional plasma fails to form outside the magnetar, and the pair plasma is sustained by relativistic Landau scattering, we consider the same
current profile as modeled by Ref.~\cite{BeloborodovThompson2007} (hereafter referred to as `BT07'):  the current is strongly localized within a bundle centered on the magnetic pole (Figure~\ref{fig:model_cartoons}). In this case, we find that the peak sensitivity to axion conversion signals is highly sensitive to the opening angle of the polar bundle. When the opening angle is large $\theta_c \gtrsim 0.2$ rad, the peak sensitivity to axions is suppressed, but spread over a wide range of frequencies. For smaller opening angle $\theta_c \lesssim 0.2$ rad, conversion takes place efficiently in the equatorial region, where the plasma distribution is well-described by the GJ model. In the alternative scenario in which collisional plasma does form, we consider a range of different arcade profiles (see Figure~\ref{fig:collisional_model_geometries}), showing how the conversion surface and radio flux depend on the underlying arcade geometry. Here, the resonant conversion is concentrated outside the arcades, where the magnetic field is assumed to be nearly potential. In this case, the plasma number density far exceeds the co-rotation density, reducing the efficiency of photon production from light axions with $m_a \lesssim 10^{-4}$ eV, but extending sensitivity to axions in the $\sim$ meV range (orders of magnitude higher than expected in the GJ model). 

\vspace{0.1in}

\noindent 2. The sensitivity to axions evolves with time. Unlike in radio pulsars, whose magnetospheres (in the closed zone) evolve over the $\sim$ kyr--Myr spindown time, magnetars' magnetospheres may evolve over much shorter timescales. Shortly after an outburst, magnetars are both highly dynamical and radiative, imposing modeling difficulties and large backgrounds. As the magnetosphere quiesces, the current density in the magnetosphere drops, magnetospheric twist is erased, and radiative backgrounds are reduced. Consequently, magnetar observations over human timescales probe a broad range of axion masses.

\vspace{0.1in}

\begin{figure*}
    \centering
    \includegraphics[width=0.4\linewidth]{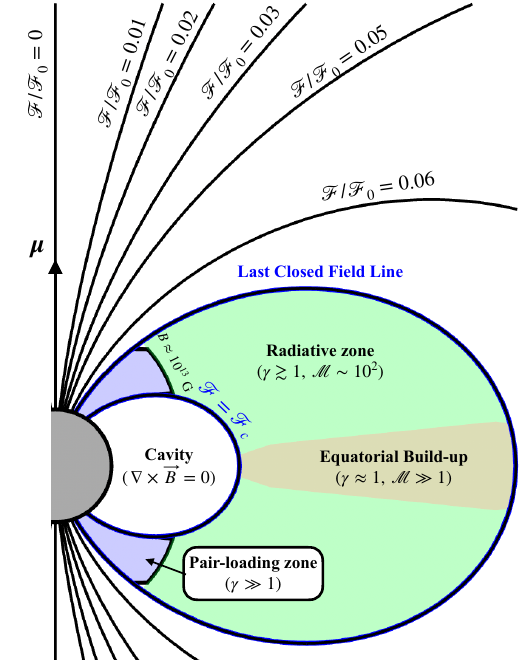} \hspace{1in} 
    \includegraphics[width=0.4\linewidth]{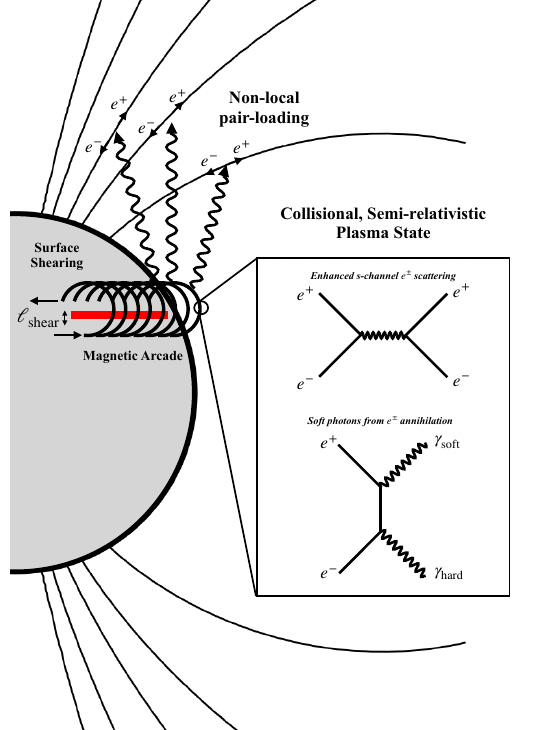}
    
    \vspace{-1em}
    
    \caption{Schematic diagrams illustrating the two models under consideration. Left panel: Relativistic Double Layer (``BT07'') Model.  Magnetospheric twist in the closed magnetosphere is confined to the j-bundle (region bounded by the `Last Closed Field Line' and {\color{black} the flux surface} `$\mathcal{F} = \mathcal{F}_c$').  The j-bundle 
    {\color{black} divides} into three main regions: the pair-loading zone (light blue, shaded), where pairs are produced by up-scattered thermal photons, the radiative zone (light green, shaded), where non-thermal X-ray emission is sourced and particle motion is governed by radiative drag forces, and the equatorial region (light orange, shaded), where the {\color{black} $e^\pm$ flow slows} and the pair density builds up. Also shown are various flux surfaces with different values of $\mathcal{F}$. Right panel:  Collisional, Transrelativistic (``TK20'') Model.  Shearing of the magnetospheric surface occurs in a region represented by the red band, with typical width $\ell_{\rm shear}$. The plasma state on magnetic field in the shear zone is determined by strong-field quantum electrodynamic (QED) processes shown in the inset. High energy photons produced in the shear zone populate the remainder of the magnetosphere through $\gamma + \gamma \to e^+ + e^-$ pair creation. 
    }
    \label{fig:model_cartoons}
\end{figure*}

To summarize, the efficiency of axion-photon conversion in magnetar magnetospheres is sensitive to the distribution of currents in the closed magnetospheres, the channels for producing the plasma required to conduct this current, and the equilibrium state into which this plasma settles. We have identified two qualitatively different models for explaining the wealth of magnetar observations that have received particular scrutiny in the literature: the relativistic double layer model (`BT07') and the collisional model (`TK20'). In the following sections, we review these models and derive analytical expressions describing their respective plasma states. In Sec.~\ref{sec:modelComp}, we examine observational evidence from specific sources that may favor one model over the other.



\section{Relativistic Double Layer} \label{sec:jbundle}

In this section we review the relativistic double layer model (‘BT07’). In canonical radio pulsars, the magnetospheric twist, which is localized to the open zone, is sustained
by a rotationally-generated longitudinal voltage. At low magnetic fields ($B \sim 10^{12}$ G), this voltage
is naturally regulated to be about $e\Phi \sim 10^7 m_e \approx 5$ TeV via the creation of $e^\pm$ pairs
sourced by curvature/synchrotron radiation emitted from seed particles~\cite{Timokhin:2015dua, Philippov2022Review}.  

Near a magnetar, which is a bright source of keV photons
(luminosity $\sim 10-100\,L_\odot$), the dominant channel for pair creation by relativistic $e^\pm$ is
inverse Compton scattering of X-rays at the first Landau resonance.  The excited electron or positron
will return to the ground Landau state by emitting a photon.  When $b \equiv B/B_{\rm Q} > 4$ this photon can sometimes directly convert to a pair; where the magnetic field is somewhat weaker, its
energy in the frame of the star	is still large enough to convert with some delay as it propagates through
the curved magnetic field\footnote{Resonant Inverse Compton scattering (RICS)
is also believed to be the dominant pair creation process for
radio pulsars with strong magnetic fields, $B\sim 10^{13}$ G~\cite{HibschmanArons2001, Timokhin:2018vdn}.}.

In this way, collisionless and relativistic $e^\pm$ plasma may be sustained anywhere in the magnetar
magnetosphere, as long as the field is smoothly twisted~\cite{BeloborodovThompson2007}.
The energy of the resonantly scattering photons drops with
increasing radius and decreasing magnetic field strength.  On field lines that extend far enough
from the star, their energy
drops below $2m_e$ and pair creation freezes out.  At even greater distances from
the magnetar, the out-flowing pair creation emits a broad and spectrally hard X-ray continuum~\cite{Beloborodov2013a, Beloborodov2013b}.

This process regulates the longitudinal voltage to be about $e\Phi \sim$ GeV, which causes the
plasma state to differ significantly from that in the open zone of a radio pulsar.
The location of the twisted field bundle is also constrained	if the
magnetar is to be a bright source of 10--100 keV X-rays.	 The magnetic field in the outer
parts of the twisted bundle must drop below $\sim 10^{13}$ G, which limits where it is anchored
on the magnetar	surface:  the polar angle must be smaller than $\theta_{\cal F} \sim 25^\circ\,B_{\rm pole,15}^{-1/6}$ where $B_{\rm pole,15} = B_{\rm pole}/10^{15}\,{\rm G}$.

Beloborodov~\cite{Beloborodov_2009} describes how twist implanted at larger $\theta_{\cal F}$ can migrate Ohmically
toward the magnetic pole, forming a `j-bundle' with an approximately circular cross-section that encloses all of the magnetospheric twist. As a result, a current-depleted `cavity' opens up around the magnetic equator.
This process offers a promising explanation for the months-delayed onset of accelerated spin-down following a bright X-ray outburst~\cite{Archibald:2014dla}; twist that is initially distributed through the magnetosphere is transferred to the polar region on the resistive timescale. Enhanced twist in the polar region leads to a higher torque exerted by the magnetic field on the star. It is worth mentioning that, in sources like the 
GC magnetar PSR J1745$-$2900, where the hard X-ray 
continuum is accompanied by small-area afterglow following the outburst, some localization of the forcing motion
near the pole is still required. Additionally, it is possible for magnetospheric twist to migrate towards the polar region without a current-depleted cavity forming in the equatorial region if, for example, the diffusion of currents is mediated by internal processes (as opposed to external Ohmic dissipation) such as magnetic reconnection ~\cite{Thompson2008III}. 

This section provides a detailed kinetic description of relativistic $e^\pm$ in	such a polar j-bundle.
{\color{black} The magnetic field lines are assumed to reach far enough from the star that outflowing
$e^\pm$ experience a strong outward radiation pressure force by cyclotron scattering keV photons,
so that the $e^\pm$ do not return to the star~\cite{Thompson:2004yg, Beloborodov2013a}.  In this sense,}
the j-bundle is analogous to the open zone in canonical radio pulsars.} 
When modeling axion-to-photon conversion, we consider a few possible values for the opening angle 
{\color{black} $\theta_c$ of the bundle and the corresponding flux surface ${\cal F}_c$.}

\subsection{J-bundle Plasma State} \label{sec:jbundle_plasma}

Below we describe the plasma state in the j-bundle by separating it into two distinct regions with qualitatively different plasma-radiation interactions. The \emph{pair-loading zone} in which kinetic energy from high-energy seed particles is reprocessed into a flow with lower average momentum, but higher multiplicity, and the \emph{radiative zone} in which particles experience strong radiative drag from the ambient photon field, and decelerate through the emission of hard X-ray photons that lie below the pair-production threshold. A depiction of these regions is shown in Fig.~\ref{fig:model_cartoons}.

\vspace{0.1in}

\emph{Pair loading zone.---}  In the ultra-strong magnetic fields found near the surfaces of NSs, electrons and positrons are effectively confined to the ground Landau state, which forces particles to flow nearly perfectly along magnetic field lines. In this limit, the dynamics are effectively one-dimensional, being governed by
\begin{align}
{\partial \epar \over \partial t} = (\boldsymbol{\nabla} \times \bfB)_{\parallel} - j_B \, .
\end{align}
In the force-free limit, \ie when plasma supply is sufficiently abundant to drive $\epar = 0$ and $\partial_t \epar = 0$, the twisted magnetic field configuration is sustained by a steady-state current density. 

The force-free limit is, in general, expected to break down in localized regions; for standard pulsars, this is known to occur in the polar caps (\ie the footprints of the open field lines), while for magnetars the breakdown may occur over much more extended regions (since the twisted magnetic field is not confined to the open field lines). 
When a sufficient current cannot be supplied, a large $\epar$ develops along the field line, accelerating particles and driving pair cascades, which in turn serve to replenish the supply of plasma, screen the electric field, and re-establish the required current density. 

The total voltage drop along the j-bundle is regulated by the interaction of accelerated $e^\pm$ pairs with the ambient photon field. Once pairs reach a critical energy, $m_e \gamma_c$, they resonantly scatter thermal photons from the magnetar surface (or X-rays produced in other regions of the magnetosphere). The resonance condition is obtained from energy and momentum conservation in the electron rest frame:

\begin{align} 
    \omega' + m_e = \gamma E_B, \quad \omega' = \beta \gamma E_B,
\end{align}
where $\omega' = \omega \gamma_c (1 - \cos \theta)$ is the photon energy in the electron rest frame. The quantities $\omega$, $\theta$, and $\gamma_c \equiv 1/\sqrt{1 - \beta_c^2}$ are the frequency of the photon, the angle its momentum makes with the magnetic field, and the electron Lorentz factor, all in the frame of the star. The quantity $\gamma$ is the Lorentz factor of the electron after it absorbs the photon (in the pre-absorption electron rest frame), and $E_B = m_e \sqrt{2 b + 1}$ is the energy of the first Landau level. Here, and below, we define $b \equiv B/B_Q$. The Lorentz factor for electrons to resonantly scatter photons is

\begin{align} \label{eqn:resonance}
	\gamma_c = {b m_e \over  \omega (1 - \beta_c \cos\theta)} \approx 10^3 \left( {B \over 10^{14} \ {\rm G}} \right) \left({\omega \over {\rm keV}} \right)^{-1}. 
\end{align}

Resonant scattering may be viewed as a two-step process: resonant absorption of a photon by an accelerated pair followed by radiative de-excitation.  When $b \gtrsim 1$,
the de-excitation photon carries away a significant fraction of the electron energy, with average energy ~\cite{Beloborodov2013b}
\begin{align} \label{eqn:egam}
	\left\langle E_\gamma \right \rangle = \gamma_c  m_e \left( 1- {1 \over \sqrt{2 b + 1} } \right) \, .
\end{align}

Resonant scattering with thermal photons gives rise to a drag force on 
outflowing $e^\pm$ pairs.  Simulations of particle motion in the ambient radiation field around a NS \cite{Beloborodov2013b}
demonstrated that the $e^\pm$ flow self-regulates to resonantly scatter photons with frequency $\omega \sim 10 \ T_{\rm NS}$ (where $T_{\rm NS}$ is the surface temperature of the NS), slightly above the main peak of the thermal spectrum $\omega_{\rm peak} \approx 3 \ T_{\rm NS}$. From~\eqref{eqn:resonance}, the self-regulated gamma factor is 
\begin{align}
    \label{eqn:gammasc}
	\gamma_{\rm sc}(b) \approx  {b m_e \over 10 T_{\rm NS
    }}.
\end{align}
{\color{black} Combining this with Eq. \eqref{eqn:egam}, one finds $\langle E_\gamma\rangle
\gg 2m_ec^2$ for $b \gg 0.2$, in which case the up-scattered photon is above threshold for
single-photon pair conversion.  The radial distance where $b$ falls below $b_0 \sim 0.2$
marks the end of the pair production zone.}


Resonant scattering creates a broad distribution of particle energies.
{\color{black} After scattering, the energy of a particle is reduced by a factor $\sim 1/\sqrt{2b+1}$ and the
secondary particles have a characteristic energy}
\begin{align} \label{eqn:gammal}
	\gamma_\ell(b) \approx {\left\langle E_\gamma \right\rangle \over 2 m_e} = {b m_e \over 20 T_{\rm NS}} \left( 1- {1 \over \sqrt{2 b + 1} } \right),
\end{align} 
where $ \left\langle E_\gamma \right\rangle$ is given in (\ref{eqn:egam}). 
{\color{black}  Further broadening in the distribution of $e^\pm$ energies results from
variations in the emission angle of the de-excitation photon and delays in pair creation.
In magnetic fields $b_0 \approx 0.2 < b < 4$,} 
several secondary particles are created with $\gamma_\ell(b) \lesssim \gamma \lesssim \gammasc(b)$. Monte Carlo (MC) simulations of particle motion and production 
performed in~\cite{Beloborodov2013b} find that the particle distribution function (of both seed and secondary particles) shows two peaks, located near $\gamma = \gammasc$ and $\gamma = \gamma_\ell$. 

{\color{black} Excitation of a two-stream instability will tend to flatten the double-peaked distribution function
that is sourced by $e^\pm$ creation.  When evaluating the electromagnetic dispersion,
we will therefore adopt a simple flat ``waterbag'' distribution,}
\begin{align} \label{eqn:distfuncPP}
    f_\pm(b, p) &= {1 \over p_+ - p_-} \left\{ \begin{matrix}
        1, & p_- \le p \le p_+; \\
        0, & {\rm otherwise}.
    \end{matrix}\right.
\end{align}
{\color{black} Here, $b$ is a proxy for position along a field line.  
Bulk motion of the plasma corresponds to taking $p_- + p_+ \neq 0$. 

In the polar j-bundle considered here, the $e^\pm$ do not return to the star and already
form a bulk outflow in the pair creation zone.  We adopt $p_+ = \sqrt{\gammasc(b)^2 - 1}$ and $p_- = \sqrt{\gamma_\ell(b)^2 - 1}$.}
Conservation of energy and momentum imply 
\begin{align} \label{eqn:conserved}
    {n_\pm(b) \bar \gamma (b) \over b} = {\rm const.}, \quad \bar \gamma(b) = {\gamma_+(b) + \gamma_-(b) \over 2},
\end{align}
where $\bar\gamma \gg 1$ represents the average gamma factor which has been computed using the waterbag distribution.
{\color{black} The constant in Eq. \eqref{eqn:conserved} is obtained by balancing the kinetic energy of the $e^\pm$
with the Ohmic losses in the circuit, which are concentrated near the base of the pair creation zone.
For net rate $\dot N_\pm$ of pair creation, j-bundle current $I$ and longitudinal potential drop $\Phi$, 
this gives $\dot N_\pm \bar\gamma m_e = I\Phi$, or equivalently~\cite{Beloborodov2013b}}
\begin{align} \label{eqn:nb0}
    n_\pm(b) \approx {j_B(b_0) \Phi b\over 2 m_e \bar \gamma(b) b_0 } \, .
\end{align}

{\color{black} The net result is that, at the top of the pair creation zone where $b \approx b_0 \approx 0.2$,
the average $e^\pm$ Lorentz factor has dropped to $\bar\gamma(b_0) \sim \gamma_{\rm sc}/2 \sim 
10\,(T_{\rm NS}/0.5~{\rm keV})^{-1}$, and the multiplicity has grown to}
\begin{align}
    \mathcal{M} \equiv \frac{e \dot{N_\pm}}{I} \approx {e \Phi \over 2 \bar \gamma(b_0) m_e} 
    = 100\Phi_9\left({T_{\rm NS}\over 0.5~{\rm keV}}\right),
\end{align}
where $\Phi_9 = e\Phi/{\rm GeV}$.
Note that overall charge neutrality $n_+ \approx n_-$ along with the constraint equation on the current, $j_B = e n_+ v_+ - e n_- v_-$, implies $n_- \gg j_B/e$.  There is equivalently a small imbalance between
the average flow speeds of positive and negative charges,
\begin{align} \label{eqn:velratio}
    {\bar \beta_- \over \bar \beta_+} = 1 - {2 \over \mathcal{M}+1}.
\end{align}

Below we describe dynamics in this ``radiative zone''.

\vspace{0.1in}

\begin{figure*}
    \centering
    \includegraphics[width=0.62\linewidth]{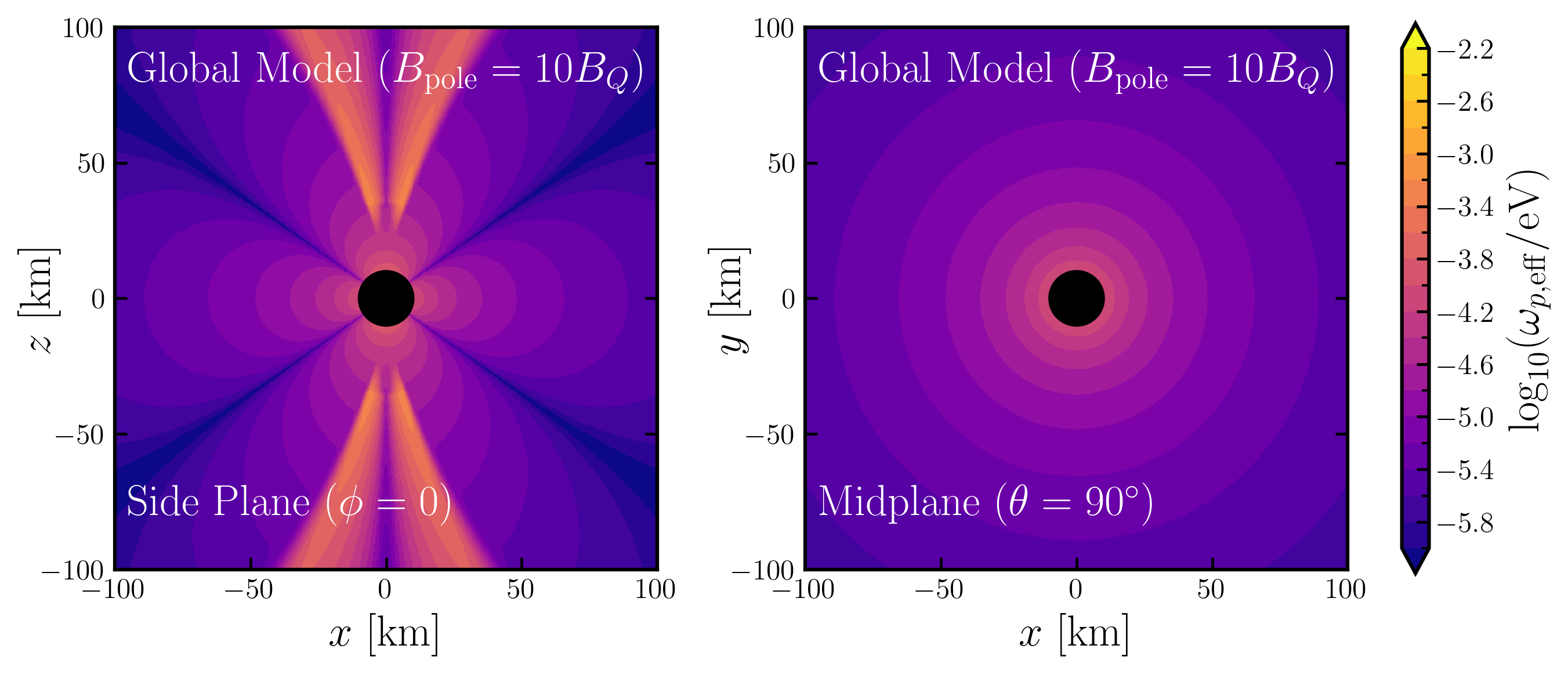}
    \includegraphics[width=0.37\linewidth]{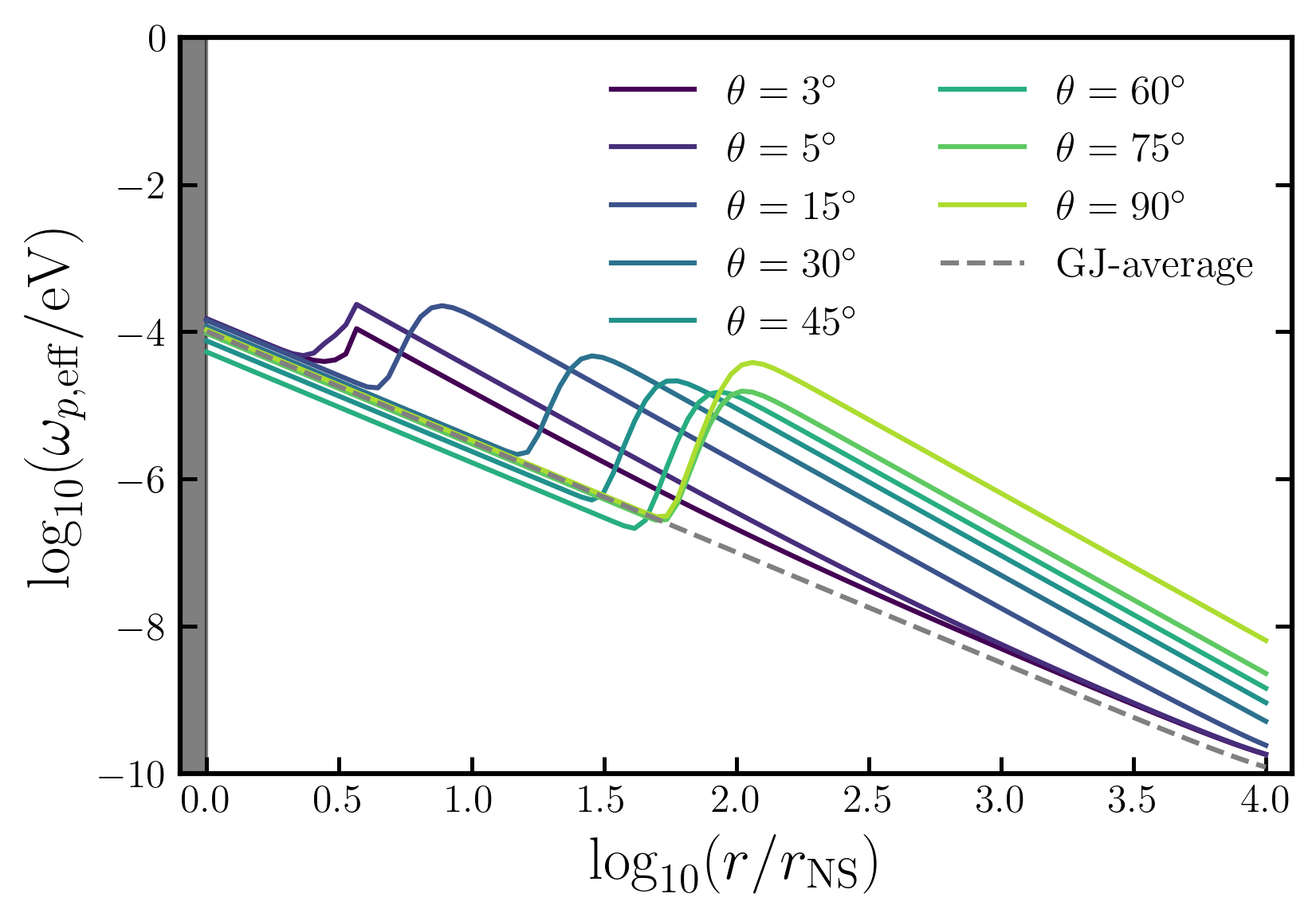}
    
    \vspace{-1em}
    
    \caption{Effective plasma frequency ($\omega_{p,\,\rm{eff}}$) profile 
    in the globally-twisted relativistic double layer magnetar model (``BT07'') with $B_{\rm pole}=10\,B_Q$, $\psi_0=1.6$, $\theta_c = 0.1$, and assuming a rotation period of $3.76\,\rm{s}$. 
    {\color{black} Left panel: vertical slice through the center of the magnetosphere.  Middle panel: midplane slice.
    Right panel: radial profile of $\omega_{p,\rm eff}$ at different polar angles.}  The gray dashed line {\color{black} shows the polar angle-averaged Goldreich-Julian (GJ) plasma profile.} 
    }
    \label{fig:plasma_profiles_global}
\end{figure*}

\emph{Radiative Zone.---} Far from the stellar surface ($r \gg R_{\rm NS}$), when $b \ll b_0$, up-scattered photons are not sufficiently energetic to pair-produce. Therefore, the number flux of particles is conserved until the pairs reach the outer part of the loop (see ``equatorial build-up'' in the left panel of Fig.~\ref{fig:model_cartoons}) where they may annihilate to photons~\cite{Beloborodov2013a}. 
In the outer parts of the loop, pairs experience a strong radiative drag from ambient X-rays,
{\color{black} and will decelerate to trans-relativistic speeds.  The equilibrium speed $\bar\beta$ depends on
the angular distribution of the X-rays, which may be modified substantially by resonant scattering.
We follow \cite{Beloborodov2013a} in taking the X-ray flow to be unperturbed and radial, which gives
$\bar\beta = |B_r/B|$ and}
\begin{align} \label{eqn:gammasat}
    \bar p(r, \theta) = m_e \left|{B_r\over B_\theta}\right| = 2 m_e \cot \theta,
\end{align}
{\color{black}assuming a mostly dipolar field.}
Radiative drag tends to  attract particles towards $p = \bar p(r,\theta)$, but electrons and positrons must maintain a small velocity mismatch in order to sustain the required current density. 


We adopt a simplified two-fluid description of the charge distribution in which the electron and positron distribution functions are broad,
\begin{align} \label{eqn:distfuncrad}
        f_\pm(p) = \left\{ \begin{matrix}
        p_{\pm}^{-1}, & 0 \le p \le p_\pm \\
        0, & {\rm otherwise}
    \end{matrix} \right.,
\end{align}
along with the condition that $(p_+ + p_-)/2 = \bar{p}$. Combining this condition with Eq.~\eqref{eqn:velratio} gives

\begin{align}
    p_\pm &= \bar{p} \pm \sqrt{\bar{p}^2 + {1 \over 2}(\mathcal{M}^2 - 1) \left(1 - \sqrt{1 + {4 \bar{p}^2 \over \mathcal{M}^2}} \right)}.
\end{align}
In the limit that $\mathcal{M} \gg \bar{p}$, $|p_\pm - \bar{p}|/\bar{p} \approx \bar{\gamma}/\mathcal{M}$.
This suggests that the most energetic particles in both the $e^+$ and $e^-$ flows have momenta very close to $\bar{p}$. 

\color{black}

\vspace{0.1in}

\subsection{Global Model} \label{sec:jbundle_global}

In this section, we build on the results of the previous section to construct a simplified global model of a magnetosphere featuring a polar j-bundle ($\boldsymbol{\nabla} \times \bfB \ne 0$) and an equatorial cavity ($\nablabold \times \bfB = 0$). Specifically, we connect the near-surface plasma regime -- shaped by pair creation and radiative drag -- with the far-field plasma behavior, which is governed by radiative drag alone.

At present, the j-bundle and cavity are assumed to be separated by a critical flux surface $\ff_c$, with $\ff < \ff_c$ corresponding to the j-bundle and $\ff > \ff_c$ to the cavity. During this phase, twist has been erased from the cavity ($\ff > \ff_c$) and we adopt the simplified twist profile
\begin{align}
    \psi(\theta) &= \psi_0 \Theta(\theta_c - \theta),
\end{align}
where $\Theta(x)$ is the Heaviside theta function and $\theta_c$ is the polar angle of the intersection of flux surface $\ff_c$ with the NS surface. 

Within the j-bundle, the plasma number density is given by
\begin{align} \label{eqn:npmjbundle}
    n_\pm(r,\theta) = \psi_0 {B(r, \theta) \sin^2 \theta \over r} \left( {e\Phi \over 2 m_e \bar{\gamma}(\max(b, b_0 ))} \right), 
\end{align}
where a nearly dipole field is assumed.
We consider a single distribution function that captures the physics in both the pair-loading and radiative zones. The momentum-space distribution in the j-bundle is taken to be a waterbag distribution, \ie 
\begin{align} \label{eqn:distfunc}
    f_\pm(b, p) &= \left\{ \begin{matrix}
        (p_+(b) - p_-(b))^{-1}, & p_-(b) \le p \le p_+(b), \\
        0, & {\rm otherwise}
    \end{matrix}\right.,
\end{align}
where
\begin{align} \label{eqn:momentum_pm}
    p_+(r, \theta) &= \max\left[\sqrt{\gammasc(b(r, \theta))^2 - 1}, \bar{p}(\theta) \right], \nonumber \\
    p_-(r, \theta) &= \sqrt{\gamma_\ell(b(r, \theta))^2 -1} \times \Theta[b(r, \theta) - b_0].
\end{align}
Here, $\gammasc(b)$ and $\gamma_\ell(b)$ are given by Eqs. (\ref{eqn:gammasc}) and (\ref{eqn:gammal}), respectively, and $\bar{p}(\theta)$ is given by (\ref{eqn:gammasat}). 
By construction, the distribution function (\ref{eqn:distfunc}) is continuous across the boundary between the pair-loading zone and the radiative zone. This follows from our definition of the edge of the pair-loading zone: $\gamma_\ell(b_0) = 1$, which means that $p_-(b_0)$ approaches zero from both sides of the boundary.

A global view of the effective plasma frequency throughout the magnetosphere is shown in Fig. \ref{fig:plasma_profiles_global} for a surface magnetic field strength $B_{\rm pole} = 10 B_Q$, a twist parameter $\psi_0 = 1.6$, a critical surface $\theta_c = 0.1$, a rotational period of $3.76$ s, and assuming aligned magnetic and rotational axes. In the left plot, one can see the sharp enhancement of the plasma frequency along the j-bundle, which allows for resonant axion-photon conversion at much higher axion masses. In the right panel, one can see that the radial plasma frequency profile at fixed $\theta$. Values are compared to that of the GJ model (dashed line), and are typically enhanced by one to two orders of magnitude, but only radial distances $r$ beyond some critical threshold $r_c(\theta)$. Note that the appearance of a local minimum in the plasma frequency implies that photons sourced for a particular range of axion masses may actually become trapped in localized under-densities -- we comment on this possibility in the future sections. 

\begin{figure*}
    \centering
    \includegraphics[width=0.325\linewidth,trim={10cm 0 0 0},clip]{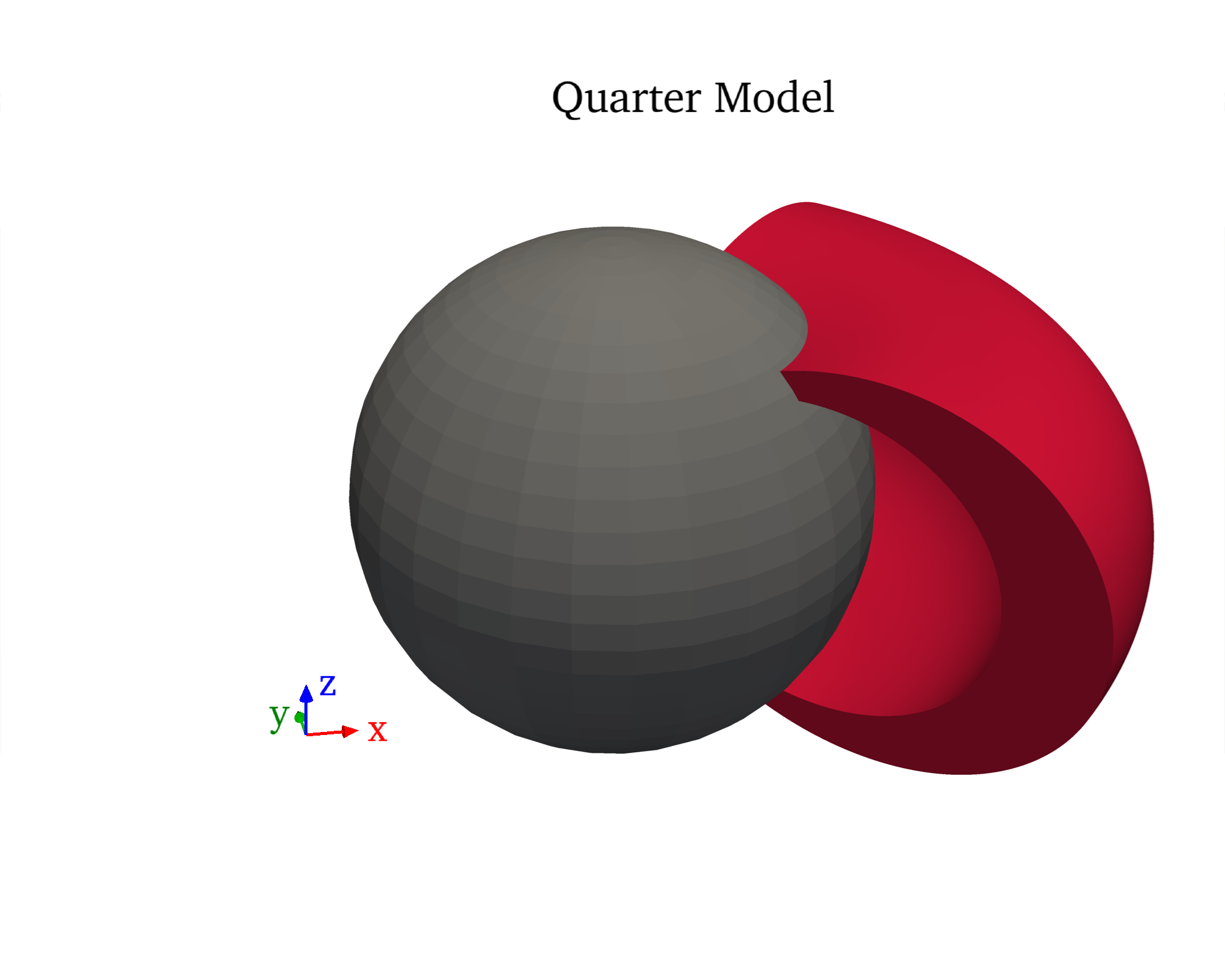}
    \includegraphics[width=0.325\linewidth,trim={10cm 0 0 0},clip]{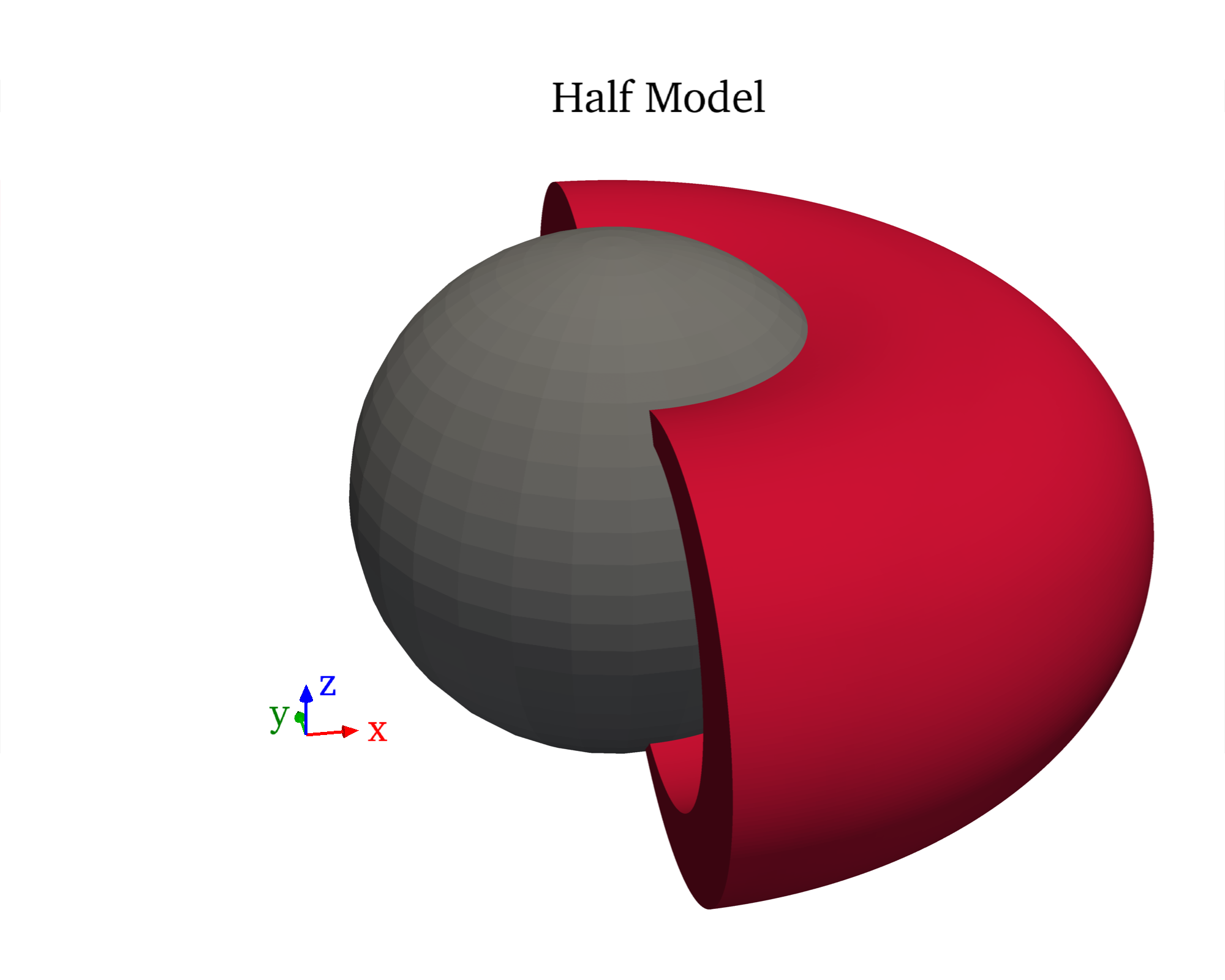}
    \includegraphics[width=0.325\linewidth,trim={10cm 0 0 0},clip]{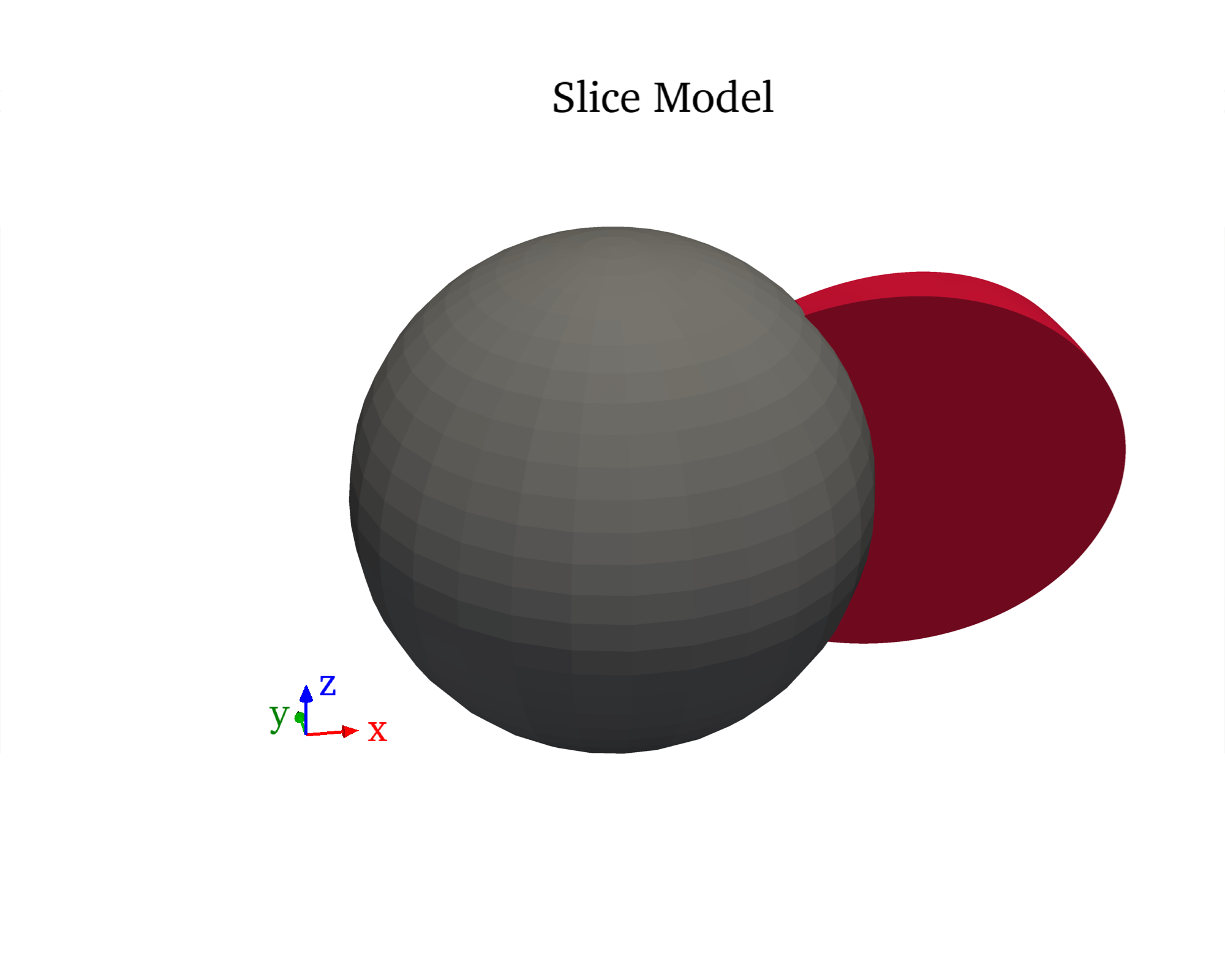}
    
    \vspace{-1em}
    
    \caption{{\color{black} Current carrying arcades (red) 
    anchored to narrow shear zones in the magnetar crust (gray).
    Electron-positron plasma supporting the current is transrelativistic and collisional when the 
    current density is high due to small-scale magnetic braiding (``TK20'' model).   These
    are referred to (left to right) as the ``quarter'', ``half'', and ``slice'' geometries. 
    Coordinate axes are shown for reference.}
    }
    \label{fig:collisional_model_geometries}
\end{figure*}

\begin{figure*}
    \centering
     \includegraphics[width=0.61\linewidth]{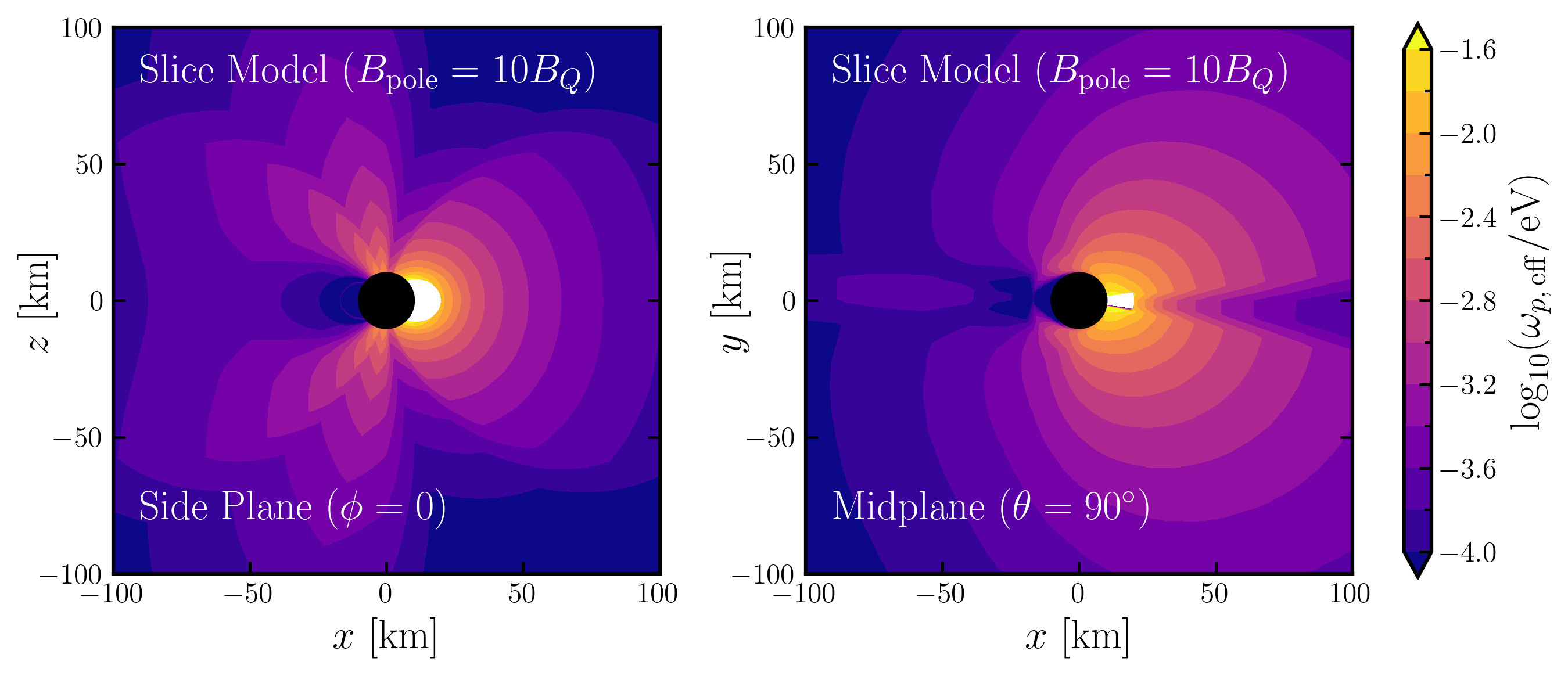}
     \includegraphics[width=0.37\linewidth]{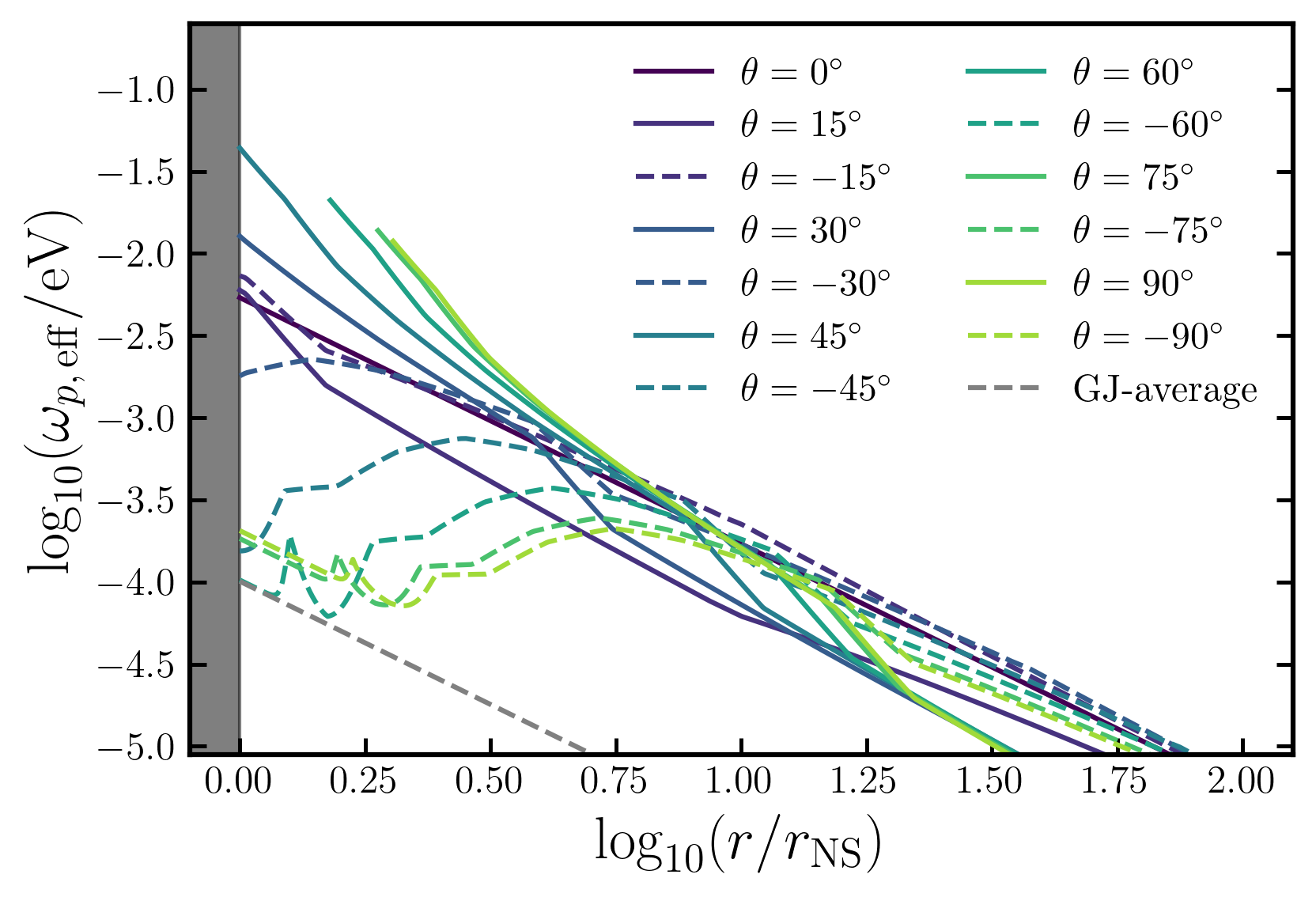}
     \includegraphics[width=0.61\linewidth]{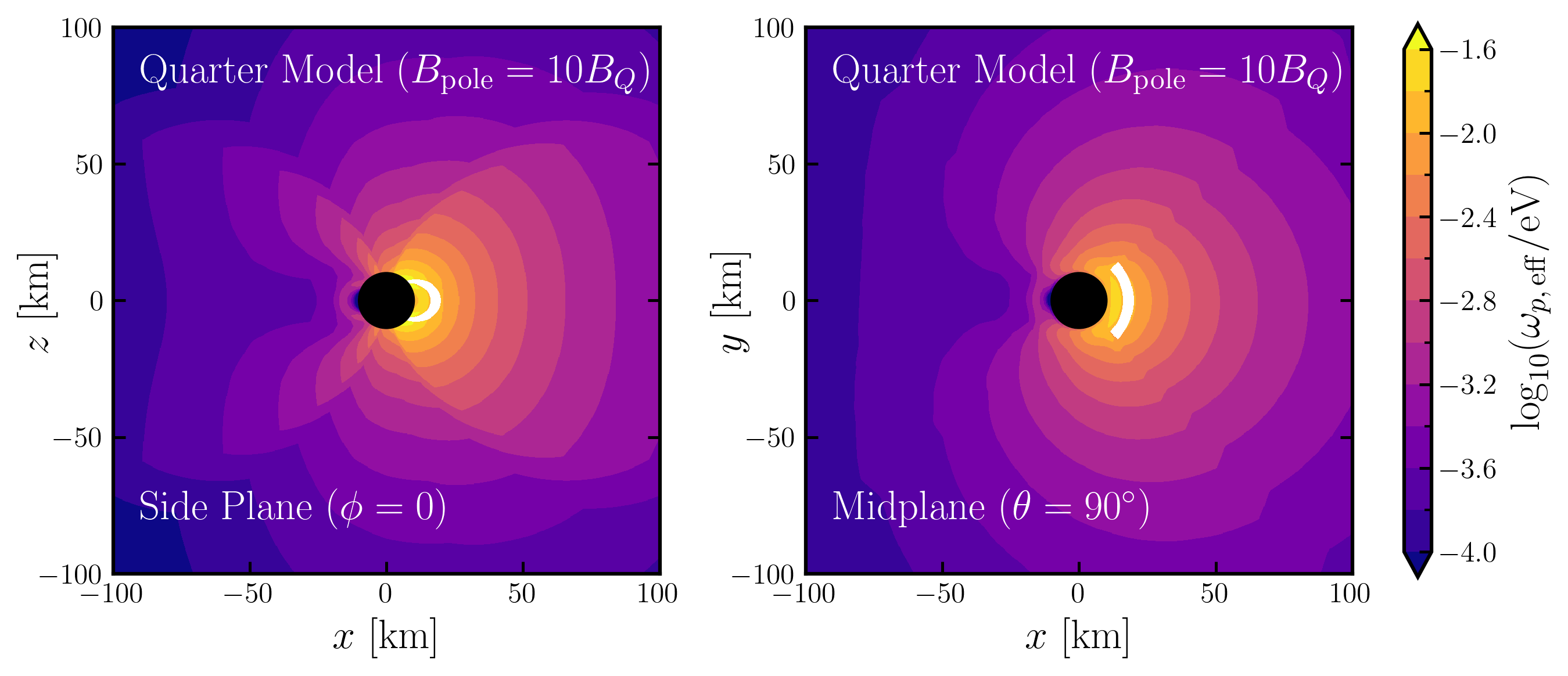}
     \includegraphics[width=0.37\linewidth]{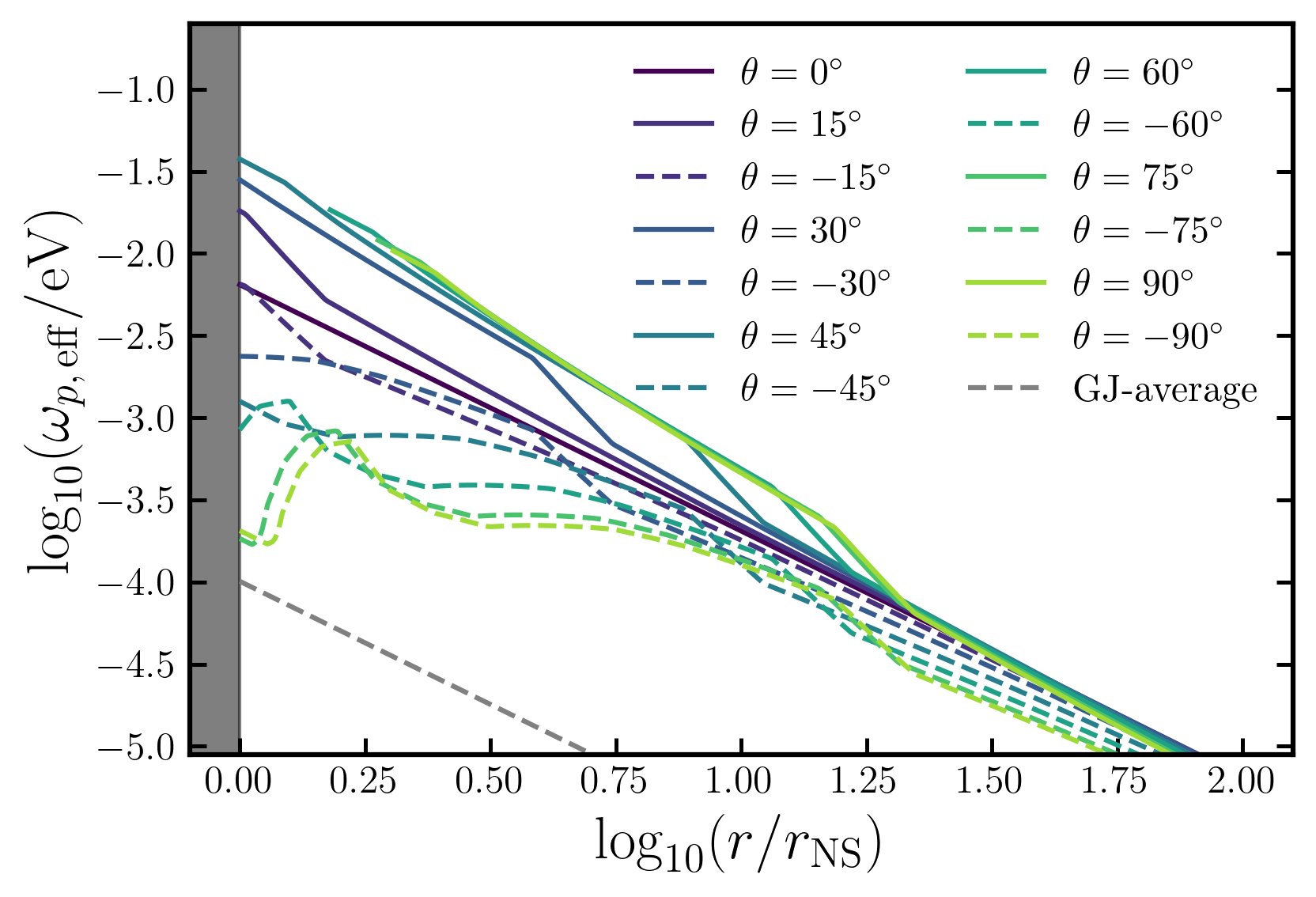}
    \includegraphics[width=0.61\linewidth]{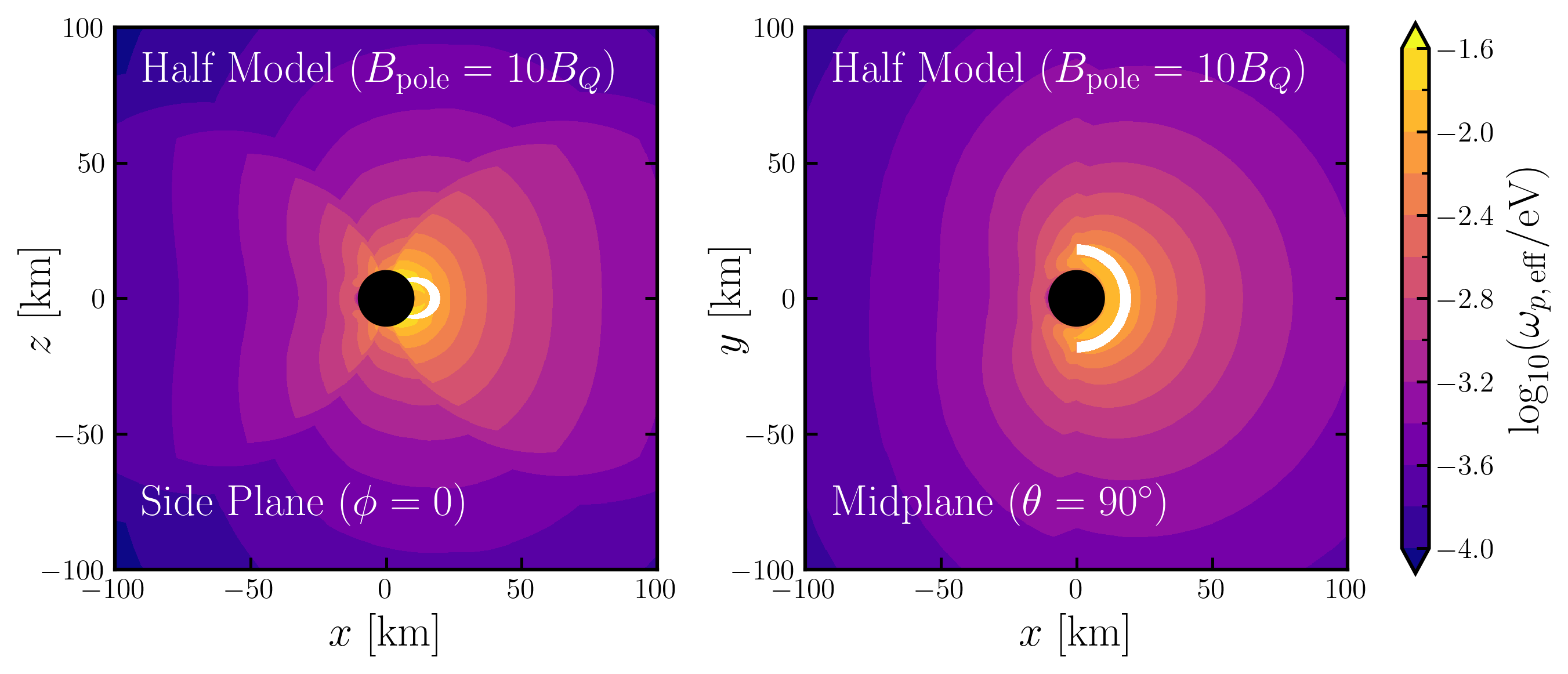}
    \includegraphics[width=0.37\linewidth]{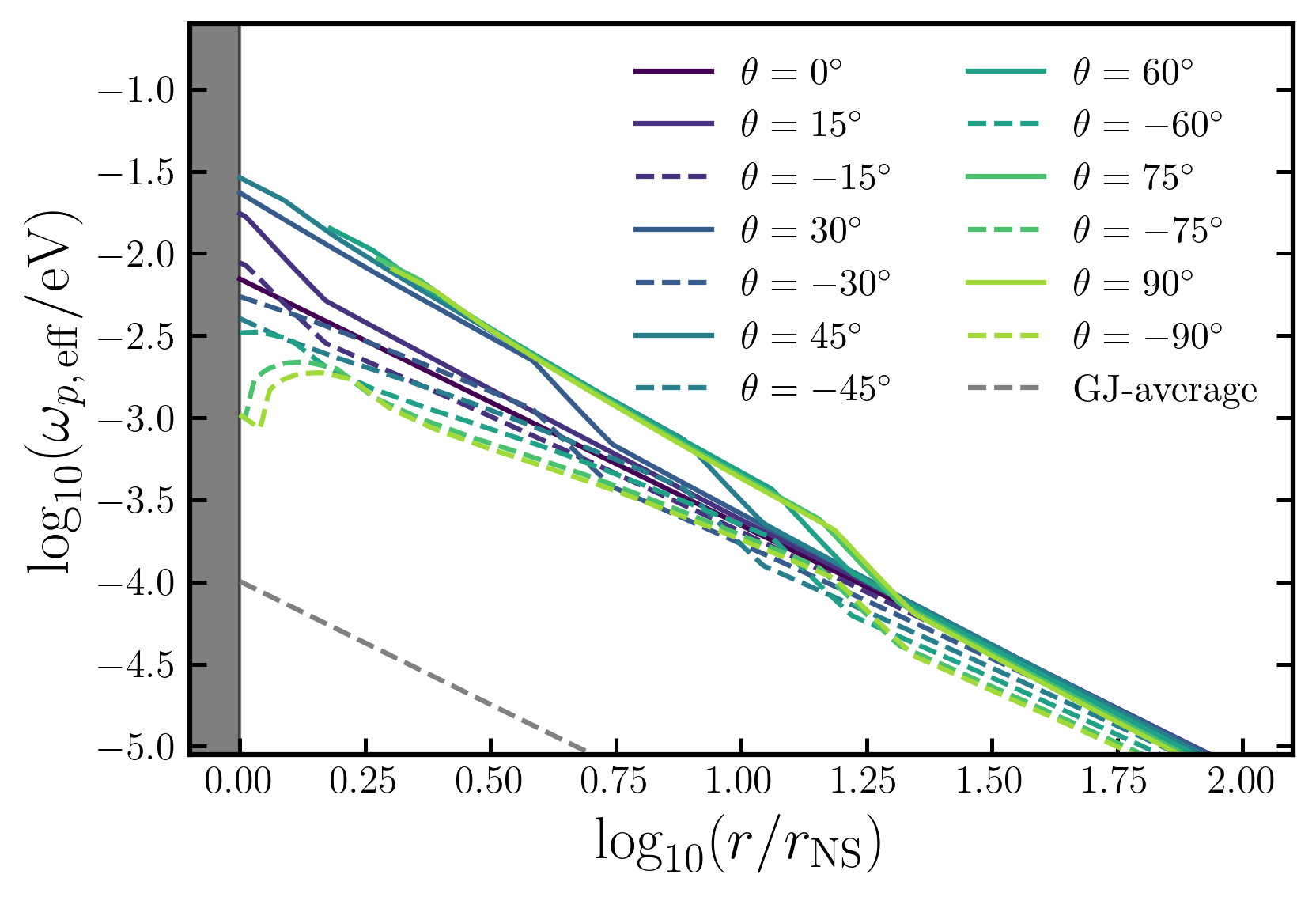}
    
    \vspace{-1em}
    
    \caption{Effective plasma frequency ($\omega_{p,\,\rm{eff}}$) profiles for three different highly-collisional magnetar arcade geometries {\color{black} (the TK20 model, as in Figure \ref{fig:collisional_model_geometries}, {\color{black} with arcade luminosity $L_0 = 10^{35}$
    erg s$^{-1}$ up to $\sim 1$ MeV}).  The
    $e^\pm$ density outside the arcades is sourced by the collision of gamma rays that are emitted from the
    arcade surfaces. Top, middle and bottom rows correspond to the slice, quarter and half arcade geometries.
    Left panel: vertical slice through the center of the magnetosphere.  Middle panel:  midplane slice.} The NS is shown in black, and the arcade regions are shown in white. {\color{black} Right column: 
    radial profile of $\omega_{p,\rm eff}$ at various polar angles.  The gray, dashed line shows the
    polar angle-averaged GJ plasma profile.  
    Profiles are not plotted where they intersect the arcades.}
    }
    \label{fig:plasma_profiles}
\end{figure*}

\section{Collisional, Dilute Pair Plasma} \label{sec:collisionalModel}

We now turn to an alternative framework for understanding the non-thermal behavior of magnetars in their relatively quiescent states, as proposed in Thompson \& Kostenko (2020) (TK20)~\cite{ThompsonKostenko2020}. The TK20 model differs significantly from the BT07 model described in the previous section -- most notably in the assumed location of the magnetic twist and the mechanism of plasma generation. Despite their differences, both models have successfully accounted for various magnetar phenomena in distinct sources. However, there is currently no consensus on which framework provides a more accurate description. In Sec.~\ref{sec:modelComp}, we examine specific observations that may help discriminate between the two models in individual cases. 

In the TK20 model, strong, localized currents once again play a central role in powering non-thermal X-ray emission. However, the underlying QED processes differ from the resonant scattering mechanism discussed in Section~\ref{sec:jbundle}. In this case, the emission originates from a collisional, trans-relativistic plasma confined to compact magnetic loops near the magnetar surface, rather than from a polar j-bundle extending far from the surface.

The possibility that magnetar currents are supported by trans-relativistic and collisional $e^\pm$ plasma is motivated most directly by a similarity between data and ab initio QED calculations
of $e^\pm$ annihilation.  The measured $> 10$ keV X-ray spectrum 
in several sources bears a strong similarity to the photon spectrum of annihilating pairs in a super-Schwinger magnetic field.  The annihilation line produced by
trans-relativistic $e^\pm$ becomes broader as the background magnetic
field approaches $B_{\rm Q}$, and when $b > 4$ the spectrum asymptotes to the same
flat distribution as ordinary bremsstrahlung \cite{KostenkoThompson2018}.
 
In this approach, constraints are relaxed on the location of	the currents flowing into the exterior of the magnetar,
thereby allowing the possibility of more extended, current-carrying arcade structures.
Fig. \ref{fig:collisional_model_geometries} shows some possible configurations in 
which the sheared magnetic field is tied to fault-like features in the crust.

An important consequence of the collisional plasma state is the appearance of a broad halo of pair 
creation outside the primary current-carrying structures, mediated by collisions between out-flowing
gamma rays, $\gamma + \gamma \rightarrow e^+ + e^-$ \cite{ZhangThompson2024}.  This is the most significant distinction 
with the double layer circuit model, where gamma rays are emitted nearly tangent to the magnetic field and secondary pair creation
is more tightly confined around the j-bundle.  In the collisional state, non-local pair creation 
is further aided by the enhancement in the cross section for photon collisions, 
by a factor $\sim b$ in the strong magnetic field \cite{KozlenkovMitrofanov1986, KostenkoThompson2018}.
The net effect is to raise the plasma frequency over a broad part of the magnetar magnetosphere.

Below, we review the key QED processes that lead to the emergence of a collisional, trans-relativistic equilibrium in magnetar magnetospheres. In particular, we highlight how the cross sections for these processes are modified in the presence of super-critical magnetic fields. For detailed derivations and further discussion, see~\cite{KostenkoThompson2018, KostenkoThompson2019}.

\subsection{Enhanced $e^+-e^-$ Collisions}

Backscattering of $e^+$ and $e^-$ introduces resistance into the magnetar circuit. 
The dynamics of electrons and positrons is essentially one-dimensional near the star;
where $B > B_{\rm Q}$, Landau excitations are kinematically forbidden when the particle motion is only mildly relativistic.

The ultrastrong magnetic field opens up an enhanced channel for backscattering via the reaction
$e^+ + e^- \rightarrow \gamma \rightarrow e^+ + e^-$.  Annihilation into a single photon is kinematically
allowed because generalized momentum $e({\bm A}_+ - {\bm A}_-)$, corresponding to a relative displacement
of the electron and positron across the magnetic field, can be interchanged with the kinetic momentum of the
gamma ray.  This enables $s$-channel scattering to proceed through annihilation into an on-shell photon.
When the intermediate photon re-converts to a pair, half the time the particle momenta are interchanged,
leading to back-scattering. The cross section for $b \gtrsim 1$ is~\cite{KostenkoThompson2019}
\begin{align}\label{eqn:sigma_pm}
    \sigma_{\pm} &= {3 \pi  e^{- 2 \gamma^2/b}\over 8 \beta \gamma^2 \alpha b} \sigma_T,
\end{align}
in the center-of-mass frame, where $\sigma_T = 8\pi (\alpha^2/m_e)^2/3$ is the Thomson cross section and $\alpha$ is the fine structure constant.  
Compared with unmagnetized pair plasma, there
is a net enhancement by a factor of $\sim (\alpha b)^{-1}$ (for $\gamma \sim 1$).

{\color{black}
\subsection{Annihilation Bremsstrahlung Cooling}

Ohmic heating of the circuit is balanced by two-photon annihilation of pairs, $e^+ + e^- \rightarrow \gamma + \gamma$.
This process acts as a calorimeter:  when $\gamma > 2$ it is possible for both photons to convert directly to pairs, thereby reducing the temperature of the plasma.  Otherwise, one annihilation photon may convert,
with the remainder of the energy released as a softer ``bremsstrahlung'' photon.
The cross section for this process is 
\begin{align} \label{eqn:sigmaann}
    {d \sigma_{\rm ann} \over d\ln \omega } = {3 \sigma_T \over b} f(\beta),
\end{align}
where
\begin{align}
        \quad  f(\beta) = {1 \over 2 \beta \gamma^2} \left[\left(\beta + {1 \over \beta} \right) \ln \left[ \gamma (1 + \beta)\right] - 1 \right] \nonumber,
\end{align}
and $\gamma = 1/\sqrt{1-\beta^2}$.  This cross section can also be obtained by multiplying
the single-photon decay cross section by the appropriate soft photon factor.}

The equilibrium pair density is obtained by balancing energy loss via this ``annihilation bremsstrahlung'' emission
with Ohmic heating in the circuit.  Using the Drude conductivity
\begin{align}\label{eqn:drude_conductivity}
    \sigma_{\rm ohm} = {e^2 \over \beta \gamma^3 m_e c \sigma_\pm},
\end{align}
one finds that $n_\pm$ must exceed $n_{\rm min} = |j|/ec$ by a multiplicity factor $\mathcal{M} \sim 15$ \citep{ThompsonKostenko2020}.

\subsection{Enhanced Two-photon Pair Creation} 

As is the case for $e^\pm$ annihilation, the presence of a strong magnetic field alters the kinematics of two-photon pair creation. 
In vacuum, energy-momentum conservation gives rise to the following kinematic threshold for two-photon pair creation:
\begin{align}\label{eqn:pair_create_two_vac}
    2 \omega_1 \omega_2 (1 - \cos\theta_{12}) \ge 4 m_e^2,
\end{align}
where $\omega_1$ and $\omega_2$ are the frequencies of the incoming photons, and $\theta_{12}$ is the angle between their three-momenta. 
In the presence of a strong magnetic field, kinetic momentum is conserved only in the direction along ${\bm B}$, 
where translational symmetry is preserved.
The kinematic threshold for pair creation is~\cite{KostenkoThompson2018}
\begin{align}\label{eqn:pair_create_two_B}
    C \equiv (\omega_1 + \omega_2)^2 - (\omega_1 \cos\theta_1 + \omega_2 \cos\theta_2)^2 \ge 4m_e^2,
\end{align}
when $b > 4$; here, $\theta_1$ and $\theta_2$ are the angles between the photon momenta and the magnetic field. 
According to Eq. ~\eqref{eqn:pair_create_two_vac}, in vacuum a low-energy photon cannot pair create unless it collides with a photon with energy far above the pair creation threshold. By contrast, in the presence of a magnetic field, a low-energy photon can pair create by colliding with a photon only slightly above the pair creation threshold. The cross-section for two-photon pair creation (valid at $b > 4$) is~\cite{KostenkoThompson2018} 
    \begin{align} \label{eqn:two_photon_cross_section}
       &|1 - \cos\theta_{12}| \sigma(\theta_1, \theta_2, b) =  \sqrt{1 - {4 \over C} } \times \nonumber \\
       &{12 \sigma_T b C^2 \sin^2 \theta_1 \sin^2\theta_2 \over \omega_1^3 \omega_2^3 [C \sin^2\theta_1 \sin^2\theta_2 + 4 (\cos\theta_1 - \cos\theta_2)^2]^2}.
    \end{align}

The modification of these fundamental QED processes leads to a qualitatively new equilibrium plasma state in twist zones, as 
we now discuss.

\subsection{Shear Zone Plasma State} \label{sec:TK20}

The Ohmically heated plasma zones described here are localized in a small part of the magnetosphere.
Diverse arguments suggest that these current carrying arcades have a sub-kilometer thickness.  These are based on
measurements of surface hotspot emission and, independently, on global models of magnetic field drift, plastic flow
and thermal evolution in magnetar crusts~\cite{GourgouliatosCumming2013, Thompson_2017}.  A natural thickness $\Delta$ for
the arcade is the crustal vertical thickness $\Delta R_\mu \sim 0.3$ km (as weighted by shear modulus).
Such a structure of horizontal length $L$ has a cross sectional area $A_{\perp,\rm NS}$ at the surface
comprising a fraction $A_{\perp,\rm NS}/4\pi R_{\rm NS}^2 \sim 0.01 (L/\pi R_{\rm NS})(\Delta/\Delta R_\mu)$ of the star.

A collisional and mildly relativistic plasma state can be sustained where
the magnetic field experiences a critical level of shear.  The strength
of the current may be characterized	in terms of a shear length $\ell_{\rm shear}
= B/|\boldnabla\times{\bm B}|$.	 The corresponding pair	density	is
$n_\pm = {\cal M}n_{\rm min} = (2/3)\alpha b {\cal M} / \sigma_{\rm T}\ell_{\rm shear}$.
Taking a characteristic thermal speed $\beta$ for the $e^\pm$, we may integrate
the rate for a $e^\pm$ to collide with a counter-flowing $e^\mp$, obtaining
\begin{equation}
\int {d\ell\over\beta}\, 2\beta\cdot\sigma_\pm(\beta){n_\pm\over 4} \sim	
{\pi\over 16} {{\cal M}\over\beta\gamma^2} {\ell\over \ell_{\rm shear}},
\end{equation}
where we have approximated $n_\pm \propto |j| \propto r^{-3}$ along the magnetic flux surface.
Although a large-scale magnetic twist is limited by hydromagnetic stability
to $\ell_{\rm shear} \gtrsim R_{\rm NS}$, small-scale braiding can enhance the current density, 
generating $k_\perp \delta B_\perp \gg B/R_{\rm NS}$ and an effective shear length $\ell_{\rm shear} \sim 1$ km.

Balancing Ohmic heating with radiative cooling (${\cal M} = 15$) 
yields the following equilibrium number density,
\begin{align}\label{eqn:number_density_shear_zone}
    n_\pm(\ell) &\approx 2 \times 10^{18} \ {\rm cm}^{-3} \left({\rm km} \over \ell_{\rm shear}\right) \ b(\ell) \nonumber \\ 
    &\approx 15\,n_{\rm min}\quad (b \gtrsim 5),
\end{align}
where we have dropped the exponential in Eq.~\eqref{eqn:sigma_pm}, which is a good approximation in the large-field limit, $b \gg 1$. An interesting point to notice is that $n_\pm/B$ is constant along a field line, which is expected in a collisionless plasma, due to the balance of forces along a field line, but not necessarily in a collision-dominated plasma. 
The corresponding plasma frequency is 
\begin{align}\label{eqn:TK20_plasma}
    \omega_{p} \approx 0.1 \ {\rm eV} \left({\rm km} \over \ell_{\rm shear}\right)^{1/2} \ b^{1/2}(\ell).
\end{align}

The collisional $e^\pm$ plasma described here is a source of annihilation bremsstrahlung
radiation that permeates much of the remainder of the magnetosphere.  The spectrum of this radiation
is slightly modified from the two-photon vacuum annihilation spectrum.  Detailed MC
simulations described in~\cite{ThompsonKostenko2020} account for (i) various annihilation
channels, with 0, 1 or 2 decay photons individually reconverting to pairs in the ultrastrong magnetic field;
(ii) scattering of low-energy photons off $e^\pm$ pairs (including the strong $u$-channel resonance
encountered when the photon energy approaches the threshold for pair creation); as well as (iii) Breit-Wheeler
pair creation.  A fit to the escaping X-ray and gamma-ray MC spectrum is \cite{ZhangThompson2024}
\begin{align} \label{eqn:annihilation_bremsstralung_spectrum}
    {d L_\gamma \over d\omega} = C\cdot  { L_0 \over \omega_0} e^{-(\omega/\omega_0)^4}
\end{align}
where $C$ is a normalization constant, $L_0$ is the total annihilation bremsstrahlung luminosity, and $\omega_0 = 1.5 m_e$.
As this spectrum extends above $\omega = m_e$, the highest energy annihilation photons are a significant source of
pairs even in zones of weak current density.

\subsection{Non-local Pair Creation and the Global Magnetosphere} \label{sec:TK20_magnetosphere}

We now summarize our approach to computing the structure of	the large-scale $e^\pm$ cloud surrounding a	magnetar,
following \citep{ZhangThompson2024}.
We find in Section \ref{sec:mixing} that the dominant contribution to the resonant $a\rightarrow\gamma$ conversion
comes not from the strongest current arcades, but from the halo of trans-relativistic pairs that
are suspended around these structures, in zones that may carry little or no current.  When the magnetar
is rotating relatively slowly (e.g. with a spin period $P = 1-10$ s) this $e^\pm$ halo is still much
denser than the baseline corotation particle population.

This global $e^\pm$ halo has been computed with a MC procedure in \cite{ZhangThompson2024}. The super-Schwinger
magnetic field acts as a capacious reservoir for pairs, for a few reasons.   First, photons of energy $\sim m_ec^2$
have a significantly lower density than $e^\pm$ in the lowest Landau state when photons and pairs are in
approximate kinetic equilibrium.   The cross section~\eqref{eqn:sigmaann} for two photon pair creation is enhanced by
a factor $b$, whereas the corresponding	annihilation cross section~\eqref{eqn:sigma_pm} is suppressed by	a factor $\sim b^{-1}$.
Equilibrium between pair creation and annihilation yields $n_\pm/n_\gamma \sim (\sigma_{\gamma\gamma}/\sigma_\pm)^{1/2} \sim b$.
Second,	positrons flowing down to the surface of the magnetar are mainly reflected, with only $\sim 1/3$ being
absorbed and annihilating \cite{ZhangThompson2024}.

The computation of the equilibrium pair density at an arbitrary position in the magnetosphere is carried out as follows.
First, an arcade structure is chosen from the sample in Fig. \ref{fig:collisional_model_geometries}. 
Photons are drawn from a spectrum~\eqref{eqn:annihilation_bremsstralung_spectrum} and emitted isotropically from the
arcade surface with uniform flux per unit surface area and total luminosity $L_0$.  Gravitational lensing significantly influences which
parts of the magnetar exterior directly receive rays from the arcade structure and is computed assuming a non-rotating
Schwarzschild geometry.  The space around the magnetar is discretized and the photon occupancy computed in each
spatial cell for a discrete set of propagation directions.   The steady state pair density is computed assuming
that pairs have time to sample the full volume of a magnetic field bundle, with density at radius $r > R_{\rm NS}$
diluted according to $n_\pm(r) = n_{\pm,\rm NS}B(r)/B_{\rm NS}$, where $n_{\pm,\rm NS}$ and $B_{\rm NS}$ are the surface number density and field strength where the bundle intersects the star.  



We compute the equilibrium density by balancing the total pair creation rate with the pair loss rate in the entirety of the tube. The main processes considered here and in~\cite{ZhangThompson2024} are: two-photon pair creation ($\gamma + \gamma \to e^+ + e^-$), pair annihilation ($e^+ + e^- \to \gamma + \gamma$) into pairs below the pair production threshold, surface annihilation, and outflow through radiative drag.

The total pair production rate in a field bundle is
\begin{align} \label{eqn:pair_creation_rate}
    &\dot N_{\gamma + \gamma\to e^+ + e^-} =\displaystyle \int_{0}^{\ell_{>4}} d\ell \displaystyle\int \ dA_\perp(\ell) \displaystyle\int d^3 p_1 d^3 p_2  \nonumber\\
    &\left[{d^3 n_{\gamma} \over d^3 \vec p}(\vec p_1){d^3 n_{\gamma} \over d^3 \vec p}(\vec p_2) |1 - \hat{p}_1 \cdot \hat{p_2}| \sigma(\vec p_1, \vec  p_2, B)\right]
\end{align}
{where $\ell_{>4}$ is the length along the tube where $B \ge 4 B_Q$,} $\vec p_1$ and $\vec p_2$ are the photon three-momenta, $d^3 n_\gamma/d^3\vec p$ is the differential number density of photons at the given location along the bundle, and the final term $|1 - \hat p_1 \cdot \hat p_2| \sigma (\vec p _1, \vec p_2, B)$ is given in~\eqref{eqn:pair_create_two_B}.

The two-photon annihilation channel of interest is the one in which both decay photons are below threshold for
immediate pair conversion (this channel is distinct from the annihilation bremsstrahlung channel in which one
pair is quickly regenerated).  The cross section 
is approximately $\sigma_{\rm ann} = 1.42 \ \sigma_T /b$~\cite{ThompsonKostenko2020}. The volumetric annihilation rate is then 
computed as
\begin{align}\label{eqn:ndot_ann}
    \dot{N}_{\rm ann} &= \displaystyle \int_{0}^{\ell_{>1}} d\ell \displaystyle\int \ dA_\perp(\ell) \ n^2 _\pm(\ell) \sigma_{\rm ann}(b(\ell)) \beta \nonumber \\
    &= A_{\perp, \rm NS} \ n^2_{\pm,\rm NS} \ \sigma_{\rm ann}(b_{\rm NS}) \ell_{>1} \beta,
\end{align}
where $\ell_{> 1}$ is the length of the field line with $B>B_Q$.
In the second line, we use the fact that $\sigma_{\rm ann} \propto B^{-1}$, and $n_\pm(\ell)/b(\ell)$, and $n_\pm(\ell) dA_{\perp}(\ell)$ are constant on a field line. 

A significant flux of particles moving in the tube can annihilate in a thin surface layer at the magnetar surface. The total annihilation rate was derived in~\cite{ZhangThompson2024} to be
\begin{align}
    \dot{N}_{\rm surf} &= {A_{\perp,\rm NS}\over 6} n_{\pm,\rm NS} \beta,
\end{align}
where $\beta \approx 0.6$ is the typical particle velocity in the trans-relativistic state. Pairs may also be evacuated from the tube due to radiation pressure from the surface. As discussed in Sec. \ref{sec:jbundle}, pairs moving along field lines that extend sufficiently far from the magnetar surface experience powerful radiative drag due to resonant scattering with thermal photons. Field lines whose foot points satisfy $\theta < \theta_{\rm max}$, where

\begin{align}
    \theta_{\rm max} = \left( {T_{\rm NS} \over m_e b_{\rm NS} \gamma (1 + \beta)} \right)^{1/6} \approx 12^\circ \left( {b_{\rm NS} \over 10}\right)^{-1/6},
\end{align}
where $T_{\rm NS} \sim$ keV is the surface temperature of the magnetar, and $\beta = 0.6$. The pair loss rate in this region is ~\cite{ZhangThompson2024}

\begin{align}
    \dot{N}_{\rm rad} &= {A_{\perp,\rm NS}\over 2} n_{\pm,\rm NS} \beta \quad (\theta < \theta_{\rm max}).
\end{align}

The loss of pairs through pair annihilation, surface annihilation, and radiation-driven outflow is balanced by two-photon pair creation ($\gamma +\gamma \to e^+ + e^-$) in the magnetosphere. Photon collision is an inherently non-local process; photons created with energies $2m_e < E < 2 m_e/\sin\theta_{\rm em}$, where $\theta_{\rm em}$ is the angle at which the photon is emitted relative to the background magnetic field, can propagate to neighboring field lines before pair producing. Inclusion of photon propagation effects and non-local pair creation can be found in~\cite{ZhangThompson2024}. The simulations consider four non-azimuthally symmetric geometries for the shear zones\footnote{ We only consider three of the four geometries present in Ref.~\cite{ZhangThompson2024}. The model not considered, the `double quarter model,' contains two arcades like the one in the `quarter model' (see Fig.~\ref{fig:collisional_model_geometries}), located diametrically opposite from one another. We neglect it because it provides almost identical axion projections to the `half model.' } with corresponding plasma frequency profiles shown in Fig.~\ref{fig:plasma_profiles}. We use the results of these simulations to predict the two-photon pair creation rate in the magnetosphere.


Finally, the equilibrium pair density at the surface of the NS is given by the solution to, 

\begin{align} \label{eqn:number_equilibrium}
    \dot{N}_{\rm ann} + \dot N_{\rm surf} + \dot N_{\rm rad} - \dot{N}_{\gamma + \gamma\to e^+ + e^-} = 0,
\end{align}
which is quadratic in $n_{\pm,\rm NS}$. Recall that to compute the density elsewhere, one simply uses the relation $n(\ell) B(\ell) = n(\rns) B_{\rm NS}$.

{\color{black}

\section{Observational Discrimination} \label{sec:modelComp}


To summarize, the BT07 (Sec.~\ref{sec:jbundle}) and TK20 (Sec.~\ref{sec:collisionalModel}) models make significantly different predictions about the location of magnetospheric twist in the magnetosphere and the state of plasma in their respective twist zones. In the BT07 model, {\color{black} currents
are concentrated around the magnetic pole; this} twisted j-bundle 
contains relativistic plasma whose dynamics are dominated by radiative drag (see Fig.~\ref{fig:model_cartoons}).
In the TK20 model, magnetospheric twist {\color{black} may be} localized to shear zones with field lines that do not extend far from the star (see Fig.~\ref{fig:model_cartoons}). These shear zones host a collisional, semi-relativistic plasma that supplies the remainder of the magnetosphere with plasma through non-local two-photon pair creation. In this section, we discuss how observations can be used to distinguish the two models and examine a few test cases where such methods have been applied. 

A key feature of magnetars that distinguishes them from their lower-field counterparts is the fact that their magnetospheres are dynamic, exhibiting transient behavior, both temporal (e.g., spin-up glitches, spin-down anti-glitches, and large variations in spin-down rate) and radiative (e.g., X-ray bursts and outbursts, giant flares, and fast radio bursts), with time scales ranging from a fraction of a second to years. Transient behavior is often temporally localized to active ``bursting phases,'' which commence with large and sudden increases in X-ray flux, and outbursts that decay on a timescale of minutes to hours and are followed by ``afterglow'' emission that decays over weeks to months. A long time (months to years) after the initial outburst, the magnetar
{\color{black} may relax} to a quiescent state, characterized by a {\color{black} 2--10 keV} X-ray spectrum that is dominated by a blackbody with a temperature in the range of $T_{\rm NS} \approx 0.5$ keV
{\color{black} and a steep power law that dominates at $\omega \gtrsim 3$ keV}~\cite{Kaspi2014}.
{\color{black} The models described in this paper are therefore most applicable to X-ray bright Anomalous X-ray Pulsars -- magnetars that have not yet been observed to emit hyper-Eddington X-ray bursts.
The more actively bursting Soft Gamma Repeaters have flatter 2--100 keV energy spectra \citep{Enoto_2017}, and by
implication stronger and more broadly distributed magnetospheric currents.}

In a search for axions, quiescent magnetars are preferred targets to their bursting counterparts due to their lower backgrounds and the fact that their global magnetospheric structures are expected to remain static over an hours-long observational window. In this section, we outline key observations of quiescent magnetars that may be used to distinguish between the BT07 and TK20 models.  



\vspace{0.1in}

\noindent \textbf{\textit{Hard X-ray Emission.---}} Non-thermal X-ray emission provides a powerful tracer for magnetospheric twist in quiescent magnetars.  {\color{black} The common detection of a hard, rising energy spectrum in the 10--70 keV band, sometimes extending to at least 100 keV \cite{Enoto_2017},
imposes strong constraints on models of the current-carrying plasma.}


In the BT07 model, hard X-ray photons are produced by resonant inverse Compton scattering in the radiative zone, where the upscattered photons are below the pair production threshold. The total Ohmic dissipation rate in the j-bundle is given $L_{\rm diss} = I \Phi$, where $I$ is the total current flowing in the loop, and $\Phi$ is the longitudinal voltage.  {\color{black} This} is approximately
\begin{align}\label{eq:ldiss}
    L_{\rm diss} \approx 6 \times 10^{35} \ {\rm erg \over  s} \ \left( {B_{\rm pole} \over 10\,B_{\rm Q}} \right) \Phi_9 \ \psi \ \theta_c^4,
\end{align}
where $B_{\rm pole}$ is the dipolar magnetic field evaluated at the pole, $\psi$ is the twist in the bundle, and $\theta_c$ is the angular extent of the j-bundle. Most of this energy is dissipated in the form of hard X-ray and gamma-ray emission produced at the edge of the radiative zone and with typical energy $E \sim 4$ MeV (see~\eqref{eqn:gammal}), and a steep spectral index~\cite{Beloborodov2013b}. 


A similar X-ray luminosity is expected from annihilation bremsstrahlung in the TK20 model. An estimate for the luminosity can be obtained by {\color{black}integrating the Ohmic dissipation $j^2/\sigma_{\rm ohm}$ along
a magnetic flux tube of surface cross section $A_{\perp,\rm NS}$, using the conductivity $\sigma_{\rm ohm}$
given in Eq. \eqref{eqn:drude_conductivity}} 
\begin{align}
    L_{\rm ann} \approx 4 \times 10^{35} \ {\rm erg \over  s} \ \left( {B_{\rm NS} \over 10 B_Q} \right)^{4/3} \left( {A_{\perp,\rm NS}/\ell_{\rm shear}^2 \over 100} \right);
\end{align}
{\color{black} here, we adopt an RMS particle velocity $\beta = 0.6$ and take a length $l \sim 
R_{\rm NS} b_{\rm NS}^{1/3}$ for the length of the emitting tube.}

A key observable that can be used to distinguish the two models is therefore the photon spectrum in the 
MeV range.  {\color{black} Resonant scattering in} the BT07 model {\color{black} can generate photons
of energy well above 1 MeV}
whereas the TK20 model predicts a {\color{black} hyper-exponential cutoff} above 
0.5--1 MeV.

Another {\color{black}observational diagnostic}
is the relative output from thermal hot spots and hard X-ray emission.  
{\color{black} For a given magnetic twist, the relativistic double layer produces a much higher {\color{black} energy 
flux in particles} impacting the surface
than does the collisional $e^\pm$ plasma state; but this effect may be compensated
if the j-bundle is confined close to the pole, where the current density $j \propto \theta^2$.  Then
the polar energy flux is
\begin{eqnarray}
F_{X,\rm pole} &=& j\cdot\Phi \simeq {cB_{\rm pole}\theta_c^2\over 4\pi R_{\rm NS}}\psi\cdot\Phi \nonumber \\
&\sim& 4\times 10^{22}\,{\rm erg\over cm^2~s}\,
\left({B_{\rm pole}\over 10\,B_{\rm Q}}\right)\left({\theta_c\over 0.1}\right)^2 
\Phi_9\psi.\nonumber \\
\end{eqnarray}
For comparison, the energy flux released by the annihilation of downmoving $e^+$ in a collisional flux bundle with
magnetic shear length $\ell_{\rm shear}$ is
\begin{eqnarray}
F_{X,\rm NS} &=& f_{\rm ann}\beta {n_\pm\over 4}\cdot 2m_e \nonumber \\
&=& 1.5\times 10^{23}\,{\rm erg\over cm^2~s}\,f_{\rm ann}
{(B_{\rm NS}/10\,B_{\rm Q})\over (\ell_{\rm shear}/{\rm km})}.\nonumber \\
\end{eqnarray}
Here, $f_{\rm ann}$ is the fraction of downmoving $e^+$ that annihilate rather than reflecting
and we have taken a multiplicity ${\cal M} = 15$.
}

Another remarkable phenomenon that has been observed in some quiescent magnetars is that of shrinking hot spots, in which the size and luminosity of hot spots are observed to decrease over months to years-long timescales (see, e.g., magnetar XTE J1810-197~\cite{Gotthelf2007, Perna2008}). 
{\color{black} This could be connected with resistive untwisting of 
a non-potential magnetic field near the magnetic pole
following an impulsive injection of twist into the magnetosphere \cite{Beloborodov_2009};
alternatively, it could reflect residual plastic flow in narrow fault-like zones in the magnetar crust
\cite{Thompson_2017}. The implications are} discussed further in Sec.~\ref{sec:gc_mag}. 

\vspace{0.1in}

\noindent \textbf{\textit{Correlation of X-rays and Radio Emission/ Spindown.---}} Spin-down measurements of magnetars carry direct information about {\color{black} the magnetic twist, and hence the current,
in the polar region.}
Another way to distinguish the two models is to search for correlations between X-ray emission and changes in the observed spindown rate, as an enhanced spindown rate implies injection of twist into the open field lines. 

A significant delay between an X-ray outburst and a change in spindown rate is a key prediction of both the BT07 and TK20 models.  
In the BT07 model, the delay is related to the resistive timescale for a global twist (imparted by the outburst) to migrate to the polar region. The delay may be more significant in the TK20 model since currents are typically injected further from the polar region. Such a delay has been observed in a number of sources such as 1E 1048.1-5937 and is on the order of months to years, consistent with the expected resistive timescale~\cite{Gavriil:2004hb}. Magnetospheric twist near the open zone may also lead to phase overlap between X-ray and radio emission, which has been observed in sources such as XTE J1810-197 \footnote{This argument assumes magnetar radio emission comes from the open field lines as in a canonical radio pulsar. The mechanism for magnetar radio emission is not currently understood, so this statement should be taken with some caution.}. Some sources such as 1E 2259+586, however, exhibit hard X-ray emission in the absence of both anomalous spindown~\cite{DibKaspi2014} and pulsed radio emission~\cite{Coe1994}, which implies a spatial separation between regions sourcing hard X-ray emission and the open field lines, as predicted by the TK20 model.


\vspace{0.1in}

\noindent \textbf{\textit{Cyclotron Absorption Features.---}} Cyclotron absorption is the direct absorption of photons with energy equal to the cyclotron frequency of charged particles in a magnetic field. Phase-dependent cyclotron absorption features are indicative of powerful localized magnetic loops, such as those described in the TK20 model, and have been observed in multiple sources. Magnetar SGR 0418+572913 has exhibited a phase-dependent cyclotron absorption line which, if interpreted as a proton cyclotron line, implies the existence of a near-surface magnetic field in the range of $2 \times 10^{14} \ {\rm G} - 10^{15} \ {\rm G}$, much stronger than its spindown-inferred dipole field of $\sim 6 \times 10^{12}$ G near the equator~\cite{Tiengo2013}. Similar features have been observed in other sources including SWIFT J1822.3-1606~\cite{Castillo2016}, and  1E 2259+586~\cite{Pizzocaro2019}, both of which infer small-scale magnetic fields with strengths larger than the dipole component.

In the following section, we examine two sources: PSR J1745--2900 (the GC magnetar) and AXP 1E 2259+586, and discuss observational support for their associations with the BT07 and TK20 models, respectively. 

\subsection{Application to Galactic Magnetars} \label{sec:observational_targets}


Below, we apply the polar j-bundle (BT07) and collisional trans-relativistic (TK20) models to observations of two specific sources: the GC magnetar PSR J1745–2900 and AXP 1E 2259+586. In the literature, PSR J1745–2900 has been analyzed in the context of both the BT07 model~\cite{CotiZelati:2017eid, Rea2020} and AXP 1E 2259+586 in the context of the TK20 model~\cite{ZhangThompson2024}. We briefly review these interpretations, while emphasizing that neither source has been conclusively described by either model. Additional data and analysis will be necessary to distinguish between them for any given object.

\subsubsection{Galactic Center Magnetar PSR J1745--2900} \label{sec:gc_mag}

The GC magnetar, PSR J1745--2900, has been identified as an ideal candidate to search for axion dark matter (see e.g.~\cite{Hook:2018iia, Leroy:2019ghm,Foster:2020pgt,Witte2021,battye2021robust, Foster:2022fxn,McDonald:2023shx}). Observations by the \emph{Chandra} and Neil Gehrels \emph{Swift} observatories localized the magnetar to {\color{black} a projected distance} of $\sim$0.1 parsec from the GC supermassive black hole, Sagittarius A*~\cite{Rea2013}. Subsequent observations of X-ray pulsations by \emph{NuSTAR} revealed a rotation period of $P = 3.76$ sec and a period derivative $\dot{P} = 6.5 \times 10^{-12}$, which, if interpreted in the context of rotational spin down, would imply a {\color{black} polar} surface magnetic field of $B_{\rm pole} = 3.3\times 10^{14}$ G $\approx 8\,B_Q$  ~\cite{Mori_2013,Rea2013}\footnote{Crucially, spin-down measurements only provide information about the strength of the dipole component of the magnetic field. Much stronger multipolar fields may exist on the magnetar surface (e.g. \cite{delima2020}). Modeling of such fields is expected to enhance axion-photon mixing, but robust modeling of multipolar fields of PSR J1745--2900 is left to future work.}. \gcmag\, is one of six magnetars observed to emit pulsed radio emission, with an approximately flat spectrum between 1.2 GHz and 291 GHz~\cite{Eatough2013, Spitler2014, Torne2017}. Although the precise mechanism of radio emission is not known, it is likely related to plasma oscillations at the corresponding frequencies (as in the case of standard pulsars); should this be the case, it would imply that axions with masses between approximately $5 \ \mu$eV and 1 meV can efficiently resonantly convert in the magnetosphere.

Following its initial outburst in April 2013, PSR J1745--2900 has exhibited spectral evolution consistent with {\color{black} emission from a shrinking hotspot
on the magnetar surface.  {\color{black} We now fit this data to
a contracting polar j-bundle in the BT07 circuit model.  This allows us
to infer the properties of the j-bundle in the source's quiescent state, and
to test the consistency of the model.  The j-bundle is defined by its
twist $\psi$ and size \cite{Beloborodov_2009}.}
The latter can be directly inferred {\color{black} from the measured temperature
and luminosity, whereas the former is independently constrained by the luminosity
and by changes in spindown rate once $\theta_c$ and $B_{\rm pole}$ are known}. 
{\color{black} In this way, we obtain fiducial parameters to be used in the
sensitivity analyses in the following sections.}

{\color{black}
The luminosity $L_X$ of the surface hotspot below an untwisting j-bundle
can be obtained from Eq.~\eqref{eq:ldiss}
in the approximation where one half of the current above the polar cap is carried by charges returning to the star.  Exchanging the angular boundary $\theta_c$ for
the blackbody radius $R_{\rm BB}  \simeq \theta_c R_{\rm NS}/2$, we have
\begin{equation} \label{eqn:luminosityarea}
L_X \simeq {1\over 2}L_{\rm diss} = 4\times 10^{33} \,{\rm erg\over s}\,
\left({B_{\rm pole}\over 10\,B_Q}\right) R_{\rm BB,km}^4 \Phi_9\psi.
\end{equation}
}
{\color{black} Note that}
the blackbody radius $R_{\rm BB} $ is defined by $A_{\rm BB}   = 4\pi R_{\rm BB}  ^2$, where $A_{\rm BB}  $ is the area of the thermal hot spot on the NS surface. {\color{black} This has been determined for PSR J1745$-$2900} in~\cite{CotiZelati:2017eid}. Eq.~\eqref{eqn:luminosityarea} assumes the twist and voltage are constant across the j-bundle, but not necessarily in time, and vanish outside the bundle. A joint simultaneous measurement of $L_X$ and $R_{\rm BB}$ would allow one to infer the product of $\psi\Phi_9$, but not uniquely $\psi$. This degeneracy can be broken by noting that high twist near the polar zone leads to enhanced spindown~\cite{Beloborodov_2009}
\begin{align} \label{eqn:dpdot}
    {\Delta \dot{P} \over \dot{P} } \approx {\psi^2 \over 2\pi} \ln\left( 
    \theta_c^2{R_{\rm LC} \over \rns} \right).
\end{align}

Therefore, a unique determination of $\psi$ can be obtained by combining X-ray measurements with information on the rate of change of spindown. The difficulty in the context of PSR J1745--2900 stems from the fact that recent archival observations are not readily available. Following the initial outburst in 2013, the X-ray luminosity was measured by the \emph{Chandra} High Resolution Camera (HRC) and the Advanced CCD Imaging Spectrometer array (ACIS-S) as well as the European Photon Imaging Cameras (EPIC) aboard \emph{XMM-Newton}. The data span about three and a half years, from April 29, 2013, until October 14, 2016. Using this information, we can attempt to extrapolate the X-ray luminosity and the evolution in the size of the hot spot  to infer $\psi \Phi_9$.

The evolution of the post-outburst X-ray luminosity (in the range 0.3--10 keV) can be well-fit by a double exponential model~\cite{CotiZelati:2017eid}, 
\begin{align}\label{eqn:luminositytime}
    L_X(t) = \displaystyle\sum_{i=1,2} A_i e^{-t/\tau_i} + L_Q,
\end{align}
where the best-fit parameters are (see Fig. \ref{fig:fits}) $A_1 \approx 2.46 \times 10^{35}$ erg s$^{-1}$, $A_2 \approx 2.04 \times10^{35}$ erg s$^{-1}$, $\tau_1 \approx 106.9$ days, and $\tau_2 \approx 444.3$ days. The quantity $L_Q$ is the quiescent luminosity. 
The absence of X-ray emission in archival \emph{Chandra} data at the position of PSR J1745--2900 indicates that $L_Q \lesssim 10^{32}$ erg s$^{-1}$~\cite{Muno2009}. We adopt $L_Q = 10^{32}$ erg s$^{-1}$, although the choice does not affect results considerably. 


The evolution of the hot spot size can be obtained 
by {\color{black} generalizing the simplest j-bundle model with constant
$\psi = O(1)$ to allow for an increase of the twist with time as the j-bundle
shrinks in size.  We assume a power-law dependence 
$\psi \Phi_9 \propto R_{\rm BB}^{-\alpha}$, or equivalently $\psi \Phi_9 \propto 
L_X^{-\alpha/(4-\alpha)}$.}
{\color{black} Applying this to the data, one finds $\alpha \simeq 1.42$ (see Fig.~\ref{fig:fits}, right panel) and}
\begin{align} \label{eqn:luminosityradiusfit}
    L_{X}(R_{\rm BB, km}) = 3.7 \times 10^{34} \, {\rm erg\over s}
    \times R_{\rm BB,km}^{2.58}\,.
\end{align}
 The fits outlined in Eqs.~\eqref{eqn:luminositytime} and \eqref{eqn:luminosityradiusfit} are shown explicitly alongside the aforementioned X-ray data in Fig.~\ref{fig:fits}.  
 {\color{black} Taking $B_{\rm pole} = 3.3\times 10^{14}$
 G (as measured from spindown) we infer from Eq.~\eqref{eqn:luminosityarea} 
 that $\psi \Phi_9 \simeq 12$ at the point in the evolution 
 where $R_{\rm BB} = 1$ km.}

{\color{black} The X-ray luminosity and hot spot size are readily extrapolated
to the present,} given that no further X-ray outbursts have been observed from PSR J1745--2900 since 2019~\cite{Rea2020}.
{\color{black} We infer}
$L_X = 1.15 \times 10^{32}$ erg s$^{-1}$ and $R_{\rm BB, km} \approx 0.1$ km.   Then the surface j-bundle radius $2R_{\rm BB}$ is only slightly larger than the polar cap radius, $\rpc = \rns \sqrt{\Omega \rns} \approx 75$ m.
Taking $R_{\rm BB} = 0.1$ km, one can infer that the twist 
has increased substantially from the earlier value, and now
$\psi\Phi_9 \approx 300$.

{\color{black} We conclude that a large combination of polar magnetic
twist $\psi$ and voltage $\Phi$ is required to fit the j-bundle model to
the cooling behavior of \gcmag.  As described in Sec. \ref{sec:jbundle},
the polar voltage is regulated by secondary $e^\pm$ pair creation.
Near the magnetar surface, the threshold for resonant scattering of
blackbody photons is first reached by $e^+$ or $e^-$ returning to the star.  
A conservative bound on the voltage is obtained by drawing target X-ray
photons from the blackbody peak, with energy $\sim 3 k_B T_{\rm NS}$.   Then
for \gcmag, we deduce $e\Phi = \gammasc m_e < bm_e^2/6 k_B T_{\rm NS} = 0.8~{\rm GeV}
(k_B T_{\rm NS}/0.5~{\rm keV})^{-1}$.  This is a robust limit, since $e^\pm$ of
this energy experience a very high rate of resonant scattering; in fact,
pairs will be mainly seeded by lower-energy particles that scatter
photons drawn from well above the peak.  (Given a resonant cross section 
$(2\pi^2e^2/m_e)\delta(\omega-eB/m_e)$, the
cumulative optical depth for an ingoing $e^+$ or $e^-$ to scatter a photon of
energy $3kT_{\rm bb}$ is $\sim 10^5(F_{X,\rm pole}/10^{23}~{\rm erg~cm^{-2}~s^{-1}})^{3/4}$.)

Next consider the implications for the polar twist $\psi$.}
Assuming the voltage drop along the bundle remains fixed, Eqs.~\eqref{eqn:luminosityarea} and~\eqref{eqn:luminosityradiusfit} imply that the twist angle $\psi \propto L_X^{-0.55}$. Therefore, from Eq.~\eqref{eqn:dpdot},  {\color{black} the increment in the} spindown rate is expected to 
{\color{black} grow with decreasing hotspot brightness as}
$\Delta \dot P/\dot P \propto L_X^{-1.1}$. 
This behavior {\color{black} may be} inconsistent with data showing constant spindown rates during {\color{black} the later stages of} X-ray dimming~\cite{Rea2020}. The 
{\color{black} spindown rate grew by a factor $\sim 2-5$ from the first timing
measurements \cite{CotiZelati:2017eid}, suggesting a twist of $\psi \approx 1-1.6$,
but a further increase is required to maintain the power-law relation between
$L_X$ and $R_{\rm BB}$ to the lowest measured flux level.}

Another challenge
introduced by large twist is that the j-bundle will be
susceptible to non-axisymmetric kink instabilities (see, e.g.,~\cite{Mahlmann_2023}).  The non-potential magnetic energy would
be released in the form of enhanced X-ray emission, which is in conflict
with the reduced $L_X$ of PSR J1745-2900.
While some compensating increase in the voltage $\Phi$ is expected
with decreasing temperature of the polar cap,
keeping $\psi$ less than unity implies much larger values of $\Phi$
than are allowed by considerations of resonant scattering.

\begin{figure*}
    \centering
    \includegraphics[width=\linewidth]{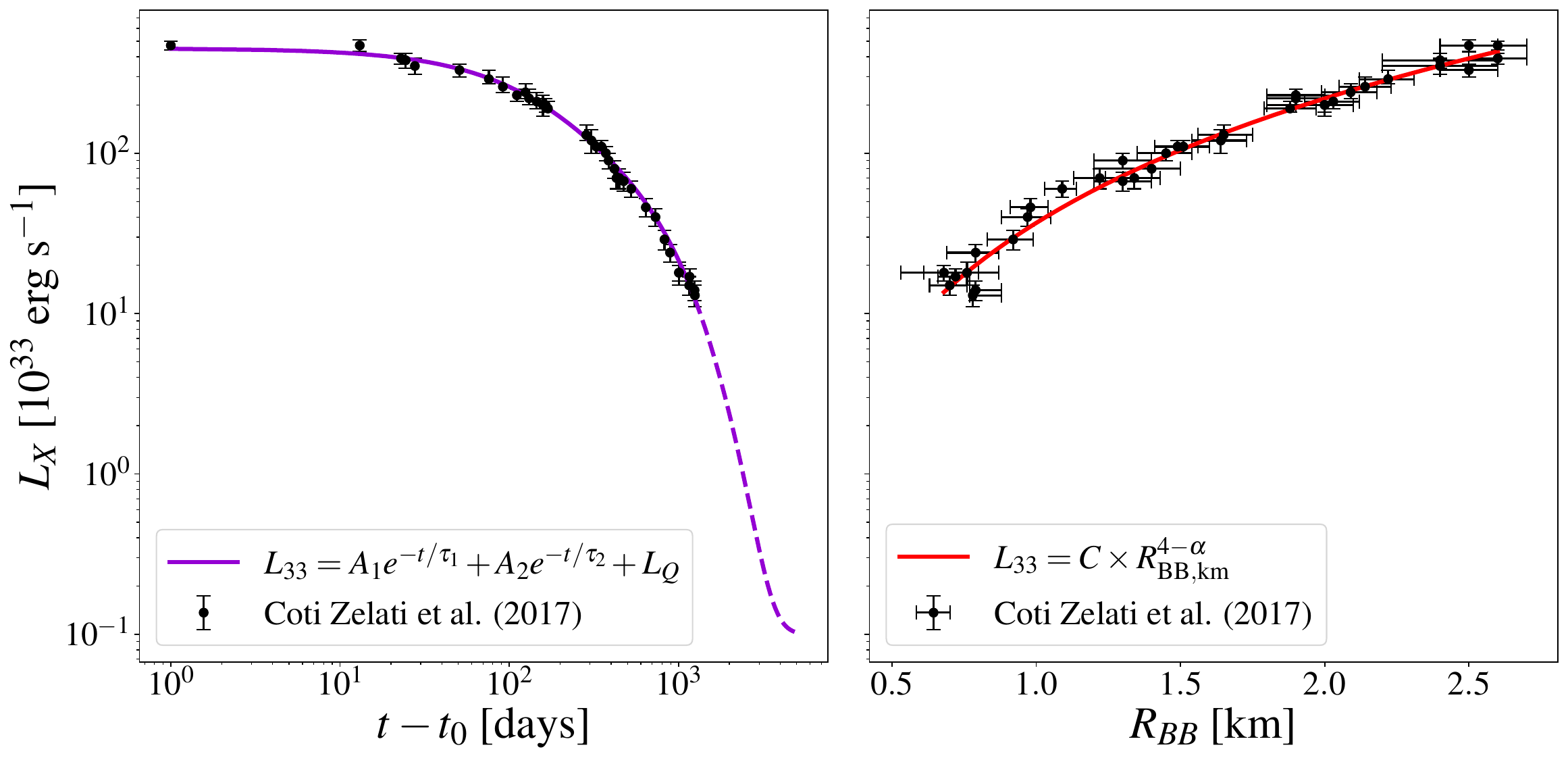}
    \caption{Fits and extrapolation of X-ray observations of GC magnetar PSR  J1745--2900. Left panel: fit of X-ray luminosity data (black points) as a function of time since outburst to a double-exponential as in~\cite{CotiZelati:2017eid}, as well as an extrapolation to present day (dashed). Right panel: {\color{black} fitted correlation between} X-ray luminosity and thermal blackbody radius (in km).}
    \label{fig:fits}
\end{figure*}

\subsubsection{AXP 1E 2259+586: Collisional, Transrelativistic Model}

The prototypical Anomalous X-ray Pulsar 1E 2259+586 was observed in supernova remnant G109.1--1.0 as a pulsating X-ray source with periodicity $P = 6.98$ sec~\cite{GregoryFahlman1980}. {\color{black} The polar magnetic field inferred from spin-down
is $B_{\rm pole} \approx 1.2 \times 10^{14}$ G.}  In addition to this large-scale dipolar magnetic field, {\color{black} there is indirect evidence that AXP 1E 2259+586 hosts stronger magnetic fields in its interior, or in the form of
higher multipoles.} This source has exhibited bursts, spin-up glitches, and spin-down anti-glitches \cite{Kaspi_2017}.
Additionally, \emph{XMM-Newton} observations revealed a phase-dependent, low-energy ($E \approx 0.7$ keV) absorption feature that can be explained by resonant proton cyclotron scattering~\cite{Pizzocaro2019}. Follow-up observations with IXPE, NICER, and \emph{XMM-Newton} confirmed this feature and modeled it as X-rays scattering off of non-relativistic protons in a plasma loop confined close to the magnetar surface and with a magnetic field greater than $10^{15}$ G~\cite{Heyl2024}. Such plasma loops share the properties of the magnetic arcades in the TK20 model, namely that they have $B \gg B_Q$ and are localized close to the surface.

AXP 1E 2259+586 also displays a hard X-ray spectrum in the 20--50 keV range that agrees remarkably with the prediction of annihilation bremsstrahlung as the source of hard X-rays~\cite{ThompsonKostenko2020}. Moreover, and as mentioned above, this source has exhibited only a few timing anomalies (glitches and anti-glitches), but otherwise has a relatively quiet spindown history~\cite{DibKaspi2014} and no observed radio emission~\cite{Coe1994}. The observation of hard X-ray emission without either radio emission or significant spindown variations suggests a spatial separation between the X-ray emitting region and the polar zone. Finally, AXP 1E 2259+586 has exhibited pulsed near-IR emission that is much brighter than expected from a blackbody with temperature $T\approx 0.5$ keV~\cite{Hulleman2001}. Coherent optical and IR emission are {\color{black} a natural consequence
of current-driven instability either in the magnetosphere or NS
atmosphere \cite{ThompsonKostenko2020}.}

To summarize, we have discussed a number of observations that can be used to distinguish the BT07 and TK20 models, and discussed particular sources whose observational characteristics are particularly consistent with one source or another. At present, there is no consensus on a unified magnetar model to describe all sources. Given the source-to-source variability, it is entirely possible that different models must be adopted to describe different sources. We leave more in-depth modeling of individual sources to future work.

\begin{figure*}
    \centering
    \includegraphics[width=0.49\linewidth]{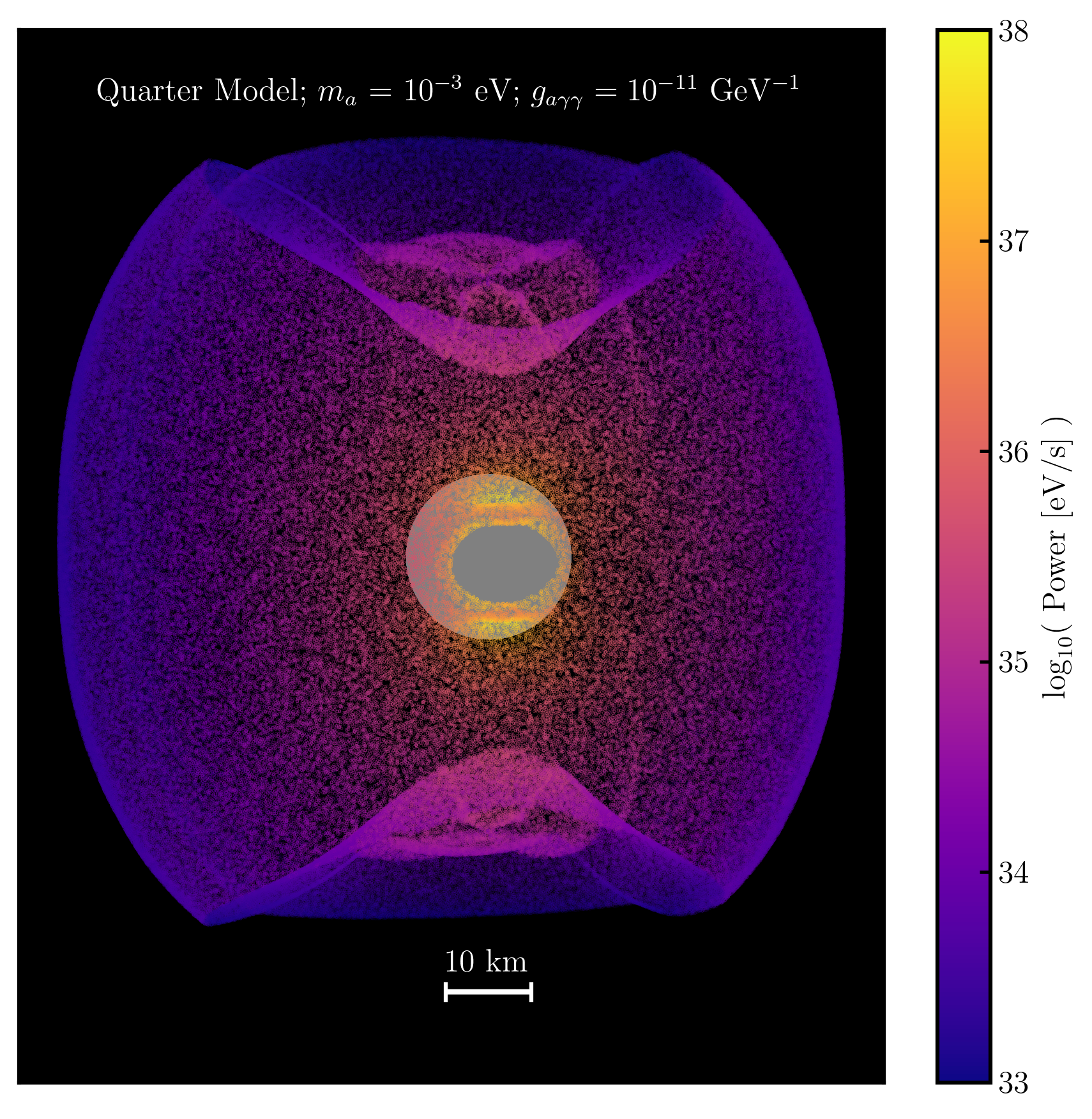}
    \includegraphics[width=0.5\linewidth]{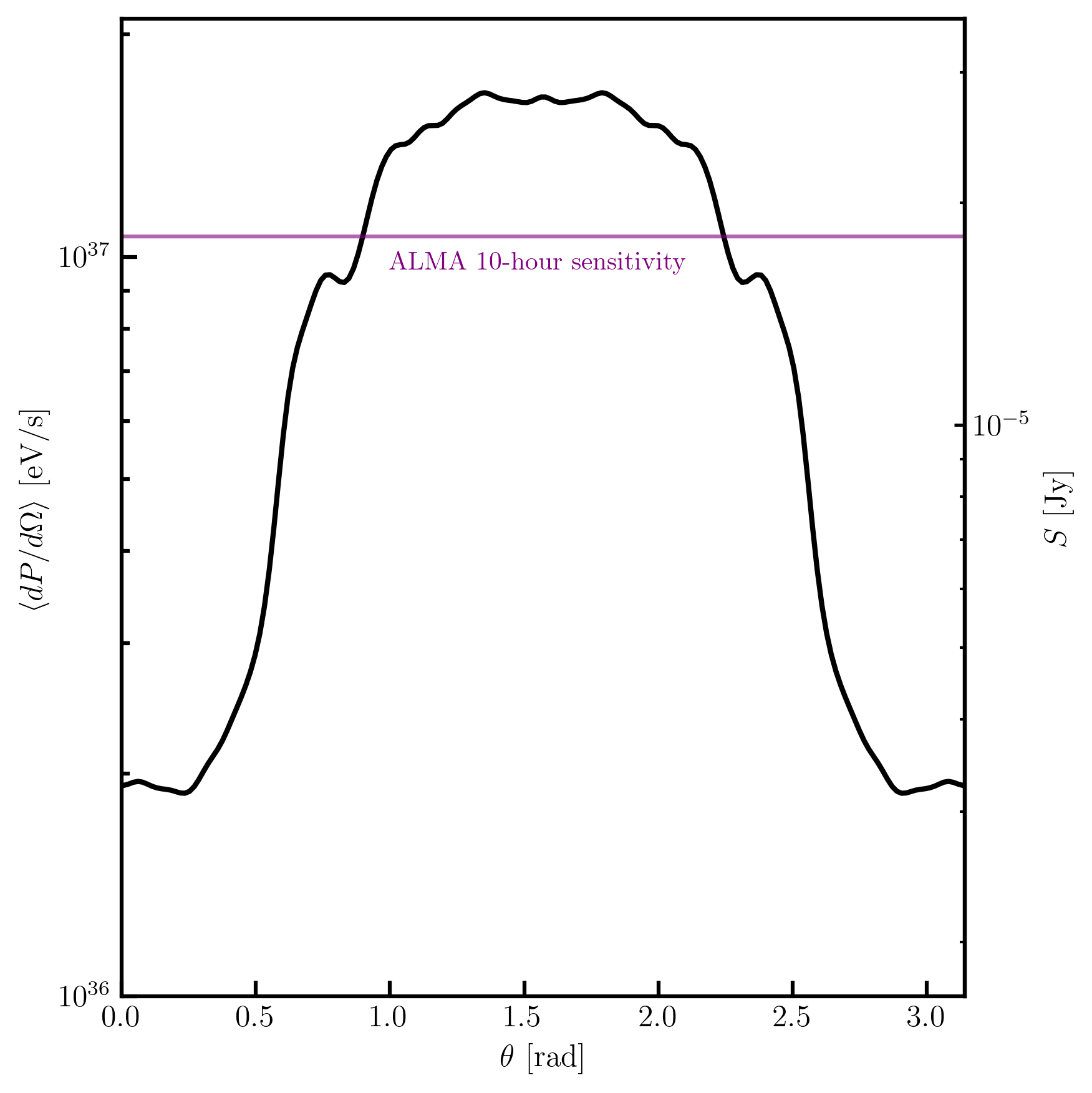}
    \vspace{-1em}
    \caption{\emph{Left Panel}:  {\color{black} Ray-traced resonant conversion surface 
    in the TK20 collisional plasma model.  View is of the quarter arcade geometry, with the
    arcade positioned on the opposite side of the magnetar.  Parameters:
    $B_{\rm pole} = 10B_Q$ (corresponding to the GC magnetar \gcmag),} $m_a = 10^{-3}\eV$, and $g_{a\gamma\gamma}=10^{-11}\, \rm{GeV}^{-1}$.  {\color{black} The color scale represents}
    the resonant conversion photon power on the unit sphere. The gray sphere represents the NS surface and photons originating {\color{black} behind} the NS are masked. The resonant conversion surface follows the shape in Fig.~\ref{fig:plasma_profiles} and the highest power regions are those {\color{black} at the rear of} the NS, converting close to the surface where magnetic fields are strongest. Resonant conversion surfaces for the other 
    {\color{black} arcade geometries} are shown in App.~\ref{sec:raytracingappendix}. \emph{Right Panel}: {\color{black} Azimuthally averaged} differential power, $\langle dP/d\Omega\rangle$, as a function of {\color{black} polar}
    angle $\theta$ for the same model and parameters as the left panel.  Maximum differential power is recorded near the midplane (i.e. $\theta\sim \pi/2$) where axion-photon conversion occurs close the NS surface. We also display the minimum sensitivity for the Atacama Large Millimeter/submillimeter Array (ALMA) in purple, assuming a ten-hour observation, and the expected signal flux density, $S$, in Jansky (Jy) using a calculated signal bandwidth of $\sim 10^{-4}\, m_a$ (see App.~\ref{sec:raytracingappendix}). Overall, ALMA could be sensitive to axion-photon conversion around the GC magnetar \gcmag\, after ten hours of observation for $m_a\sim 10^{-3}\eV$ if PSR J1745--2900's magnetosphere has a similar structure to the quarter model. 
    }
    \vspace{-1em}
    \label{fig:quarter-resonant-surface}
\end{figure*}

\section{Axion-Photon Mixing} \label{sec:mixing}

We now return to the question at hand: namely, how do the expected charge distributions found in the near-field regime of magnetars, which are believed to differ markedly from the baseline GJ model, influence the electromagnetic signal produced from the conversion of in-falling axion dark matter. Below, we outline the procedure taken in this work to compute the characteristic flux and the properties of the spectral line produced from the mixing in these environments. This section relies heavily on the formalism and code developed in~\cite{Witte2021,McDonald:2023shx,Tjemsland:2023vvc}, which has been used to compute observables arising from axion-photon mixing around standard pulsars (see e.g.~\cite{Foster:2022fxn,noordhuis2023novel}).

In order to compute the electromagnetic signal, we begin by fully defining the magnetic field structure and plasma distribution around the NS of interest. Assuming a quasi-stationary solution (valid for NSs), the total electromagnetic power sourced by axions flowing out of the magnetosphere is given by
\begin{equation}\label{eq:powerout}
    \mathcal{P} = \int d^3 {\bf x} \, d^3 {\bf k}_\gamma \, E_\gamma \, \mathcal{C}[{\bf k}_\gamma, {\bf x} ]
\end{equation}
where 
\begin{multline}\label{eq:Collision}
 \mathcal{C}[{\bf k}_\gamma, {\bf x}] = \frac{1}{\partial_{k_0} \mathcal{H}} \int \frac{d^3 {\bf k}_a}{(2\pi)^3 2 E_a} (2\pi)^4 \, \delta(E_\gamma({\bf k}_\gamma, {\bf x}) - E_a({\bf k}_a)) \\ 
 \times \delta^{(3)}({\bf k}_\gamma - {\bf k}_a) \, |\mathcal{M}_{a\rightarrow\gamma}|^2 \, f_a({\bf k}_a, {\bf x}) \, .
\end{multline}
Here, we have defined ${\bf k}_\gamma/{\bf k}_a$ and $E_\gamma/E_a$ as the photon/axion momentum and energy, $\mathcal{M}_{a\rightarrow\gamma}$ as the axion-photon conversion matrix element (where we have implicitly assumed that photons are only sourced at resonance), and $f_a$ as the axion phase space distribution.
{\color{black} The dispersion relations of electromagnetic modes can be obtained from the 
effective Hamiltonian $\mathcal{H}$ of the photon by solving for the roots, i.e., $\mathcal{H}=0$.}
In the strongly magnetized limit, $\mathcal{H}$ is given by~\cite{PhysRevE.64.027401,Witte2021,millar2021axionphotonUPDATED,McDonald:2023shx}
\begin{equation} \label{eqn:Heff}
    \mathcal{H}({\bf x},{\bf k}) = g^{\mu\nu}k_\mu k_\nu + (\omega^2 - k_{\parallel}^2) \left< \frac{\omega_p^2}{\tilde{\omega}^2 \gamma^3} \right> \, .
\end{equation}
Here, $k_{\parallel} = k \cdot B / \sqrt{B \cdot B}$; {\color{black} the photon frequency $\omega = u^\mu k_\mu$
 is measured in} the local inertial frame, which we take to be the pulsar rest frame,
\color{black} corresponding to}  $u = (\sqrt{-g_{tt}^{-1}}, 0, 0, 0)$.  {\color{black} Also, the factor}
\begin{align} \label{eqn:wpeff}
    \left\langle\frac{\omega_p^2}{\tilde{\omega}^2 \gamma^3} \right\rangle &= \displaystyle\sum_s \displaystyle\int_{-\infty}^\infty \frac{\omega_{p, s}^2}{ (\omega - k_{\parallel} v_{\parallel})^2 \gamma_s^3} f_{s, \parallel} (p_\parallel) \ dp_\parallel,
\end{align}
where $f_{s,\parallel}$, $p_{\parallel}$, $v_{\parallel}$, and $\gamma_s$ are the distribution function, the parallel momentum, velocity, and boost factor of {\color{black} charged particle} species $s$, and $\tilde{\omega} \equiv \omega - k_{\parallel} v_{\parallel}$. Note that in the infinite magnetic field limit, particles are forced to flow along magnetic field lines, and thus $p_\perp = 0$, or equivalently $p_{\parallel} = p$. 

{\color{black} When considering}
the BT07 model, we use the distribution function defined in and below Eq.~\eqref{eqn:distfunc} and evaluate~\eqref{eqn:wpeff} analytically. In the TK20 model, the plasma is trans-relativistic ($\gamma \sim 1.25$), so we neglect the factor of $\left\langle \gamma^{-3} \right\rangle$. {We neglect strong-field Euler-Heisenberg corrections to the photon Hamiltonian in Eq.~\eqref{eqn:Heff}. 

In practice, axions {\color{black} accreted around the magnetar} are only semi-relativistic, and thus we approximate $\tilde{\omega} \approx \omega$.  This simplification allows one to re-define an `effective plasma frequency'
\begin{equation}
    \omega_{p,{\rm eff}}^2 \equiv  \displaystyle\sum_s \omega_{p, s}^2 \displaystyle\int_{-\infty}^\infty \frac{f_{s, \parallel} (p_\parallel)}{ \gamma_s^3}  \ dp_\parallel \, ;    
\end{equation}
{\color{black} the computations of the axion-photon conversion surface and conversion probability are
readily generalized from the case of a static, non-relativistic plasma (discussed in e.g. ~\cite{Witte2021,McDonald:2023shx}) to the case of relativistic streaming by substituting $\omega_p \rightarrow \omega_{p, {\rm eff}}$.}

Resonantly enhanced mixing occurs when $k_\mu^\gamma = k_\mu^a$. For a fixed axion momentum ${\bf k}$, one can define a 2+1 dimensional resonant conversion surface $\Sigma_{{\bf k}}$, defined by 
\begin{equation}
    \Sigma_{{\bf k}} = {{\bf x} : E_\gamma({\bf k},{\bf x}) - E_a({\bf k}) = 0} \, ,
\end{equation}
where $E_a$ is the axion energy and $ E_\gamma({\bf k},{\bf x}) = \frac{1}{2} \left( k^2 +  \omega_{p,{\rm eff}}^2 + \sqrt{ \omega_{p,{\rm eff}}^4 + k^4 + 2 k^2  \omega_{p,{\rm eff}}^2 (1 - 2 \cos^2 \theta_B) }\right) $ is the associated energy of the sourced electromagnetic mode.
{\color{black} Here,} $\theta_B$ is the angle between the wave vector and the magnetic field, and the functional dependence of ${\bf x}$ is understood to enter implicitly in  both $ \omega_{p,{\rm eff}}$ and $\theta_B$. This definition allows one to write Eq.~\eqref{eq:powerout} as
\begin{equation}\label{eq:power2d}
    \mathcal{P} = \int d^3{\bf k} \int d\Sigma_{\bf k} \, \cdot \vec{v}_a \, P_{a\gamma} \, E_a \, f_a \, ,
\end{equation}
where we have defined the conversion probability $P_{a\gamma}$ as
\begin{equation}
    P_{a\gamma} \equiv \frac{\pi |\mathcal{M}_{a\rightarrow\gamma}|^2}{E_\gamma \partial_{k_0}\mathcal{H} |\vec{v}_a \cdot \nabla_x E_\gamma|} \, .
\end{equation}

The conversion probability {\color{black} in strongly magnetized plasma}  has recently been computed using kinetic theory~\cite{McDonald:2023ohd} and wave optics~\cite{McDonald:2024uuh}, with the results yielding excellent agreement with independently run numerical simulations~\cite{Gines:2024ekm}.  The result in the non-adiabatic limit ($P_{a\gamma} \ll 1$) is given by
\begin{equation} \label{eqn:conversion_probability_full} 
    P_{a\gamma} = \frac{\pi}{2}\frac{\omega^4 g_{a\gamma}^2 B_0^2 \sin^2\theta_B}{\omega_{p,{\rm eff}}^2 (\omega_{p,{\rm eff}}^2 - 2 E_\gamma^2) \cos^2\theta_B + \omega^4} \frac{1}{|\vec{v}_p \cdot \nabla E_\gamma|} \, ,
\end{equation}
where $B_0$ is the local magnetic field strength,  and $\vec{v}_p \equiv \vec{k} / \omega$ is the phase velocity. All quantities in this expression should be understood to be evaluated on resonance. Eq.~\eqref{eq:power2d} can then be computed via a MC integration procedure outlined in~\cite{Witte2021,McDonald:2023shx};  for completeness, we provide a brief discussion of this procedure below.

Assuming the asymptotic axion phase space distribution to be uniform and isotropic \footnote{Inhomogeneous and anisotropic distributions can also be dealt with, however tend to require modifications to the MC sampling procedure -- see e.g.~\cite{Witte:2022cjj} for an example where this is applied to NS-minicluster encounters.}, i.e.
\begin{equation}
    {\rm lim}_{|\vec{x}| \rightarrow \infty} f_a(\vec{x},\vec{k}) = \frac{n_{a, \infty}}{(\pi k_0^2)^{3/2}} e^{- |\vec{k}|^2 / k_0^2} \, ,
\end{equation}
where $n_{a, \infty}$ is the axion number density far from the NS, {\color{black}the conversion probability can be computed by:} $(i)$ drawing a sample from the asymptotic axion speed distribution, $(ii)$ drawing a random direction on a two-sphere which characterizes the direction of the local axion velocity (note that fixing the local direction and asymptotic velocity fixes $\vec{k}$ at any point $\vec{x}$, implying this sampling procedure pre-selects a two-dimensional surface of the foliation defined by $\Sigma_{\bf k}$), $(iii)$ sampling a point on the two-dimensional surface $\Sigma_{\bf k}$, $(iv)$ expressing the local axion phase space as
\begin{equation}
    f_a(x,k) = \frac{n_{a, \infty}}{(\pi k_0^2)^{3/2}} \, {\rm exp} \left[\frac{-1}{k_0^2} \left(k^2 - \frac{2 G M m_a^2}{|\vec{x}|} \right) \right] \, ,
\end{equation}
and, finally, $(v)$ averaging the weighted samples. 

In general, one is often interested not in the total radiated power but rather in the flux as observable on Earth. The latter can be obtained by using techniques from geometric ray tracing, where the evolution of the photon world lines can be obtained by solving the coupled differential equations:
\begin{equation}
\frac{dx^\mu}{d\lambda} = \frac{\partial \mathcal{H}}{\partial k_\mu} \hspace{.5cm} \frac{dk_\mu}{d\lambda} = -\frac{\partial \mathcal{H}}{\partial x^\mu} \, .
\end{equation}
The power radiated in a direction $(\theta, \phi)$ can then be obtained by averaging the weighted samples that asymptotically enter a small bin $(\theta \pm \delta \theta, \phi \pm \delta \phi)$. The ray tracing procedure naturally allows one to account for the non-linear propagation of photons in the highly magnetized and relativistic background, include dispersive effects arising from the time evolution of the background plasma (which serve to modify the line profile), and incorporate the possibility of resonant cyclotron absorption, which can in some cases lead to a strong suppression of the radiated power (see e.g.~\cite{Witte2021} for further discussion). 

It is worth highlighting that the above procedure is valid as long as photon production remains relatively inefficient. For strong magnetic fields and couplings near the current upper limits, one can find scenarios in which this condition is no longer met. Should this occur, the mixing can modify the initial axion phase space in a highly non-trivial way, and a more detailed MC integration approach is required~\cite{Tjemsland:2023vvc}.  

\subsection{Comment on Small-Scale Structures} \label{eqn:small_scales}

The resonant enhancement in axion-photon mixing seen in Eq.~\eqref{eqn:conversion_probability_full} is valid when the background magnetic field and plasma frequency vary slowly compared to the wavelength of the axion/photon field. This assumption is well justified in the majority of the magnetosphere, where the magnetic field and current density vary over length scales that are on the order of the NS radius. Regions containing magnetospheric twist, however, may develop large spatial gradients in the current density, which would significantly reduce the axion-to-photon conversion probability. An example of this can be seen in the open zones of radio pulsars, where the current density can vary on scales much smaller than the NS radius due to either inhomogeneous pair creation in the polar cap region~\cite{RudermanSutherland1975, Philippov2020, Cruz:2020vfm, Cruz2021, Bransgrove2023, Benacek:2024lat, Chernoglazov:2024rvo} or due to the excitation of high-wavenumber tearing modes during magnetic reconnection~\cite{Thompson:2021oey, Thompson:2021jin}.

The excitation of small-scale currents may also arise within magnetar twist zones. In general, magnetic twist is expected to exhibit a spectrum, $B_\perp(k) \propto k^{-\alpha}$, where $B_\perp$ represents the non-potential component of the magnetic field and $\alpha$ is the spectral index. For $\alpha < 1$, as in the case of Alfv\'enic turbulence, the current spectrum $\tilde{\jmath}(k) \sim k B_\perp(k)$ increases with wavenumber, indicating a buildup of structure at small scales (see Appendix \ref{sec:conversion_turbulent}). In the collisional model, small-scale current overdensities can further enhance collisionality, which may modify the results of Sec.~\ref{sec:collisionalModel}. Another possible source of small-scale magnetic structure is streaming instabilities. For example, in the relativistic double layer model (Sec.~\ref{sec:jbundle}), plasma in the pair production zone is susceptible to such instabilities, which can lead to the formation of fluctuations at the plasma scale~\cite{BeloborodovThompson2007, Beloborodov2013a, Beloborodov2013b}.

The presence of small-scale fluctuations may invalidate the assumption of a slowly varying background and render the WKB approximation inapplicable in the derivation of Eq.~\eqref{eqn:conversion_probability_full}. As a result, prior studies of axion-photon mixing in NS magnetospheres have typically excluded the open field line region and focused solely on conversion within the closed zone (see, e.g.,~\cite{McDonald:2023shx}; although, it is worth highlighting that the open field lines contribute volumetrically at a very small level in the near-field regime, and thus cutting these regions induces negligible corrections). We adopt a similar approach in our analysis of axion-photon mixing in magnetar magnetospheres and do not consider axion masses for which axion-photon conversion would occur near the twisted arcades in the collisional model, as discussed in the following section.


\begin{figure*}
    \centering
    \includegraphics[width=0.49\linewidth]{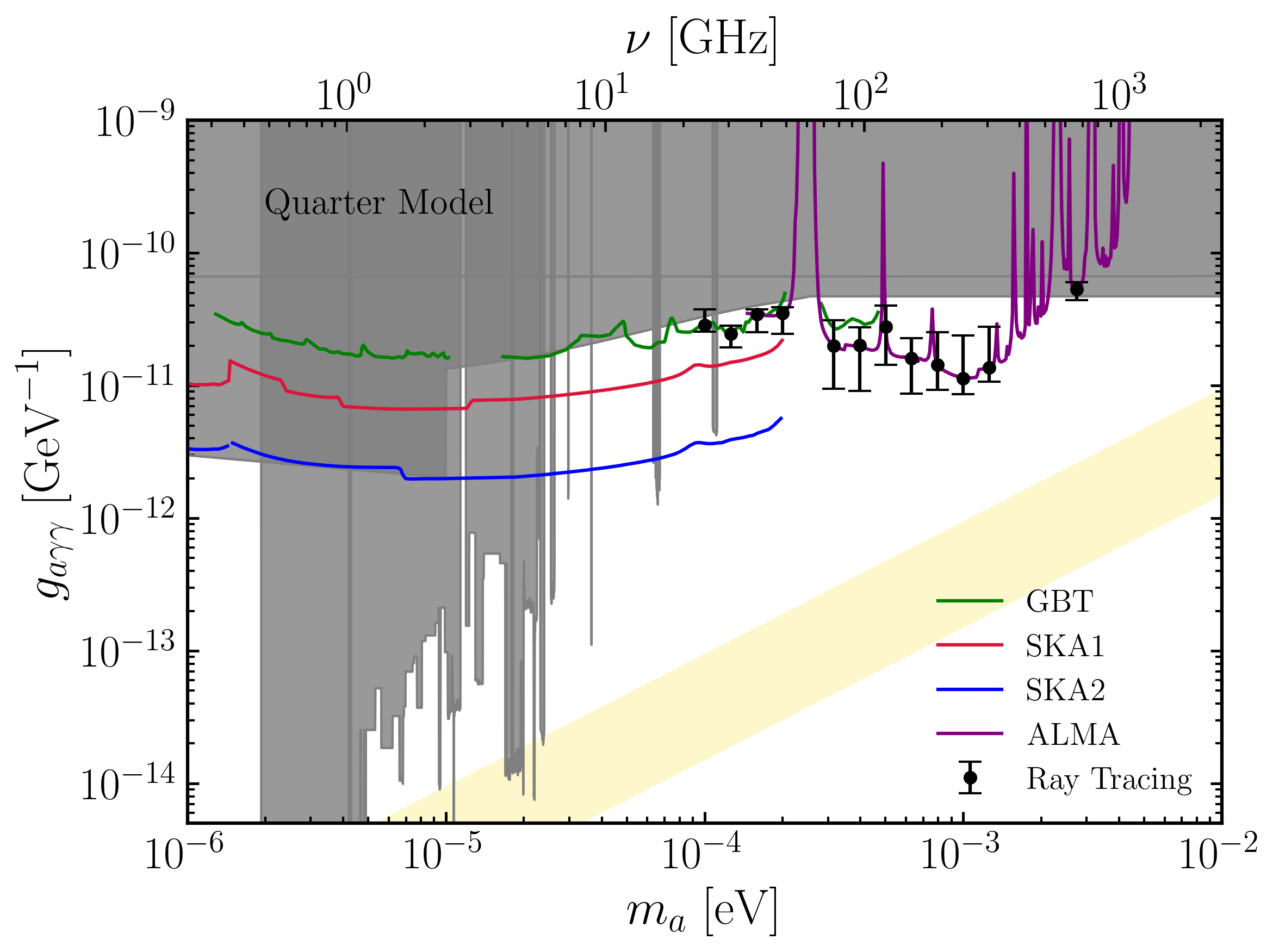}
    \includegraphics[width=0.49\linewidth]{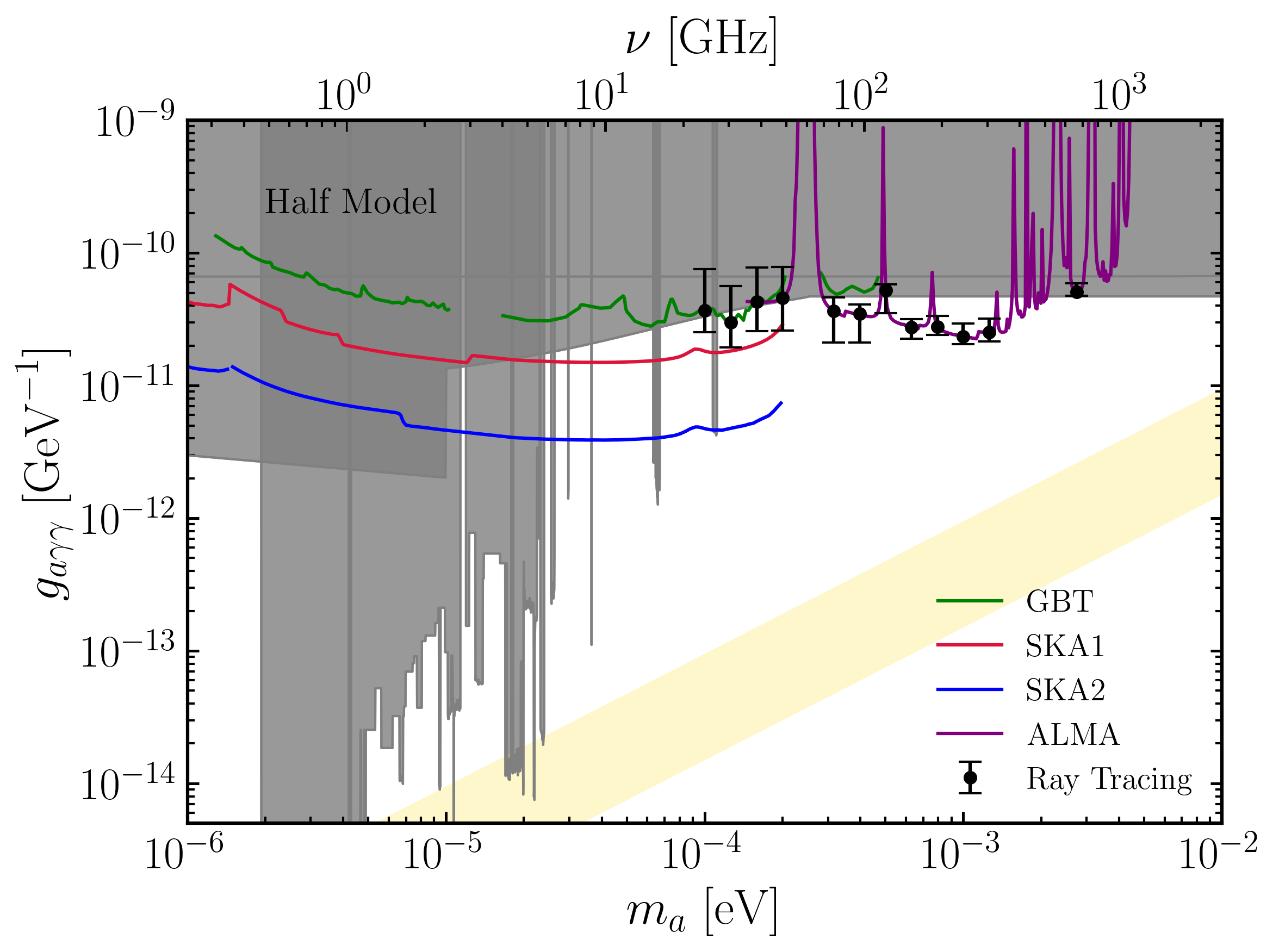}
    \includegraphics[width=0.49\linewidth]{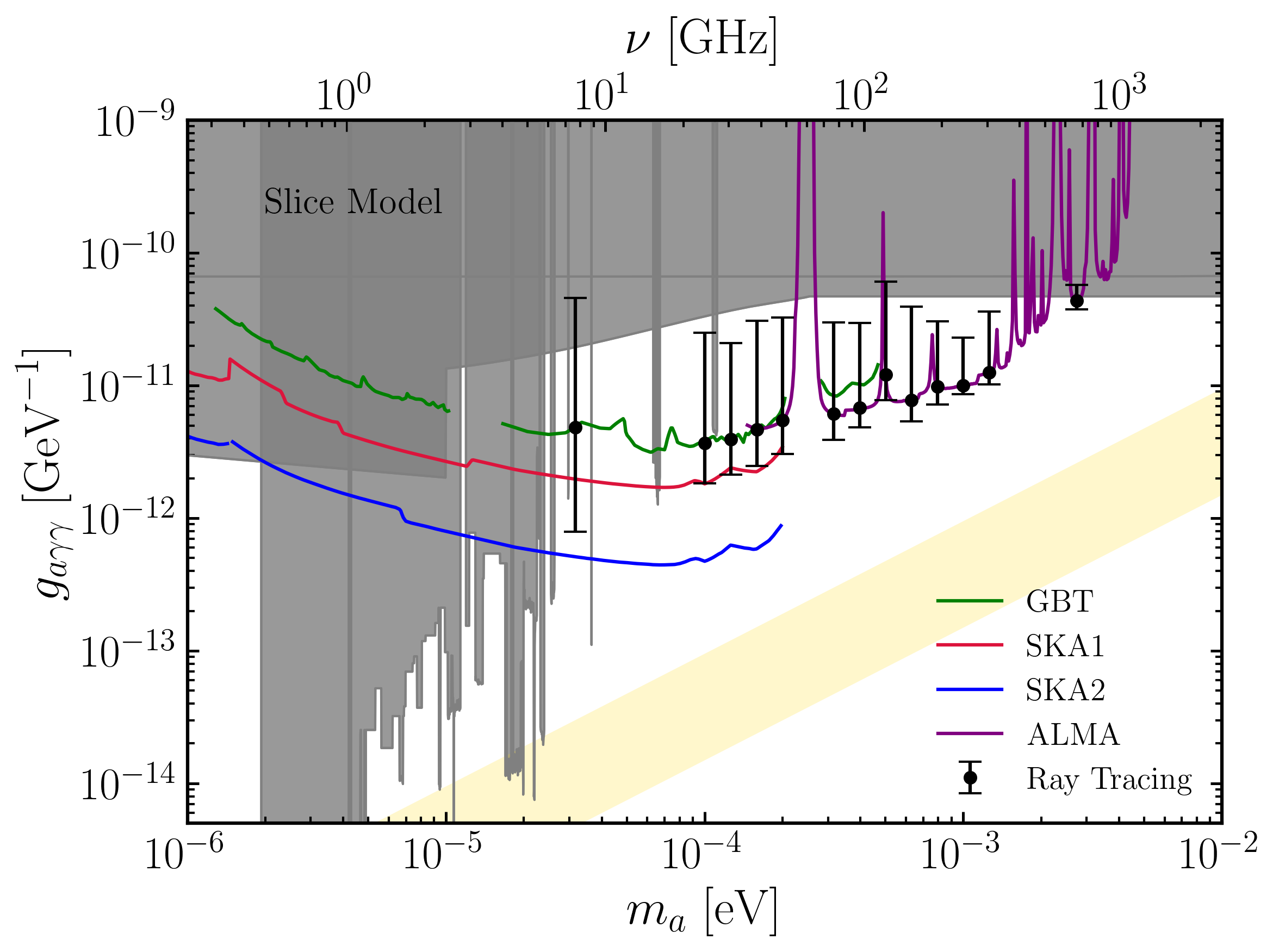}
    \includegraphics[width=0.49\linewidth]{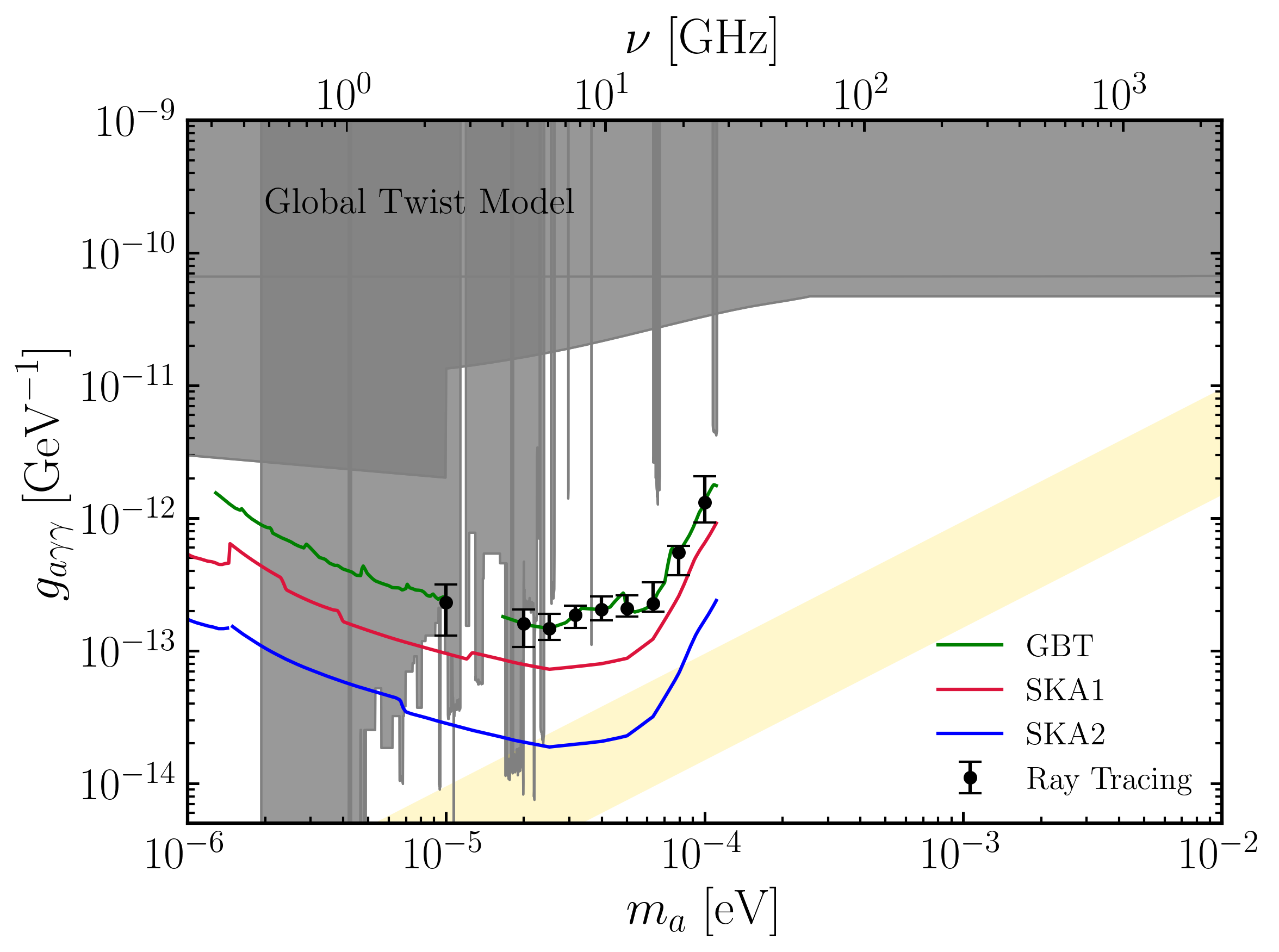}

    \vspace{-1em}
    \caption{\emph{Upper Left Panel}: {\color{black} Sensitivity to radio and microwave lines produced by the resonant conversion of axions (of mass $m_a$ and coupling $g_{a\gamma\gamma}$) in the magnetosphere of the GC magnetar PSR J1745$-$2900.   We assume $B_{\rm pole} = 10\,B_Q$ and consider measurements by four facilities:} the Green Bank Telescope (GBT), the Square Kilometre Array in Phases One and Two (SKA1 and SKA2), and the Atacama Large Millimeter/submillimeter Array (ALMA). {\color{black} Datapoints show the result of extensive
    MC ray-tracing simulations;  error bars represent the 1-$\sigma$ containment region of the
    distribution of photon flux, averaged over viewing angle.  The best sensitivity of ALMA and GBT is
    adopted for each value of $m_a$.}  
    To fill in the rest of $m_a$ space, we log-linearly extrapolate the differential fluxes from the simulations and then calculate sensitivities for each telescope. All sensitivities assume an integration time of ten hours and a bandwidth of $\sim10^{-4}\, m_a$.
    {\color{black} See the Appendices for further details of the calibration by ray-tracing simulations}.
    The yellow band roughly highlights the parameter space of the QCD axion, while the gray regions have been excluded by existing axion searches~\cite{Sikivie1983,DePanfilis:1987,Hagmann:1990,Hagmann:1998cb,Asztalos:2001tf,Asztalos:2009yp,Du:2018uak,Braine2020,Bradley2003,Bradley2004,Shokair2014,HAYSTAC,Zhong2018,Backes_2021,mcallister2017organ,QUAX:2020adt,Choi_2021,Alvarez_Melcon_2021,Anastassopoulos2017,noordhuis2023novel,Ruz:2024gkl,Dolan:2022kul}. \emph{Upper Right Panel}: Same as upper left panel but using the half arcade geometry. \emph{Lower Left Panel}: Same as upper row plots but using the slice geometry. \emph{Lower Right Panel}: Same as the other panels but assuming the global {\color{black} j-bundle} model with the same parameters as in Fig.~\ref{fig:plasma_profiles_global}. In all cases, we find that current and future-generation ground-based telescopes are uniquely sensitive to axions with feeble couplings to Standard Model photons and will be able probe currently unconstrained parameter space. However, the specific sensitivities and mass ranges vary depending on the magnetosphere model assumed. }
    \label{fig:sensitivities}
    \vspace{-1em}
\end{figure*}

\vspace{0.1in}

\section{Sensitivity Projections: \gcmag}\label{sec:sense}

\begin{figure*}
    \centering
    \includegraphics[width=0.99\linewidth]{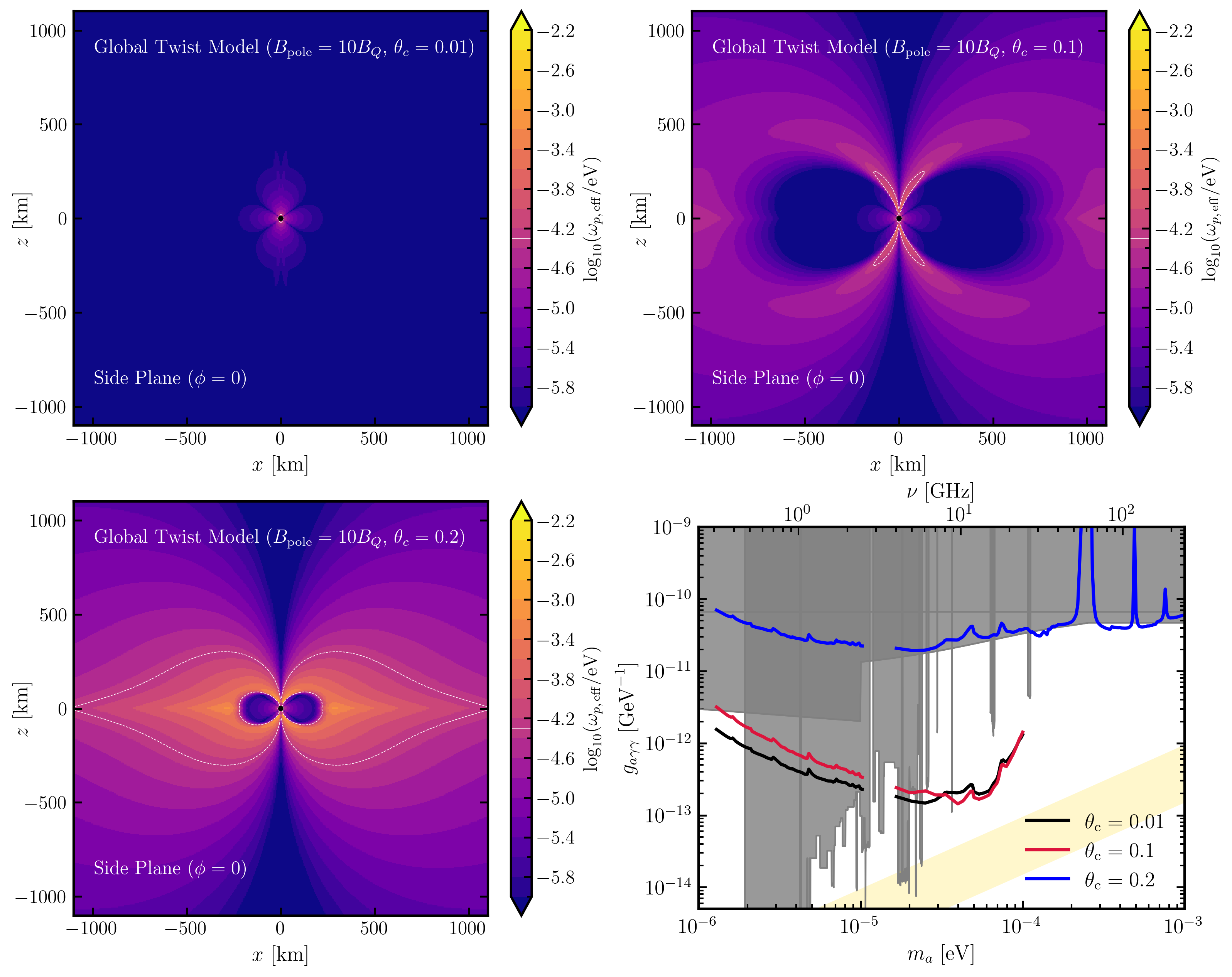}    

    \vspace{-1em}
    \caption{\emph{Top Left Panel}: {\color{black} Profile of $\omega_{p,\rm eff}$ in the global
    twist model, viewed from the side, for $B_{\rm pole} = 10B_Q$ and j-bundle surface opening 
    angle $\theta_c = 0.01$
    (as derived for the quiescent state of \gcmag\, in Sec.~\ref{sec:gc_mag}).  This profile bears some
    similarity to the baseline GJ distribution in a rotating magnetosphere.  Dashed white line:
    contour where $\omega_{p,\rm eff} = 5\times 10^{-5}$ eV, here located close to the NS surface.}
    \emph{Upper Right Panel}: Same as upper left panel but now {\color{black} for $\theta_c = 0.1$.  The plasma density in the polar region now greatly
    exceeds GJ and the contour $\omega_{p,\rm eff} = 5\times 10^{-5}$ extends much further from the magnetar surface.} \emph{Lower Left Panel}: Same as the upper panels but now {\color{black} for} $\theta_c=0.2$ (close to the expected maximum value for \gcmag\, given the fits in Sec.~\ref{sec:gc_mag}). 
    {\color{black} The low-current cavity is now pushed much close to the NS surface and the contour
    $\omega_{p,\rm eff} = 5\times 10^{-5}$ bifurcates into two surfaces.}
    Resonant axion-photon conversion in the inner cavity is effectively``shielded" by the outer j-bundle, {except} in narrow, pole regions. \emph{Lower Right Panel}:
    Interpolated ray-traced median sensitivities for the global twist model {\color{black} and various}
    opening angles $\theta_c$, {\color{black} otherwise adopting} the same magnetar parameters as Fig.~\ref{fig:sensitivities}. {\color{black} Sensitivity is weakened at
    the highest $\theta_c = 0.2$ because the j-bundle suppresses the escape of photons emitted close to 
    the magnetar surface; the escaping photons tend to be emitted at greater radii, where the conversion
    probability is suppressed.}
    }
    \label{fig:global_sensitivities_varying_thetacrit}
    \vspace{-2em}
\end{figure*}

\begin{figure}
    \centering
    \includegraphics[width=1.0\linewidth]{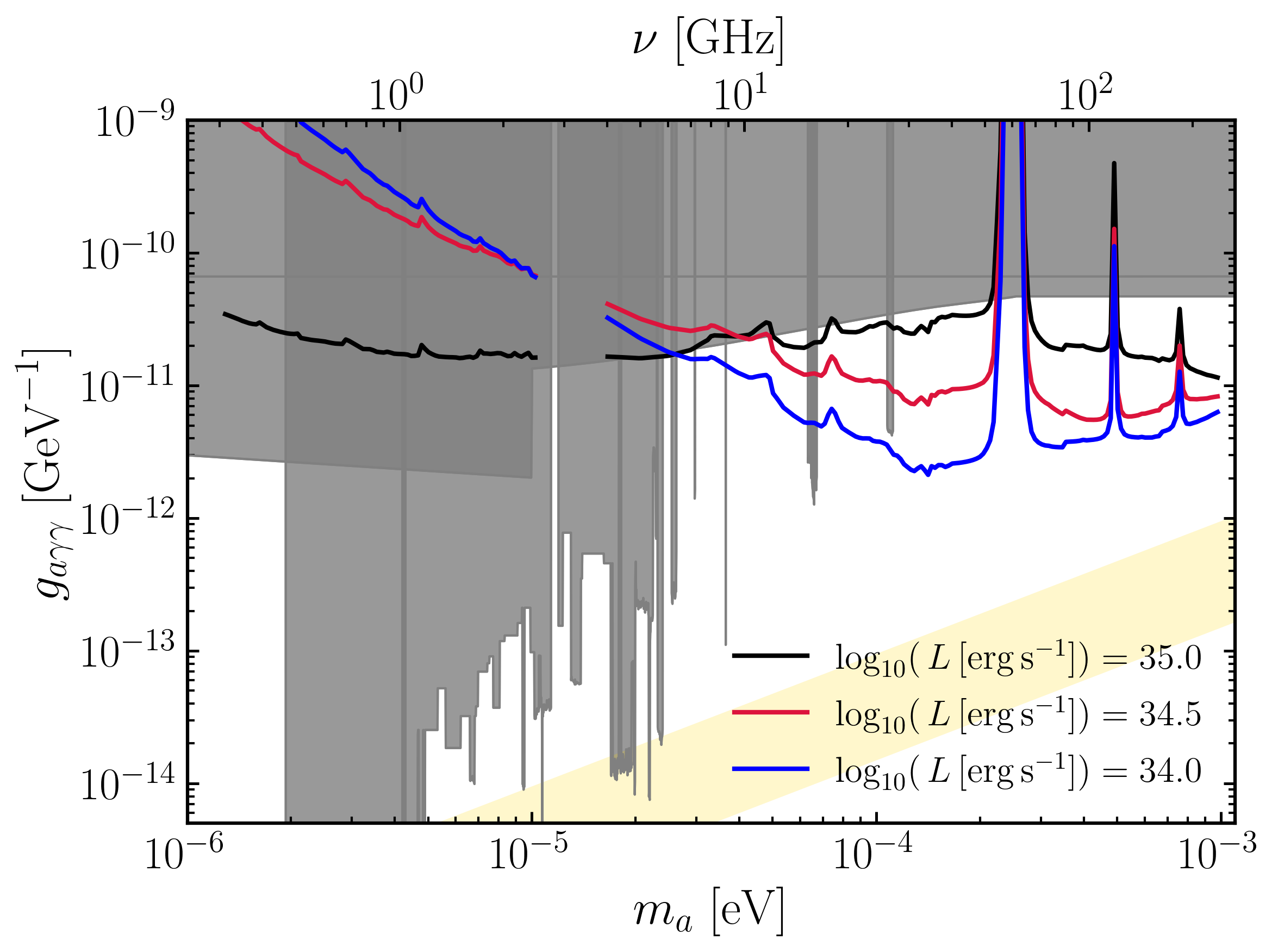}

    \vspace{-1em}
    \caption{Interpolated ray-traced sensitivities {\color{black} in the collisional TK20 model 
    for various values of the arcade gamma-ray luminosity $L$.  Quarter arcade geometry is adopted and
    other magnetar parameters
    and telescope sensistivies are the same as in Fig.~\ref{fig:sensitivities}.  Lowering $L$ results
    in lower $\omega_p$ due to a suppressed rate of photon collisions.} {\color{black} For} 
    axion masses in the range $10^{-4}\eV\lesssim m_a \lesssim 10^{-3}\eV$, lower {\color{black} $L$ pushes
    the conversion closer to the magnetar, where $B$ is stronger and the conversion probability is
    enhanced.}
    The $\omega_p$ profiles have different slopes at low frequency, 
    resulting in differing sensitivities at low $m_a$. }
    \label{fig:quarter_sensitivities_varying_luminosity}
    \vspace{-1em}
\end{figure}

In this section, we revisit sensitivity projections using the j-bundle (BT07) model and the collisional (TK20) model. We focus on using the GC magnetar \gcmag\, as a target, as its close proximity to the GC guarantees a large flux of ambient axion dark matter\footnote{Although we previously discussed AXP 1E 2259+586 in the context of the TK20 model, we do not include it in our analysis here due to its unfavorable location within the dark matter halo. Since \gcmag\, has not been definitively associated with either the BT07 or TK20 model, we consider both frameworks when modeling \gcmag.}. 

Many studies attempting to probe axion-photon conversion in magnetar magnetospheres have assumed the plasma state is described by the GJ model~\cite{Leroy:2019ghm, Witte2021, Hook:2018iia, Huang:2018lxq}. In the GJ model, magnetospheric twist arises from the breakdown of co-rotation of the magnetosphere at the light cylinder, located at a distance $R_{\rm LC} = c/\Omega$ from the center of the NS. The twist is localized to the polar cap region, which makes up a very small fraction of the NS surface and has a negligible impact on axion-induced photon production. Though applicable to lower-field pulsars, the GJ model is expected to be a poor descriptor of the plasma state in magnetar magnetospheres. Ref.~\cite{McDonald:2023shx} presents a more accurate treatment, in which the magnetar magnetosphere is described by a globally twisted dipole with number density roughly compatible with Eq.~\eqref{eqn:number_density_shear_zone}, and including a global multiplicity factor $\lambda$ which serves to account for the fact that the plasma is not charge separated (the adopted values are taken to be $\lambda = 1$ and $\lambda = 20$, the latter being chosen to reproduce the characteristic plasma frequencies identified in~\cite{ThompsonKostenko2020}) -- note that the assumption of uniformity is adopted for simplicity, with more complicated twist geometries likely falling between these more limiting cases.  While this approach is useful to get a rough handle on how enhanced charge densities in magnetar magnetospheres alter the expected signal, it is neither fully self-consistent, nor does it capture a number of interesting features that naturally arise in the BT07 and TK20 models discussed here (such as non-local pair production, streaming plasma, broad distribution functions, and spatial inhomogeneities). Below, we outline the modeling assumptions adopted in each framework and use them to estimate the corresponding sensitivity to axions.

\subsection{Magnetosphere Model Parameters}

In the BT07 model, three parameters (apart from the dipole magnetic field strength, the rotation period, and the misalignment angle)  are required to describe the magnetospheric plasma in the inner magnetosphere (see Sec.~\ref{sec:jbundle_global}): the surface temperature, $T_{\rm NS}$, the polar angle corresponding to the boundary of the j-bundle, $\theta_c$, and the twist within the bundle, $\psi$ \footnote{In general, the twist may be non-uniform within the bundle, but we assume it to be constant for simplicity.}. The first X-ray observations of PSR J1745-2900 using \emph{Swift} and \emph{NuSTAR} show a spectrum associated with a blackbody with temperature $T_{\rm NS} = 0.956^{+0.015}_{-0.017}$ keV~\cite{Mori_2013}. In Sec.~\ref{sec:gc_mag}, we discussed how X-ray observations can be used to infer {\color{black} the properties of the j-bundle in
a more recent quiescent state: radius} $R \approx 0.1$ km, corresponding to an opening angle $\theta_c \approx 0.01$ rad, and a twist angle $\psi \approx 1.6$ rad. 
 We compute the axion sensitivity assuming these values, but also explore the impact of varying the opening angle by considering additional values $\theta_c = 0.1$ and $0.2$ rad.

While the BT07 model describes large-scale twist located near the polar region, the TK20 model is governed by smaller-scale current structures that need not extend far from the surface. 
We estimate the sensitivity using the three {\color{black} trial arcade geometries shown in 
Fig.~\ref{fig:collisional_model_geometries}.} Another key parameter in determining the magnetospheric plasma distribution is the total annihilation bremsstrahlung luminosity, $L_0$, defined in and below Eq.~\eqref{eqn:annihilation_bremsstralung_spectrum}. Determining $L_0$ observationally requires multi-wavelength X-ray and gamma-ray observations up to $ \sim$ MeV photon energies. In the absence of such data, we adopt a fiducial value for the luminosity $L_0 = 10^{35}~{\rm erg~s^{-1}}$ but also calculate axion-photon coupling sensitivities for two other fiducial luminosities: $L_0 = 10^{34}$,
$10^{34.5}~{\rm erg~s^{-1}}$. We report the axion-photon coupling sensitivities for the TK20 models in Fig.~\ref{fig:sensitivities} assuming a fiducial luminosity of $10^{35}\ergpersecond$; however, for the sake of completeness, we also report sensitivities using alternative luminosities in Fig.~\ref{fig:quarter_sensitivities_varying_luminosity}.

{\color{black} The plasma density in the j-bundle is defined by Eqs.
(\ref{eqn:npmjbundle})--(\ref{eqn:momentum_pm}) along with the model parameters
described above.  Stellar rotation cannot be neglected when determining
the particle density outside the j-bundle, since the current density is
assumed to vanish in the equatorial cavity at $\theta > \theta_c$.
A similar effect occurs in the collisional TK20 model, where the pair
density is determined by the procedure described in Section \ref{sec:TK20_magnetosphere}.  The 0.5--1 MeV gamma rays emitted by the 
arcade do not reach a part of the magnetosphere antipodal to the arcade,
where $\gamma-\gamma$ pair creation is suppressed.  We therefore} add a minimal number density, $n_{\rm GJ} = \rho_{\rm GJ}/e$ throughout the magnetosphere, which gives an equilibrium plasma frequency of
\begin{align}\label{eqn:omegaptot}
    \omega_{p} = \sqrt{\displaystyle{\omega_{p, \rm GJ}^2 + \omega_{p,\rm eff}^2}},
\end{align}
where $\omega_{p, \rm GJ}^2 = e^2 n_{\rm GJ}/m_e$  and $\omega_{p, \rm eff}$ is as given in Eq.~\eqref{eqn:wpeff}.

\subsection{Signal Computation}

{\color{black}Combining the magnetospheric plasma distribution (\ref{eqn:omegaptot})
with} the ray-tracing procedure outlined in Sec.~\ref{sec:mixing}, we derive the expected flux density for different viewing angles using the following formula:
\begin{align} \label{eqn:S_sig}
    S = {1 \over D^2 \Delta \nu } {d P \over d\Omega},
\end{align}
where $D$ is the distance to the NS, and the differential power should be evaluated along the line-of-sight from the pulsar to the Earth. We use a galactocentric distance $D = 8.178\kpc$~\cite{GravityCollab2019}, with a offset from the GC black hole Sagittarius A* of $\approx 0.097\pc$~\cite{Rea2013,Rea2020}\footnote{Technically, this is the inferred two-dimensional projected distance. Should the line of sight distance be significantly larger, our approach may lead to an overestimation of the dark matter density; nevertheless, adopting this value serves as useful benchmark in understanding future capabilities.}. {\color{black} The orientation of the magnetar
spin and magnetic moment are not known, and so} 
we average the differential power over the azimuthal 
direction, weighted by $\sin\theta$ where $\theta$ is the 
magnetic polar angle, and compute the 15.7\%, 50\%, and 84.3\% containment regions. At each viewing angle, we also compute the width of the photon spectra, $\Delta \nu$, which we use to calculate the differential flux density (we find that for most axion masses and models we test, $\Delta\nu \sim 10^{-4}\,m_a$). 

For illustration, we show an example of a resonant conversion surface, as well as the phase-resolved differential power produced from axion-photon mixing, in Fig.~\ref{fig:quarter-resonant-surface}. On the left panel, we show the surface of resonant axion-photon conversion assuming a collisional magnetosphere with the quarter arcade geometry as well as the axion parameters $m_a=10^{-3}\eV$ and $g_{a\gamma\gamma}=10^{-11}\GeV^{-1}$. The photon rays on the resonant surface are colored by their asymptotic power on the unit sphere. The highest power regions are {\color{black} on the side of the NS opposite to the
emitting arcade}, where the resonant surface intersects the NS surface. There, the magnetic fields are strongest and axion-photon conversion is greatest. This is also reflected in the right panel,
{\color{black} where the rotation-averaged differential power is seen
to peak near the magnetic equator.}

\subsection{Observation and Backgrounds}

Observations of a given magnetar are most sensitive to axions whose mass coincides with the effective plasma frequency close to the stellar surface. Examining Figs. \ref{fig:plasma_profiles_global} and \ref{fig:plasma_profiles}, the sources considered in this work are most sensitive to axions with mass in the range $10^{-5}$ eV $\lesssim m_a \lesssim 10^{-2}$ eV. This mass range is best probed with sensitive radio and millimeter telescopes. In this section, we compute the sensitivity of current radio and millimeter telescopes, the Green Bank Telescope (GBT)~\cite{Jewell2004,White2022,GBTTelescope2025} and the Atacama
Large Millimeter/submillimeter Array (ALMA)~\cite{Wootten2009,almatechnicalhandbook}, and the projected sensitivities of the upcoming Square Kilometer Array (SKA)~\cite{Dewdney2009,Braun2019}. 

To project our sensitivities, we use the system equivalent flux densities (SEFD) for each telescope array and use the radiometer equation to derive the minimum observable photon line flux density ($S_{\rm min}$):
\begin{equation}
    S_{\rm min} = \frac{\rm{SEFD}}{\sqrt{2 \Delta \nu \,\Delta t_{\rm obs}}}.
    \label{eq:Smintelescope}
\end{equation}
Here, $\Delta \nu$ is the axion signal bandwidth, which we compute using our ray-tracing code for different viewing angles, and $\Delta t_{\rm obs}$ is the total observation time (our sensitivities assume $\Delta t_{\rm obs}=10$ hours). For ALMA, we use the sensitivities calculator tool \texttt{analysisUtils}~\cite{analysisUtilsALMA}. For GBT, we use the SEFDs listed in the Proposer's Guide~\cite{GBTTelescope2025} assuming average weather conditions and the 12-meter telescope array. For SKA, we project the sensitivity to future array's phase-1 and phase-2 observational programs (i.e. SKA1 and SKA2) based on their anticipated performance~\cite{Braun2019}. Individual SEFDs are shown in App.~\ref{sec:telescope_appendices}. 

To derive the minimum observable axion-photon coupling using a given telescope, we compare Eq.~\eqref{eq:Smintelescope} to the signal flux density in Eq.~\eqref{eqn:S_sig}, and use the relation $S \propto \gagg^2$. These {\color{black} quantities} 
then form the axion-photon coupling sensitivities and uncertainties we report in Figs.~\ref{fig:sensitivities}, \ref{fig:global_sensitivities_varying_thetacrit}, and \ref{fig:quarter_sensitivities_varying_luminosity}.

An important consideration when calculating the expected resonant conversion signals from the GC magnetar is extinction by dust in the galactic plane. Three-dimensional dust maps exist~\cite{Green_2019} but we conservatively model the expected extinction assuming all emission originates at infinity as in~\cite{Roy:2023omw}. As a result, the expected dust extinction of the resonant conversion signal $\exp\left(-\int_0^s ds' \,n_d \, \sigma\right)$ becomes $10^{-0.44 A_\lambda}$ where $n_d$ is the dust density, $\sigma$ is the wavelength-dependent dust-photon cross section, and $A_\lambda$ is the galactic dust extinction at wavelength $\lambda$. Using the galactic dust models in ~\cite{1998ApJ...500..525S,doug_2011} and the extinction curve from~\cite{Gao_2013}, we find that our expected resonant conversion flux from the GC is multiplied by an attenuation factor $\lesssim 0.90$ for axion masses $m_a \gtrsim 4\times 10^{-3}$ eV. As a result, we ignore all resonant conversion in our sensitivities for greater axion masses $m_a \gtrsim 4\times 10^{-3}$ eV to neglect the effects of GC dust extinction. 

We note that there are significant uncertainties regarding the precise dark matter densities and velocity dispersions expected in the inner halo. Using Jeans analyses to calculate equilibrium inner velocity dispersions will likely not capture the disequilibrium caused by mergers like the Large Magellanic Cloud and Sagittarius and other sources of systematic uncertainties (see \cite{Laporte2018,Ou2025} for examples of how recent mergers have disrupted the galactic disk). Additionally, N-Body simulations of Milky Way-like galaxies are often not able to finely resolve the inner $\sim 0.1$ pc due to resolution limits (e.g. \cite{Hopkins2018}), although new simulation methods are beginning to resolve the inner accretion disks around central supermassive black holes as well as the larger-scale structural properties of Milky Way-like galaxies~\cite{Hopkins2024a,Hopkins2024b,Hopkins2024c}. To be conservative, we assume a dark matter velocity dispersion of 
$220\kms$ and calculate the dark matter density in the GC using the best-fit generalized NFW profile\footnote{The best fit inner density slope for the generalised NFW profile in \cite{Ou2025} is $\approx -0.9$, which is similar to the traditional NFW inner slope of $-1$.} 
in~\cite{Ou2025} evaluated at a galactocentric distance of 0.097~pc~\cite{Rea2013,Rea2020}. We note that assuming an Einasto profile \cite{Ou2025} or cored Burkert profile \cite{Nesti2013} would decrease our assumed inner densities by a factor of $\sim 10^4-10^5$, resulting in weaker axion-photon coupling sensitivities by a factor of $\sim 10^2 - 10^{2.5}$. On the other hand, assuming a spiked inner dark matter density profile, as hypothesized for regions close to the Milky Way's supermassive black hole Sagittarius A* \cite{Lacroix2018}, would enhance our assumed densities by a factor of $\sim 10^4$, resulting in sensitivities to axion-photon couplings two orders of magnitude lower than our current projections. 

\subsection{Results} \label{sec:results}

The sensitivities of observations of \gcmag\, to axion dark matter are shown in Figs.~\ref{fig:sensitivities}, \ref{fig:global_sensitivities_varying_thetacrit}, and \ref{fig:quarter_sensitivities_varying_luminosity}. The analysis in Fig.~\ref{fig:sensitivities} includes four different magnetar magnetosphere geometries across a broad range of axion masses, including the sensitivities for the telescopes GBT, ALMA, and SKA (phases 1 and 2). We calculate all sensitivities assuming a total observation time of $10$ hours and use the detailed ray-tracing procedure in Sec.~\ref{sec:mixing} to calculate differential fluxes as a function of viewing angles, as in Fig.~\ref{fig:quarter-resonant-surface}. The 1-$\sigma$ containment regions for the individual simulations run are shown as error bars in each panel of Fig.~\ref{fig:sensitivities}, where we calculate the differential power percentiles based on the distribution of rotation-averaged differential power vs. viewing angle. We then log-linearly interpolate these rotation-averaged differential power values as a function of axion mass to generate the full sensitivity curves.

In Fig.~\ref{fig:sensitivities}, the quarter, half, and slice magnetosphere geometries for the collisional TK20 model all display the most competitive sensitivities in the region of $10^{-4}\eV\lesssim m_a\lesssim 10^{-3}\eV$. For these axion masses, ten hours of observation time would provide the leading sensitivities to such dark matter candidates. Sensitivities drop off at greater $m_a$ because of the loss of sensitivity of the telescope (ALMA) and because the resonant conversion surface has a low area despite being close to the NS surface. The slice geometry 
{\color{black} stands out} in that it provides a competitive sensitivity at lower axion masses $m_a \lesssim 3\times 10^{-5}\eV$, although the viewing-angle dependent scatter is more severe. 
{\color{black} In this configuration, it turns out that the
resonant surface extends close to the NS surface at low $\omega_p$, although} only in narrow regions as shown in Fig.~\ref{fig:plasma_profiles}. (See App.~\ref{sec:raytracingappendix} for an example surface.) 

The final model shown in Fig.~\ref{fig:sensitivities} is the BT07 globally twisted j-bundle,
{\color{black} with opening angle $\theta_c = 0.01$, appropriate to \gcmag\, in its present quiescent
state.  Here, the surface plasma frequency, which is
determined by the low corotation particle density, does not rise above} $\omega_p\sim10^{-4}\eV$, too low to be observed by ALMA. The improved sensitivity in the $10^{-5}-10^{-4}$ eV mass range is a direct result of the fact that the conversion surface is closer to the star (implying a larger magnetic field, and thus more efficient axion-photon conversion). We predict for SKA phase 2, that {\color{black} observations
of \gcmag} could be sensitive to QCD axions in the mass range $2\times 10^{-5}\eV\lesssim m_a \lesssim 8\times 10^{-5}\eV$. 


{\color{black} This model assumes that most of the magnetosphere of \gcmag\, has relaxed close to the baseline
GJ particle density in its quiescent state; the results are therefore consistent} with previous analyses of GJ-like magnetars with similar magnetic field strengths \cite{Witte2021,McDonald:2023shx}. 
{\color{black} Note that the j-bundle fit to the X-ray data of \gcmag\, predicts larger $\theta_c$ 
at higher fluxes (up to $\theta_c \sim 0.25$), a configuration that might be repeated during later outbursts.}
We therefore calculate the median sensitivities for the global twist model assuming different opening angle values in Fig.~\ref{fig:global_sensitivities_varying_thetacrit}, specifically $\theta_c = 0.01,\, 0.1,\, 0.2$. 

{\color{black} This increase in j-bundle opening angle can generate}
significantly different plasma profiles, and thus different axion-photon conversion sensitivities. In Fig.~\ref{fig:global_sensitivities_varying_thetacrit}, we plot the plasma profiles for various values of $\theta_c$, as well as the projected median sensitivities from ALMA and GBT (bottom right). For small  $\theta_c$, the plasma closely resembles a GJ plasma. For larger $\theta_c$, one sees an enhancement in the plasma density, and the appearance of local plasma under-densities, which effectively trap low energy photons, blocking out near-field resonant conversion. {\color{black} The net effect is that} increasing $\theta_c$ leads to a decrease in sensitivity, but broadens the sensitivity to much higher $m_a$ than would be possible in the GJ limit. 

{\color{black} Our results highlight an interesting feature of the global j-bundle model with large
opening angle:  the inner resonant conversion surface is surrounded by a high-density plasma ``wall''.}
This could have 
significant implications for the observability of axions. 
{\color{black} Low-frequency photons deposited inside the cavity may be channeled through holes in this
plasma wall, thereby producing bright spots across the sky.}
In order to avoid overstating sensitivity to current radio observations, we adopt the procedure 
of {\color{black} excising any conversion photons that intercept a high-density plasma zone.}


One of the systematic uncertainties entering the analysis of the collisional TK20 model arises from the unknown total luminosity of annihilation bremsstrahlung photons. In our ray-tracing simulations (and in the sensitivity projections shown in Fig.~\ref{fig:sensitivities}), we adopt a fiducial luminosity of $L_0 = 10^{35}\ergpersecond$. However, lower luminosities would result in lower density magnetospheres and thus lower plasma frequencies. In Fig.~\ref{fig:quarter_sensitivities_varying_luminosity}, we calculate the median sensitivities for the quarter arcade structure, using the same procedure as in Fig.~\ref{fig:sensitivities},
but now with $L_0 = 10^{34.5}\ergpersecond$ and $L_0 = 10^{34}\ergpersecond$, while
{\color{black} adopting the same} SEFDs for ALMA and GBT.  {\color{black} We find that the conversion
of axions with masses $m_a \lesssim 10^{-3}$ eV is pushed closer to the NS, where the magnetic field
is stronger.}
As a result, lowering {\color{black} the dissipation in the magnetospheric circuit} increases the sensitivity to heavier axions and reduces the sensitivity to very light axions (which convert much further from the star). This analysis suggests that one may optimistically be able to probe axion-photon couplings down to $\approx 2\times 10^{-12}\GeV^{-1}$ for axion masses $m_a \gtrsim 10^{-4}$ eV.  {\color{black} Lowering $L_0$ also has the effect of restricting
the range of $m_a$ that can convert, this being below} $\sim 10^{-3}\eV$ for $L \lesssim 10^{34.5}\ergpersecond$, {\color{black} as compared with
$\sim 5\times 10^{-3}$ eV for the fiducial luminosity.  Spectral measurements around 1 MeV are needed
to accurately calibrate the gamma ray output.}


\vspace{0.1in}

\section{Conclusion}\label{sec:conclude}

In this paper, we have revisited {\color{black} the constraints on} axion dark matter {\color{black} that may be derived from} observations of magnetars {\color{black} in the radio and microwave bands.  We have
adopted up-to-date models of the plasma state around a magnetar.  These
models have been developed already in some detail but are based on simple mechanisms
of $e^\pm$ multiplication in super-Schwinger magnetic fields.  They are
motivated and constrained by the remarkable X-ray, infrared, radio, and spindown phenomenology of magnetars.  Independent of the underlying mechanism, it is clear
that the current flow around a magnetar must depart dramatically from the predictions of the standard pulsar model.  One expects significant modifications
to the brightness and frequency range of the emission line arising from 
resonant axion-photon conversion.}

{\color{black} The two models considered apply to relatively quiescent
magnetars and involve different plasma states and mechanisms of pair creation.  
The first model (which we term the BT07 model) applies in a situation where
the current density is not too high, the $e^\pm$ flow is collisionless and
relativistic, and pair creation is sustained by scattering of keV photons
at the Landau resonance.  The relativistic pairs source a hard X-ray 
continuum similar to that observed from most quiescent magnetars if the
current is concentrated in a narrow j-bundle near the magnetic dipole axis.
We consider j-bundles of various angular widths.  The resonant conversion of axions is
concentrated outside the j-bundle, where the plasma density may not much
exceed the baseline GJ value.

The second approach (termed the TK20 model) is based on the possibility of
enhanced current density in zones of strong magnetic shear.  In this situation,
the $e^\pm$ gas can relax to a collisional, transrelativistic, but still dilute
state.  Ohmic heating of the circuit is mediated by repeated $e^+-e^-$
annihilation and compensated by the emission of a broad spectrum of annihilation
X-rays.  Here, the location of the currents is less constrained; we consider
various arcade structures tied to narrow crustal slip zones.  In contrast
with the BT07 model, much of the remaining magnetosphere is bathed with
$0.5-1$ MeV photons that collide and form an extended cloud of pairs.
As a result, the plasma density and resonant conversion frequency are
generally higher than in the BT07 model.

It should be emphasized that all magnetars appear to be variable on month-to-year 
timescales, whether or not they emit brief (0.1--1000 s) and ultraluminous X-ray bursts.  
This variability has been ascribed to a combination of continued crustal flow and
resistive evolution in the magnetosphere.
As a result, the plasma state around a given source is not expected to be fixed.
In many respects, magnetar states of lower luminosity and plasma density turn out
to be more favorable for the detection of a microwave axion conversion line.

We summarize our main conclusions here:}

\begin{itemize}
    \item {Radio and sub-mm observations of} magnetars {\emph{may}} be sensitive to axions {\color{black}
    with mass} in the meV range. The highest $m_a$ that can be probed by resonant conversion to photons is set by the plasma frequency near the magnetar surface. In the GJ model, the surface plasma frequency of \gcmag\, is about $\omega_{p}(\rns) \approx 0.1$ meV.  {\color{black} This rises to about 30 meV
    in the TK20 model when the arcade gamma-ray luminosity is $L_0 \sim 10^{35}$ erg s$^{-1}$.}
    We show that dedicated ALMA observations of \gcmag\, could be sensitive to meV-scale axions with couplings as low as $\sim 2 \times 10^{-12}$ GeV$^{-1}$ if the TK20 model is correct {\color{black} and
    $L_0$ is as low as $\sim 10^{34}$ erg s$^{-1}$; the constraint weakens with increasing
    luminosity.  Measurements of the gamma ray spectrum around 1 MeV are key to testing
    the plasma state.}  
    
    
    \item Since the the magnetospheric plasma state evolves over long 
    timescales, the sensitivity to axions varies accordingly with observing epoch. Following an outburst, the magnetosphere becomes highly twisted, resulting in enhanced plasma densities. This environment enhances sensitivity to higher-mass axions, although it also increases radiative background levels. As the magnetosphere untwists
    {\color{black} and the rate of plastic flow diminishes}, the plasma density drops, {\color{black} shifting the sensitivity to lower 
    $m_a$.}


    \item {\color{black} In a situation where the BT07 model applies -- e.g. the magnetic twist
    is too weak to support collisional $e^\pm$ plasma --} future observations of \gcmag\, with the SKA may yield sensitivity to the QCD axion.  This conclusion is consistent with previous studies~\cite{Hook:2018iia,Battye:2019aco,Witte:2021arp} that employed the GJ model for the magnetar magnetosphere. 
    {\color{black} If instead this magnetar forms stronger currents supported by
    transrelativistic $e^\pm$,} ALMA observations yield sensitivity to axion-like particles in the meV-range, but not to the QCD axion. This conclusion is consistent with the approximations made in~\cite{McDonald:2023shx}.
    
\end{itemize}

Another possibility we do not consider here is a `hybrid' model. While we have treated BT07 and TK20 as distinct models, a realistic magnetar magnetosphere may exhibit features of both. For example, a magnetosphere could simultaneously support {\color{black} 
magnetic arcades with dense $e^\pm$ plasma in combination with
zones of weaker twist (on the opposing side of the star
\cite{ZhangThompson2024}) where a relativistic double layer structure would emerge.  A magnetar in this 
configuration would combine sensitivity to meV-mass axions with a sensitivity to QCD axions approaching
that of the 
pure j-bundle -- even though the hard X-ray continuum would arise from the collisional plasma and
not from resonant scattering in the polar zone.}
{\color{black}  Considerable variability may be expected
in the density of the reflecting plasma wall forming on extended field lines, this being sensitive to the rate
of pair creation near the magnetic dipole axis.}
A more comprehensive treatment of these hybrid magnetospheric configurations is left to future work.

Future observations of the GC may yield a population of sources like \gcmag\, that both have favorable magnetic and plasma conditions for axion conversion and are located in a particularly dense region of the dark matter halo. {\color{black} The baseline GJ plasma frequency may decrease only slowly with time as a magnetar
spins down, as $\omega_p \propto B^{1/2}t^{-1/4}$;  this means that older magnetars with magnetic fields frozen 
into their superconducting
cores can still be viable sites for
resonant conversion.}  Finally, in addition to converting dark matter axions into narrow radio signals, magnetars may themselves be prolific axion factories. The production efficiency varies significantly between the BT07 and TK20 models. In the former, there is a relatively low plasma density, which admits regions with $\epar \ne 0$, particularly in the pair production zone, which can source axions as in~\cite{Prabhu2021, noordhuis2023novel}.  We defer a dedicated analysis of axion production from magnetar magnetospheres to future work.

\vspace{0.1in}

\section{Acknowledgments}

The authors would like to acknowledge helpful conversations and feedback from Christopher Dessert, Dylan Folsom, Yonatan Kahn, Mariia Khelashvili, and Mariangela Lisanti. SR is supported by the National Science Foundation~(NSF), under Award Number AST 2307789, and by the Department of Energy~(DOE), under Award Number DE-SC0007968. AP is supported by the Princeton Center for Theoretical Science. CT~acknowledges the support of the Natural Sciences and Engineering Research Council of Canada, funding reference number RGPIN-2023-04612, and of the Simons Foundation (MP-SCMPS-00001470).  SJW acknowledges support from a Royal Society University Research Fellowship (URF-R1-231065). This article/publication is based upon work from COST Action COSMIC WISPers CA21106, supported by COST (European Cooperation in Science and Technology). This work is also supported by the Deutsche Forschungsgemeinschaft under Germany’s Excellence Strategy—EXC 2121 “Quantum Universe”—390833306. The work of C.B.~was supported in part by NASA through the NASA Hubble Fellowship Program grant HST-HF2-51451.001-A awarded by the Space Telescope Science Institute, which is operated by the Association of Universities for Research in Astronomy, Inc., for NASA, under contract NAS5-26555, as well as by the European Research Council under grant 742104.

The ray-tracing simulations presented in this article were performed on computational resources managed and supported by Princeton Research Computing, a consortium of groups including the Princeton Institute for Computational Science and Engineering (PICSciE) and Research Computing at Princeton University.


\bibliography{magnetardm}

\begin{thebibliography}{168}%
\makeatletter
\providecommand \@ifxundefined [1]{%
 \@ifx{#1\undefined}
}%
\providecommand \@ifnum [1]{%
 \ifnum #1\expandafter \@firstoftwo
 \else \expandafter \@secondoftwo
 \fi
}%
\providecommand \@ifx [1]{%
 \ifx #1\expandafter \@firstoftwo
 \else \expandafter \@secondoftwo
 \fi
}%
\providecommand \natexlab [1]{#1}%
\providecommand \enquote  [1]{``#1''}%
\providecommand \bibnamefont  [1]{#1}%
\providecommand \bibfnamefont [1]{#1}%
\providecommand \citenamefont [1]{#1}%
\providecommand \href@noop [0]{\@secondoftwo}%
\providecommand \href [0]{\begingroup \@sanitize@url \@href}%
\providecommand \@href[1]{\@@startlink{#1}\@@href}%
\providecommand \@@href[1]{\endgroup#1\@@endlink}%
\providecommand \@sanitize@url [0]{\catcode `\\12\catcode `\$12\catcode
  `\&12\catcode `\#12\catcode `\^12\catcode `\_12\catcode `\%12\relax}%
\providecommand \@@startlink[1]{}%
\providecommand \@@endlink[0]{}%
\providecommand \url  [0]{\begingroup\@sanitize@url \@url }%
\providecommand \@url [1]{\endgroup\@href {#1}{\urlprefix }}%
\providecommand \urlprefix  [0]{URL }%
\providecommand \Eprint [0]{\href }%
\providecommand \doibase [0]{http://dx.doi.org/}%
\providecommand \selectlanguage [0]{\@gobble}%
\providecommand \bibinfo  [0]{\@secondoftwo}%
\providecommand \bibfield  [0]{\@secondoftwo}%
\providecommand \translation [1]{[#1]}%
\providecommand \BibitemOpen [0]{}%
\providecommand \bibitemStop [0]{}%
\providecommand \bibitemNoStop [0]{.\EOS\space}%
\providecommand \EOS [0]{\spacefactor3000\relax}%
\providecommand \BibitemShut  [1]{\csname bibitem#1\endcsname}%
\let\auto@bib@innerbib\@empty
\bibitem [{\citenamefont {Peccei}\ and\ \citenamefont
  {Quinn}(1977{\natexlab{a}})}]{PQ1}%
  \BibitemOpen
  \bibfield  {author} {\bibinfo {author} {\bibfnamefont {R.~D.}\ \bibnamefont
  {Peccei}}\ and\ \bibinfo {author} {\bibfnamefont {H.~R.}\ \bibnamefont
  {Quinn}},\ }\href {\doibase 10.1103/PhysRevLett.38.1440} {\bibfield
  {journal} {\bibinfo  {journal} {Phys. Rev. Lett.}\ }\textbf {\bibinfo
  {volume} {38}},\ \bibinfo {pages} {1440} (\bibinfo {year}
  {1977}{\natexlab{a}})}\BibitemShut {NoStop}%
\bibitem [{\citenamefont {Peccei}\ and\ \citenamefont
  {Quinn}(1977{\natexlab{b}})}]{PQ2}%
  \BibitemOpen
  \bibfield  {author} {\bibinfo {author} {\bibfnamefont {R.~D.}\ \bibnamefont
  {Peccei}}\ and\ \bibinfo {author} {\bibfnamefont {H.~R.}\ \bibnamefont
  {Quinn}},\ }\href {\doibase 10.1103/PhysRevD.16.1791} {\bibfield  {journal}
  {\bibinfo  {journal} {Phys. Rev. D}\ }\textbf {\bibinfo {volume} {16}},\
  \bibinfo {pages} {1791} (\bibinfo {year} {1977}{\natexlab{b}})}\BibitemShut
  {NoStop}%
\bibitem [{\citenamefont {Weinberg}(1978)}]{WeinbergAxion}%
  \BibitemOpen
  \bibfield  {author} {\bibinfo {author} {\bibfnamefont {S.}~\bibnamefont
  {Weinberg}},\ }\href {\doibase 10.1103/PhysRevLett.40.223} {\bibfield
  {journal} {\bibinfo  {journal} {Phys. Rev. Lett.}\ }\textbf {\bibinfo
  {volume} {40}},\ \bibinfo {pages} {223} (\bibinfo {year} {1978})}\BibitemShut
  {NoStop}%
\bibitem [{\citenamefont {Wilczek}(1978)}]{WilczekAxion}%
  \BibitemOpen
  \bibfield  {author} {\bibinfo {author} {\bibfnamefont {F.}~\bibnamefont
  {Wilczek}},\ }\href {\doibase 10.1103/PhysRevLett.40.279} {\bibfield
  {journal} {\bibinfo  {journal} {Phys. Rev. Lett.}\ }\textbf {\bibinfo
  {volume} {40}},\ \bibinfo {pages} {279} (\bibinfo {year} {1978})}\BibitemShut
  {NoStop}%
\bibitem [{\citenamefont {Preskill}\ \emph {et~al.}(1983)\citenamefont
  {Preskill}, \citenamefont {Wise},\ and\ \citenamefont
  {Wilczek}}]{PRESKILL1983127}%
  \BibitemOpen
  \bibfield  {author} {\bibinfo {author} {\bibfnamefont {J.}~\bibnamefont
  {Preskill}}, \bibinfo {author} {\bibfnamefont {M.~B.}\ \bibnamefont {Wise}},
  \ and\ \bibinfo {author} {\bibfnamefont {F.}~\bibnamefont {Wilczek}},\ }\href
  {\doibase https://doi.org/10.1016/0370-2693(83)90637-8} {\bibfield  {journal}
  {\bibinfo  {journal} {Physics Letters B}\ }\textbf {\bibinfo {volume}
  {120}},\ \bibinfo {pages} {127 } (\bibinfo {year} {1983})}\BibitemShut
  {NoStop}%
\bibitem [{\citenamefont {Abbott}\ and\ \citenamefont
  {Sikivie}(1983)}]{Abbott1982}%
  \BibitemOpen
  \bibfield  {author} {\bibinfo {author} {\bibfnamefont {L.}~\bibnamefont
  {Abbott}}\ and\ \bibinfo {author} {\bibfnamefont {P.}~\bibnamefont
  {Sikivie}},\ }\href {\doibase 10.1016/0370-2693(83)90638-X} {\bibfield
  {journal} {\bibinfo  {journal} {Phys. Lett. B}\ }\textbf {\bibinfo {volume}
  {120}},\ \bibinfo {pages} {133} (\bibinfo {year} {1983})}\BibitemShut
  {NoStop}%
\bibitem [{\citenamefont {Dine}\ and\ \citenamefont
  {Fischler}(1983)}]{Fischler1982}%
  \BibitemOpen
  \bibfield  {author} {\bibinfo {author} {\bibfnamefont {M.}~\bibnamefont
  {Dine}}\ and\ \bibinfo {author} {\bibfnamefont {W.}~\bibnamefont
  {Fischler}},\ }\href {\doibase 10.1016/0370-2693(83)90639-1} {\bibfield
  {journal} {\bibinfo  {journal} {Phys. Lett. B}\ }\textbf {\bibinfo {volume}
  {120}},\ \bibinfo {pages} {137} (\bibinfo {year} {1983})}\BibitemShut
  {NoStop}%
\bibitem [{\citenamefont {Witte}\ \emph
  {et~al.}(2021{\natexlab{a}})\citenamefont {Witte}, \citenamefont {Noordhuis},
  \citenamefont {Edwards},\ and\ \citenamefont {Weniger}}]{Witte:2021arp}%
  \BibitemOpen
  \bibfield  {author} {\bibinfo {author} {\bibfnamefont {S.~J.}\ \bibnamefont
  {Witte}}, \bibinfo {author} {\bibfnamefont {D.}~\bibnamefont {Noordhuis}},
  \bibinfo {author} {\bibfnamefont {T.~D.~P.}\ \bibnamefont {Edwards}}, \ and\
  \bibinfo {author} {\bibfnamefont {C.}~\bibnamefont {Weniger}},\ }\href
  {\doibase 10.1103/PhysRevD.104.103030} {\bibfield  {journal} {\bibinfo
  {journal} {Phys. Rev. D}\ }\textbf {\bibinfo {volume} {104}},\ \bibinfo
  {pages} {103030} (\bibinfo {year} {2021}{\natexlab{a}})},\ \Eprint
  {http://arxiv.org/abs/2104.07670} {arXiv:2104.07670 [hep-ph]} \BibitemShut
  {NoStop}%
\bibitem [{\citenamefont {{Millar}}\ \emph {et~al.}(2021)\citenamefont
  {{Millar}}, \citenamefont {{Baum}}, \citenamefont {{Lawson}},\ and\
  \citenamefont {{Marsh}}}]{millar2021axionphotonUPDATED}%
  \BibitemOpen
  \bibfield  {author} {\bibinfo {author} {\bibfnamefont {A.~J.}\ \bibnamefont
  {{Millar}}}, \bibinfo {author} {\bibfnamefont {S.}~\bibnamefont {{Baum}}},
  \bibinfo {author} {\bibfnamefont {M.}~\bibnamefont {{Lawson}}}, \ and\
  \bibinfo {author} {\bibfnamefont {M.~C.~D.}\ \bibnamefont {{Marsh}}},\ }\href
  {\doibase 10.1088/1475-7516/2021/11/013} {\bibfield  {journal} {\bibinfo
  {journal} {\jcap}\ }\textbf {\bibinfo {volume} {2021}},\ \bibinfo {eid} {013}
  (\bibinfo {year} {2021})},\ \Eprint {http://arxiv.org/abs/2107.07399}
  {arXiv:2107.07399 [hep-ph]} \BibitemShut {NoStop}%
\bibitem [{\citenamefont {McDonald}\ \emph {et~al.}(2023)\citenamefont
  {McDonald}, \citenamefont {Garbrecht},\ and\ \citenamefont
  {Millington}}]{McDonald:2023ohd}%
  \BibitemOpen
  \bibfield  {author} {\bibinfo {author} {\bibfnamefont {J.~I.}\ \bibnamefont
  {McDonald}}, \bibinfo {author} {\bibfnamefont {B.}~\bibnamefont {Garbrecht}},
  \ and\ \bibinfo {author} {\bibfnamefont {P.}~\bibnamefont {Millington}},\
  }\href {\doibase 10.1088/1475-7516/2023/12/031} {\bibfield  {journal}
  {\bibinfo  {journal} {JCAP}\ }\textbf {\bibinfo {volume} {12}},\ \bibinfo
  {pages} {031} (\bibinfo {year} {2023})},\ \Eprint
  {http://arxiv.org/abs/2307.11812} {arXiv:2307.11812 [hep-ph]} \BibitemShut
  {NoStop}%
\bibitem [{\citenamefont {Gin\'es}\ \emph {et~al.}(2024)\citenamefont
  {Gin\'es}, \citenamefont {Noordhuis}, \citenamefont {Weniger},\ and\
  \citenamefont {Witte}}]{Gines:2024ekm}%
  \BibitemOpen
  \bibfield  {author} {\bibinfo {author} {\bibfnamefont {E.~U.}\ \bibnamefont
  {Gin\'es}}, \bibinfo {author} {\bibfnamefont {D.}~\bibnamefont {Noordhuis}},
  \bibinfo {author} {\bibfnamefont {C.}~\bibnamefont {Weniger}}, \ and\
  \bibinfo {author} {\bibfnamefont {S.~J.}\ \bibnamefont {Witte}},\ }\href
  {\doibase 10.1103/PhysRevD.110.083007} {\bibfield  {journal} {\bibinfo
  {journal} {Phys. Rev. D}\ }\textbf {\bibinfo {volume} {110}},\ \bibinfo
  {pages} {083007} (\bibinfo {year} {2024})},\ \Eprint
  {http://arxiv.org/abs/2405.08865} {arXiv:2405.08865 [hep-ph]} \BibitemShut
  {NoStop}%
\bibitem [{\citenamefont {McDonald}\ and\ \citenamefont
  {Millington}(2024)}]{McDonald:2024uuh}%
  \BibitemOpen
  \bibfield  {author} {\bibinfo {author} {\bibfnamefont {J.~I.}\ \bibnamefont
  {McDonald}}\ and\ \bibinfo {author} {\bibfnamefont {P.}~\bibnamefont
  {Millington}},\ }\href {\doibase 10.1088/1475-7516/2024/09/072} {\bibfield
  {journal} {\bibinfo  {journal} {JCAP}\ }\textbf {\bibinfo {volume} {09}},\
  \bibinfo {pages} {072} (\bibinfo {year} {2024})},\ \Eprint
  {http://arxiv.org/abs/2407.11192} {arXiv:2407.11192 [hep-ph]} \BibitemShut
  {NoStop}%
\bibitem [{\citenamefont {Pshirkov}\ and\ \citenamefont
  {Popov}(2009)}]{Pshirkov:2007st}%
  \BibitemOpen
  \bibfield  {author} {\bibinfo {author} {\bibfnamefont {M.~S.}\ \bibnamefont
  {Pshirkov}}\ and\ \bibinfo {author} {\bibfnamefont {S.~B.}\ \bibnamefont
  {Popov}},\ }\href {\doibase 10.1134/S1063776109030030} {\bibfield  {journal}
  {\bibinfo  {journal} {J. Exp. Theor. Phys.}\ }\textbf {\bibinfo {volume}
  {108}},\ \bibinfo {pages} {384} (\bibinfo {year} {2009})},\ \Eprint
  {http://arxiv.org/abs/0711.1264} {arXiv:0711.1264 [astro-ph]} \BibitemShut
  {NoStop}%
\bibitem [{\citenamefont {Huang}\ \emph {et~al.}(2018)\citenamefont {Huang},
  \citenamefont {Kadota}, \citenamefont {Sekiguchi},\ and\ \citenamefont
  {Tashiro}}]{Huang:2018lxq}%
  \BibitemOpen
  \bibfield  {author} {\bibinfo {author} {\bibfnamefont {F.~P.}\ \bibnamefont
  {Huang}}, \bibinfo {author} {\bibfnamefont {K.}~\bibnamefont {Kadota}},
  \bibinfo {author} {\bibfnamefont {T.}~\bibnamefont {Sekiguchi}}, \ and\
  \bibinfo {author} {\bibfnamefont {H.}~\bibnamefont {Tashiro}},\ }\href
  {\doibase 10.1103/PhysRevD.97.123001} {\bibfield  {journal} {\bibinfo
  {journal} {Phys. Rev. D}\ }\textbf {\bibinfo {volume} {97}},\ \bibinfo
  {pages} {123001} (\bibinfo {year} {2018})},\ \Eprint
  {http://arxiv.org/abs/1803.08230} {arXiv:1803.08230 [hep-ph]} \BibitemShut
  {NoStop}%
\bibitem [{\citenamefont {Hook}\ \emph {et~al.}(2018)\citenamefont {Hook},
  \citenamefont {Kahn}, \citenamefont {Safdi},\ and\ \citenamefont
  {Sun}}]{Hook:2018iia}%
  \BibitemOpen
  \bibfield  {author} {\bibinfo {author} {\bibfnamefont {A.}~\bibnamefont
  {Hook}}, \bibinfo {author} {\bibfnamefont {Y.}~\bibnamefont {Kahn}}, \bibinfo
  {author} {\bibfnamefont {B.~R.}\ \bibnamefont {Safdi}}, \ and\ \bibinfo
  {author} {\bibfnamefont {Z.}~\bibnamefont {Sun}},\ }\href {\doibase
  10.1103/PhysRevLett.121.241102} {\bibfield  {journal} {\bibinfo  {journal}
  {Phys. Rev. Lett.}\ }\textbf {\bibinfo {volume} {121}},\ \bibinfo {pages}
  {241102} (\bibinfo {year} {2018})},\ \Eprint
  {http://arxiv.org/abs/1804.03145} {arXiv:1804.03145 [hep-ph]} \BibitemShut
  {NoStop}%
\bibitem [{\citenamefont {Safdi}\ \emph {et~al.}(2019)\citenamefont {Safdi},
  \citenamefont {Sun},\ and\ \citenamefont {Chen}}]{Safdi:2018oeu}%
  \BibitemOpen
  \bibfield  {author} {\bibinfo {author} {\bibfnamefont {B.~R.}\ \bibnamefont
  {Safdi}}, \bibinfo {author} {\bibfnamefont {Z.}~\bibnamefont {Sun}}, \ and\
  \bibinfo {author} {\bibfnamefont {A.~Y.}\ \bibnamefont {Chen}},\ }\href
  {\doibase 10.1103/PhysRevD.99.123021} {\bibfield  {journal} {\bibinfo
  {journal} {Phys. Rev. D}\ }\textbf {\bibinfo {volume} {99}},\ \bibinfo
  {pages} {123021} (\bibinfo {year} {2019})},\ \Eprint
  {http://arxiv.org/abs/1811.01020} {arXiv:1811.01020 [astro-ph.CO]}
  \BibitemShut {NoStop}%
\bibitem [{\citenamefont {Battye}\ \emph {et~al.}(2020)\citenamefont {Battye},
  \citenamefont {Garbrecht}, \citenamefont {McDonald}, \citenamefont {Pace},\
  and\ \citenamefont {Srinivasan}}]{Battye:2019aco}%
  \BibitemOpen
  \bibfield  {author} {\bibinfo {author} {\bibfnamefont {R.~A.}\ \bibnamefont
  {Battye}}, \bibinfo {author} {\bibfnamefont {B.}~\bibnamefont {Garbrecht}},
  \bibinfo {author} {\bibfnamefont {J.~I.}\ \bibnamefont {McDonald}}, \bibinfo
  {author} {\bibfnamefont {F.}~\bibnamefont {Pace}}, \ and\ \bibinfo {author}
  {\bibfnamefont {S.}~\bibnamefont {Srinivasan}},\ }\href {\doibase
  10.1103/PhysRevD.102.023504} {\bibfield  {journal} {\bibinfo  {journal}
  {Phys. Rev. D}\ }\textbf {\bibinfo {volume} {102}},\ \bibinfo {pages}
  {023504} (\bibinfo {year} {2020})},\ \Eprint
  {http://arxiv.org/abs/1910.11907} {arXiv:1910.11907 [astro-ph.CO]}
  \BibitemShut {NoStop}%
\bibitem [{\citenamefont {Leroy}\ \emph {et~al.}(2020)\citenamefont {Leroy},
  \citenamefont {Chianese}, \citenamefont {Edwards},\ and\ \citenamefont
  {Weniger}}]{Leroy:2019ghm}%
  \BibitemOpen
  \bibfield  {author} {\bibinfo {author} {\bibfnamefont {M.}~\bibnamefont
  {Leroy}}, \bibinfo {author} {\bibfnamefont {M.}~\bibnamefont {Chianese}},
  \bibinfo {author} {\bibfnamefont {T.~D.~P.}\ \bibnamefont {Edwards}}, \ and\
  \bibinfo {author} {\bibfnamefont {C.}~\bibnamefont {Weniger}},\ }\href
  {\doibase 10.1103/PhysRevD.101.123003} {\bibfield  {journal} {\bibinfo
  {journal} {Phys. Rev. D}\ }\textbf {\bibinfo {volume} {101}},\ \bibinfo
  {pages} {123003} (\bibinfo {year} {2020})},\ \Eprint
  {http://arxiv.org/abs/1912.08815} {arXiv:1912.08815 [hep-ph]} \BibitemShut
  {NoStop}%
\bibitem [{\citenamefont {Foster}\ \emph {et~al.}(2020)\citenamefont {Foster}
  \emph {et~al.}}]{Foster:2020pgt}%
  \BibitemOpen
  \bibfield  {author} {\bibinfo {author} {\bibfnamefont {J.~W.}\ \bibnamefont
  {Foster}} \emph {et~al.},\ }\href {\doibase 10.1103/PhysRevLett.125.171301}
  {\bibfield  {journal} {\bibinfo  {journal} {Phys. Rev. Lett.}\ }\textbf
  {\bibinfo {volume} {125}},\ \bibinfo {pages} {171301} (\bibinfo {year}
  {2020})},\ \Eprint {http://arxiv.org/abs/2004.00011} {arXiv:2004.00011
  [astro-ph.CO]} \BibitemShut {NoStop}%
\bibitem [{\citenamefont {Prabhu}\ and\ \citenamefont
  {Rapidis}(2020)}]{Prabhu:2020yif}%
  \BibitemOpen
  \bibfield  {author} {\bibinfo {author} {\bibfnamefont {A.}~\bibnamefont
  {Prabhu}}\ and\ \bibinfo {author} {\bibfnamefont {N.~M.}\ \bibnamefont
  {Rapidis}},\ }\href {\doibase 10.1088/1475-7516/2020/10/054} {\bibfield
  {journal} {\bibinfo  {journal} {JCAP}\ }\textbf {\bibinfo {volume} {10}},\
  \bibinfo {pages} {054} (\bibinfo {year} {2020})},\ \Eprint
  {http://arxiv.org/abs/2005.03700} {arXiv:2005.03700 [astro-ph.CO]}
  \BibitemShut {NoStop}%
\bibitem [{\citenamefont {Buckley}\ \emph {et~al.}(2021)\citenamefont
  {Buckley}, \citenamefont {Dev}, \citenamefont {Ferrer},\ and\ \citenamefont
  {Huang}}]{Buckley:2020fmh}%
  \BibitemOpen
  \bibfield  {author} {\bibinfo {author} {\bibfnamefont {J.~H.}\ \bibnamefont
  {Buckley}}, \bibinfo {author} {\bibfnamefont {P.~S.~B.}\ \bibnamefont {Dev}},
  \bibinfo {author} {\bibfnamefont {F.}~\bibnamefont {Ferrer}}, \ and\ \bibinfo
  {author} {\bibfnamefont {F.~P.}\ \bibnamefont {Huang}},\ }\href {\doibase
  10.1103/PhysRevD.103.043015} {\bibfield  {journal} {\bibinfo  {journal}
  {Phys. Rev. D}\ }\textbf {\bibinfo {volume} {103}},\ \bibinfo {pages}
  {043015} (\bibinfo {year} {2021})},\ \Eprint
  {http://arxiv.org/abs/2004.06486} {arXiv:2004.06486 [astro-ph.HE]}
  \BibitemShut {NoStop}%
\bibitem [{\citenamefont {Battye}\ \emph
  {et~al.}(2021{\natexlab{a}})\citenamefont {Battye}, \citenamefont
  {Garbrecht}, \citenamefont {McDonald},\ and\ \citenamefont
  {Srinivasan}}]{Battye:2021xvt}%
  \BibitemOpen
  \bibfield  {author} {\bibinfo {author} {\bibfnamefont {R.~A.}\ \bibnamefont
  {Battye}}, \bibinfo {author} {\bibfnamefont {B.}~\bibnamefont {Garbrecht}},
  \bibinfo {author} {\bibfnamefont {J.~I.}\ \bibnamefont {McDonald}}, \ and\
  \bibinfo {author} {\bibfnamefont {S.}~\bibnamefont {Srinivasan}},\ }\href
  {\doibase 10.1007/JHEP09(2021)105} {\bibfield  {journal} {\bibinfo  {journal}
  {JHEP}\ }\textbf {\bibinfo {volume} {09}},\ \bibinfo {pages} {105} (\bibinfo
  {year} {2021}{\natexlab{a}})},\ \Eprint {http://arxiv.org/abs/2104.08290}
  {arXiv:2104.08290 [hep-ph]} \BibitemShut {NoStop}%
\bibitem [{\citenamefont {Battye}\ \emph
  {et~al.}(2021{\natexlab{b}})\citenamefont {Battye}, \citenamefont {Darling},
  \citenamefont {McDonald},\ and\ \citenamefont
  {Srinivasan}}]{battye2021robust}%
  \BibitemOpen
  \bibfield  {author} {\bibinfo {author} {\bibfnamefont {R.~A.}\ \bibnamefont
  {Battye}}, \bibinfo {author} {\bibfnamefont {J.}~\bibnamefont {Darling}},
  \bibinfo {author} {\bibfnamefont {J.}~\bibnamefont {McDonald}}, \ and\
  \bibinfo {author} {\bibfnamefont {S.}~\bibnamefont {Srinivasan}},\
  }\href@noop {} {\enquote {\bibinfo {title} {Towards robust constraints on
  axion dark matter using psr j1745-2900},}\ } (\bibinfo {year}
  {2021}{\natexlab{b}}),\ \Eprint {http://arxiv.org/abs/2107.01225}
  {arXiv:2107.01225 [astro-ph.CO]} \BibitemShut {NoStop}%
\bibitem [{\citenamefont {Nurmi}\ \emph {et~al.}(2021)\citenamefont {Nurmi},
  \citenamefont {Schiappacasse},\ and\ \citenamefont
  {Yanagida}}]{Nurmi:2021xds}%
  \BibitemOpen
  \bibfield  {author} {\bibinfo {author} {\bibfnamefont {S.}~\bibnamefont
  {Nurmi}}, \bibinfo {author} {\bibfnamefont {E.~D.}\ \bibnamefont
  {Schiappacasse}}, \ and\ \bibinfo {author} {\bibfnamefont {T.~T.}\
  \bibnamefont {Yanagida}},\ }\href@noop {} {\  (\bibinfo {year} {2021})},\
  \Eprint {http://arxiv.org/abs/2102.05680} {arXiv:2102.05680 [hep-ph]}
  \BibitemShut {NoStop}%
\bibitem [{\citenamefont {Foster}\ \emph {et~al.}(2022)\citenamefont {Foster},
  \citenamefont {Witte}, \citenamefont {Lawson}, \citenamefont {Linden},
  \citenamefont {Gajjar}, \citenamefont {Weniger},\ and\ \citenamefont
  {Safdi}}]{Foster:2022fxn}%
  \BibitemOpen
  \bibfield  {author} {\bibinfo {author} {\bibfnamefont {J.~W.}\ \bibnamefont
  {Foster}}, \bibinfo {author} {\bibfnamefont {S.~J.}\ \bibnamefont {Witte}},
  \bibinfo {author} {\bibfnamefont {M.}~\bibnamefont {Lawson}}, \bibinfo
  {author} {\bibfnamefont {T.}~\bibnamefont {Linden}}, \bibinfo {author}
  {\bibfnamefont {V.}~\bibnamefont {Gajjar}}, \bibinfo {author} {\bibfnamefont
  {C.}~\bibnamefont {Weniger}}, \ and\ \bibinfo {author} {\bibfnamefont
  {B.~R.}\ \bibnamefont {Safdi}},\ }\href {\doibase
  10.1103/PhysRevLett.129.251102} {\bibfield  {journal} {\bibinfo  {journal}
  {Phys. Rev. Lett.}\ }\textbf {\bibinfo {volume} {129}},\ \bibinfo {pages}
  {251102} (\bibinfo {year} {2022})},\ \Eprint
  {http://arxiv.org/abs/2202.08274} {arXiv:2202.08274 [astro-ph.CO]}
  \BibitemShut {NoStop}%
\bibitem [{\citenamefont {Witte}\ \emph {et~al.}(2023)\citenamefont {Witte},
  \citenamefont {Baum}, \citenamefont {Lawson}, \citenamefont {Marsh},
  \citenamefont {Millar},\ and\ \citenamefont {Salinas}}]{Witte:2022cjj}%
  \BibitemOpen
  \bibfield  {author} {\bibinfo {author} {\bibfnamefont {S.~J.}\ \bibnamefont
  {Witte}}, \bibinfo {author} {\bibfnamefont {S.}~\bibnamefont {Baum}},
  \bibinfo {author} {\bibfnamefont {M.}~\bibnamefont {Lawson}}, \bibinfo
  {author} {\bibfnamefont {M.~C.~D.}\ \bibnamefont {Marsh}}, \bibinfo {author}
  {\bibfnamefont {A.~J.}\ \bibnamefont {Millar}}, \ and\ \bibinfo {author}
  {\bibfnamefont {G.}~\bibnamefont {Salinas}},\ }\href {\doibase
  10.1103/PhysRevD.107.063013} {\bibfield  {journal} {\bibinfo  {journal}
  {Phys. Rev. D}\ }\textbf {\bibinfo {volume} {107}},\ \bibinfo {pages}
  {063013} (\bibinfo {year} {2023})},\ \Eprint
  {http://arxiv.org/abs/2212.08079} {arXiv:2212.08079 [hep-ph]} \BibitemShut
  {NoStop}%
\bibitem [{\citenamefont {Battye}\ \emph {et~al.}(2023)\citenamefont {Battye},
  \citenamefont {Keith}, \citenamefont {McDonald}, \citenamefont {Srinivasan},
  \citenamefont {Stappers},\ and\ \citenamefont {Weltevrede}}]{Battye:2023oac}%
  \BibitemOpen
  \bibfield  {author} {\bibinfo {author} {\bibfnamefont {R.~A.}\ \bibnamefont
  {Battye}}, \bibinfo {author} {\bibfnamefont {M.~J.}\ \bibnamefont {Keith}},
  \bibinfo {author} {\bibfnamefont {J.~I.}\ \bibnamefont {McDonald}}, \bibinfo
  {author} {\bibfnamefont {S.}~\bibnamefont {Srinivasan}}, \bibinfo {author}
  {\bibfnamefont {B.~W.}\ \bibnamefont {Stappers}}, \ and\ \bibinfo {author}
  {\bibfnamefont {P.}~\bibnamefont {Weltevrede}},\ }\href {\doibase
  10.1103/PhysRevD.108.063001} {\bibfield  {journal} {\bibinfo  {journal}
  {Phys. Rev. D}\ }\textbf {\bibinfo {volume} {108}},\ \bibinfo {pages}
  {063001} (\bibinfo {year} {2023})},\ \Eprint
  {http://arxiv.org/abs/2303.11792} {arXiv:2303.11792 [astro-ph.CO]}
  \BibitemShut {NoStop}%
\bibitem [{\citenamefont {McDonald}\ and\ \citenamefont
  {Witte}(2023)}]{McDonald:2023shx}%
  \BibitemOpen
  \bibfield  {author} {\bibinfo {author} {\bibfnamefont {J.~I.}\ \bibnamefont
  {McDonald}}\ and\ \bibinfo {author} {\bibfnamefont {S.~J.}\ \bibnamefont
  {Witte}},\ }\href {\doibase 10.1103/PhysRevD.108.103021} {\bibfield
  {journal} {\bibinfo  {journal} {Phys. Rev. D}\ }\textbf {\bibinfo {volume}
  {108}},\ \bibinfo {pages} {103021} (\bibinfo {year} {2023})},\ \Eprint
  {http://arxiv.org/abs/2309.08655} {arXiv:2309.08655 [hep-ph]} \BibitemShut
  {NoStop}%
\bibitem [{\citenamefont {Noordhuis}\ \emph {et~al.}(2023)\citenamefont
  {Noordhuis}, \citenamefont {Prabhu}, \citenamefont {Witte}, \citenamefont
  {Chen}, \citenamefont {Cruz},\ and\ \citenamefont
  {Weniger}}]{noordhuis2023novel}%
  \BibitemOpen
  \bibfield  {author} {\bibinfo {author} {\bibfnamefont {D.}~\bibnamefont
  {Noordhuis}}, \bibinfo {author} {\bibfnamefont {A.}~\bibnamefont {Prabhu}},
  \bibinfo {author} {\bibfnamefont {S.~J.}\ \bibnamefont {Witte}}, \bibinfo
  {author} {\bibfnamefont {A.~Y.}\ \bibnamefont {Chen}}, \bibinfo {author}
  {\bibfnamefont {F.}~\bibnamefont {Cruz}}, \ and\ \bibinfo {author}
  {\bibfnamefont {C.}~\bibnamefont {Weniger}},\ }\href {\doibase
  10.1103/PhysRevLett.131.111004} {\bibfield  {journal} {\bibinfo  {journal}
  {Phys. Rev. Lett.}\ }\textbf {\bibinfo {volume} {131}},\ \bibinfo {pages}
  {111004} (\bibinfo {year} {2023})},\ \Eprint
  {http://arxiv.org/abs/2209.09917} {arXiv:2209.09917 [hep-ph]} \BibitemShut
  {NoStop}%
\bibitem [{\citenamefont {Xue}\ \emph {et~al.}(2023)\citenamefont {Xue},
  \citenamefont {Lee}, \citenamefont {Gao},\ and\ \citenamefont
  {Xu}}]{Xue:2023ejt}%
  \BibitemOpen
  \bibfield  {author} {\bibinfo {author} {\bibfnamefont {Z.~H.}\ \bibnamefont
  {Xue}}, \bibinfo {author} {\bibfnamefont {K.~J.}\ \bibnamefont {Lee}},
  \bibinfo {author} {\bibfnamefont {X.~D.}\ \bibnamefont {Gao}}, \ and\
  \bibinfo {author} {\bibfnamefont {R.~X.}\ \bibnamefont {Xu}},\ }\href
  {\doibase 10.1103/PhysRevD.108.083009} {\bibfield  {journal} {\bibinfo
  {journal} {Phys. Rev. D}\ }\textbf {\bibinfo {volume} {108}},\ \bibinfo
  {pages} {083009} (\bibinfo {year} {2023})},\ \Eprint
  {http://arxiv.org/abs/2310.06660} {arXiv:2310.06660 [astro-ph.HE]}
  \BibitemShut {NoStop}%
\bibitem [{\citenamefont {Tjemsland}\ \emph {et~al.}(2024)\citenamefont
  {Tjemsland}, \citenamefont {McDonald},\ and\ \citenamefont
  {Witte}}]{Tjemsland:2023vvc}%
  \BibitemOpen
  \bibfield  {author} {\bibinfo {author} {\bibfnamefont {J.}~\bibnamefont
  {Tjemsland}}, \bibinfo {author} {\bibfnamefont {J.}~\bibnamefont {McDonald}},
  \ and\ \bibinfo {author} {\bibfnamefont {S.~J.}\ \bibnamefont {Witte}},\
  }\href {\doibase 10.1103/PhysRevD.109.023015} {\bibfield  {journal} {\bibinfo
   {journal} {Phys. Rev. D}\ }\textbf {\bibinfo {volume} {109}},\ \bibinfo
  {pages} {023015} (\bibinfo {year} {2024})},\ \Eprint
  {http://arxiv.org/abs/2310.18403} {arXiv:2310.18403 [hep-ph]} \BibitemShut
  {NoStop}%
\bibitem [{\citenamefont {Noordhuis}\ \emph {et~al.}(2024)\citenamefont
  {Noordhuis}, \citenamefont {Prabhu}, \citenamefont {Weniger},\ and\
  \citenamefont {Witte}}]{Noordhuis:2023wid}%
  \BibitemOpen
  \bibfield  {author} {\bibinfo {author} {\bibfnamefont {D.}~\bibnamefont
  {Noordhuis}}, \bibinfo {author} {\bibfnamefont {A.}~\bibnamefont {Prabhu}},
  \bibinfo {author} {\bibfnamefont {C.}~\bibnamefont {Weniger}}, \ and\
  \bibinfo {author} {\bibfnamefont {S.~J.}\ \bibnamefont {Witte}},\ }\href
  {\doibase 10.1103/PhysRevX.14.041015} {\bibfield  {journal} {\bibinfo
  {journal} {Phys. Rev. X}\ }\textbf {\bibinfo {volume} {14}},\ \bibinfo
  {pages} {041015} (\bibinfo {year} {2024})},\ \Eprint
  {http://arxiv.org/abs/2307.11811} {arXiv:2307.11811 [hep-ph]} \BibitemShut
  {NoStop}%
\bibitem [{\citenamefont {Caputo}\ \emph {et~al.}(2024)\citenamefont {Caputo},
  \citenamefont {Witte}, \citenamefont {Philippov},\ and\ \citenamefont
  {Jacobson}}]{Caputo:2023cpv}%
  \BibitemOpen
  \bibfield  {author} {\bibinfo {author} {\bibfnamefont {A.}~\bibnamefont
  {Caputo}}, \bibinfo {author} {\bibfnamefont {S.~J.}\ \bibnamefont {Witte}},
  \bibinfo {author} {\bibfnamefont {A.~A.}\ \bibnamefont {Philippov}}, \ and\
  \bibinfo {author} {\bibfnamefont {T.}~\bibnamefont {Jacobson}},\ }\href
  {\doibase 10.1103/PhysRevLett.133.161001} {\bibfield  {journal} {\bibinfo
  {journal} {Phys. Rev. Lett.}\ }\textbf {\bibinfo {volume} {133}},\ \bibinfo
  {pages} {161001} (\bibinfo {year} {2024})},\ \Eprint
  {http://arxiv.org/abs/2311.14795} {arXiv:2311.14795 [hep-ph]} \BibitemShut
  {NoStop}%
\bibitem [{\citenamefont {Witte}\ \emph {et~al.}(2024)\citenamefont {Witte},
  \citenamefont {Noordhuis}, \citenamefont {Prabhu},\ and\ \citenamefont
  {Weniger}}]{Witte:2024akb}%
  \BibitemOpen
  \bibfield  {author} {\bibinfo {author} {\bibfnamefont {S.}~\bibnamefont
  {Witte}}, \bibinfo {author} {\bibfnamefont {D.}~\bibnamefont {Noordhuis}},
  \bibinfo {author} {\bibfnamefont {A.}~\bibnamefont {Prabhu}}, \ and\ \bibinfo
  {author} {\bibfnamefont {C.}~\bibnamefont {Weniger}},\ }\href {\doibase
  10.22323/1.454.0022} {\bibfield  {journal} {\bibinfo  {journal} {PoS}\
  }\textbf {\bibinfo {volume} {COSMICWISPers}},\ \bibinfo {pages} {022}
  (\bibinfo {year} {2024})}\BibitemShut {NoStop}%
\bibitem [{\citenamefont {Kouvaris}\ \emph {et~al.}(2024)\citenamefont
  {Kouvaris}, \citenamefont {Liu},\ and\ \citenamefont
  {Lyu}}]{Kouvaris:2022guf}%
  \BibitemOpen
  \bibfield  {author} {\bibinfo {author} {\bibfnamefont {C.}~\bibnamefont
  {Kouvaris}}, \bibinfo {author} {\bibfnamefont {T.}~\bibnamefont {Liu}}, \
  and\ \bibinfo {author} {\bibfnamefont {K.-F.}\ \bibnamefont {Lyu}},\ }\href
  {\doibase 10.1103/PhysRevD.109.023008} {\bibfield  {journal} {\bibinfo
  {journal} {Phys. Rev. D}\ }\textbf {\bibinfo {volume} {109}},\ \bibinfo
  {pages} {023008} (\bibinfo {year} {2024})},\ \Eprint
  {http://arxiv.org/abs/2202.11096} {arXiv:2202.11096 [astro-ph.HE]}
  \BibitemShut {NoStop}%
\bibitem [{\citenamefont {Maseizik}\ and\ \citenamefont
  {Sigl}(2024)}]{Maseizik:2024qly}%
  \BibitemOpen
  \bibfield  {author} {\bibinfo {author} {\bibfnamefont {D.}~\bibnamefont
  {Maseizik}}\ and\ \bibinfo {author} {\bibfnamefont {G.}~\bibnamefont
  {Sigl}},\ }\href@noop {} {\  (\bibinfo {year} {2024})},\ \Eprint
  {http://arxiv.org/abs/2404.07908} {arXiv:2404.07908 [astro-ph.CO]}
  \BibitemShut {NoStop}%
\bibitem [{\citenamefont {Song}\ \emph {et~al.}(2025)\citenamefont {Song},
  \citenamefont {Su},\ and\ \citenamefont {Wu}}]{Song:2024rru}%
  \BibitemOpen
  \bibfield  {author} {\bibinfo {author} {\bibfnamefont {N.}~\bibnamefont
  {Song}}, \bibinfo {author} {\bibfnamefont {L.}~\bibnamefont {Su}}, \ and\
  \bibinfo {author} {\bibfnamefont {L.}~\bibnamefont {Wu}},\ }\href {\doibase
  10.1103/PhysRevD.111.043025} {\bibfield  {journal} {\bibinfo  {journal}
  {Phys. Rev. D}\ }\textbf {\bibinfo {volume} {111}},\ \bibinfo {pages}
  {043025} (\bibinfo {year} {2025})},\ \Eprint
  {http://arxiv.org/abs/2402.15144} {arXiv:2402.15144 [hep-ph]} \BibitemShut
  {NoStop}%
\bibitem [{\citenamefont {Witte}\ \emph
  {et~al.}(2021{\natexlab{b}})\citenamefont {Witte}, \citenamefont {Noordhuis},
  \citenamefont {Edwards},\ and\ \citenamefont {Weniger}}]{Witte2021}%
  \BibitemOpen
  \bibfield  {author} {\bibinfo {author} {\bibfnamefont {S.~J.}\ \bibnamefont
  {Witte}}, \bibinfo {author} {\bibfnamefont {D.}~\bibnamefont {Noordhuis}},
  \bibinfo {author} {\bibfnamefont {T.~D.}\ \bibnamefont {Edwards}}, \ and\
  \bibinfo {author} {\bibfnamefont {C.}~\bibnamefont {Weniger}},\ }\href
  {\doibase 10.1103/physrevd.104.103030} {\bibfield  {journal} {\bibinfo
  {journal} {Physical Review D}\ }\textbf {\bibinfo {volume} {104}} (\bibinfo
  {year} {2021}{\natexlab{b}}),\ 10.1103/physrevd.104.103030}\BibitemShut
  {NoStop}%
\bibitem [{\citenamefont {Battye}\ \emph {et~al.}(2022)\citenamefont {Battye},
  \citenamefont {Darling}, \citenamefont {McDonald},\ and\ \citenamefont
  {Srinivasan}}]{Battye:2021yue}%
  \BibitemOpen
  \bibfield  {author} {\bibinfo {author} {\bibfnamefont {R.~A.}\ \bibnamefont
  {Battye}}, \bibinfo {author} {\bibfnamefont {J.}~\bibnamefont {Darling}},
  \bibinfo {author} {\bibfnamefont {J.~I.}\ \bibnamefont {McDonald}}, \ and\
  \bibinfo {author} {\bibfnamefont {S.}~\bibnamefont {Srinivasan}},\ }\href
  {\doibase 10.1103/PhysRevD.105.L021305} {\bibfield  {journal} {\bibinfo
  {journal} {Phys. Rev. D}\ }\textbf {\bibinfo {volume} {105}},\ \bibinfo
  {pages} {L021305} (\bibinfo {year} {2022})},\ \Eprint
  {http://arxiv.org/abs/2107.01225} {arXiv:2107.01225 [astro-ph.CO]}
  \BibitemShut {NoStop}%
\bibitem [{\citenamefont {Prabhu}(2023)}]{Prabhu:2023cgb}%
  \BibitemOpen
  \bibfield  {author} {\bibinfo {author} {\bibfnamefont {A.}~\bibnamefont
  {Prabhu}},\ }\href {\doibase 10.3847/2041-8213/acc7a7} {\bibfield  {journal}
  {\bibinfo  {journal} {Astrophys. J. Lett.}\ }\textbf {\bibinfo {volume}
  {946}},\ \bibinfo {pages} {L52} (\bibinfo {year} {2023})},\ \Eprint
  {http://arxiv.org/abs/2302.11645} {arXiv:2302.11645 [astro-ph.HE]}
  \BibitemShut {NoStop}%
\bibitem [{\citenamefont {{Goldreich}}\ and\ \citenamefont
  {{Julian}}(1969)}]{GoldreichJulian1969}%
  \BibitemOpen
  \bibfield  {author} {\bibinfo {author} {\bibfnamefont {P.}~\bibnamefont
  {{Goldreich}}}\ and\ \bibinfo {author} {\bibfnamefont {W.~H.}\ \bibnamefont
  {{Julian}}},\ }\href {\doibase 10.1086/150119} {\bibfield  {journal}
  {\bibinfo  {journal} {\apj}\ }\textbf {\bibinfo {volume} {157}},\ \bibinfo
  {pages} {869} (\bibinfo {year} {1969})}\BibitemShut {NoStop}%
\bibitem [{\citenamefont {{Woods}}\ and\ \citenamefont
  {{Thompson}}(2006)}]{WoodsThompson2006}%
  \BibitemOpen
  \bibfield  {author} {\bibinfo {author} {\bibfnamefont {P.~M.}\ \bibnamefont
  {{Woods}}}\ and\ \bibinfo {author} {\bibfnamefont {C.}~\bibnamefont
  {{Thompson}}},\ }in\ \href {\doibase 10.48550/arXiv.astro-ph/0406133} {\emph
  {\bibinfo {booktitle} {Compact stellar X-ray sources}}},\ Vol.~\bibinfo
  {volume} {39},\ \bibinfo {editor} {edited by\ \bibinfo {editor}
  {\bibfnamefont {W.~H.~G.}\ \bibnamefont {{Lewin}}}\ and\ \bibinfo {editor}
  {\bibfnamefont {M.}~\bibnamefont {{van der Klis}}}}\ (\bibinfo {year}
  {2006})\ pp.\ \bibinfo {pages} {547--586}\BibitemShut {NoStop}%
\bibitem [{\citenamefont {{Thompson}}\ \emph {et~al.}(2000)\citenamefont
  {{Thompson}}, \citenamefont {{Duncan}}, \citenamefont {{Woods}},
  \citenamefont {{Kouveliotou}}, \citenamefont {{Finger}},\ and\ \citenamefont
  {{van Paradijs}}}]{Thompson2000}%
  \BibitemOpen
  \bibfield  {author} {\bibinfo {author} {\bibfnamefont {C.}~\bibnamefont
  {{Thompson}}}, \bibinfo {author} {\bibfnamefont {R.~C.}\ \bibnamefont
  {{Duncan}}}, \bibinfo {author} {\bibfnamefont {P.~M.}\ \bibnamefont
  {{Woods}}}, \bibinfo {author} {\bibfnamefont {C.}~\bibnamefont
  {{Kouveliotou}}}, \bibinfo {author} {\bibfnamefont {M.~H.}\ \bibnamefont
  {{Finger}}}, \ and\ \bibinfo {author} {\bibfnamefont {J.}~\bibnamefont {{van
  Paradijs}}},\ }\href {\doibase 10.1086/317072} {\bibfield  {journal}
  {\bibinfo  {journal} {\apj}\ }\textbf {\bibinfo {volume} {543}},\ \bibinfo
  {pages} {340} (\bibinfo {year} {2000})},\ \Eprint
  {http://arxiv.org/abs/astro-ph/9908086} {arXiv:astro-ph/9908086 [astro-ph]}
  \BibitemShut {NoStop}%
\bibitem [{\citenamefont {Kaspi}\ and\ \citenamefont
  {Beloborodov}(2017)}]{Kaspi_2017}%
  \BibitemOpen
  \bibfield  {author} {\bibinfo {author} {\bibfnamefont {V.~M.}\ \bibnamefont
  {Kaspi}}\ and\ \bibinfo {author} {\bibfnamefont {A.~M.}\ \bibnamefont
  {Beloborodov}},\ }\href {\doibase 10.1146/annurev-astro-081915-023329}
  {\bibfield  {journal} {\bibinfo  {journal} {Annual Review of Astronomy and
  Astrophysics}\ }\textbf {\bibinfo {volume} {55}},\ \bibinfo {pages}
  {261–301} (\bibinfo {year} {2017})}\BibitemShut {NoStop}%
\bibitem [{\citenamefont {Kuiper}\ \emph {et~al.}(2006)\citenamefont {Kuiper},
  \citenamefont {Hermsen}, \citenamefont {den Hartog},\ and\ \citenamefont
  {Collmar}}]{Kuiper_2006}%
  \BibitemOpen
  \bibfield  {author} {\bibinfo {author} {\bibfnamefont {L.}~\bibnamefont
  {Kuiper}}, \bibinfo {author} {\bibfnamefont {W.}~\bibnamefont {Hermsen}},
  \bibinfo {author} {\bibfnamefont {P.~R.}\ \bibnamefont {den Hartog}}, \ and\
  \bibinfo {author} {\bibfnamefont {W.}~\bibnamefont {Collmar}},\ }\href
  {\doibase 10.1086/504317} {\bibfield  {journal} {\bibinfo  {journal} {The
  Astrophysical Journal}\ }\textbf {\bibinfo {volume} {645}},\ \bibinfo {pages}
  {556–575} (\bibinfo {year} {2006})}\BibitemShut {NoStop}%
\bibitem [{\citenamefont {{G{\"o}tz}}\ \emph {et~al.}(2006)\citenamefont
  {{G{\"o}tz}}, \citenamefont {{Mereghetti}}, \citenamefont {{Tiengo}},\ and\
  \citenamefont {{Esposito}}}]{Gotz2006}%
  \BibitemOpen
  \bibfield  {author} {\bibinfo {author} {\bibfnamefont {D.}~\bibnamefont
  {{G{\"o}tz}}}, \bibinfo {author} {\bibfnamefont {S.}~\bibnamefont
  {{Mereghetti}}}, \bibinfo {author} {\bibfnamefont {A.}~\bibnamefont
  {{Tiengo}}}, \ and\ \bibinfo {author} {\bibfnamefont {P.}~\bibnamefont
  {{Esposito}}},\ }\href {\doibase 10.1051/0004-6361:20064870} {\bibfield
  {journal} {\bibinfo  {journal} {\aap}\ }\textbf {\bibinfo {volume} {449}},\
  \bibinfo {pages} {L31} (\bibinfo {year} {2006})},\ \Eprint
  {http://arxiv.org/abs/astro-ph/0602359} {arXiv:astro-ph/0602359 [astro-ph]}
  \BibitemShut {NoStop}%
\bibitem [{\citenamefont {Enoto}\ \emph {et~al.}(2017)\citenamefont {Enoto},
  \citenamefont {Shibata}, \citenamefont {Kitaguchi}, \citenamefont {Suwa},
  \citenamefont {Uchide}, \citenamefont {Nishioka}, \citenamefont {Kisaka},
  \citenamefont {Nakano}, \citenamefont {Murakami},\ and\ \citenamefont
  {Makishima}}]{Enoto_2017}%
  \BibitemOpen
  \bibfield  {author} {\bibinfo {author} {\bibfnamefont {T.}~\bibnamefont
  {Enoto}}, \bibinfo {author} {\bibfnamefont {S.}~\bibnamefont {Shibata}},
  \bibinfo {author} {\bibfnamefont {T.}~\bibnamefont {Kitaguchi}}, \bibinfo
  {author} {\bibfnamefont {Y.}~\bibnamefont {Suwa}}, \bibinfo {author}
  {\bibfnamefont {T.}~\bibnamefont {Uchide}}, \bibinfo {author} {\bibfnamefont
  {H.}~\bibnamefont {Nishioka}}, \bibinfo {author} {\bibfnamefont
  {S.}~\bibnamefont {Kisaka}}, \bibinfo {author} {\bibfnamefont
  {T.}~\bibnamefont {Nakano}}, \bibinfo {author} {\bibfnamefont
  {H.}~\bibnamefont {Murakami}}, \ and\ \bibinfo {author} {\bibfnamefont
  {K.}~\bibnamefont {Makishima}},\ }\href {\doibase 10.3847/1538-4365/aa6f0a}
  {\bibfield  {journal} {\bibinfo  {journal} {The Astrophysical Journal
  Supplement Series}\ }\textbf {\bibinfo {volume} {231}},\ \bibinfo {pages} {8}
  (\bibinfo {year} {2017})}\BibitemShut {NoStop}%
\bibitem [{\citenamefont {{Dib}}\ and\ \citenamefont
  {{Kaspi}}(2014)}]{DibKaspi2014}%
  \BibitemOpen
  \bibfield  {author} {\bibinfo {author} {\bibfnamefont {R.}~\bibnamefont
  {{Dib}}}\ and\ \bibinfo {author} {\bibfnamefont {V.~M.}\ \bibnamefont
  {{Kaspi}}},\ }\href {\doibase 10.1088/0004-637X/784/1/37} {\bibfield
  {journal} {\bibinfo  {journal} {\apj}\ }\textbf {\bibinfo {volume} {784}},\
  \bibinfo {eid} {37} (\bibinfo {year} {2014})},\ \Eprint
  {http://arxiv.org/abs/1401.3085} {arXiv:1401.3085 [astro-ph.HE]} \BibitemShut
  {NoStop}%
\bibitem [{\citenamefont {{Hulleman}}\ \emph {et~al.}(2001)\citenamefont
  {{Hulleman}}, \citenamefont {{Tennant}}, \citenamefont {{van Kerkwijk}},
  \citenamefont {{Kulkarni}}, \citenamefont {{Kouveliotou}},\ and\
  \citenamefont {{Patel}}}]{Hulleman2001}%
  \BibitemOpen
  \bibfield  {author} {\bibinfo {author} {\bibfnamefont {F.}~\bibnamefont
  {{Hulleman}}}, \bibinfo {author} {\bibfnamefont {A.~F.}\ \bibnamefont
  {{Tennant}}}, \bibinfo {author} {\bibfnamefont {M.~H.}\ \bibnamefont {{van
  Kerkwijk}}}, \bibinfo {author} {\bibfnamefont {S.~R.}\ \bibnamefont
  {{Kulkarni}}}, \bibinfo {author} {\bibfnamefont {C.}~\bibnamefont
  {{Kouveliotou}}}, \ and\ \bibinfo {author} {\bibfnamefont {S.~K.}\
  \bibnamefont {{Patel}}},\ }\href {\doibase 10.1086/338478} {\bibfield
  {journal} {\bibinfo  {journal} {\apjl}\ }\textbf {\bibinfo {volume} {563}},\
  \bibinfo {pages} {L49} (\bibinfo {year} {2001})},\ \Eprint
  {http://arxiv.org/abs/astro-ph/0110172} {arXiv:astro-ph/0110172 [astro-ph]}
  \BibitemShut {NoStop}%
\bibitem [{\citenamefont {Durant}\ and\ \citenamefont {van
  Kerkwijk}(2006)}]{Durant:2006qd}%
  \BibitemOpen
  \bibfield  {author} {\bibinfo {author} {\bibfnamefont {M.}~\bibnamefont
  {Durant}}\ and\ \bibinfo {author} {\bibfnamefont {M.~H.}\ \bibnamefont {van
  Kerkwijk}},\ }\href {\doibase 10.1086/505740} {\bibfield  {journal} {\bibinfo
   {journal} {Astrophys. J.}\ }\textbf {\bibinfo {volume} {648}},\ \bibinfo
  {pages} {534} (\bibinfo {year} {2006})},\ \Eprint
  {http://arxiv.org/abs/astro-ph/0605330} {arXiv:astro-ph/0605330} \BibitemShut
  {NoStop}%
\bibitem [{\citenamefont {Beloborodov}\ and\ \citenamefont
  {Li}(2016)}]{Beloborodov:2016mmx}%
  \BibitemOpen
  \bibfield  {author} {\bibinfo {author} {\bibfnamefont {A.~M.}\ \bibnamefont
  {Beloborodov}}\ and\ \bibinfo {author} {\bibfnamefont {X.}~\bibnamefont
  {Li}},\ }\href {\doibase 10.3847/1538-4357/833/2/261} {\bibfield  {journal}
  {\bibinfo  {journal} {Astrophys. J.}\ }\textbf {\bibinfo {volume} {833}},\
  \bibinfo {pages} {261} (\bibinfo {year} {2016})},\ \Eprint
  {http://arxiv.org/abs/1605.09077} {arXiv:1605.09077 [astro-ph.HE]}
  \BibitemShut {NoStop}%
\bibitem [{\citenamefont {{Beloborodov}}\ and\ \citenamefont
  {{Thompson}}(2007)}]{BeloborodovThompson2007}%
  \BibitemOpen
  \bibfield  {author} {\bibinfo {author} {\bibfnamefont {A.~M.}\ \bibnamefont
  {{Beloborodov}}}\ and\ \bibinfo {author} {\bibfnamefont {C.}~\bibnamefont
  {{Thompson}}},\ }\href {\doibase 10.1086/508917} {\bibfield  {journal}
  {\bibinfo  {journal} {\apj}\ }\textbf {\bibinfo {volume} {657}},\ \bibinfo
  {pages} {967} (\bibinfo {year} {2007})},\ \Eprint
  {http://arxiv.org/abs/astro-ph/0602417} {arXiv:astro-ph/0602417 [astro-ph]}
  \BibitemShut {NoStop}%
\bibitem [{\citenamefont {Thompson}\ and\ \citenamefont
  {Kostenko}(2020)}]{ThompsonKostenko2020}%
  \BibitemOpen
  \bibfield  {author} {\bibinfo {author} {\bibfnamefont {C.}~\bibnamefont
  {Thompson}}\ and\ \bibinfo {author} {\bibfnamefont {A.}~\bibnamefont
  {Kostenko}},\ }\href {\doibase 10.3847/1538-4357/abbe87} {\bibfield
  {journal} {\bibinfo  {journal} {The Astrophysical Journal}\ }\textbf
  {\bibinfo {volume} {904}},\ \bibinfo {pages} {184} (\bibinfo {year}
  {2020})}\BibitemShut {NoStop}%
\bibitem [{\citenamefont {Zhang}\ and\ \citenamefont
  {Thompson}(2024)}]{ZhangThompson2024}%
  \BibitemOpen
  \bibfield  {author} {\bibinfo {author} {\bibfnamefont {J.}~\bibnamefont
  {Zhang}}\ and\ \bibinfo {author} {\bibfnamefont {C.}~\bibnamefont
  {Thompson}},\ }\href@noop {} {\  (\bibinfo {year} {2024})},\ \Eprint
  {http://arxiv.org/abs/2407.08810} {arXiv:2407.08810 [astro-ph.HE]}
  \BibitemShut {NoStop}%
\bibitem [{\citenamefont {{Kostenko}}\ and\ \citenamefont
  {{Thompson}}(2018)}]{KostenkoThompson2018}%
  \BibitemOpen
  \bibfield  {author} {\bibinfo {author} {\bibfnamefont {A.}~\bibnamefont
  {{Kostenko}}}\ and\ \bibinfo {author} {\bibfnamefont {C.}~\bibnamefont
  {{Thompson}}},\ }\href {\doibase 10.3847/1538-4357/aae0ef} {\bibfield
  {journal} {\bibinfo  {journal} {\apj}\ }\textbf {\bibinfo {volume} {869}},\
  \bibinfo {eid} {44} (\bibinfo {year} {2018})},\ \Eprint
  {http://arxiv.org/abs/1904.03324} {arXiv:1904.03324 [astro-ph.HE]}
  \BibitemShut {NoStop}%
\bibitem [{\citenamefont {{Thompson}}\ and\ \citenamefont
  {{Duncan}}(1995)}]{ThompsonDuncan1995}%
  \BibitemOpen
  \bibfield  {author} {\bibinfo {author} {\bibfnamefont {C.}~\bibnamefont
  {{Thompson}}}\ and\ \bibinfo {author} {\bibfnamefont {R.~C.}\ \bibnamefont
  {{Duncan}}},\ }\href {\doibase 10.1093/mnras/275.2.255} {\bibfield  {journal}
  {\bibinfo  {journal} {\mnras}\ }\textbf {\bibinfo {volume} {275}},\ \bibinfo
  {pages} {255} (\bibinfo {year} {1995})}\BibitemShut {NoStop}%
\bibitem [{\citenamefont {{Thompson}}\ and\ \citenamefont
  {{Duncan}}(2001)}]{ThompsonDuncan2001}%
  \BibitemOpen
  \bibfield  {author} {\bibinfo {author} {\bibfnamefont {C.}~\bibnamefont
  {{Thompson}}}\ and\ \bibinfo {author} {\bibfnamefont {R.~C.}\ \bibnamefont
  {{Duncan}}},\ }\href {\doibase 10.1086/323256} {\bibfield  {journal}
  {\bibinfo  {journal} {\apj}\ }\textbf {\bibinfo {volume} {561}},\ \bibinfo
  {pages} {980} (\bibinfo {year} {2001})},\ \Eprint
  {http://arxiv.org/abs/astro-ph/0110675} {arXiv:astro-ph/0110675 [astro-ph]}
  \BibitemShut {NoStop}%
\bibitem [{\citenamefont {Thompson}\ \emph {et~al.}(2017)\citenamefont
  {Thompson}, \citenamefont {Yang},\ and\ \citenamefont
  {Ortiz}}]{Thompson_2017}%
  \BibitemOpen
  \bibfield  {author} {\bibinfo {author} {\bibfnamefont {C.}~\bibnamefont
  {Thompson}}, \bibinfo {author} {\bibfnamefont {H.}~\bibnamefont {Yang}}, \
  and\ \bibinfo {author} {\bibfnamefont {N.}~\bibnamefont {Ortiz}},\ }\href
  {\doibase 10.3847/1538-4357/aa6c30} {\bibfield  {journal} {\bibinfo
  {journal} {The Astrophysical Journal}\ }\textbf {\bibinfo {volume} {841}},\
  \bibinfo {pages} {54} (\bibinfo {year} {2017})}\BibitemShut {NoStop}%
\bibitem [{\citenamefont {{Parfrey}}\ \emph {et~al.}(2012)\citenamefont
  {{Parfrey}}, \citenamefont {{Beloborodov}},\ and\ \citenamefont
  {{Hui}}}]{Parfrey2012}%
  \BibitemOpen
  \bibfield  {author} {\bibinfo {author} {\bibfnamefont {K.}~\bibnamefont
  {{Parfrey}}}, \bibinfo {author} {\bibfnamefont {A.~M.}\ \bibnamefont
  {{Beloborodov}}}, \ and\ \bibinfo {author} {\bibfnamefont {L.}~\bibnamefont
  {{Hui}}},\ }\href {\doibase 10.1088/2041-8205/754/1/L12} {\bibfield
  {journal} {\bibinfo  {journal} {\apjl}\ }\textbf {\bibinfo {volume} {754}},\
  \bibinfo {eid} {L12} (\bibinfo {year} {2012})},\ \Eprint
  {http://arxiv.org/abs/1201.3635} {arXiv:1201.3635 [astro-ph.HE]} \BibitemShut
  {NoStop}%
\bibitem [{\citenamefont {{Parfrey}}\ \emph {et~al.}(2013)\citenamefont
  {{Parfrey}}, \citenamefont {{Beloborodov}},\ and\ \citenamefont
  {{Hui}}}]{Parfrey2013}%
  \BibitemOpen
  \bibfield  {author} {\bibinfo {author} {\bibfnamefont {K.}~\bibnamefont
  {{Parfrey}}}, \bibinfo {author} {\bibfnamefont {A.~M.}\ \bibnamefont
  {{Beloborodov}}}, \ and\ \bibinfo {author} {\bibfnamefont {L.}~\bibnamefont
  {{Hui}}},\ }\href {\doibase 10.1088/0004-637X/774/2/92} {\bibfield  {journal}
  {\bibinfo  {journal} {\apj}\ }\textbf {\bibinfo {volume} {774}},\ \bibinfo
  {eid} {92} (\bibinfo {year} {2013})},\ \Eprint
  {http://arxiv.org/abs/1306.4335} {arXiv:1306.4335 [astro-ph.HE]} \BibitemShut
  {NoStop}%
\bibitem [{\citenamefont {Mahlmann}\ \emph {et~al.}(2019)\citenamefont
  {Mahlmann}, \citenamefont {Akg\"un}, \citenamefont {Pons}, \citenamefont
  {Aloy},\ and\ \citenamefont {Cerd\'a-Dur\'an}}]{Mahlmann:2019arj}%
  \BibitemOpen
  \bibfield  {author} {\bibinfo {author} {\bibfnamefont {J.~F.}\ \bibnamefont
  {Mahlmann}}, \bibinfo {author} {\bibfnamefont {T.}~\bibnamefont {Akg\"un}},
  \bibinfo {author} {\bibfnamefont {J.~A.}\ \bibnamefont {Pons}}, \bibinfo
  {author} {\bibfnamefont {M.~A.}\ \bibnamefont {Aloy}}, \ and\ \bibinfo
  {author} {\bibfnamefont {P.}~\bibnamefont {Cerd\'a-Dur\'an}},\ }\href
  {\doibase 10.1093/mnras/stz2729} {\bibfield  {journal} {\bibinfo  {journal}
  {Mon. Not. Roy. Astron. Soc.}\ }\textbf {\bibinfo {volume} {490}},\ \bibinfo
  {pages} {4858} (\bibinfo {year} {2019})},\ \Eprint
  {http://arxiv.org/abs/1908.00010} {arXiv:1908.00010 [astro-ph.HE]}
  \BibitemShut {NoStop}%
\bibitem [{\citenamefont {Mahlmann}\ \emph {et~al.}(2023)\citenamefont
  {Mahlmann}, \citenamefont {Philippov}, \citenamefont {Mewes}, \citenamefont
  {Ripperda}, \citenamefont {Most},\ and\ \citenamefont
  {Sironi}}]{Mahlmann_2023}%
  \BibitemOpen
  \bibfield  {author} {\bibinfo {author} {\bibfnamefont {J.~F.}\ \bibnamefont
  {Mahlmann}}, \bibinfo {author} {\bibfnamefont {A.~A.}\ \bibnamefont
  {Philippov}}, \bibinfo {author} {\bibfnamefont {V.}~\bibnamefont {Mewes}},
  \bibinfo {author} {\bibfnamefont {B.}~\bibnamefont {Ripperda}}, \bibinfo
  {author} {\bibfnamefont {E.~R.}\ \bibnamefont {Most}}, \ and\ \bibinfo
  {author} {\bibfnamefont {L.}~\bibnamefont {Sironi}},\ }\href {\doibase
  10.3847/2041-8213/accada} {\bibfield  {journal} {\bibinfo  {journal} {The
  Astrophysical Journal Letters}\ }\textbf {\bibinfo {volume} {947}},\ \bibinfo
  {pages} {L34} (\bibinfo {year} {2023})}\BibitemShut {NoStop}%
\bibitem [{\citenamefont {Beloborodov}(2009)}]{Beloborodov_2009}%
  \BibitemOpen
  \bibfield  {author} {\bibinfo {author} {\bibfnamefont {A.~M.}\ \bibnamefont
  {Beloborodov}},\ }\href {\doibase 10.1088/0004-637X/703/1/1044} {\bibfield
  {journal} {\bibinfo  {journal} {The Astrophysical Journal}\ }\textbf
  {\bibinfo {volume} {703}},\ \bibinfo {pages} {1044} (\bibinfo {year}
  {2009})}\BibitemShut {NoStop}%
\bibitem [{\citenamefont {{G{\"o}{\v{g}}{\"u}{\c{S}} }}\ \emph
  {et~al.}(1999)\citenamefont {{G{\"o}{\v{g}}{\"u}{\c{S}} }}, \citenamefont
  {{Woods}}, \citenamefont {{Kouveliotou}}, \citenamefont {{van Paradijs}},
  \citenamefont {{Briggs}}, \citenamefont {{Duncan}},\ and\ \citenamefont
  {{Thompson}}}]{Gogus1999}%
  \BibitemOpen
  \bibfield  {author} {\bibinfo {author} {\bibfnamefont {E.}~\bibnamefont
  {{G{\"o}{\v{g}}{\"u}{\c{S}} }}}, \bibinfo {author} {\bibfnamefont {P.~M.}\
  \bibnamefont {{Woods}}}, \bibinfo {author} {\bibfnamefont {C.}~\bibnamefont
  {{Kouveliotou}}}, \bibinfo {author} {\bibfnamefont {J.}~\bibnamefont {{van
  Paradijs}}}, \bibinfo {author} {\bibfnamefont {M.~S.}\ \bibnamefont
  {{Briggs}}}, \bibinfo {author} {\bibfnamefont {R.~C.}\ \bibnamefont
  {{Duncan}}}, \ and\ \bibinfo {author} {\bibfnamefont {C.}~\bibnamefont
  {{Thompson}}},\ }\href {\doibase 10.1086/312380} {\bibfield  {journal}
  {\bibinfo  {journal} {\apjl}\ }\textbf {\bibinfo {volume} {526}},\ \bibinfo
  {pages} {L93} (\bibinfo {year} {1999})},\ \Eprint
  {http://arxiv.org/abs/astro-ph/9910062} {arXiv:astro-ph/9910062 [astro-ph]}
  \BibitemShut {NoStop}%
\bibitem [{\citenamefont {{G{\"o}{\v{g}}{\"u}{\c{s}}}}\ \emph
  {et~al.}(2000)\citenamefont {{G{\"o}{\v{g}}{\"u}{\c{s}}}}, \citenamefont
  {{Woods}}, \citenamefont {{Kouveliotou}}, \citenamefont {{van Paradijs}},
  \citenamefont {{Briggs}}, \citenamefont {{Duncan}},\ and\ \citenamefont
  {{Thompson}}}]{Gogus2000}%
  \BibitemOpen
  \bibfield  {author} {\bibinfo {author} {\bibfnamefont {E.}~\bibnamefont
  {{G{\"o}{\v{g}}{\"u}{\c{s}}}}}, \bibinfo {author} {\bibfnamefont {P.~M.}\
  \bibnamefont {{Woods}}}, \bibinfo {author} {\bibfnamefont {C.}~\bibnamefont
  {{Kouveliotou}}}, \bibinfo {author} {\bibfnamefont {J.}~\bibnamefont {{van
  Paradijs}}}, \bibinfo {author} {\bibfnamefont {M.~S.}\ \bibnamefont
  {{Briggs}}}, \bibinfo {author} {\bibfnamefont {R.~C.}\ \bibnamefont
  {{Duncan}}}, \ and\ \bibinfo {author} {\bibfnamefont {C.}~\bibnamefont
  {{Thompson}}},\ }\href {\doibase 10.1086/312583} {\bibfield  {journal}
  {\bibinfo  {journal} {\apjl}\ }\textbf {\bibinfo {volume} {532}},\ \bibinfo
  {pages} {L121} (\bibinfo {year} {2000})},\ \Eprint
  {http://arxiv.org/abs/astro-ph/0002181} {arXiv:astro-ph/0002181 [astro-ph]}
  \BibitemShut {NoStop}%
\bibitem [{\citenamefont {Rea}\ \emph {et~al.}(2009)\citenamefont {Rea},
  \citenamefont {Israel}, \citenamefont {Turolla}, \citenamefont {Esposito},
  \citenamefont {Mereghetti}, \citenamefont {Götz}, \citenamefont {Zane},
  \citenamefont {Tiengo}, \citenamefont {Hurley}, \citenamefont {Feroci},
  \citenamefont {Still}, \citenamefont {Yershov}, \citenamefont {Winkler},
  \citenamefont {Perna}, \citenamefont {Bernardini}, \citenamefont {Ubertini},
  \citenamefont {Stella}, \citenamefont {Campana}, \citenamefont {Van
  Der~Klis},\ and\ \citenamefont {Woods}}]{Rea2009}%
  \BibitemOpen
  \bibfield  {author} {\bibinfo {author} {\bibfnamefont {N.}~\bibnamefont
  {Rea}}, \bibinfo {author} {\bibfnamefont {G.~L.}\ \bibnamefont {Israel}},
  \bibinfo {author} {\bibfnamefont {R.}~\bibnamefont {Turolla}}, \bibinfo
  {author} {\bibfnamefont {P.}~\bibnamefont {Esposito}}, \bibinfo {author}
  {\bibfnamefont {S.}~\bibnamefont {Mereghetti}}, \bibinfo {author}
  {\bibfnamefont {D.}~\bibnamefont {Götz}}, \bibinfo {author} {\bibfnamefont
  {S.}~\bibnamefont {Zane}}, \bibinfo {author} {\bibfnamefont {A.}~\bibnamefont
  {Tiengo}}, \bibinfo {author} {\bibfnamefont {K.}~\bibnamefont {Hurley}},
  \bibinfo {author} {\bibfnamefont {M.}~\bibnamefont {Feroci}}, \bibinfo
  {author} {\bibfnamefont {M.}~\bibnamefont {Still}}, \bibinfo {author}
  {\bibfnamefont {V.}~\bibnamefont {Yershov}}, \bibinfo {author} {\bibfnamefont
  {C.}~\bibnamefont {Winkler}}, \bibinfo {author} {\bibfnamefont
  {R.}~\bibnamefont {Perna}}, \bibinfo {author} {\bibfnamefont
  {F.}~\bibnamefont {Bernardini}}, \bibinfo {author} {\bibfnamefont
  {P.}~\bibnamefont {Ubertini}}, \bibinfo {author} {\bibfnamefont
  {L.}~\bibnamefont {Stella}}, \bibinfo {author} {\bibfnamefont
  {S.}~\bibnamefont {Campana}}, \bibinfo {author} {\bibfnamefont
  {M.}~\bibnamefont {Van Der~Klis}}, \ and\ \bibinfo {author} {\bibfnamefont
  {P.}~\bibnamefont {Woods}},\ }\href {\doibase
  10.1111/j.1365-2966.2009.14920.x} {\bibfield  {journal} {\bibinfo  {journal}
  {Monthly Notices of the Royal Astronomical Society}\ }\textbf {\bibinfo
  {volume} {396}},\ \bibinfo {pages} {2419} (\bibinfo {year} {2009})},\ \Eprint
  {http://arxiv.org/abs/https://academic.oup.com/mnras/article-pdf/396/4/2419/18440830/mnras0396-2419.pdf}
  {https://academic.oup.com/mnras/article-pdf/396/4/2419/18440830/mnras0396-2419.pdf}
  \BibitemShut {NoStop}%
\bibitem [{\citenamefont {Rea}\ and\ \citenamefont
  {Esposito}(2010)}]{Rea_2010}%
  \BibitemOpen
  \bibfield  {author} {\bibinfo {author} {\bibfnamefont {N.}~\bibnamefont
  {Rea}}\ and\ \bibinfo {author} {\bibfnamefont {P.}~\bibnamefont {Esposito}},\
  }\enquote {\bibinfo {title} {Magnetar outbursts: an observational review},}\
  in\ \href {\doibase 10.1007/978-3-642-17251-9_21} {\emph {\bibinfo
  {booktitle} {High-Energy Emission from Pulsars and their Systems}}}\
  (\bibinfo  {publisher} {Springer Berlin Heidelberg},\ \bibinfo {year}
  {2010})\ p.\ \bibinfo {pages} {247–273}\BibitemShut {NoStop}%
\bibitem [{\citenamefont {{Esposito}}\ \emph {et~al.}(2020)\citenamefont
  {{Esposito}}, \citenamefont {{Rea}}, \citenamefont {{Borghese}},
  \citenamefont {{Coti Zelati}}, \citenamefont {{Vigan{\`o}}}, \citenamefont
  {{Israel}}, \citenamefont {{Tiengo}}, \citenamefont {{Ridolfi}},
  \citenamefont {{Possenti}}, \citenamefont {{Burgay}}, \citenamefont
  {{G{\"o}tz}}, \citenamefont {{Pintore}}, \citenamefont {{Stella}},
  \citenamefont {{Dehman}}, \citenamefont {{Ronchi}}, \citenamefont
  {{Campana}}, \citenamefont {{Garcia-Garcia}}, \citenamefont {{Graber}},
  \citenamefont {{Mereghetti}}, \citenamefont {{Perna}}, \citenamefont
  {{Rodr{\'\i}guez Castillo}}, \citenamefont {{Turolla}},\ and\ \citenamefont
  {{Zane}}}]{Esposito2020}%
  \BibitemOpen
  \bibfield  {author} {\bibinfo {author} {\bibfnamefont {P.}~\bibnamefont
  {{Esposito}}}, \bibinfo {author} {\bibfnamefont {N.}~\bibnamefont {{Rea}}},
  \bibinfo {author} {\bibfnamefont {A.}~\bibnamefont {{Borghese}}}, \bibinfo
  {author} {\bibfnamefont {F.}~\bibnamefont {{Coti Zelati}}}, \bibinfo {author}
  {\bibfnamefont {D.}~\bibnamefont {{Vigan{\`o}}}}, \bibinfo {author}
  {\bibfnamefont {G.~L.}\ \bibnamefont {{Israel}}}, \bibinfo {author}
  {\bibfnamefont {A.}~\bibnamefont {{Tiengo}}}, \bibinfo {author}
  {\bibfnamefont {A.}~\bibnamefont {{Ridolfi}}}, \bibinfo {author}
  {\bibfnamefont {A.}~\bibnamefont {{Possenti}}}, \bibinfo {author}
  {\bibfnamefont {M.}~\bibnamefont {{Burgay}}}, \bibinfo {author}
  {\bibfnamefont {D.}~\bibnamefont {{G{\"o}tz}}}, \bibinfo {author}
  {\bibfnamefont {F.}~\bibnamefont {{Pintore}}}, \bibinfo {author}
  {\bibfnamefont {L.}~\bibnamefont {{Stella}}}, \bibinfo {author}
  {\bibfnamefont {C.}~\bibnamefont {{Dehman}}}, \bibinfo {author}
  {\bibfnamefont {M.}~\bibnamefont {{Ronchi}}}, \bibinfo {author}
  {\bibfnamefont {S.}~\bibnamefont {{Campana}}}, \bibinfo {author}
  {\bibfnamefont {A.}~\bibnamefont {{Garcia-Garcia}}}, \bibinfo {author}
  {\bibfnamefont {V.}~\bibnamefont {{Graber}}}, \bibinfo {author}
  {\bibfnamefont {S.}~\bibnamefont {{Mereghetti}}}, \bibinfo {author}
  {\bibfnamefont {R.}~\bibnamefont {{Perna}}}, \bibinfo {author} {\bibfnamefont
  {G.~A.}\ \bibnamefont {{Rodr{\'\i}guez Castillo}}}, \bibinfo {author}
  {\bibfnamefont {R.}~\bibnamefont {{Turolla}}}, \ and\ \bibinfo {author}
  {\bibfnamefont {S.}~\bibnamefont {{Zane}}},\ }\href {\doibase
  10.3847/2041-8213/ab9742} {\bibfield  {journal} {\bibinfo  {journal} {\apjl}\
  }\textbf {\bibinfo {volume} {896}},\ \bibinfo {eid} {L30} (\bibinfo {year}
  {2020})},\ \Eprint {http://arxiv.org/abs/2004.04083} {arXiv:2004.04083
  [astro-ph.HE]} \BibitemShut {NoStop}%
\bibitem [{\citenamefont {Bochenek}(2020)}]{Bochenek2020}%
  \BibitemOpen
  \bibfield  {author} {\bibinfo {author} {\bibfnamefont {C.~D.~o.}\
  \bibnamefont {Bochenek}},\ }\href {\doibase 10.1038/s41586-020-2872-x}
  {\bibfield  {journal} {\bibinfo  {journal} {Nature}\ }\textbf {\bibinfo
  {volume} {587}},\ \bibinfo {pages} {59–62} (\bibinfo {year}
  {2020})}\BibitemShut {NoStop}%
\bibitem [{\citenamefont {Amiri}\ \emph {et~al.}(2018)\citenamefont {Amiri}
  \emph {et~al.}}]{CHIME1}%
  \BibitemOpen
  \bibfield  {author} {\bibinfo {author} {\bibfnamefont {M.}~\bibnamefont
  {Amiri}} \emph {et~al.} (\bibinfo {collaboration} {CHIME/FRB}),\ }\href
  {\doibase 10.3847/1538-4357/aad188} {\bibfield  {journal} {\bibinfo
  {journal} {The Astrophysical Journal}\ }\textbf {\bibinfo {volume} {863}},\
  \bibinfo {pages} {48} (\bibinfo {year} {2018})}\BibitemShut {NoStop}%
\bibitem [{\citenamefont {Amiri}\ \emph {et~al.}(2019)\citenamefont {Amiri}
  \emph {et~al.}}]{CHIME2}%
  \BibitemOpen
  \bibfield  {author} {\bibinfo {author} {\bibfnamefont {M.}~\bibnamefont
  {Amiri}} \emph {et~al.} (\bibinfo {collaboration} {CHIME/FRB}),\ }\href
  {\doibase 10.1038/s41586-018-0867-7} {\bibfield  {journal} {\bibinfo
  {journal} {Nature}\ }\textbf {\bibinfo {volume} {566}},\ \bibinfo {pages}
  {230} (\bibinfo {year} {2019})},\ \Eprint {http://arxiv.org/abs/1901.04524}
  {arXiv:1901.04524 [astro-ph.HE]} \BibitemShut {NoStop}%
\bibitem [{\citenamefont {Woods}\ \emph {et~al.}(2004)\citenamefont {Woods},
  \citenamefont {Kaspi}, \citenamefont {Thompson}, \citenamefont {Gavriil},
  \citenamefont {Marshall}, \citenamefont {Chakrabarty}, \citenamefont
  {Flanagan}, \citenamefont {Heyl},\ and\ \citenamefont
  {Hernquist}}]{Woods:2003si}%
  \BibitemOpen
  \bibfield  {author} {\bibinfo {author} {\bibfnamefont {P.~M.}\ \bibnamefont
  {Woods}}, \bibinfo {author} {\bibfnamefont {V.~M.}\ \bibnamefont {Kaspi}},
  \bibinfo {author} {\bibfnamefont {C.}~\bibnamefont {Thompson}}, \bibinfo
  {author} {\bibfnamefont {F.~P.}\ \bibnamefont {Gavriil}}, \bibinfo {author}
  {\bibfnamefont {H.~L.}\ \bibnamefont {Marshall}}, \bibinfo {author}
  {\bibfnamefont {D.}~\bibnamefont {Chakrabarty}}, \bibinfo {author}
  {\bibfnamefont {K.}~\bibnamefont {Flanagan}}, \bibinfo {author}
  {\bibfnamefont {J.}~\bibnamefont {Heyl}}, \ and\ \bibinfo {author}
  {\bibfnamefont {L.}~\bibnamefont {Hernquist}},\ }\href {\doibase
  10.1086/382233} {\bibfield  {journal} {\bibinfo  {journal} {Astrophys. J.}\
  }\textbf {\bibinfo {volume} {605}},\ \bibinfo {pages} {378} (\bibinfo {year}
  {2004})},\ \Eprint {http://arxiv.org/abs/astro-ph/0310575}
  {arXiv:astro-ph/0310575} \BibitemShut {NoStop}%
\bibitem [{\citenamefont {{Lyubarsky}}\ \emph {et~al.}(2002)\citenamefont
  {{Lyubarsky}}, \citenamefont {{Eichler}},\ and\ \citenamefont
  {{Thompson}}}]{LyubarskyEichlerThompson2002}%
  \BibitemOpen
  \bibfield  {author} {\bibinfo {author} {\bibfnamefont {Y.}~\bibnamefont
  {{Lyubarsky}}}, \bibinfo {author} {\bibfnamefont {D.}~\bibnamefont
  {{Eichler}}}, \ and\ \bibinfo {author} {\bibfnamefont {C.}~\bibnamefont
  {{Thompson}}},\ }\href {\doibase 10.1086/345402} {\bibfield  {journal}
  {\bibinfo  {journal} {\apjl}\ }\textbf {\bibinfo {volume} {580}},\ \bibinfo
  {pages} {L69} (\bibinfo {year} {2002})},\ \Eprint
  {http://arxiv.org/abs/astro-ph/0211110} {arXiv:astro-ph/0211110 [astro-ph]}
  \BibitemShut {NoStop}%
\bibitem [{\citenamefont {Li}\ and\ \citenamefont
  {Beloborodov}(2015)}]{Li:2015epa}%
  \BibitemOpen
  \bibfield  {author} {\bibinfo {author} {\bibfnamefont {X.}~\bibnamefont
  {Li}}\ and\ \bibinfo {author} {\bibfnamefont {A.~M.}\ \bibnamefont
  {Beloborodov}},\ }\href {\doibase 10.1088/0004-637X/815/1/25} {\bibfield
  {journal} {\bibinfo  {journal} {Astrophys. J.}\ }\textbf {\bibinfo {volume}
  {815}},\ \bibinfo {pages} {25} (\bibinfo {year} {2015})},\ \Eprint
  {http://arxiv.org/abs/1505.03465} {arXiv:1505.03465 [astro-ph.HE]}
  \BibitemShut {NoStop}%
\bibitem [{\citenamefont {{Thompson}}\ \emph {et~al.}(2002)\citenamefont
  {{Thompson}}, \citenamefont {{Lyutikov}},\ and\ \citenamefont
  {{Kulkarni}}}]{TLK02}%
  \BibitemOpen
  \bibfield  {author} {\bibinfo {author} {\bibfnamefont {C.}~\bibnamefont
  {{Thompson}}}, \bibinfo {author} {\bibfnamefont {M.}~\bibnamefont
  {{Lyutikov}}}, \ and\ \bibinfo {author} {\bibfnamefont {S.~R.}\ \bibnamefont
  {{Kulkarni}}},\ }\href {\doibase 10.1086/340586} {\bibfield  {journal}
  {\bibinfo  {journal} {\apj}\ }\textbf {\bibinfo {volume} {574}},\ \bibinfo
  {pages} {332} (\bibinfo {year} {2002})},\ \Eprint
  {http://arxiv.org/abs/astro-ph/0110677} {arXiv:astro-ph/0110677 [astro-ph]}
  \BibitemShut {NoStop}%
\bibitem [{\citenamefont {{Goldreich}}\ and\ \citenamefont
  {{Reisenegger}}(1992)}]{GoldreichReisenegger1992}%
  \BibitemOpen
  \bibfield  {author} {\bibinfo {author} {\bibfnamefont {P.}~\bibnamefont
  {{Goldreich}}}\ and\ \bibinfo {author} {\bibfnamefont {A.}~\bibnamefont
  {{Reisenegger}}},\ }\href {\doibase 10.1086/171646} {\bibfield  {journal}
  {\bibinfo  {journal} {\apj}\ }\textbf {\bibinfo {volume} {395}},\ \bibinfo
  {pages} {250} (\bibinfo {year} {1992})}\BibitemShut {NoStop}%
\bibitem [{\citenamefont {Cumming}\ \emph {et~al.}(2004)\citenamefont
  {Cumming}, \citenamefont {Arras},\ and\ \citenamefont
  {Zweibel}}]{Cumming:2004mf}%
  \BibitemOpen
  \bibfield  {author} {\bibinfo {author} {\bibfnamefont {A.}~\bibnamefont
  {Cumming}}, \bibinfo {author} {\bibfnamefont {P.}~\bibnamefont {Arras}}, \
  and\ \bibinfo {author} {\bibfnamefont {E.~G.}\ \bibnamefont {Zweibel}},\
  }\href {\doibase 10.1086/421324} {\bibfield  {journal} {\bibinfo  {journal}
  {Astrophys. J.}\ }\textbf {\bibinfo {volume} {609}},\ \bibinfo {pages} {999}
  (\bibinfo {year} {2004})},\ \Eprint {http://arxiv.org/abs/astro-ph/0402392}
  {arXiv:astro-ph/0402392} \BibitemShut {NoStop}%
\bibitem [{\citenamefont {Gourgouliatos}\ and\ \citenamefont
  {Cumming}(2013)}]{GourgouliatosCumming2013}%
  \BibitemOpen
  \bibfield  {author} {\bibinfo {author} {\bibfnamefont {K.~N.}\ \bibnamefont
  {Gourgouliatos}}\ and\ \bibinfo {author} {\bibfnamefont {A.}~\bibnamefont
  {Cumming}},\ }\href {\doibase 10.1093/mnras/stt2300} {\bibfield  {journal}
  {\bibinfo  {journal} {Monthly Notices of the Royal Astronomical Society}\
  }\textbf {\bibinfo {volume} {438}},\ \bibinfo {pages} {1618} (\bibinfo {year}
  {2013})},\ \Eprint
  {http://arxiv.org/abs/https://academic.oup.com/mnras/article-pdf/438/2/1618/18501517/stt2300.pdf}
  {https://academic.oup.com/mnras/article-pdf/438/2/1618/18501517/stt2300.pdf}
  \BibitemShut {NoStop}%
\bibitem [{\citenamefont {Timokhin}\ and\ \citenamefont
  {Harding}(2015)}]{Timokhin:2015dua}%
  \BibitemOpen
  \bibfield  {author} {\bibinfo {author} {\bibfnamefont {A.~N.}\ \bibnamefont
  {Timokhin}}\ and\ \bibinfo {author} {\bibfnamefont {A.~K.}\ \bibnamefont
  {Harding}},\ }\href {\doibase 10.1088/0004-637X/810/2/144} {\bibfield
  {journal} {\bibinfo  {journal} {Astrophys. J.}\ }\textbf {\bibinfo {volume}
  {810}},\ \bibinfo {pages} {144} (\bibinfo {year} {2015})},\ \Eprint
  {http://arxiv.org/abs/1504.02194} {arXiv:1504.02194 [astro-ph.HE]}
  \BibitemShut {NoStop}%
\bibitem [{\citenamefont {{Philippov}}\ and\ \citenamefont
  {{Kramer}}(2022)}]{Philippov2022Review}%
  \BibitemOpen
  \bibfield  {author} {\bibinfo {author} {\bibfnamefont {A.}~\bibnamefont
  {{Philippov}}}\ and\ \bibinfo {author} {\bibfnamefont {M.}~\bibnamefont
  {{Kramer}}},\ }\href {\doibase 10.1146/annurev-astro-052920-112338}
  {\bibfield  {journal} {\bibinfo  {journal} {Annual Review of Astron and
  Astrophys}\ }\textbf {\bibinfo {volume} {60}},\ \bibinfo {pages} {495}
  (\bibinfo {year} {2022})}\BibitemShut {NoStop}%
\bibitem [{\citenamefont {{Hibschman}}\ and\ \citenamefont
  {{Arons}}(2001)}]{HibschmanArons2001}%
  \BibitemOpen
  \bibfield  {author} {\bibinfo {author} {\bibfnamefont {J.~A.}\ \bibnamefont
  {{Hibschman}}}\ and\ \bibinfo {author} {\bibfnamefont {J.}~\bibnamefont
  {{Arons}}},\ }\href {\doibase 10.1086/321378} {\bibfield  {journal} {\bibinfo
   {journal} {\apj}\ }\textbf {\bibinfo {volume} {554}},\ \bibinfo {pages}
  {624} (\bibinfo {year} {2001})},\ \Eprint
  {http://arxiv.org/abs/astro-ph/0102175} {arXiv:astro-ph/0102175 [astro-ph]}
  \BibitemShut {NoStop}%
\bibitem [{\citenamefont {Timokhin}\ and\ \citenamefont
  {Harding}(2019)}]{Timokhin:2018vdn}%
  \BibitemOpen
  \bibfield  {author} {\bibinfo {author} {\bibfnamefont {A.~N.}\ \bibnamefont
  {Timokhin}}\ and\ \bibinfo {author} {\bibfnamefont {A.~K.}\ \bibnamefont
  {Harding}},\ }\href {\doibase 10.3847/1538-4357/aaf050} {\bibfield  {journal}
  {\bibinfo  {journal} {Astrophys. J.}\ }\textbf {\bibinfo {volume} {871}},\
  \bibinfo {pages} {12} (\bibinfo {year} {2019})},\ \Eprint
  {http://arxiv.org/abs/1803.08924} {arXiv:1803.08924 [astro-ph.HE]}
  \BibitemShut {NoStop}%
\bibitem [{\citenamefont
  {{Beloborodov}}(2013{\natexlab{a}})}]{Beloborodov2013a}%
  \BibitemOpen
  \bibfield  {author} {\bibinfo {author} {\bibfnamefont {A.~M.}\ \bibnamefont
  {{Beloborodov}}},\ }\href {\doibase 10.1088/0004-637X/777/2/114} {\bibfield
  {journal} {\bibinfo  {journal} {\apj}\ }\textbf {\bibinfo {volume} {777}},\
  \bibinfo {eid} {114} (\bibinfo {year} {2013}{\natexlab{a}})},\ \Eprint
  {http://arxiv.org/abs/1209.4063} {arXiv:1209.4063 [astro-ph.HE]} \BibitemShut
  {NoStop}%
\bibitem [{\citenamefont
  {{Beloborodov}}(2013{\natexlab{b}})}]{Beloborodov2013b}%
  \BibitemOpen
  \bibfield  {author} {\bibinfo {author} {\bibfnamefont {A.~M.}\ \bibnamefont
  {{Beloborodov}}},\ }\href {\doibase 10.1088/0004-637X/762/1/13} {\bibfield
  {journal} {\bibinfo  {journal} {\apj}\ }\textbf {\bibinfo {volume} {762}},\
  \bibinfo {eid} {13} (\bibinfo {year} {2013}{\natexlab{b}})},\ \Eprint
  {http://arxiv.org/abs/1201.0664} {arXiv:1201.0664 [astro-ph.HE]} \BibitemShut
  {NoStop}%
\bibitem [{\citenamefont {Archibald}\ \emph {et~al.}(2015)\citenamefont
  {Archibald}, \citenamefont {Kaspi}, \citenamefont {Ng}, \citenamefont
  {Scholz}, \citenamefont {Beardmore}, \citenamefont {Gehrels},\ and\
  \citenamefont {Kennea}}]{Archibald:2014dla}%
  \BibitemOpen
  \bibfield  {author} {\bibinfo {author} {\bibfnamefont {R.~F.}\ \bibnamefont
  {Archibald}}, \bibinfo {author} {\bibfnamefont {V.~M.}\ \bibnamefont
  {Kaspi}}, \bibinfo {author} {\bibfnamefont {C.~Y.}\ \bibnamefont {Ng}},
  \bibinfo {author} {\bibfnamefont {P.}~\bibnamefont {Scholz}}, \bibinfo
  {author} {\bibfnamefont {A.~P.}\ \bibnamefont {Beardmore}}, \bibinfo {author}
  {\bibfnamefont {N.}~\bibnamefont {Gehrels}}, \ and\ \bibinfo {author}
  {\bibfnamefont {J.~A.}\ \bibnamefont {Kennea}},\ }\href {\doibase
  10.1088/0004-637X/800/1/33} {\bibfield  {journal} {\bibinfo  {journal}
  {Astrophys. J.}\ }\textbf {\bibinfo {volume} {800}},\ \bibinfo {pages} {33}
  (\bibinfo {year} {2015})},\ \Eprint {http://arxiv.org/abs/1412.2780}
  {arXiv:1412.2780 [astro-ph.HE]} \BibitemShut {NoStop}%
\bibitem [{\citenamefont {{Thompson}}(2008)}]{Thompson2008III}%
  \BibitemOpen
  \bibfield  {author} {\bibinfo {author} {\bibfnamefont {C.}~\bibnamefont
  {{Thompson}}},\ }\href {\doibase 10.1086/592263} {\bibfield  {journal}
  {\bibinfo  {journal} {\apj}\ }\textbf {\bibinfo {volume} {688}},\ \bibinfo
  {pages} {1258} (\bibinfo {year} {2008})},\ \Eprint
  {http://arxiv.org/abs/0802.2571} {arXiv:0802.2571 [astro-ph]} \BibitemShut
  {NoStop}%
\bibitem [{\citenamefont {Thompson}\ and\ \citenamefont
  {Beloborodov}(2005)}]{Thompson:2004yg}%
  \BibitemOpen
  \bibfield  {author} {\bibinfo {author} {\bibfnamefont {C.}~\bibnamefont
  {Thompson}}\ and\ \bibinfo {author} {\bibfnamefont {A.~M.}\ \bibnamefont
  {Beloborodov}},\ }\href {\doibase 10.1086/432245} {\bibfield  {journal}
  {\bibinfo  {journal} {Astrophys. J.}\ }\textbf {\bibinfo {volume} {634}},\
  \bibinfo {pages} {565} (\bibinfo {year} {2005})},\ \Eprint
  {http://arxiv.org/abs/astro-ph/0408538} {arXiv:astro-ph/0408538} \BibitemShut
  {NoStop}%
\bibitem [{\citenamefont {{Kozlenkov}}\ and\ \citenamefont
  {{Mitrofanov}}(1986)}]{KozlenkovMitrofanov1986}%
  \BibitemOpen
  \bibfield  {author} {\bibinfo {author} {\bibfnamefont {A.~A.}\ \bibnamefont
  {{Kozlenkov}}}\ and\ \bibinfo {author} {\bibfnamefont {I.~G.}\ \bibnamefont
  {{Mitrofanov}}},\ }\href@noop {} {\bibfield  {journal} {\bibinfo  {journal}
  {Soviet Journal of Experimental and Theoretical Physics}\ }\textbf {\bibinfo
  {volume} {64}},\ \bibinfo {pages} {1173} (\bibinfo {year}
  {1986})}\BibitemShut {NoStop}%
\bibitem [{\citenamefont {{Kostenko}}\ and\ \citenamefont
  {{Thompson}}(2019)}]{KostenkoThompson2019}%
  \BibitemOpen
  \bibfield  {author} {\bibinfo {author} {\bibfnamefont {A.}~\bibnamefont
  {{Kostenko}}}\ and\ \bibinfo {author} {\bibfnamefont {C.}~\bibnamefont
  {{Thompson}}},\ }\href {\doibase 10.3847/1538-4357/aae82e} {\bibfield
  {journal} {\bibinfo  {journal} {\apj}\ }\textbf {\bibinfo {volume} {875}},\
  \bibinfo {eid} {23} (\bibinfo {year} {2019})},\ \Eprint
  {http://arxiv.org/abs/1904.03325} {arXiv:1904.03325 [astro-ph.HE]}
  \BibitemShut {NoStop}%
\bibitem [{\citenamefont {{Olausen}}\ and\ \citenamefont
  {{Kaspi}}(2014)}]{Kaspi2014}%
  \BibitemOpen
  \bibfield  {author} {\bibinfo {author} {\bibfnamefont {S.~A.}\ \bibnamefont
  {{Olausen}}}\ and\ \bibinfo {author} {\bibfnamefont {V.~M.}\ \bibnamefont
  {{Kaspi}}},\ }\href {\doibase 10.1088/0067-0049/212/1/6} {\bibfield
  {journal} {\bibinfo  {journal} {ApJS}\ }\textbf {\bibinfo {volume} {212}},\
  \bibinfo {eid} {6} (\bibinfo {year} {2014})},\ \Eprint
  {http://arxiv.org/abs/1309.4167} {arXiv:1309.4167 [astro-ph.HE]} \BibitemShut
  {NoStop}%
\bibitem [{\citenamefont {{Gotthelf}}\ and\ \citenamefont
  {{Halpern}}(2007)}]{Gotthelf2007}%
  \BibitemOpen
  \bibfield  {author} {\bibinfo {author} {\bibfnamefont {E.~V.}\ \bibnamefont
  {{Gotthelf}}}\ and\ \bibinfo {author} {\bibfnamefont {J.~P.}\ \bibnamefont
  {{Halpern}}},\ }\href {\doibase 10.1007/s10509-007-9327-9} {\bibfield
  {journal} {\bibinfo  {journal} {\apss}\ }\textbf {\bibinfo {volume} {308}},\
  \bibinfo {pages} {79} (\bibinfo {year} {2007})},\ \Eprint
  {http://arxiv.org/abs/astro-ph/0608473} {arXiv:astro-ph/0608473 [astro-ph]}
  \BibitemShut {NoStop}%
\bibitem [{\citenamefont {{Perna}}\ and\ \citenamefont
  {{Gotthelf}}(2008)}]{Perna2008}%
  \BibitemOpen
  \bibfield  {author} {\bibinfo {author} {\bibfnamefont {R.}~\bibnamefont
  {{Perna}}}\ and\ \bibinfo {author} {\bibfnamefont {E.~V.}\ \bibnamefont
  {{Gotthelf}}},\ }\href {\doibase 10.1086/588211} {\bibfield  {journal}
  {\bibinfo  {journal} {\apj}\ }\textbf {\bibinfo {volume} {681}},\ \bibinfo
  {pages} {522} (\bibinfo {year} {2008})},\ \Eprint
  {http://arxiv.org/abs/0803.2042} {arXiv:0803.2042 [astro-ph]} \BibitemShut
  {NoStop}%
\bibitem [{\citenamefont {Gavriil}\ and\ \citenamefont
  {Kaspi}(2004)}]{Gavriil:2004hb}%
  \BibitemOpen
  \bibfield  {author} {\bibinfo {author} {\bibfnamefont {F.~P.}\ \bibnamefont
  {Gavriil}}\ and\ \bibinfo {author} {\bibfnamefont {V.~M.}\ \bibnamefont
  {Kaspi}},\ }\href {\doibase 10.1086/422751} {\bibfield  {journal} {\bibinfo
  {journal} {Astrophys. J. Lett.}\ }\textbf {\bibinfo {volume} {609}},\
  \bibinfo {pages} {L67} (\bibinfo {year} {2004})},\ \Eprint
  {http://arxiv.org/abs/astro-ph/0404113} {arXiv:astro-ph/0404113} \BibitemShut
  {NoStop}%
\bibitem [{\citenamefont {{Coe}}\ \emph {et~al.}(1994)\citenamefont {{Coe}},
  \citenamefont {{Jones}},\ and\ \citenamefont {{Lehto}}}]{Coe1994}%
  \BibitemOpen
  \bibfield  {author} {\bibinfo {author} {\bibfnamefont {M.~J.}\ \bibnamefont
  {{Coe}}}, \bibinfo {author} {\bibfnamefont {L.~R.}\ \bibnamefont {{Jones}}},
  \ and\ \bibinfo {author} {\bibfnamefont {H.}~\bibnamefont {{Lehto}}},\ }\href
  {\doibase 10.1093/mnras/270.1.178} {\bibfield  {journal} {\bibinfo  {journal}
  {\mnras}\ }\textbf {\bibinfo {volume} {270}},\ \bibinfo {pages} {178}
  (\bibinfo {year} {1994})}\BibitemShut {NoStop}%
\bibitem [{\citenamefont {{Tiengo}}\ \emph {et~al.}(2013)\citenamefont
  {{Tiengo}}, \citenamefont {{Esposito}}, \citenamefont {{Mereghetti}},
  \citenamefont {{Turolla}}, \citenamefont {{Nobili}}, \citenamefont
  {{Gastaldello}}, \citenamefont {{G{\"o}tz}}, \citenamefont {{Israel}},
  \citenamefont {{Rea}}, \citenamefont {{Stella}}, \citenamefont {{Zane}},\
  and\ \citenamefont {{Bignami}}}]{Tiengo2013}%
  \BibitemOpen
  \bibfield  {author} {\bibinfo {author} {\bibfnamefont {A.}~\bibnamefont
  {{Tiengo}}}, \bibinfo {author} {\bibfnamefont {P.}~\bibnamefont
  {{Esposito}}}, \bibinfo {author} {\bibfnamefont {S.}~\bibnamefont
  {{Mereghetti}}}, \bibinfo {author} {\bibfnamefont {R.}~\bibnamefont
  {{Turolla}}}, \bibinfo {author} {\bibfnamefont {L.}~\bibnamefont {{Nobili}}},
  \bibinfo {author} {\bibfnamefont {F.}~\bibnamefont {{Gastaldello}}}, \bibinfo
  {author} {\bibfnamefont {D.}~\bibnamefont {{G{\"o}tz}}}, \bibinfo {author}
  {\bibfnamefont {G.~L.}\ \bibnamefont {{Israel}}}, \bibinfo {author}
  {\bibfnamefont {N.}~\bibnamefont {{Rea}}}, \bibinfo {author} {\bibfnamefont
  {L.}~\bibnamefont {{Stella}}}, \bibinfo {author} {\bibfnamefont
  {S.}~\bibnamefont {{Zane}}}, \ and\ \bibinfo {author} {\bibfnamefont {G.~F.}\
  \bibnamefont {{Bignami}}},\ }\href {\doibase 10.1038/nature12386} {\bibfield
  {journal} {\bibinfo  {journal} {\nat}\ }\textbf {\bibinfo {volume} {500}},\
  \bibinfo {pages} {312} (\bibinfo {year} {2013})},\ \Eprint
  {http://arxiv.org/abs/1308.4987} {arXiv:1308.4987 [astro-ph.HE]} \BibitemShut
  {NoStop}%
\bibitem [{\citenamefont {{Rodr{\'\i}guez Castillo}}\ \emph
  {et~al.}(2016)\citenamefont {{Rodr{\'\i}guez Castillo}}, \citenamefont
  {{Israel}}, \citenamefont {{Tiengo}}, \citenamefont {{Salvetti}},
  \citenamefont {{Turolla}}, \citenamefont {{Zane}}, \citenamefont {{Rea}},
  \citenamefont {{Esposito}}, \citenamefont {{Mereghetti}}, \citenamefont
  {{Perna}}, \citenamefont {{Stella}}, \citenamefont {{Pons}}, \citenamefont
  {{Campana}}, \citenamefont {{G{\"o}tz}},\ and\ \citenamefont
  {{Motta}}}]{Castillo2016}%
  \BibitemOpen
  \bibfield  {author} {\bibinfo {author} {\bibfnamefont {G.~A.}\ \bibnamefont
  {{Rodr{\'\i}guez Castillo}}}, \bibinfo {author} {\bibfnamefont {G.~L.}\
  \bibnamefont {{Israel}}}, \bibinfo {author} {\bibfnamefont {A.}~\bibnamefont
  {{Tiengo}}}, \bibinfo {author} {\bibfnamefont {D.}~\bibnamefont
  {{Salvetti}}}, \bibinfo {author} {\bibfnamefont {R.}~\bibnamefont
  {{Turolla}}}, \bibinfo {author} {\bibfnamefont {S.}~\bibnamefont {{Zane}}},
  \bibinfo {author} {\bibfnamefont {N.}~\bibnamefont {{Rea}}}, \bibinfo
  {author} {\bibfnamefont {P.}~\bibnamefont {{Esposito}}}, \bibinfo {author}
  {\bibfnamefont {S.}~\bibnamefont {{Mereghetti}}}, \bibinfo {author}
  {\bibfnamefont {R.}~\bibnamefont {{Perna}}}, \bibinfo {author} {\bibfnamefont
  {L.}~\bibnamefont {{Stella}}}, \bibinfo {author} {\bibfnamefont {J.~A.}\
  \bibnamefont {{Pons}}}, \bibinfo {author} {\bibfnamefont {S.}~\bibnamefont
  {{Campana}}}, \bibinfo {author} {\bibfnamefont {D.}~\bibnamefont
  {{G{\"o}tz}}}, \ and\ \bibinfo {author} {\bibfnamefont {S.}~\bibnamefont
  {{Motta}}},\ }\href {\doibase 10.1093/mnras/stv2490} {\bibfield  {journal}
  {\bibinfo  {journal} {\mnras}\ }\textbf {\bibinfo {volume} {456}},\ \bibinfo
  {pages} {4145} (\bibinfo {year} {2016})},\ \Eprint
  {http://arxiv.org/abs/1510.09157} {arXiv:1510.09157 [astro-ph.HE]}
  \BibitemShut {NoStop}%
\bibitem [{\citenamefont {{Pizzocaro}}\ \emph {et~al.}(2019)\citenamefont
  {{Pizzocaro}}, \citenamefont {{Tiengo}}, \citenamefont {{Mereghetti}},
  \citenamefont {{Turolla}}, \citenamefont {{Esposito}}, \citenamefont
  {{Stella}}, \citenamefont {{Zane}}, \citenamefont {{Rea}}, \citenamefont
  {{Coti Zelati}},\ and\ \citenamefont {{Israel}}}]{Pizzocaro2019}%
  \BibitemOpen
  \bibfield  {author} {\bibinfo {author} {\bibfnamefont {D.}~\bibnamefont
  {{Pizzocaro}}}, \bibinfo {author} {\bibfnamefont {A.}~\bibnamefont
  {{Tiengo}}}, \bibinfo {author} {\bibfnamefont {S.}~\bibnamefont
  {{Mereghetti}}}, \bibinfo {author} {\bibfnamefont {R.}~\bibnamefont
  {{Turolla}}}, \bibinfo {author} {\bibfnamefont {P.}~\bibnamefont
  {{Esposito}}}, \bibinfo {author} {\bibfnamefont {L.}~\bibnamefont
  {{Stella}}}, \bibinfo {author} {\bibfnamefont {S.}~\bibnamefont {{Zane}}},
  \bibinfo {author} {\bibfnamefont {N.}~\bibnamefont {{Rea}}}, \bibinfo
  {author} {\bibfnamefont {F.}~\bibnamefont {{Coti Zelati}}}, \ and\ \bibinfo
  {author} {\bibfnamefont {G.}~\bibnamefont {{Israel}}},\ }\href {\doibase
  10.1051/0004-6361/201834784} {\bibfield  {journal} {\bibinfo  {journal}
  {\aap}\ }\textbf {\bibinfo {volume} {626}},\ \bibinfo {eid} {A39} (\bibinfo
  {year} {2019})},\ \Eprint {http://arxiv.org/abs/1904.07553} {arXiv:1904.07553
  [astro-ph.HE]} \BibitemShut {NoStop}%
\bibitem [{\citenamefont {Coti~Zelati}\ \emph {et~al.}(2017)\citenamefont
  {Coti~Zelati} \emph {et~al.}}]{CotiZelati:2017eid}%
  \BibitemOpen
  \bibfield  {author} {\bibinfo {author} {\bibfnamefont {F.}~\bibnamefont
  {Coti~Zelati}} \emph {et~al.},\ }\href {\doibase 10.1093/mnras/stx1700}
  {\bibfield  {journal} {\bibinfo  {journal} {Mon. Not. Roy. Astron. Soc.}\
  }\textbf {\bibinfo {volume} {471}},\ \bibinfo {pages} {1819} (\bibinfo {year}
  {2017})},\ \Eprint {http://arxiv.org/abs/1707.01514} {arXiv:1707.01514
  [astro-ph.HE]} \BibitemShut {NoStop}%
\bibitem [{\citenamefont {Rea}\ \emph {et~al.}(2020)\citenamefont {Rea},
  \citenamefont {Zelati}, \citenamefont {Viganò}, \citenamefont {Papitto},
  \citenamefont {Baganoff}, \citenamefont {Borghese}, \citenamefont {Campana},
  \citenamefont {Esposito}, \citenamefont {Haggard}, \citenamefont {Israel},
  \citenamefont {Mereghetti}, \citenamefont {Mignani}, \citenamefont {Perna},
  \citenamefont {Pons}, \citenamefont {Ponti}, \citenamefont {Stella},
  \citenamefont {Torres}, \citenamefont {Turolla},\ and\ \citenamefont
  {Zane}}]{Rea2020}%
  \BibitemOpen
  \bibfield  {author} {\bibinfo {author} {\bibfnamefont {N.}~\bibnamefont
  {Rea}}, \bibinfo {author} {\bibfnamefont {F.~C.}\ \bibnamefont {Zelati}},
  \bibinfo {author} {\bibfnamefont {D.}~\bibnamefont {Viganò}}, \bibinfo
  {author} {\bibfnamefont {A.}~\bibnamefont {Papitto}}, \bibinfo {author}
  {\bibfnamefont {F.}~\bibnamefont {Baganoff}}, \bibinfo {author}
  {\bibfnamefont {A.}~\bibnamefont {Borghese}}, \bibinfo {author}
  {\bibfnamefont {S.}~\bibnamefont {Campana}}, \bibinfo {author} {\bibfnamefont
  {P.}~\bibnamefont {Esposito}}, \bibinfo {author} {\bibfnamefont
  {D.}~\bibnamefont {Haggard}}, \bibinfo {author} {\bibfnamefont {G.~L.}\
  \bibnamefont {Israel}}, \bibinfo {author} {\bibfnamefont {S.}~\bibnamefont
  {Mereghetti}}, \bibinfo {author} {\bibfnamefont {R.~P.}\ \bibnamefont
  {Mignani}}, \bibinfo {author} {\bibfnamefont {R.}~\bibnamefont {Perna}},
  \bibinfo {author} {\bibfnamefont {J.~A.}\ \bibnamefont {Pons}}, \bibinfo
  {author} {\bibfnamefont {G.}~\bibnamefont {Ponti}}, \bibinfo {author}
  {\bibfnamefont {L.}~\bibnamefont {Stella}}, \bibinfo {author} {\bibfnamefont
  {D.~F.}\ \bibnamefont {Torres}}, \bibinfo {author} {\bibfnamefont
  {R.}~\bibnamefont {Turolla}}, \ and\ \bibinfo {author} {\bibfnamefont
  {S.}~\bibnamefont {Zane}},\ }\href {\doibase 10.3847/1538-4357/AB8387}
  {\bibfield  {journal} {\bibinfo  {journal} {The Astrophysical Journal}\
  }\textbf {\bibinfo {volume} {894}},\ \bibinfo {pages} {159} (\bibinfo {year}
  {2020})}\BibitemShut {NoStop}%
\bibitem [{\citenamefont {{Rea}}\ \emph {et~al.}(2013)\citenamefont {{Rea}},
  \citenamefont {{Esposito}}, \citenamefont {{Pons}}, \citenamefont
  {{Turolla}}, \citenamefont {{Torres}}, \citenamefont {{Israel}},
  \citenamefont {{Possenti}}, \citenamefont {{Burgay}}, \citenamefont
  {{Vigan{\`o}}}, \citenamefont {{Papitto}}, \citenamefont {{Perna}},
  \citenamefont {{Stella}}, \citenamefont {{Ponti}}, \citenamefont
  {{Baganoff}}, \citenamefont {{Haggard}}, \citenamefont {{Camero-Arranz}},
  \citenamefont {{Zane}}, \citenamefont {{Minter}}, \citenamefont
  {{Mereghetti}}, \citenamefont {{Tiengo}}, \citenamefont {{Sch{\"o}del}},
  \citenamefont {{Feroci}}, \citenamefont {{Mignani}},\ and\ \citenamefont
  {{G{\"o}tz}}}]{Rea2013}%
  \BibitemOpen
  \bibfield  {author} {\bibinfo {author} {\bibfnamefont {N.}~\bibnamefont
  {{Rea}}}, \bibinfo {author} {\bibfnamefont {P.}~\bibnamefont {{Esposito}}},
  \bibinfo {author} {\bibfnamefont {J.~A.}\ \bibnamefont {{Pons}}}, \bibinfo
  {author} {\bibfnamefont {R.}~\bibnamefont {{Turolla}}}, \bibinfo {author}
  {\bibfnamefont {D.~F.}\ \bibnamefont {{Torres}}}, \bibinfo {author}
  {\bibfnamefont {G.~L.}\ \bibnamefont {{Israel}}}, \bibinfo {author}
  {\bibfnamefont {A.}~\bibnamefont {{Possenti}}}, \bibinfo {author}
  {\bibfnamefont {M.}~\bibnamefont {{Burgay}}}, \bibinfo {author}
  {\bibfnamefont {D.}~\bibnamefont {{Vigan{\`o}}}}, \bibinfo {author}
  {\bibfnamefont {A.}~\bibnamefont {{Papitto}}}, \bibinfo {author}
  {\bibfnamefont {R.}~\bibnamefont {{Perna}}}, \bibinfo {author} {\bibfnamefont
  {L.}~\bibnamefont {{Stella}}}, \bibinfo {author} {\bibfnamefont
  {G.}~\bibnamefont {{Ponti}}}, \bibinfo {author} {\bibfnamefont {F.~K.}\
  \bibnamefont {{Baganoff}}}, \bibinfo {author} {\bibfnamefont
  {D.}~\bibnamefont {{Haggard}}}, \bibinfo {author} {\bibfnamefont
  {A.}~\bibnamefont {{Camero-Arranz}}}, \bibinfo {author} {\bibfnamefont
  {S.}~\bibnamefont {{Zane}}}, \bibinfo {author} {\bibfnamefont
  {A.}~\bibnamefont {{Minter}}}, \bibinfo {author} {\bibfnamefont
  {S.}~\bibnamefont {{Mereghetti}}}, \bibinfo {author} {\bibfnamefont
  {A.}~\bibnamefont {{Tiengo}}}, \bibinfo {author} {\bibfnamefont
  {R.}~\bibnamefont {{Sch{\"o}del}}}, \bibinfo {author} {\bibfnamefont
  {M.}~\bibnamefont {{Feroci}}}, \bibinfo {author} {\bibfnamefont
  {R.}~\bibnamefont {{Mignani}}}, \ and\ \bibinfo {author} {\bibfnamefont
  {D.}~\bibnamefont {{G{\"o}tz}}},\ }\href {\doibase
  10.1088/2041-8205/775/2/L34} {\bibfield  {journal} {\bibinfo  {journal}
  {\apjl}\ }\textbf {\bibinfo {volume} {775}},\ \bibinfo {eid} {L34} (\bibinfo
  {year} {2013})},\ \Eprint {http://arxiv.org/abs/1307.6331} {arXiv:1307.6331
  [astro-ph.GA]} \BibitemShut {NoStop}%
\bibitem [{\citenamefont {Mori}\ \emph {et~al.}(2013)\citenamefont {Mori},
  \citenamefont {Gotthelf}, \citenamefont {Zhang}, \citenamefont {An},
  \citenamefont {Baganoff}, \citenamefont {Barrière}, \citenamefont
  {Beloborodov}, \citenamefont {Boggs}, \citenamefont {Christensen},
  \citenamefont {Craig}, \citenamefont {Dufour}, \citenamefont {Grefenstette},
  \citenamefont {Hailey}, \citenamefont {Harrison}, \citenamefont {Hong},
  \citenamefont {Kaspi}, \citenamefont {Kennea}, \citenamefont {Madsen},
  \citenamefont {Markwardt}, \citenamefont {Nynka}, \citenamefont {Stern},
  \citenamefont {Tomsick},\ and\ \citenamefont {Zhang}}]{Mori_2013}%
  \BibitemOpen
  \bibfield  {author} {\bibinfo {author} {\bibfnamefont {K.}~\bibnamefont
  {Mori}}, \bibinfo {author} {\bibfnamefont {E.~V.}\ \bibnamefont {Gotthelf}},
  \bibinfo {author} {\bibfnamefont {S.}~\bibnamefont {Zhang}}, \bibinfo
  {author} {\bibfnamefont {H.}~\bibnamefont {An}}, \bibinfo {author}
  {\bibfnamefont {F.~K.}\ \bibnamefont {Baganoff}}, \bibinfo {author}
  {\bibfnamefont {N.~M.}\ \bibnamefont {Barrière}}, \bibinfo {author}
  {\bibfnamefont {A.~M.}\ \bibnamefont {Beloborodov}}, \bibinfo {author}
  {\bibfnamefont {S.~E.}\ \bibnamefont {Boggs}}, \bibinfo {author}
  {\bibfnamefont {F.~E.}\ \bibnamefont {Christensen}}, \bibinfo {author}
  {\bibfnamefont {W.~W.}\ \bibnamefont {Craig}}, \bibinfo {author}
  {\bibfnamefont {F.}~\bibnamefont {Dufour}}, \bibinfo {author} {\bibfnamefont
  {B.~W.}\ \bibnamefont {Grefenstette}}, \bibinfo {author} {\bibfnamefont
  {C.~J.}\ \bibnamefont {Hailey}}, \bibinfo {author} {\bibfnamefont {F.~A.}\
  \bibnamefont {Harrison}}, \bibinfo {author} {\bibfnamefont {J.}~\bibnamefont
  {Hong}}, \bibinfo {author} {\bibfnamefont {V.~M.}\ \bibnamefont {Kaspi}},
  \bibinfo {author} {\bibfnamefont {J.~A.}\ \bibnamefont {Kennea}}, \bibinfo
  {author} {\bibfnamefont {K.~K.}\ \bibnamefont {Madsen}}, \bibinfo {author}
  {\bibfnamefont {C.~B.}\ \bibnamefont {Markwardt}}, \bibinfo {author}
  {\bibfnamefont {M.}~\bibnamefont {Nynka}}, \bibinfo {author} {\bibfnamefont
  {D.}~\bibnamefont {Stern}}, \bibinfo {author} {\bibfnamefont {J.~A.}\
  \bibnamefont {Tomsick}}, \ and\ \bibinfo {author} {\bibfnamefont {W.~W.}\
  \bibnamefont {Zhang}},\ }\href {\doibase 10.1088/2041-8205/770/2/l23}
  {\bibfield  {journal} {\bibinfo  {journal} {The Astrophysical Journal}\
  }\textbf {\bibinfo {volume} {770}},\ \bibinfo {pages} {L23} (\bibinfo {year}
  {2013})}\BibitemShut {NoStop}%
\bibitem [{\citenamefont {{de Lima}}\ \emph {et~al.}(2020)\citenamefont {{de
  Lima}}, \citenamefont {{Coelho}}, \citenamefont {{Pereira}}, \citenamefont
  {{Rodrigues}},\ and\ \citenamefont {{Rueda}}}]{delima2020}%
  \BibitemOpen
  \bibfield  {author} {\bibinfo {author} {\bibfnamefont {R.~C.~R.}\
  \bibnamefont {{de Lima}}}, \bibinfo {author} {\bibfnamefont {J.~G.}\
  \bibnamefont {{Coelho}}}, \bibinfo {author} {\bibfnamefont {J.~P.}\
  \bibnamefont {{Pereira}}}, \bibinfo {author} {\bibfnamefont {C.~V.}\
  \bibnamefont {{Rodrigues}}}, \ and\ \bibinfo {author} {\bibfnamefont {J.~A.}\
  \bibnamefont {{Rueda}}},\ }\href {\doibase 10.3847/1538-4357/ab65f4}
  {\bibfield  {journal} {\bibinfo  {journal} {\apj}\ }\textbf {\bibinfo
  {volume} {889}},\ \bibinfo {eid} {165} (\bibinfo {year} {2020})},\ \Eprint
  {http://arxiv.org/abs/1912.12336} {arXiv:1912.12336 [astro-ph.SR]}
  \BibitemShut {NoStop}%
\bibitem [{\citenamefont {{Eatough}}\ \emph {et~al.}(2013)\citenamefont
  {{Eatough}}, \citenamefont {{Falcke}}, \citenamefont {{Karuppusamy}},
  \citenamefont {{Lee}}, \citenamefont {{Champion}}, \citenamefont {{Keane}},
  \citenamefont {{Desvignes}}, \citenamefont {{Schnitzeler}}, \citenamefont
  {{Spitler}}, \citenamefont {{Kramer}}, \citenamefont {{Klein}}, \citenamefont
  {{Bassa}}, \citenamefont {{Bower}}, \citenamefont {{Brunthaler}},
  \citenamefont {{Cognard}}, \citenamefont {{Deller}}, \citenamefont
  {{Demorest}}, \citenamefont {{Freire}}, \citenamefont {{Kraus}},
  \citenamefont {{Lyne}}, \citenamefont {{Noutsos}}, \citenamefont
  {{Stappers}},\ and\ \citenamefont {{Wex}}}]{Eatough2013}%
  \BibitemOpen
  \bibfield  {author} {\bibinfo {author} {\bibfnamefont {R.~P.}\ \bibnamefont
  {{Eatough}}}, \bibinfo {author} {\bibfnamefont {H.}~\bibnamefont {{Falcke}}},
  \bibinfo {author} {\bibfnamefont {R.}~\bibnamefont {{Karuppusamy}}}, \bibinfo
  {author} {\bibfnamefont {K.~J.}\ \bibnamefont {{Lee}}}, \bibinfo {author}
  {\bibfnamefont {D.~J.}\ \bibnamefont {{Champion}}}, \bibinfo {author}
  {\bibfnamefont {E.~F.}\ \bibnamefont {{Keane}}}, \bibinfo {author}
  {\bibfnamefont {G.}~\bibnamefont {{Desvignes}}}, \bibinfo {author}
  {\bibfnamefont {D.~H.~F.~M.}\ \bibnamefont {{Schnitzeler}}}, \bibinfo
  {author} {\bibfnamefont {L.~G.}\ \bibnamefont {{Spitler}}}, \bibinfo {author}
  {\bibfnamefont {M.}~\bibnamefont {{Kramer}}}, \bibinfo {author}
  {\bibfnamefont {B.}~\bibnamefont {{Klein}}}, \bibinfo {author} {\bibfnamefont
  {C.}~\bibnamefont {{Bassa}}}, \bibinfo {author} {\bibfnamefont {G.~C.}\
  \bibnamefont {{Bower}}}, \bibinfo {author} {\bibfnamefont {A.}~\bibnamefont
  {{Brunthaler}}}, \bibinfo {author} {\bibfnamefont {I.}~\bibnamefont
  {{Cognard}}}, \bibinfo {author} {\bibfnamefont {A.~T.}\ \bibnamefont
  {{Deller}}}, \bibinfo {author} {\bibfnamefont {P.~B.}\ \bibnamefont
  {{Demorest}}}, \bibinfo {author} {\bibfnamefont {P.~C.~C.}\ \bibnamefont
  {{Freire}}}, \bibinfo {author} {\bibfnamefont {A.}~\bibnamefont {{Kraus}}},
  \bibinfo {author} {\bibfnamefont {A.~G.}\ \bibnamefont {{Lyne}}}, \bibinfo
  {author} {\bibfnamefont {A.}~\bibnamefont {{Noutsos}}}, \bibinfo {author}
  {\bibfnamefont {B.}~\bibnamefont {{Stappers}}}, \ and\ \bibinfo {author}
  {\bibfnamefont {N.}~\bibnamefont {{Wex}}},\ }\href {\doibase
  10.1038/nature12499} {\bibfield  {journal} {\bibinfo  {journal} {\nat}\
  }\textbf {\bibinfo {volume} {501}},\ \bibinfo {pages} {391} (\bibinfo {year}
  {2013})},\ \Eprint {http://arxiv.org/abs/1308.3147} {arXiv:1308.3147
  [astro-ph.GA]} \BibitemShut {NoStop}%
\bibitem [{\citenamefont {Spitler}\ \emph {et~al.}(2014)\citenamefont {Spitler}
  \emph {et~al.}}]{Spitler2014}%
  \BibitemOpen
  \bibfield  {author} {\bibinfo {author} {\bibfnamefont {L.~G.}\ \bibnamefont
  {Spitler}} \emph {et~al.},\ }\href {\doibase 10.1088/0004-637x/790/2/101}
  {\bibfield  {journal} {\bibinfo  {journal} {The Astrophysical Journal}\
  }\textbf {\bibinfo {volume} {790}},\ \bibinfo {pages} {101} (\bibinfo {year}
  {2014})}\BibitemShut {NoStop}%
\bibitem [{\citenamefont {{Torne}}\ \emph {et~al.}(2017)\citenamefont
  {{Torne}}, \citenamefont {{Desvignes}}, \citenamefont {{Eatough}},
  \citenamefont {{Karuppusamy}}, \citenamefont {{Paubert}}, \citenamefont
  {{Kramer}}, \citenamefont {{Cognard}}, \citenamefont {{Champion}},\ and\
  \citenamefont {{Spitler}}}]{Torne2017}%
  \BibitemOpen
  \bibfield  {author} {\bibinfo {author} {\bibfnamefont {P.}~\bibnamefont
  {{Torne}}}, \bibinfo {author} {\bibfnamefont {G.}~\bibnamefont
  {{Desvignes}}}, \bibinfo {author} {\bibfnamefont {R.~P.}\ \bibnamefont
  {{Eatough}}}, \bibinfo {author} {\bibfnamefont {R.}~\bibnamefont
  {{Karuppusamy}}}, \bibinfo {author} {\bibfnamefont {G.}~\bibnamefont
  {{Paubert}}}, \bibinfo {author} {\bibfnamefont {M.}~\bibnamefont {{Kramer}}},
  \bibinfo {author} {\bibfnamefont {I.}~\bibnamefont {{Cognard}}}, \bibinfo
  {author} {\bibfnamefont {D.~J.}\ \bibnamefont {{Champion}}}, \ and\ \bibinfo
  {author} {\bibfnamefont {L.~G.}\ \bibnamefont {{Spitler}}},\ }\href {\doibase
  10.1093/mnras/stw2757} {\bibfield  {journal} {\bibinfo  {journal} {\mnras}\
  }\textbf {\bibinfo {volume} {465}},\ \bibinfo {pages} {242} (\bibinfo {year}
  {2017})},\ \Eprint {http://arxiv.org/abs/1610.07616} {arXiv:1610.07616
  [astro-ph.HE]} \BibitemShut {NoStop}%
\bibitem [{\citenamefont {{Muno}}\ \emph {et~al.}(2009)\citenamefont {{Muno}},
  \citenamefont {{Bauer}}, \citenamefont {{Baganoff}}, \citenamefont
  {{Bandyopadhyay}}, \citenamefont {{Bower}}, \citenamefont {{Brandt}},
  \citenamefont {{Broos}}, \citenamefont {{Cotera}}, \citenamefont
  {{Eikenberry}}, \citenamefont {{Garmire}}, \citenamefont {{Hyman}},
  \citenamefont {{Kassim}}, \citenamefont {{Lang}}, \citenamefont {{Lazio}},
  \citenamefont {{Law}}, \citenamefont {{Mauerhan}}, \citenamefont {{Morris}},
  \citenamefont {{Nagata}}, \citenamefont {{Nishiyama}}, \citenamefont
  {{Park}}, \citenamefont {{Ram{\`\i}rez}}, \citenamefont {{Stolovy}},
  \citenamefont {{Wijnands}}, \citenamefont {{Wang}}, \citenamefont {{Wang}},\
  and\ \citenamefont {{Yusef-Zadeh}}}]{Muno2009}%
  \BibitemOpen
  \bibfield  {author} {\bibinfo {author} {\bibfnamefont {M.~P.}\ \bibnamefont
  {{Muno}}}, \bibinfo {author} {\bibfnamefont {F.~E.}\ \bibnamefont {{Bauer}}},
  \bibinfo {author} {\bibfnamefont {F.~K.}\ \bibnamefont {{Baganoff}}},
  \bibinfo {author} {\bibfnamefont {R.~M.}\ \bibnamefont {{Bandyopadhyay}}},
  \bibinfo {author} {\bibfnamefont {G.~C.}\ \bibnamefont {{Bower}}}, \bibinfo
  {author} {\bibfnamefont {W.~N.}\ \bibnamefont {{Brandt}}}, \bibinfo {author}
  {\bibfnamefont {P.~S.}\ \bibnamefont {{Broos}}}, \bibinfo {author}
  {\bibfnamefont {A.}~\bibnamefont {{Cotera}}}, \bibinfo {author}
  {\bibfnamefont {S.~S.}\ \bibnamefont {{Eikenberry}}}, \bibinfo {author}
  {\bibfnamefont {G.~P.}\ \bibnamefont {{Garmire}}}, \bibinfo {author}
  {\bibfnamefont {S.~D.}\ \bibnamefont {{Hyman}}}, \bibinfo {author}
  {\bibfnamefont {N.~E.}\ \bibnamefont {{Kassim}}}, \bibinfo {author}
  {\bibfnamefont {C.~C.}\ \bibnamefont {{Lang}}}, \bibinfo {author}
  {\bibfnamefont {T.~J.~W.}\ \bibnamefont {{Lazio}}}, \bibinfo {author}
  {\bibfnamefont {C.}~\bibnamefont {{Law}}}, \bibinfo {author} {\bibfnamefont
  {J.~C.}\ \bibnamefont {{Mauerhan}}}, \bibinfo {author} {\bibfnamefont
  {M.~R.}\ \bibnamefont {{Morris}}}, \bibinfo {author} {\bibfnamefont
  {T.}~\bibnamefont {{Nagata}}}, \bibinfo {author} {\bibfnamefont
  {S.}~\bibnamefont {{Nishiyama}}}, \bibinfo {author} {\bibfnamefont
  {S.}~\bibnamefont {{Park}}}, \bibinfo {author} {\bibfnamefont {S.~V.}\
  \bibnamefont {{Ram{\`\i}rez}}}, \bibinfo {author} {\bibfnamefont {S.~R.}\
  \bibnamefont {{Stolovy}}}, \bibinfo {author} {\bibfnamefont {R.}~\bibnamefont
  {{Wijnands}}}, \bibinfo {author} {\bibfnamefont {Q.~D.}\ \bibnamefont
  {{Wang}}}, \bibinfo {author} {\bibfnamefont {Z.}~\bibnamefont {{Wang}}}, \
  and\ \bibinfo {author} {\bibfnamefont {F.}~\bibnamefont {{Yusef-Zadeh}}},\
  }\href {\doibase 10.1088/0067-0049/181/1/110} {\bibfield  {journal} {\bibinfo
   {journal} {\apjs}\ }\textbf {\bibinfo {volume} {181}},\ \bibinfo {pages}
  {110} (\bibinfo {year} {2009})},\ \Eprint {http://arxiv.org/abs/0809.1105}
  {arXiv:0809.1105 [astro-ph]} \BibitemShut {NoStop}%
\bibitem [{\citenamefont {{Gregory}}\ and\ \citenamefont
  {{Fahlman}}(1980)}]{GregoryFahlman1980}%
  \BibitemOpen
  \bibfield  {author} {\bibinfo {author} {\bibfnamefont {P.~C.}\ \bibnamefont
  {{Gregory}}}\ and\ \bibinfo {author} {\bibfnamefont {G.~G.}\ \bibnamefont
  {{Fahlman}}},\ }\href {\doibase 10.1038/287805a0} {\bibfield  {journal}
  {\bibinfo  {journal} {\nat}\ }\textbf {\bibinfo {volume} {287}},\ \bibinfo
  {pages} {805} (\bibinfo {year} {1980})}\BibitemShut {NoStop}%
\bibitem [{\citenamefont {{Heyl}}\ \emph {et~al.}(2024)\citenamefont {{Heyl}},
  \citenamefont {{Taverna}}, \citenamefont {{Turolla}} \emph
  {et~al.}}]{Heyl2024}%
  \BibitemOpen
  \bibfield  {author} {\bibinfo {author} {\bibfnamefont {J.}~\bibnamefont
  {{Heyl}}}, \bibinfo {author} {\bibfnamefont {R.}~\bibnamefont {{Taverna}}},
  \bibinfo {author} {\bibfnamefont {R.}~\bibnamefont {{Turolla}}},  \emph
  {et~al.},\ }\href {\doibase 10.1093/mnras/stad3680} {\bibfield  {journal}
  {\bibinfo  {journal} {\mnras}\ }\textbf {\bibinfo {volume} {527}},\ \bibinfo
  {pages} {12219} (\bibinfo {year} {2024})},\ \Eprint
  {http://arxiv.org/abs/2311.03637} {arXiv:2311.03637 [astro-ph.HE]}
  \BibitemShut {NoStop}%
\bibitem [{\citenamefont {Gedalin}\ and\ \citenamefont
  {Melrose}(2001)}]{PhysRevE.64.027401}%
  \BibitemOpen
  \bibfield  {author} {\bibinfo {author} {\bibfnamefont {M.}~\bibnamefont
  {Gedalin}}\ and\ \bibinfo {author} {\bibfnamefont {D.~B.}\ \bibnamefont
  {Melrose}},\ }\href {\doibase 10.1103/PhysRevE.64.027401} {\bibfield
  {journal} {\bibinfo  {journal} {Phys. Rev. E}\ }\textbf {\bibinfo {volume}
  {64}},\ \bibinfo {pages} {027401} (\bibinfo {year} {2001})}\BibitemShut
  {NoStop}%
\bibitem [{\citenamefont {{Ruderman}}\ and\ \citenamefont
  {{Sutherland}}(1975)}]{RudermanSutherland1975}%
  \BibitemOpen
  \bibfield  {author} {\bibinfo {author} {\bibfnamefont {M.~A.}\ \bibnamefont
  {{Ruderman}}}\ and\ \bibinfo {author} {\bibfnamefont {P.~G.}\ \bibnamefont
  {{Sutherland}}},\ }\href {\doibase 10.1086/153393} {\bibfield  {journal}
  {\bibinfo  {journal} {\apj}\ }\textbf {\bibinfo {volume} {196}},\ \bibinfo
  {pages} {51} (\bibinfo {year} {1975})}\BibitemShut {NoStop}%
\bibitem [{\citenamefont {{Philippov}}\ \emph {et~al.}(2020)\citenamefont
  {{Philippov}}, \citenamefont {{Timokhin}},\ and\ \citenamefont
  {{Spitkovsky}}}]{Philippov2020}%
  \BibitemOpen
  \bibfield  {author} {\bibinfo {author} {\bibfnamefont {A.}~\bibnamefont
  {{Philippov}}}, \bibinfo {author} {\bibfnamefont {A.}~\bibnamefont
  {{Timokhin}}}, \ and\ \bibinfo {author} {\bibfnamefont {A.}~\bibnamefont
  {{Spitkovsky}}},\ }\href {\doibase 10.1103/PhysRevLett.124.245101} {\bibfield
   {journal} {\bibinfo  {journal} {\prl}\ }\textbf {\bibinfo {volume} {124}},\
  \bibinfo {eid} {245101} (\bibinfo {year} {2020})},\ \Eprint
  {http://arxiv.org/abs/2001.02236} {arXiv:2001.02236 [astro-ph.HE]}
  \BibitemShut {NoStop}%
\bibitem [{\citenamefont {Cruz}\ \emph {et~al.}(2021)\citenamefont {Cruz},
  \citenamefont {Grismayer},\ and\ \citenamefont {Silva}}]{Cruz:2020vfm}%
  \BibitemOpen
  \bibfield  {author} {\bibinfo {author} {\bibfnamefont {F.}~\bibnamefont
  {Cruz}}, \bibinfo {author} {\bibfnamefont {T.}~\bibnamefont {Grismayer}}, \
  and\ \bibinfo {author} {\bibfnamefont {L.~O.}\ \bibnamefont {Silva}},\ }\href
  {\doibase 10.3847/1538-4357/abd2c0} {\bibfield  {journal} {\bibinfo
  {journal} {Astrophys. J.}\ }\textbf {\bibinfo {volume} {908}},\ \bibinfo
  {pages} {149} (\bibinfo {year} {2021})},\ \Eprint
  {http://arxiv.org/abs/2012.05587} {arXiv:2012.05587 [astro-ph.HE]}
  \BibitemShut {NoStop}%
\bibitem [{\citenamefont {{Cruz}}\ \emph {et~al.}(2021)\citenamefont {{Cruz}},
  \citenamefont {{Grismayer}}, \citenamefont {{Chen}}, \citenamefont
  {{Spitkovsky}},\ and\ \citenamefont {{Silva}}}]{Cruz2021}%
  \BibitemOpen
  \bibfield  {author} {\bibinfo {author} {\bibfnamefont {F.}~\bibnamefont
  {{Cruz}}}, \bibinfo {author} {\bibfnamefont {T.}~\bibnamefont {{Grismayer}}},
  \bibinfo {author} {\bibfnamefont {A.~Y.}\ \bibnamefont {{Chen}}}, \bibinfo
  {author} {\bibfnamefont {A.}~\bibnamefont {{Spitkovsky}}}, \ and\ \bibinfo
  {author} {\bibfnamefont {L.~O.}\ \bibnamefont {{Silva}}},\ }\href {\doibase
  10.3847/2041-8213/ac2157} {\bibfield  {journal} {\bibinfo  {journal}
  {Astrophysical Journal Letters}\ }\textbf {\bibinfo {volume} {919}},\
  \bibinfo {eid} {L4} (\bibinfo {year} {2021})},\ \Eprint
  {http://arxiv.org/abs/2108.11702} {arXiv:2108.11702 [astro-ph.HE]}
  \BibitemShut {NoStop}%
\bibitem [{\citenamefont {Bransgrove}\ \emph {et~al.}(2023)\citenamefont
  {Bransgrove}, \citenamefont {Beloborodov},\ and\ \citenamefont
  {Levin}}]{Bransgrove2023}%
  \BibitemOpen
  \bibfield  {author} {\bibinfo {author} {\bibfnamefont {A.}~\bibnamefont
  {Bransgrove}}, \bibinfo {author} {\bibfnamefont {A.~M.}\ \bibnamefont
  {Beloborodov}}, \ and\ \bibinfo {author} {\bibfnamefont {Y.}~\bibnamefont
  {Levin}},\ }\href {\doibase 10.3847/2041-8213/ad0556} {\bibfield  {journal}
  {\bibinfo  {journal} {The Astrophysical Journal Letters}\ }\textbf {\bibinfo
  {volume} {958}},\ \bibinfo {pages} {L9} (\bibinfo {year} {2023})}\BibitemShut
  {NoStop}%
\bibitem [{\citenamefont {Ben\'a\v{c}ek}\ \emph {et~al.}(2024)\citenamefont
  {Ben\'a\v{c}ek}, \citenamefont {Timokhin}, \citenamefont {Mu\~noz},
  \citenamefont {Jessner}, \citenamefont {Rievajov\'a}, \citenamefont {Pohl},\
  and\ \citenamefont {B\"uchner}}]{Benacek:2024lat}%
  \BibitemOpen
  \bibfield  {author} {\bibinfo {author} {\bibfnamefont {J.}~\bibnamefont
  {Ben\'a\v{c}ek}}, \bibinfo {author} {\bibfnamefont {A.}~\bibnamefont
  {Timokhin}}, \bibinfo {author} {\bibfnamefont {P.~A.}\ \bibnamefont
  {Mu\~noz}}, \bibinfo {author} {\bibfnamefont {A.}~\bibnamefont {Jessner}},
  \bibinfo {author} {\bibfnamefont {T.}~\bibnamefont {Rievajov\'a}}, \bibinfo
  {author} {\bibfnamefont {M.}~\bibnamefont {Pohl}}, \ and\ \bibinfo {author}
  {\bibfnamefont {J.}~\bibnamefont {B\"uchner}},\ }\href {\doibase
  10.1051/0004-6361/202450949} {\bibfield  {journal} {\bibinfo  {journal}
  {Astron. Astrophys.}\ }\textbf {\bibinfo {volume} {691}},\ \bibinfo {pages}
  {A137} (\bibinfo {year} {2024})},\ \Eprint {http://arxiv.org/abs/2405.20866}
  {arXiv:2405.20866 [astro-ph.HE]} \BibitemShut {NoStop}%
\bibitem [{\citenamefont {Chernoglazov}\ \emph {et~al.}(2024)\citenamefont
  {Chernoglazov}, \citenamefont {Philippov},\ and\ \citenamefont
  {Timokhin}}]{Chernoglazov:2024rvo}%
  \BibitemOpen
  \bibfield  {author} {\bibinfo {author} {\bibfnamefont {A.}~\bibnamefont
  {Chernoglazov}}, \bibinfo {author} {\bibfnamefont {A.}~\bibnamefont
  {Philippov}}, \ and\ \bibinfo {author} {\bibfnamefont {A.}~\bibnamefont
  {Timokhin}},\ }\href {\doibase 10.3847/2041-8213/ad7e24} {\bibfield
  {journal} {\bibinfo  {journal} {Astrophys. J. Lett.}\ }\textbf {\bibinfo
  {volume} {974}},\ \bibinfo {pages} {L32} (\bibinfo {year} {2024})},\ \Eprint
  {http://arxiv.org/abs/2409.15409} {arXiv:2409.15409 [astro-ph.HE]}
  \BibitemShut {NoStop}%
\bibitem [{\citenamefont {Thompson}(2022{\natexlab{a}})}]{Thompson:2021oey}%
  \BibitemOpen
  \bibfield  {author} {\bibinfo {author} {\bibfnamefont {C.}~\bibnamefont
  {Thompson}},\ }\href {\doibase 10.3847/1538-4357/ac501f} {\bibfield
  {journal} {\bibinfo  {journal} {Astrophys. J.}\ }\textbf {\bibinfo {volume}
  {933}},\ \bibinfo {pages} {231} (\bibinfo {year} {2022}{\natexlab{a}})},\
  \Eprint {http://arxiv.org/abs/2111.01958} {arXiv:2111.01958 [astro-ph.HE]}
  \BibitemShut {NoStop}%
\bibitem [{\citenamefont {Thompson}(2022{\natexlab{b}})}]{Thompson:2021jin}%
  \BibitemOpen
  \bibfield  {author} {\bibinfo {author} {\bibfnamefont {C.}~\bibnamefont
  {Thompson}},\ }\href {\doibase 10.3847/1538-4357/ac51d4} {\bibfield
  {journal} {\bibinfo  {journal} {Astrophys. J.}\ }\textbf {\bibinfo {volume}
  {933}},\ \bibinfo {pages} {232} (\bibinfo {year} {2022}{\natexlab{b}})},\
  \Eprint {http://arxiv.org/abs/2111.01959} {arXiv:2111.01959 [astro-ph.HE]}
  \BibitemShut {NoStop}%
\bibitem [{\citenamefont {Sikivie}(1983)}]{Sikivie1983}%
  \BibitemOpen
  \bibfield  {author} {\bibinfo {author} {\bibfnamefont {P.}~\bibnamefont
  {Sikivie}},\ }\href {\doibase 10.1103/PhysRevLett.51.1415} {\bibfield
  {journal} {\bibinfo  {journal} {Phys. Rev. Lett.}\ }\textbf {\bibinfo
  {volume} {51}},\ \bibinfo {pages} {1415} (\bibinfo {year}
  {1983})}\BibitemShut {NoStop}%
\bibitem [{\citenamefont {DePanfilis}\ \emph {et~al.}(1987)\citenamefont
  {DePanfilis} \emph {et~al.}}]{DePanfilis:1987}%
  \BibitemOpen
  \bibfield  {author} {\bibinfo {author} {\bibfnamefont {S.}~\bibnamefont
  {DePanfilis}} \emph {et~al.},\ }\href {\doibase 10.1103/PhysRevLett.59.839}
  {\bibfield  {journal} {\bibinfo  {journal} {Phys. Rev. Lett.}\ }\textbf
  {\bibinfo {volume} {59}},\ \bibinfo {pages} {839} (\bibinfo {year}
  {1987})}\BibitemShut {NoStop}%
\bibitem [{\citenamefont {Hagmann}\ \emph {et~al.}(1990)\citenamefont
  {Hagmann}, \citenamefont {Sikivie}, \citenamefont {Sullivan},\ and\
  \citenamefont {Tanner}}]{Hagmann:1990}%
  \BibitemOpen
  \bibfield  {author} {\bibinfo {author} {\bibfnamefont {C.}~\bibnamefont
  {Hagmann}}, \bibinfo {author} {\bibfnamefont {P.}~\bibnamefont {Sikivie}},
  \bibinfo {author} {\bibfnamefont {N.~S.}\ \bibnamefont {Sullivan}}, \ and\
  \bibinfo {author} {\bibfnamefont {D.~B.}\ \bibnamefont {Tanner}},\ }\href
  {\doibase 10.1103/PhysRevD.42.1297} {\bibfield  {journal} {\bibinfo
  {journal} {Phys. Rev. D}\ }\textbf {\bibinfo {volume} {42}},\ \bibinfo
  {pages} {1297} (\bibinfo {year} {1990})}\BibitemShut {NoStop}%
\bibitem [{\citenamefont {Hagmann}\ \emph {et~al.}(1998)\citenamefont {Hagmann}
  \emph {et~al.}}]{Hagmann:1998cb}%
  \BibitemOpen
  \bibfield  {author} {\bibinfo {author} {\bibfnamefont {C.}~\bibnamefont
  {Hagmann}} \emph {et~al.} (\bibinfo {collaboration} {ADMX}),\ }\href
  {\doibase 10.1103/PhysRevLett.80.2043} {\bibfield  {journal} {\bibinfo
  {journal} {Phys. Rev. Lett.}\ }\textbf {\bibinfo {volume} {80}},\ \bibinfo
  {pages} {2043} (\bibinfo {year} {1998})},\ \Eprint
  {http://arxiv.org/abs/astro-ph/9801286} {arXiv:astro-ph/9801286} \BibitemShut
  {NoStop}%
\bibitem [{\citenamefont {Asztalos}\ \emph {et~al.}(2001)\citenamefont
  {Asztalos} \emph {et~al.}}]{Asztalos:2001tf}%
  \BibitemOpen
  \bibfield  {author} {\bibinfo {author} {\bibfnamefont {S.~J.}\ \bibnamefont
  {Asztalos}} \emph {et~al.} (\bibinfo {collaboration} {ADMX}),\ }\href
  {\doibase 10.1103/PhysRevD.64.092003} {\bibfield  {journal} {\bibinfo
  {journal} {Phys. Rev. D}\ }\textbf {\bibinfo {volume} {64}},\ \bibinfo
  {pages} {092003} (\bibinfo {year} {2001})}\BibitemShut {NoStop}%
\bibitem [{\citenamefont {Asztalos}\ \emph {et~al.}(2010)\citenamefont
  {Asztalos} \emph {et~al.}}]{Asztalos:2009yp}%
  \BibitemOpen
  \bibfield  {author} {\bibinfo {author} {\bibfnamefont {S.~J.}\ \bibnamefont
  {Asztalos}} \emph {et~al.} (\bibinfo {collaboration} {ADMX}),\ }\href
  {\doibase 10.1103/PhysRevLett.104.041301} {\bibfield  {journal} {\bibinfo
  {journal} {Phys. Rev. Lett.}\ }\textbf {\bibinfo {volume} {104}},\ \bibinfo
  {pages} {041301} (\bibinfo {year} {2010})},\ \Eprint
  {http://arxiv.org/abs/0910.5914} {arXiv:0910.5914 [astro-ph.CO]} \BibitemShut
  {NoStop}%
\bibitem [{\citenamefont {Du}\ \emph {et~al.}(2018)\citenamefont {Du} \emph
  {et~al.}}]{Du:2018uak}%
  \BibitemOpen
  \bibfield  {author} {\bibinfo {author} {\bibfnamefont {N.}~\bibnamefont {Du}}
  \emph {et~al.} (\bibinfo {collaboration} {ADMX}),\ }\href {\doibase
  10.1103/PhysRevLett.120.151301} {\bibfield  {journal} {\bibinfo  {journal}
  {Phys. Rev. Lett.}\ }\textbf {\bibinfo {volume} {120}},\ \bibinfo {pages}
  {151301} (\bibinfo {year} {2018})},\ \Eprint
  {http://arxiv.org/abs/1804.05750} {arXiv:1804.05750 [hep-ex]} \BibitemShut
  {NoStop}%
\bibitem [{\citenamefont {Braine}\ \emph {et~al.}(2020)\citenamefont {Braine}
  \emph {et~al.}}]{Braine2020}%
  \BibitemOpen
  \bibfield  {author} {\bibinfo {author} {\bibfnamefont {T.}~\bibnamefont
  {Braine}} \emph {et~al.} (\bibinfo {collaboration} {ADMX}),\ }\href {\doibase
  10.1103/physrevlett.124.101303} {\bibfield  {journal} {\bibinfo  {journal}
  {Physical Review Letters}\ }\textbf {\bibinfo {volume} {124}} (\bibinfo
  {year} {2020}),\ 10.1103/physrevlett.124.101303}\BibitemShut {NoStop}%
\bibitem [{\citenamefont {Bradley}\ \emph {et~al.}(2003)\citenamefont {Bradley}
  \emph {et~al.}}]{Bradley2003}%
  \BibitemOpen
  \bibfield  {author} {\bibinfo {author} {\bibfnamefont {R.}~\bibnamefont
  {Bradley}} \emph {et~al.},\ }\href {\doibase 10.1103/RevModPhys.75.777}
  {\bibfield  {journal} {\bibinfo  {journal} {Rev. Mod. Phys.}\ }\textbf
  {\bibinfo {volume} {75}},\ \bibinfo {pages} {777} (\bibinfo {year}
  {2003})}\BibitemShut {NoStop}%
\bibitem [{\citenamefont {Asztalos}\ \emph {et~al.}(2004)\citenamefont
  {Asztalos} \emph {et~al.}}]{Bradley2004}%
  \BibitemOpen
  \bibfield  {author} {\bibinfo {author} {\bibfnamefont {S.~J.}\ \bibnamefont
  {Asztalos}} \emph {et~al.},\ }\href {\doibase 10.1103/PhysRevD.69.011101}
  {\bibfield  {journal} {\bibinfo  {journal} {Phys. Rev. D}\ }\textbf {\bibinfo
  {volume} {69}},\ \bibinfo {pages} {011101} (\bibinfo {year}
  {2004})}\BibitemShut {NoStop}%
\bibitem [{\citenamefont {Shokair}\ \emph {et~al.}(2014)\citenamefont {Shokair}
  \emph {et~al.}}]{Shokair2014}%
  \BibitemOpen
  \bibfield  {author} {\bibinfo {author} {\bibfnamefont {T.~M.}\ \bibnamefont
  {Shokair}} \emph {et~al.},\ }\href {\doibase 10.1142/s0217751x14430040}
  {\bibfield  {journal} {\bibinfo  {journal} {International Journal of Modern
  Physics A}\ }\textbf {\bibinfo {volume} {29}},\ \bibinfo {pages} {1443004}
  (\bibinfo {year} {2014})}\BibitemShut {NoStop}%
\bibitem [{\citenamefont {Brubaker}\ \emph {et~al.}(2017)\citenamefont
  {Brubaker} \emph {et~al.}}]{HAYSTAC}%
  \BibitemOpen
  \bibfield  {author} {\bibinfo {author} {\bibfnamefont {B.~M.}\ \bibnamefont
  {Brubaker}} \emph {et~al.} (\bibinfo {collaboration} {HAYSTAC}),\ }\href
  {\doibase 10.1103/PhysRevLett.118.061302} {\bibfield  {journal} {\bibinfo
  {journal} {Phys. Rev. Lett.}\ }\textbf {\bibinfo {volume} {118}},\ \bibinfo
  {pages} {061302} (\bibinfo {year} {2017})}\BibitemShut {NoStop}%
\bibitem [{\citenamefont {Zhong}\ \emph {et~al.}(2018)\citenamefont {Zhong}
  \emph {et~al.}}]{Zhong2018}%
  \BibitemOpen
  \bibfield  {author} {\bibinfo {author} {\bibfnamefont {L.}~\bibnamefont
  {Zhong}} \emph {et~al.} (\bibinfo {collaboration} {HAYSTAC}),\ }\href
  {\doibase 10.1103/physrevd.97.092001} {\bibfield  {journal} {\bibinfo
  {journal} {Physical Review D}\ }\textbf {\bibinfo {volume} {97}} (\bibinfo
  {year} {2018}),\ 10.1103/physrevd.97.092001}\BibitemShut {NoStop}%
\bibitem [{\citenamefont {Backes}\ \emph {et~al.}(2021)\citenamefont {Backes}
  \emph {et~al.}}]{Backes_2021}%
  \BibitemOpen
  \bibfield  {author} {\bibinfo {author} {\bibfnamefont {K.~M.}\ \bibnamefont
  {Backes}} \emph {et~al.} (\bibinfo {collaboration} {HAYSTAC}),\ }\href
  {\doibase 10.1038/s41586-021-03226-7} {\bibfield  {journal} {\bibinfo
  {journal} {Nature}\ }\textbf {\bibinfo {volume} {590}},\ \bibinfo {pages}
  {238} (\bibinfo {year} {2021})}\BibitemShut {NoStop}%
\bibitem [{\citenamefont {McAllister}\ \emph {et~al.}(2017)\citenamefont
  {McAllister} \emph {et~al.}}]{mcallister2017organ}%
  \BibitemOpen
  \bibfield  {author} {\bibinfo {author} {\bibfnamefont {B.~T.}\ \bibnamefont
  {McAllister}} \emph {et~al.},\ }\href@noop {} {\enquote {\bibinfo {title}
  {The organ experiment: An axion haloscope above 15 ghz},}\ } (\bibinfo {year}
  {2017}),\ \Eprint {http://arxiv.org/abs/1706.00209} {arXiv:1706.00209
  [physics.ins-det]} \BibitemShut {NoStop}%
\bibitem [{\citenamefont {Crescini}\ \emph {et~al.}(2020)\citenamefont
  {Crescini} \emph {et~al.}}]{QUAX:2020adt}%
  \BibitemOpen
  \bibfield  {author} {\bibinfo {author} {\bibfnamefont {N.}~\bibnamefont
  {Crescini}} \emph {et~al.} (\bibinfo {collaboration} {QUAX}),\ }\href
  {\doibase 10.1103/PhysRevLett.124.171801} {\bibfield  {journal} {\bibinfo
  {journal} {Phys. Rev. Lett.}\ }\textbf {\bibinfo {volume} {124}},\ \bibinfo
  {pages} {171801} (\bibinfo {year} {2020})},\ \Eprint
  {http://arxiv.org/abs/2001.08940} {arXiv:2001.08940 [hep-ex]} \BibitemShut
  {NoStop}%
\bibitem [{\citenamefont {Choi}\ \emph {et~al.}(2021)\citenamefont {Choi},
  \citenamefont {Ahn}, \citenamefont {Ko}, \citenamefont {Lee},\ and\
  \citenamefont {Semertzidis}}]{Choi_2021}%
  \BibitemOpen
  \bibfield  {author} {\bibinfo {author} {\bibfnamefont {J.}~\bibnamefont
  {Choi}}, \bibinfo {author} {\bibfnamefont {S.}~\bibnamefont {Ahn}}, \bibinfo
  {author} {\bibfnamefont {B.}~\bibnamefont {Ko}}, \bibinfo {author}
  {\bibfnamefont {S.}~\bibnamefont {Lee}}, \ and\ \bibinfo {author}
  {\bibfnamefont {Y.}~\bibnamefont {Semertzidis}},\ }\href {\doibase
  10.1016/j.nima.2021.165667} {\bibfield  {journal} {\bibinfo  {journal}
  {NIM-A}\ }\textbf {\bibinfo {volume} {1013}},\ \bibinfo {pages} {165667}
  (\bibinfo {year} {2021})}\BibitemShut {NoStop}%
\bibitem [{\citenamefont {{\'{A}}lvarez-Melc{\'{o}}n}\ \emph
  {et~al.}(2021)\citenamefont {{\'{A}}lvarez-Melc{\'{o}}n} \emph
  {et~al.}}]{Alvarez_Melcon_2021}%
  \BibitemOpen
  \bibfield  {author} {\bibinfo {author} {\bibfnamefont {A.}~\bibnamefont
  {{\'{A}}lvarez-Melc{\'{o}}n}} \emph {et~al.},\ }\href {\doibase
  10.1007/jhep10(2021)075} {\bibfield  {journal} {\bibinfo  {journal} {JHEP}\
  }\textbf {\bibinfo {volume} {2021}} (\bibinfo {year} {2021}),\
  10.1007/jhep10(2021)075}\BibitemShut {NoStop}%
\bibitem [{\citenamefont {Anastassopoulos}\ \emph {et~al.}(2017)\citenamefont
  {Anastassopoulos} \emph {et~al.}}]{Anastassopoulos2017}%
  \BibitemOpen
  \bibfield  {author} {\bibinfo {author} {\bibfnamefont {V.}~\bibnamefont
  {Anastassopoulos}} \emph {et~al.} (\bibinfo {collaboration} {CAST}),\ }\href
  {\doibase 10.1038/nphys4109} {\bibfield  {journal} {\bibinfo  {journal}
  {Nature Physics}\ }\textbf {\bibinfo {volume} {13}},\ \bibinfo {pages}
  {584–590} (\bibinfo {year} {2017})}\BibitemShut {NoStop}%
\bibitem [{\citenamefont {Ruz}\ \emph {et~al.}(2024)\citenamefont {Ruz} \emph
  {et~al.}}]{Ruz:2024gkl}%
  \BibitemOpen
  \bibfield  {author} {\bibinfo {author} {\bibfnamefont {J.}~\bibnamefont
  {Ruz}} \emph {et~al.},\ }\href@noop {} {\  (\bibinfo {year} {2024})},\
  \Eprint {http://arxiv.org/abs/2407.03828} {arXiv:2407.03828 [astro-ph.CO]}
  \BibitemShut {NoStop}%
\bibitem [{\citenamefont {Dolan}\ \emph {et~al.}(2022)\citenamefont {Dolan},
  \citenamefont {Hiskens},\ and\ \citenamefont {Volkas}}]{Dolan:2022kul}%
  \BibitemOpen
  \bibfield  {author} {\bibinfo {author} {\bibfnamefont {M.~J.}\ \bibnamefont
  {Dolan}}, \bibinfo {author} {\bibfnamefont {F.~J.}\ \bibnamefont {Hiskens}},
  \ and\ \bibinfo {author} {\bibfnamefont {R.~R.}\ \bibnamefont {Volkas}},\
  }\href {\doibase 10.1088/1475-7516/2022/10/096} {\bibfield  {journal}
  {\bibinfo  {journal} {JCAP}\ }\textbf {\bibinfo {volume} {10}},\ \bibinfo
  {pages} {096} (\bibinfo {year} {2022})},\ \Eprint
  {http://arxiv.org/abs/2207.03102} {arXiv:2207.03102 [hep-ph]} \BibitemShut
  {NoStop}%
\bibitem [{\citenamefont {Abuter}\ \emph {et~al.}(2019)\citenamefont {Abuter},
  \citenamefont {Amorim}, \citenamefont {Bauböck}, \citenamefont {Berger},
  \citenamefont {Bonnet}, \citenamefont {Brandner}, \citenamefont {Clénet},
  \citenamefont {Foresto}, \citenamefont {Zeeuw}, \citenamefont {Dexter},
  \citenamefont {Duvert}, \citenamefont {Eckart}, \citenamefont {Eisenhauer},
  \citenamefont {Schreiber}, \citenamefont {Garcia}, \citenamefont {Gao},
  \citenamefont {Gendron}, \citenamefont {Genzel}, \citenamefont {Gerhard},
  \citenamefont {Gillessen}, \citenamefont {Habibi}, \citenamefont {Haubois},
  \citenamefont {Henning}, \citenamefont {Hippler}, \citenamefont {Horrobin},
  \citenamefont {Jiménez-Rosales}, \citenamefont {Jocou}, \citenamefont
  {Kervella}, \citenamefont {Lacour}, \citenamefont {Lapeyrère}, \citenamefont
  {Bouquin}, \citenamefont {Léna}, \citenamefont {Ott}, \citenamefont
  {Paumard}, \citenamefont {Perraut}, \citenamefont {Perrin}, \citenamefont
  {Pfuhl}, \citenamefont {Rabien}, \citenamefont {Coira}, \citenamefont
  {Rousset}, \citenamefont {Scheithauer}, \citenamefont {Sternberg},
  \citenamefont {Straub}, \citenamefont {Straubmeier}, \citenamefont {Sturm},
  \citenamefont {Tacconi}, \citenamefont {Vincent}, \citenamefont {Fellenberg},
  \citenamefont {Waisberg}, \citenamefont {Widmann}, \citenamefont {Wieprecht},
  \citenamefont {Wiezorrek}, \citenamefont {Woillez},\ and\ \citenamefont
  {Yazici}}]{GravityCollab2019}%
  \BibitemOpen
  \bibfield  {author} {\bibinfo {author} {\bibfnamefont {R.}~\bibnamefont
  {Abuter}}, \bibinfo {author} {\bibfnamefont {A.}~\bibnamefont {Amorim}},
  \bibinfo {author} {\bibfnamefont {M.}~\bibnamefont {Bauböck}}, \bibinfo
  {author} {\bibfnamefont {J.~P.}\ \bibnamefont {Berger}}, \bibinfo {author}
  {\bibfnamefont {H.}~\bibnamefont {Bonnet}}, \bibinfo {author} {\bibfnamefont
  {W.}~\bibnamefont {Brandner}}, \bibinfo {author} {\bibfnamefont
  {Y.}~\bibnamefont {Clénet}}, \bibinfo {author} {\bibfnamefont {V.~C.~D.}\
  \bibnamefont {Foresto}}, \bibinfo {author} {\bibfnamefont {P.~T.~D.}\
  \bibnamefont {Zeeuw}}, \bibinfo {author} {\bibfnamefont {J.}~\bibnamefont
  {Dexter}}, \bibinfo {author} {\bibfnamefont {G.}~\bibnamefont {Duvert}},
  \bibinfo {author} {\bibfnamefont {A.}~\bibnamefont {Eckart}}, \bibinfo
  {author} {\bibfnamefont {F.}~\bibnamefont {Eisenhauer}}, \bibinfo {author}
  {\bibfnamefont {N.~M.~F.}\ \bibnamefont {Schreiber}}, \bibinfo {author}
  {\bibfnamefont {P.}~\bibnamefont {Garcia}}, \bibinfo {author} {\bibfnamefont
  {F.}~\bibnamefont {Gao}}, \bibinfo {author} {\bibfnamefont {E.}~\bibnamefont
  {Gendron}}, \bibinfo {author} {\bibfnamefont {R.}~\bibnamefont {Genzel}},
  \bibinfo {author} {\bibfnamefont {O.}~\bibnamefont {Gerhard}}, \bibinfo
  {author} {\bibfnamefont {S.}~\bibnamefont {Gillessen}}, \bibinfo {author}
  {\bibfnamefont {M.}~\bibnamefont {Habibi}}, \bibinfo {author} {\bibfnamefont
  {X.}~\bibnamefont {Haubois}}, \bibinfo {author} {\bibfnamefont
  {T.}~\bibnamefont {Henning}}, \bibinfo {author} {\bibfnamefont
  {S.}~\bibnamefont {Hippler}}, \bibinfo {author} {\bibfnamefont
  {M.}~\bibnamefont {Horrobin}}, \bibinfo {author} {\bibfnamefont
  {A.}~\bibnamefont {Jiménez-Rosales}}, \bibinfo {author} {\bibfnamefont
  {L.}~\bibnamefont {Jocou}}, \bibinfo {author} {\bibfnamefont
  {P.}~\bibnamefont {Kervella}}, \bibinfo {author} {\bibfnamefont
  {S.}~\bibnamefont {Lacour}}, \bibinfo {author} {\bibfnamefont
  {V.}~\bibnamefont {Lapeyrère}}, \bibinfo {author} {\bibfnamefont {J.~B.~L.}\
  \bibnamefont {Bouquin}}, \bibinfo {author} {\bibfnamefont {P.}~\bibnamefont
  {Léna}}, \bibinfo {author} {\bibfnamefont {T.}~\bibnamefont {Ott}}, \bibinfo
  {author} {\bibfnamefont {T.}~\bibnamefont {Paumard}}, \bibinfo {author}
  {\bibfnamefont {K.}~\bibnamefont {Perraut}}, \bibinfo {author} {\bibfnamefont
  {G.}~\bibnamefont {Perrin}}, \bibinfo {author} {\bibfnamefont
  {O.}~\bibnamefont {Pfuhl}}, \bibinfo {author} {\bibfnamefont
  {S.}~\bibnamefont {Rabien}}, \bibinfo {author} {\bibfnamefont {G.~R.}\
  \bibnamefont {Coira}}, \bibinfo {author} {\bibfnamefont {G.}~\bibnamefont
  {Rousset}}, \bibinfo {author} {\bibfnamefont {S.}~\bibnamefont
  {Scheithauer}}, \bibinfo {author} {\bibfnamefont {A.}~\bibnamefont
  {Sternberg}}, \bibinfo {author} {\bibfnamefont {O.}~\bibnamefont {Straub}},
  \bibinfo {author} {\bibfnamefont {C.}~\bibnamefont {Straubmeier}}, \bibinfo
  {author} {\bibfnamefont {E.}~\bibnamefont {Sturm}}, \bibinfo {author}
  {\bibfnamefont {L.~J.}\ \bibnamefont {Tacconi}}, \bibinfo {author}
  {\bibfnamefont {F.}~\bibnamefont {Vincent}}, \bibinfo {author} {\bibfnamefont
  {S.~V.}\ \bibnamefont {Fellenberg}}, \bibinfo {author} {\bibfnamefont
  {I.}~\bibnamefont {Waisberg}}, \bibinfo {author} {\bibfnamefont
  {F.}~\bibnamefont {Widmann}}, \bibinfo {author} {\bibfnamefont
  {E.}~\bibnamefont {Wieprecht}}, \bibinfo {author} {\bibfnamefont
  {E.}~\bibnamefont {Wiezorrek}}, \bibinfo {author} {\bibfnamefont
  {J.}~\bibnamefont {Woillez}}, \ and\ \bibinfo {author} {\bibfnamefont
  {S.}~\bibnamefont {Yazici}},\ }\href {\doibase 10.1051/0004-6361/201935656}
  {\bibfield  {journal} {\bibinfo  {journal} {Astronomy \& Astrophysics}\
  }\textbf {\bibinfo {volume} {625}},\ \bibinfo {pages} {L10} (\bibinfo {year}
  {2019})}\BibitemShut {NoStop}%
\bibitem [{\citenamefont {Jewell}\ and\ \citenamefont
  {Prestage}(2004)}]{Jewell2004}%
  \BibitemOpen
  \bibfield  {author} {\bibinfo {author} {\bibfnamefont {P.~R.}\ \bibnamefont
  {Jewell}}\ and\ \bibinfo {author} {\bibfnamefont {R.~M.}\ \bibnamefont
  {Prestage}},\ }\href {\doibase 10.1117/12.550631} {\bibfield  {journal}
  {\bibinfo  {journal} {https://doi.org/10.1117/12.550631}\ }\textbf {\bibinfo
  {volume} {5489}},\ \bibinfo {pages} {312} (\bibinfo {year}
  {2004})}\BibitemShut {NoStop}%
\bibitem [{\citenamefont {White}\ \emph {et~al.}(2022)\citenamefont {White},
  \citenamefont {Ghigo}, \citenamefont {Prestage}, \citenamefont {Frayer},
  \citenamefont {Maddalena}, \citenamefont {Wu}, \citenamefont {Brandt},
  \citenamefont {Egan}, \citenamefont {Nelson},\ and\ \citenamefont
  {Ray}}]{White2022}%
  \BibitemOpen
  \bibfield  {author} {\bibinfo {author} {\bibfnamefont {E.}~\bibnamefont
  {White}}, \bibinfo {author} {\bibfnamefont {F.~D.}\ \bibnamefont {Ghigo}},
  \bibinfo {author} {\bibfnamefont {R.~M.}\ \bibnamefont {Prestage}}, \bibinfo
  {author} {\bibfnamefont {D.~T.}\ \bibnamefont {Frayer}}, \bibinfo {author}
  {\bibfnamefont {R.~J.}\ \bibnamefont {Maddalena}}, \bibinfo {author}
  {\bibfnamefont {P.~T.}\ \bibnamefont {Wu}}, \bibinfo {author} {\bibfnamefont
  {J.~J.}\ \bibnamefont {Brandt}}, \bibinfo {author} {\bibfnamefont
  {D.}~\bibnamefont {Egan}}, \bibinfo {author} {\bibfnamefont {J.~D.}\
  \bibnamefont {Nelson}}, \ and\ \bibinfo {author} {\bibfnamefont
  {J.}~\bibnamefont {Ray}},\ }\href {\doibase 10.1051/0004-6361/202141936}
  {\bibfield  {journal} {\bibinfo  {journal} {Astronomy \& Astrophysics}\
  }\textbf {\bibinfo {volume} {659}},\ \bibinfo {pages} {A113} (\bibinfo {year}
  {2022})}\BibitemShut {NoStop}%
\bibitem [{\citenamefont {Collaboration}(2025)}]{GBTTelescope2025}%
  \BibitemOpen
  \bibfield  {author} {\bibinfo {author} {\bibfnamefont {G.~B.~T.}\
  \bibnamefont {Collaboration}},\ }\href@noop {} {\  (\bibinfo {year}
  {2025})}\BibitemShut {NoStop}%
\bibitem [{\citenamefont {Wootten}\ and\ \citenamefont
  {Thompson}(2009)}]{Wootten2009}%
  \BibitemOpen
  \bibfield  {author} {\bibinfo {author} {\bibfnamefont {A.}~\bibnamefont
  {Wootten}}\ and\ \bibinfo {author} {\bibfnamefont {A.~R.}\ \bibnamefont
  {Thompson}},\ }\href {\doibase 10.1109/JPROC.2009.2020572} {\bibfield
  {journal} {\bibinfo  {journal} {Proceedings of the IEEE}\ }\textbf {\bibinfo
  {volume} {97}},\ \bibinfo {pages} {1463} (\bibinfo {year}
  {2009})}\BibitemShut {NoStop}%
\bibitem [{\citenamefont {Collaboration}()}]{almatechnicalhandbook}%
  \BibitemOpen
  \bibfield  {author} {\bibinfo {author} {\bibfnamefont {A.}~\bibnamefont
  {Collaboration}},\ }\href {https://doi.org/10.5281/zenodo.4511521} {\
  }\BibitemShut {NoStop}%
\bibitem [{\citenamefont {Dewdney}\ \emph {et~al.}(2009)\citenamefont
  {Dewdney}, \citenamefont {Hall}, \citenamefont {Schilizzi},\ and\
  \citenamefont {Lazio}}]{Dewdney2009}%
  \BibitemOpen
  \bibfield  {author} {\bibinfo {author} {\bibfnamefont {P.~E.}\ \bibnamefont
  {Dewdney}}, \bibinfo {author} {\bibfnamefont {P.~J.}\ \bibnamefont {Hall}},
  \bibinfo {author} {\bibfnamefont {R.~T.}\ \bibnamefont {Schilizzi}}, \ and\
  \bibinfo {author} {\bibfnamefont {T.~J.~L.}\ \bibnamefont {Lazio}},\ }\href
  {\doibase 10.1109/JPROC.2009.2021005} {\bibfield  {journal} {\bibinfo
  {journal} {Proceedings of the IEEE}\ }\textbf {\bibinfo {volume} {97}},\
  \bibinfo {pages} {1482} (\bibinfo {year} {2009})}\BibitemShut {NoStop}%
\bibitem [{\citenamefont {Braun}\ \emph {et~al.}(2019)\citenamefont {Braun},
  \citenamefont {Bonaldi}, \citenamefont {Bourke}, \citenamefont {Keane},\ and\
  \citenamefont {Wagg}}]{Braun2019}%
  \BibitemOpen
  \bibfield  {author} {\bibinfo {author} {\bibfnamefont {R.}~\bibnamefont
  {Braun}}, \bibinfo {author} {\bibfnamefont {A.}~\bibnamefont {Bonaldi}},
  \bibinfo {author} {\bibfnamefont {T.}~\bibnamefont {Bourke}}, \bibinfo
  {author} {\bibfnamefont {E.}~\bibnamefont {Keane}}, \ and\ \bibinfo {author}
  {\bibfnamefont {J.}~\bibnamefont {Wagg}},\ }\href
  {https://arxiv.org/abs/1912.12699v1} {\  (\bibinfo {year}
  {2019})}\BibitemShut {NoStop}%
\bibitem [{\citenamefont {Hunter}\ \emph {et~al.}()\citenamefont {Hunter},
  \citenamefont {Petry}, \citenamefont {Barkats}, \citenamefont {Corder},\ and\
  \citenamefont {Indebetouw}}]{analysisUtilsALMA}%
  \BibitemOpen
  \bibfield  {author} {\bibinfo {author} {\bibfnamefont {T.~R.}\ \bibnamefont
  {Hunter}}, \bibinfo {author} {\bibfnamefont {D.}~\bibnamefont {Petry}},
  \bibinfo {author} {\bibfnamefont {D.}~\bibnamefont {Barkats}}, \bibinfo
  {author} {\bibfnamefont {S.}~\bibnamefont {Corder}}, \ and\ \bibinfo {author}
  {\bibfnamefont {R.}~\bibnamefont {Indebetouw}},\ }\href {\doibase
  10.5281/ZENODO.7502160} {\ 10.5281/ZENODO.7502160}\BibitemShut {NoStop}%
\bibitem [{\citenamefont {Green}\ \emph {et~al.}(2019)\citenamefont {Green},
  \citenamefont {Schlafly}, \citenamefont {Zucker}, \citenamefont {Speagle},\
  and\ \citenamefont {Finkbeiner}}]{Green_2019}%
  \BibitemOpen
  \bibfield  {author} {\bibinfo {author} {\bibfnamefont {G.~M.}\ \bibnamefont
  {Green}}, \bibinfo {author} {\bibfnamefont {E.}~\bibnamefont {Schlafly}},
  \bibinfo {author} {\bibfnamefont {C.}~\bibnamefont {Zucker}}, \bibinfo
  {author} {\bibfnamefont {J.~S.}\ \bibnamefont {Speagle}}, \ and\ \bibinfo
  {author} {\bibfnamefont {D.}~\bibnamefont {Finkbeiner}},\ }\href {\doibase
  10.3847/1538-4357/ab5362} {\bibfield  {journal} {\bibinfo  {journal} {The
  Astrophysical Journal}\ }\textbf {\bibinfo {volume} {887}},\ \bibinfo {pages}
  {93} (\bibinfo {year} {2019})}\BibitemShut {NoStop}%
\bibitem [{\citenamefont {Roy}\ \emph {et~al.}(2025)\citenamefont {Roy},
  \citenamefont {Blanco}, \citenamefont {Dessert}, \citenamefont {Prabhu},\
  and\ \citenamefont {Temim}}]{Roy:2023omw}%
  \BibitemOpen
  \bibfield  {author} {\bibinfo {author} {\bibfnamefont {S.}~\bibnamefont
  {Roy}}, \bibinfo {author} {\bibfnamefont {C.}~\bibnamefont {Blanco}},
  \bibinfo {author} {\bibfnamefont {C.}~\bibnamefont {Dessert}}, \bibinfo
  {author} {\bibfnamefont {A.}~\bibnamefont {Prabhu}}, \ and\ \bibinfo {author}
  {\bibfnamefont {T.}~\bibnamefont {Temim}},\ }\href {\doibase
  10.1103/PhysRevLett.134.071003} {\bibfield  {journal} {\bibinfo  {journal}
  {Phys. Rev. Lett.}\ }\textbf {\bibinfo {volume} {134}},\ \bibinfo {pages}
  {071003} (\bibinfo {year} {2025})},\ \Eprint
  {http://arxiv.org/abs/2311.04987} {arXiv:2311.04987 [hep-ph]} \BibitemShut
  {NoStop}%
\bibitem [{\citenamefont {{Schlegel}}\ \emph {et~al.}(1998)\citenamefont
  {{Schlegel}}, \citenamefont {{Finkbeiner}},\ and\ \citenamefont
  {{Davis}}}]{1998ApJ...500..525S}%
  \BibitemOpen
  \bibfield  {author} {\bibinfo {author} {\bibfnamefont {D.~J.}\ \bibnamefont
  {{Schlegel}}}, \bibinfo {author} {\bibfnamefont {D.~P.}\ \bibnamefont
  {{Finkbeiner}}}, \ and\ \bibinfo {author} {\bibfnamefont {M.}~\bibnamefont
  {{Davis}}},\ }\href {\doibase 10.1086/305772} {\bibfield  {journal} {\bibinfo
   {journal} {\apj}\ }\textbf {\bibinfo {volume} {500}},\ \bibinfo {pages}
  {525} (\bibinfo {year} {1998})},\ \Eprint
  {http://arxiv.org/abs/astro-ph/9710327} {arXiv:astro-ph/9710327 [astro-ph]}
  \BibitemShut {NoStop}%
\bibitem [{\citenamefont {Schlafly}\ and\ \citenamefont
  {Finkbeiner}(2011)}]{doug_2011}%
  \BibitemOpen
  \bibfield  {author} {\bibinfo {author} {\bibfnamefont {E.~F.}\ \bibnamefont
  {Schlafly}}\ and\ \bibinfo {author} {\bibfnamefont {D.~P.}\ \bibnamefont
  {Finkbeiner}},\ }\href {\doibase 10.1088/0004-637x/737/2/103} {\bibfield
  {journal} {\bibinfo  {journal} {The Astrophysical Journal}\ }\textbf
  {\bibinfo {volume} {737}},\ \bibinfo {pages} {103} (\bibinfo {year}
  {2011})}\BibitemShut {NoStop}%
\bibitem [{\citenamefont {Gao}\ \emph {et~al.}(2013)\citenamefont {Gao},
  \citenamefont {Li},\ and\ \citenamefont {Jiang}}]{Gao_2013}%
  \BibitemOpen
  \bibfield  {author} {\bibinfo {author} {\bibfnamefont {J.}~\bibnamefont
  {Gao}}, \bibinfo {author} {\bibfnamefont {A.}~\bibnamefont {Li}}, \ and\
  \bibinfo {author} {\bibfnamefont {B.~W.}\ \bibnamefont {Jiang}},\ }\href
  {\doibase 10.5047/eps.2013.05.016} {\bibfield  {journal} {\bibinfo  {journal}
  {Earth, Planets and Space}\ }\textbf {\bibinfo {volume} {65}},\ \bibinfo
  {pages} {1127} (\bibinfo {year} {2013})}\BibitemShut {NoStop}%
\bibitem [{\citenamefont {Laporte}\ \emph {et~al.}(2018)\citenamefont
  {Laporte}, \citenamefont {Johnston}, \citenamefont {Gómez}, \citenamefont
  {Garavito-Camargo},\ and\ \citenamefont {Besla}}]{Laporte2018}%
  \BibitemOpen
  \bibfield  {author} {\bibinfo {author} {\bibfnamefont {C.~F.}\ \bibnamefont
  {Laporte}}, \bibinfo {author} {\bibfnamefont {K.~V.}\ \bibnamefont
  {Johnston}}, \bibinfo {author} {\bibfnamefont {F.~A.}\ \bibnamefont
  {Gómez}}, \bibinfo {author} {\bibfnamefont {N.}~\bibnamefont
  {Garavito-Camargo}}, \ and\ \bibinfo {author} {\bibfnamefont
  {G.}~\bibnamefont {Besla}},\ }\href {\doibase 10.1093/MNRAS/STY1574}
  {\bibfield  {journal} {\bibinfo  {journal} {Monthly Notices of the Royal
  Astronomical Society}\ }\textbf {\bibinfo {volume} {481}},\ \bibinfo {pages}
  {286} (\bibinfo {year} {2018})}\BibitemShut {NoStop}%
\bibitem [{\citenamefont {Ou}\ \emph {et~al.}(2025)\citenamefont {Ou},
  \citenamefont {Necib}, \citenamefont {Wetzel}, \citenamefont {Frebel},
  \citenamefont {Bailin},\ and\ \citenamefont {Oeur}}]{Ou2025}%
  \BibitemOpen
  \bibfield  {author} {\bibinfo {author} {\bibfnamefont {X.}~\bibnamefont
  {Ou}}, \bibinfo {author} {\bibfnamefont {L.}~\bibnamefont {Necib}}, \bibinfo
  {author} {\bibfnamefont {A.}~\bibnamefont {Wetzel}}, \bibinfo {author}
  {\bibfnamefont {A.}~\bibnamefont {Frebel}}, \bibinfo {author} {\bibfnamefont
  {J.}~\bibnamefont {Bailin}}, \ and\ \bibinfo {author} {\bibfnamefont
  {M.}~\bibnamefont {Oeur}},\ }\href {https://arxiv.org/abs/2503.05877v1} {\
  (\bibinfo {year} {2025})}\BibitemShut {NoStop}%
\bibitem [{\citenamefont {Hopkins}\ \emph {et~al.}(2018)\citenamefont
  {Hopkins}, \citenamefont {Wetzel}, \citenamefont {Kereš}, \citenamefont
  {Faucher-Giguère}, \citenamefont {Quataert}, \citenamefont {Boylan-Kolchin},
  \citenamefont {Murray}, \citenamefont {Hayward}, \citenamefont
  {Garrison-Kimmel}, \citenamefont {Hummels}, \citenamefont {Feldmann},
  \citenamefont {Torrey}, \citenamefont {Ma}, \citenamefont {Anglés-Alcázar},
  \citenamefont {Su}, \citenamefont {Orr}, \citenamefont {Schmitz},
  \citenamefont {Escala}, \citenamefont {Sanderson}, \citenamefont
  {Grudi´cgrudi´c}, \citenamefont {Hafen}, \citenamefont {Kim}, \citenamefont
  {Fitts}, \citenamefont {Bullock}, \citenamefont {Wheeler}, \citenamefont
  {Chan}, \citenamefont {Elbert},\ and\ \citenamefont
  {Narayanan}}]{Hopkins2018}%
  \BibitemOpen
  \bibfield  {author} {\bibinfo {author} {\bibfnamefont {P.~F.}\ \bibnamefont
  {Hopkins}}, \bibinfo {author} {\bibfnamefont {A.}~\bibnamefont {Wetzel}},
  \bibinfo {author} {\bibfnamefont {D.}~\bibnamefont {Kereš}}, \bibinfo
  {author} {\bibfnamefont {C.-A.}\ \bibnamefont {Faucher-Giguère}}, \bibinfo
  {author} {\bibfnamefont {E.}~\bibnamefont {Quataert}}, \bibinfo {author}
  {\bibfnamefont {M.}~\bibnamefont {Boylan-Kolchin}}, \bibinfo {author}
  {\bibfnamefont {N.}~\bibnamefont {Murray}}, \bibinfo {author} {\bibfnamefont
  {C.~C.}\ \bibnamefont {Hayward}}, \bibinfo {author} {\bibfnamefont
  {S.}~\bibnamefont {Garrison-Kimmel}}, \bibinfo {author} {\bibfnamefont
  {C.}~\bibnamefont {Hummels}}, \bibinfo {author} {\bibfnamefont
  {R.}~\bibnamefont {Feldmann}}, \bibinfo {author} {\bibfnamefont
  {P.}~\bibnamefont {Torrey}}, \bibinfo {author} {\bibfnamefont
  {X.}~\bibnamefont {Ma}}, \bibinfo {author} {\bibfnamefont {D.}~\bibnamefont
  {Anglés-Alcázar}}, \bibinfo {author} {\bibfnamefont {K.-Y.}\ \bibnamefont
  {Su}}, \bibinfo {author} {\bibfnamefont {M.}~\bibnamefont {Orr}}, \bibinfo
  {author} {\bibfnamefont {D.}~\bibnamefont {Schmitz}}, \bibinfo {author}
  {\bibfnamefont {I.}~\bibnamefont {Escala}}, \bibinfo {author} {\bibfnamefont
  {R.}~\bibnamefont {Sanderson}}, \bibinfo {author} {\bibfnamefont {M.~Y.}\
  \bibnamefont {Grudi´cgrudi´c}}, \bibinfo {author} {\bibfnamefont
  {Z.}~\bibnamefont {Hafen}}, \bibinfo {author} {\bibfnamefont {J.-H.}\
  \bibnamefont {Kim}}, \bibinfo {author} {\bibfnamefont {A.}~\bibnamefont
  {Fitts}}, \bibinfo {author} {\bibfnamefont {J.~S.}\ \bibnamefont {Bullock}},
  \bibinfo {author} {\bibfnamefont {C.}~\bibnamefont {Wheeler}}, \bibinfo
  {author} {\bibfnamefont {T.~K.}\ \bibnamefont {Chan}}, \bibinfo {author}
  {\bibfnamefont {O.~D.}\ \bibnamefont {Elbert}}, \ and\ \bibinfo {author}
  {\bibfnamefont {D.}~\bibnamefont {Narayanan}},\ }\href@noop {} {\bibfield
  {journal} {\bibinfo  {journal} {Mon. Not. R. Astron. Soc}\ }\textbf {\bibinfo
  {volume} {000}},\ \bibinfo {pages} {0} (\bibinfo {year} {2018})}\BibitemShut
  {NoStop}%
\bibitem [{\citenamefont {Hopkins}\ \emph
  {et~al.}(2024{\natexlab{a}})\citenamefont {Hopkins}, \citenamefont {Grudic},
  \citenamefont {Kremer}, \citenamefont {Offner}, \citenamefont {Guszejnov},\
  and\ \citenamefont {Rosen}}]{Hopkins2024a}%
  \BibitemOpen
  \bibfield  {author} {\bibinfo {author} {\bibfnamefont {P.~F.}\ \bibnamefont
  {Hopkins}}, \bibinfo {author} {\bibfnamefont {M.~Y.}\ \bibnamefont {Grudic}},
  \bibinfo {author} {\bibfnamefont {K.}~\bibnamefont {Kremer}}, \bibinfo
  {author} {\bibfnamefont {S.~S.~R.}\ \bibnamefont {Offner}}, \bibinfo {author}
  {\bibfnamefont {D.}~\bibnamefont {Guszejnov}}, \ and\ \bibinfo {author}
  {\bibfnamefont {A.~L.}\ \bibnamefont {Rosen}},\ }\href {\doibase
  10.33232/001C.122857} {\bibfield  {journal} {\bibinfo  {journal} {The Open
  Journal of Astrophysics}\ }\textbf {\bibinfo {volume} {7}} (\bibinfo {year}
  {2024}{\natexlab{a}}),\ 10.33232/001C.122857}\BibitemShut {NoStop}%
\bibitem [{\citenamefont {Hopkins}\ \emph
  {et~al.}(2024{\natexlab{b}})\citenamefont {Hopkins}, \citenamefont {Squire},
  \citenamefont {Su}, \citenamefont {Steinwandel}, \citenamefont {Kremer},
  \citenamefont {Shi}, \citenamefont {Grudić}, \citenamefont {Wellons},
  \citenamefont {Faucher-Giguère}, \citenamefont {Anglés-Alcázar},
  \citenamefont {Murray},\ and\ \citenamefont {Quataert}}]{Hopkins2024b}%
  \BibitemOpen
  \bibfield  {author} {\bibinfo {author} {\bibfnamefont {P.~F.}\ \bibnamefont
  {Hopkins}}, \bibinfo {author} {\bibfnamefont {J.}~\bibnamefont {Squire}},
  \bibinfo {author} {\bibfnamefont {K.~Y.}\ \bibnamefont {Su}}, \bibinfo
  {author} {\bibfnamefont {U.~P.}\ \bibnamefont {Steinwandel}}, \bibinfo
  {author} {\bibfnamefont {K.}~\bibnamefont {Kremer}}, \bibinfo {author}
  {\bibfnamefont {Y.}~\bibnamefont {Shi}}, \bibinfo {author} {\bibfnamefont
  {M.~Y.}\ \bibnamefont {Grudić}}, \bibinfo {author} {\bibfnamefont
  {S.}~\bibnamefont {Wellons}}, \bibinfo {author} {\bibfnamefont {C.~A.}\
  \bibnamefont {Faucher-Giguère}}, \bibinfo {author} {\bibfnamefont
  {D.}~\bibnamefont {Anglés-Alcázar}}, \bibinfo {author} {\bibfnamefont
  {N.}~\bibnamefont {Murray}}, \ and\ \bibinfo {author} {\bibfnamefont
  {E.}~\bibnamefont {Quataert}},\ }\href {\doibase 10.21105/ASTRO.2310.04506}
  {\bibfield  {journal} {\bibinfo  {journal} {Open Journal of Astrophysics}\
  }\textbf {\bibinfo {volume} {7}} (\bibinfo {year} {2024}{\natexlab{b}}),\
  10.21105/ASTRO.2310.04506}\BibitemShut {NoStop}%
\bibitem [{\citenamefont {Hopkins}\ \emph
  {et~al.}(2024{\natexlab{c}})\citenamefont {Hopkins}, \citenamefont {Grudić},
  \citenamefont {Su}, \citenamefont {Wellons}, \citenamefont
  {Anglés-Alcázar}, \citenamefont {Steinwandel}, \citenamefont {Guszejnov},
  \citenamefont {Murray}, \citenamefont {Faucher-Giguère}, \citenamefont
  {Quataert},\ and\ \citenamefont {Kereš}}]{Hopkins2024c}%
  \BibitemOpen
  \bibfield  {author} {\bibinfo {author} {\bibfnamefont {P.~F.}\ \bibnamefont
  {Hopkins}}, \bibinfo {author} {\bibfnamefont {M.~Y.}\ \bibnamefont
  {Grudić}}, \bibinfo {author} {\bibfnamefont {K.~Y.}\ \bibnamefont {Su}},
  \bibinfo {author} {\bibfnamefont {S.}~\bibnamefont {Wellons}}, \bibinfo
  {author} {\bibfnamefont {D.}~\bibnamefont {Anglés-Alcázar}}, \bibinfo
  {author} {\bibfnamefont {U.~P.}\ \bibnamefont {Steinwandel}}, \bibinfo
  {author} {\bibfnamefont {D.}~\bibnamefont {Guszejnov}}, \bibinfo {author}
  {\bibfnamefont {N.}~\bibnamefont {Murray}}, \bibinfo {author} {\bibfnamefont
  {C.~A.}\ \bibnamefont {Faucher-Giguère}}, \bibinfo {author} {\bibfnamefont
  {E.}~\bibnamefont {Quataert}}, \ and\ \bibinfo {author} {\bibfnamefont
  {D.}~\bibnamefont {Kereš}},\ }\href {\doibase 10.21105/astro.2309.13115}
  {\bibfield  {journal} {\bibinfo  {journal} {Open Journal of Astrophysics}\
  }\textbf {\bibinfo {volume} {7}},\ \bibinfo {pages} {18} (\bibinfo {year}
  {2024}{\natexlab{c}})}\BibitemShut {NoStop}%
\bibitem [{\citenamefont {Nesti}\ and\ \citenamefont
  {Salucci}(2013)}]{Nesti2013}%
  \BibitemOpen
  \bibfield  {author} {\bibinfo {author} {\bibfnamefont {F.}~\bibnamefont
  {Nesti}}\ and\ \bibinfo {author} {\bibfnamefont {P.}~\bibnamefont
  {Salucci}},\ }\href {\doibase 10.1088/1475-7516/2013/07/016} {\bibfield
  {journal} {\bibinfo  {journal} {Journal of Cosmology and Astroparticle
  Physics}\ }\textbf {\bibinfo {volume} {2013}} (\bibinfo {year} {2013}),\
  10.1088/1475-7516/2013/07/016}\BibitemShut {NoStop}%
\bibitem [{\citenamefont {Lacroix}(2018)}]{Lacroix2018}%
  \BibitemOpen
  \bibfield  {author} {\bibinfo {author} {\bibfnamefont {T.}~\bibnamefont
  {Lacroix}},\ }\href {\doibase 10.1051/0004-6361/201832652} {\bibfield
  {journal} {\bibinfo  {journal} {Astronomy \& Astrophysics}\ }\textbf
  {\bibinfo {volume} {619}},\ \bibinfo {pages} {A46} (\bibinfo {year}
  {2018})}\BibitemShut {NoStop}%
\bibitem [{\citenamefont {Prabhu}(2021)}]{Prabhu2021}%
  \BibitemOpen
  \bibfield  {author} {\bibinfo {author} {\bibfnamefont {A.}~\bibnamefont
  {Prabhu}},\ }\href {\doibase 10.1103/physrevd.104.055038} {\bibfield
  {journal} {\bibinfo  {journal} {Physical Review D}\ }\textbf {\bibinfo
  {volume} {104}} (\bibinfo {year} {2021}),\
  10.1103/physrevd.104.055038}\BibitemShut {NoStop}%
\bibitem [{\citenamefont {Heisenberg}\ and\ \citenamefont
  {Euler}(1936)}]{heisenberg1936folgerungen}%
  \BibitemOpen
  \bibfield  {author} {\bibinfo {author} {\bibfnamefont {W.}~\bibnamefont
  {Heisenberg}}\ and\ \bibinfo {author} {\bibfnamefont {H.}~\bibnamefont
  {Euler}},\ }\href@noop {} {\bibfield  {journal} {\bibinfo  {journal}
  {Zeitschrift f{\"u}r Physik}\ }\textbf {\bibinfo {volume} {98}},\ \bibinfo
  {pages} {714} (\bibinfo {year} {1936})}\BibitemShut {NoStop}%
\bibitem [{\citenamefont {Medvedev}(2023)}]{Medvedev:2023wzw}%
  \BibitemOpen
  \bibfield  {author} {\bibinfo {author} {\bibfnamefont {M.~V.}\ \bibnamefont
  {Medvedev}},\ }\href {\doibase 10.1063/5.0160628} {\bibfield  {journal}
  {\bibinfo  {journal} {Phys. Plasmas}\ }\textbf {\bibinfo {volume} {30}},\
  \bibinfo {pages} {092112} (\bibinfo {year} {2023})},\ \Eprint
  {http://arxiv.org/abs/2309.07316} {arXiv:2309.07316 [physics.plasm-ph]}
  \BibitemShut {NoStop}%
\bibitem [{\citenamefont {{Beskin}}\ \emph {et~al.}(1993)\citenamefont
  {{Beskin}}, \citenamefont {{Gurevich}},\ and\ \citenamefont
  {{Istomin}}}]{Gurevich1993}%
  \BibitemOpen
  \bibfield  {author} {\bibinfo {author} {\bibfnamefont {V.~S.}\ \bibnamefont
  {{Beskin}}}, \bibinfo {author} {\bibfnamefont {A.~V.}\ \bibnamefont
  {{Gurevich}}}, \ and\ \bibinfo {author} {\bibfnamefont {Y.~N.}\ \bibnamefont
  {{Istomin}}},\ }\href@noop {} {\emph {\bibinfo {title} {{Physics of the
  pulsar magnetosphere}}}}\ (\bibinfo {year} {1993})\BibitemShut {NoStop}%
\bibitem [{\citenamefont {Brahma}\ \emph {et~al.}(2023)\citenamefont {Brahma},
  \citenamefont {Berlin},\ and\ \citenamefont {Schutz}}]{Brahma:2023zcw}%
  \BibitemOpen
  \bibfield  {author} {\bibinfo {author} {\bibfnamefont {N.}~\bibnamefont
  {Brahma}}, \bibinfo {author} {\bibfnamefont {A.}~\bibnamefont {Berlin}}, \
  and\ \bibinfo {author} {\bibfnamefont {K.}~\bibnamefont {Schutz}},\ }\href
  {\doibase 10.1103/PhysRevD.108.095045} {\bibfield  {journal} {\bibinfo
  {journal} {Phys. Rev. D}\ }\textbf {\bibinfo {volume} {108}},\ \bibinfo
  {pages} {095045} (\bibinfo {year} {2023})},\ \Eprint
  {http://arxiv.org/abs/2308.08586} {arXiv:2308.08586 [hep-ph]} \BibitemShut
  {NoStop}%
\bibitem [{\citenamefont {{Sridhar}}\ and\ \citenamefont
  {{Goldreich}}(1994)}]{GoldreichSridhar1}%
  \BibitemOpen
  \bibfield  {author} {\bibinfo {author} {\bibfnamefont {S.}~\bibnamefont
  {{Sridhar}}}\ and\ \bibinfo {author} {\bibfnamefont {P.}~\bibnamefont
  {{Goldreich}}},\ }\href {\doibase 10.1086/174600} {\bibfield  {journal}
  {\bibinfo  {journal} {\apj}\ }\textbf {\bibinfo {volume} {432}},\ \bibinfo
  {pages} {612} (\bibinfo {year} {1994})}\BibitemShut {NoStop}%
\bibitem [{\citenamefont {{Goldreich}}\ and\ \citenamefont
  {{Sridhar}}(1995)}]{GoldreichSridhar2}%
  \BibitemOpen
  \bibfield  {author} {\bibinfo {author} {\bibfnamefont {P.}~\bibnamefont
  {{Goldreich}}}\ and\ \bibinfo {author} {\bibfnamefont {S.}~\bibnamefont
  {{Sridhar}}},\ }\href {\doibase 10.1086/175121} {\bibfield  {journal}
  {\bibinfo  {journal} {\apj}\ }\textbf {\bibinfo {volume} {438}},\ \bibinfo
  {pages} {763} (\bibinfo {year} {1995})}\BibitemShut {NoStop}%
\end{thebibliography}%

\clearpage





\appendix

\section{Axion-photon Conversion in Super-critical Magnetic Fields}\label{sec:supercritical_conversion}

The Lagrangian describing axion-photon mixing in-medium is

\begin{align}
    \mathcal{L} \supset -{1\over 4} &\gagg a F_{\mu \nu}{F}^{\mu\nu} - A_\mu J^\mu + \nonumber\\
    & {1\over 2} \partial_\mu a \partial^\mu a - {1 \over 2} m_a^2 a^2 - {1 \over 4} F_{\mu \nu} \tilde{F}^{\mu \nu} a,
\end{align}
where $F^{\mu \nu}$ is the electromagnetic tensor, $\tilde{F}^{\mu \nu} = \epsilon^{\mu\nu\alpha\beta} F_{\alpha \beta}/2$, $\epsilon$ is the Levi-Civita symbol, $J^\mu = (\phi, \vec{A})$ is the matter four-current, $\phi$ the electrostatic potential, and $\vec{A}$ the vector potential. The axion field, mass, and coupling constant are represented by $a$, $m_a$, and $\gagg$, respectively. In a strong background magnetic field, the QED Lagrangian receives quantum corrections. The renormalized one-loop QED Lagrangian was computed exactly by Euler and Heisenberg~\cite{heisenberg1936folgerungen},

\begin{align}
    \mathcal{L}_{\rm EH} &= -{1\over 8\pi^2} \displaystyle\int {d \eta \over \eta^3} e^{- \eta m_e^2} \nonumber \\
    &\left[ {e^2 a b \eta^2 \over \tanh(e b \eta) \tanh(e a \eta)} + {e^2 \eta^2\over 3} (a^2 -b^2) - 1\right],
\end{align}
where $a$ and $b$ are related to Lorentz invariant quantities $\mathcal{F} = F^2/4$, $\mathcal{G} = F \tilde{F}/4$, according to

\begin{align}
    a = \sqrt{ \sqrt{\mathcal{F}^2 + \mathcal{G}^2} - \mathcal{F}}, \quad b = \sqrt{ \sqrt{\mathcal{F}^2 + \mathcal{G}^2} + \mathcal{F}}.
\end{align}

The Lagrangian leads to the equations of axion electrodynamics 

\begin{gather}
    \vec{\nabla} \cdot \vec{D} = - \gagg \nabla a \cdot \vec{B}, \label {eqn:gauss} \\
    \vec{\nabla} \cdot \vec{B} = 0, \label{eqn:monopole} \\
    \vec{\nabla} \times \vec{E} = -{\partial \vec{B} \over \partial t}, \label{eqn:faraday} \\
    \vec{\nabla} \times \vec{H} = {\partial \vec{D} \over \partial t} + \gagg\left( \vec{B} \partial_t a + \vec{\nabla} a \times \vec{E}\right), \label{eqn:ampere} \\
    \left(\partial_t^2 -\nabla^2 + m_a^2 \right) a = \gagg \vec{E} \cdot \vec{B}, \label{eqn:klein-gordon}
\end{gather}
where $D$ and $H$ are the displacement and magnetizing fields, respectively. They are defined by the constitutive relations, $D_i = \varepsilon_{ij} E_i$ and $B_i = \mu_{ij} H_j $, where $\varepsilon$ and $\mu$ are the permittivity and permeability tensors, respectively. These tensors receive contributions from both the surrounding medium (in this case, defined by the four-current), as well as QED vacuum polarization,

\begin{eqnarray}
    \varepsilon_{ij} &=& \delta_{ij} + \chi_{ij}^{\rm plasma} + \chi_{ij}^{\rm vac} \\ (\mu^{-1})_{ij}  &=& \delta_{ij} + \eta_{ij}^{\rm plasma} + \eta_{ij}^{\rm vac} \, , \\ \nonumber
\end{eqnarray}
where $\chi$ ($\xi$) represents the electric (magnetic) susceptibility. The vacuum susceptibilities were derived explicitly in~\cite{Medvedev:2023wzw},
    \begin{eqnarray}\label{eqn:chivac}
        \chi_{ij}^{\rm vac} &=& - C_\delta \delta_{ij} + C_\epsilon b_i b_j \\ \eta_{ij}^{\rm vac} &=& - C_\delta \delta_{ij} - C_\mu b_i b_j \,
    \end{eqnarray}
where $\vec{b} = \vec{B}/B_c$ and $B_c = m_e^2/e \approx 4.4\times 10^{13}$ G is the Schwinger critical field strength. For magnetic fields of arbitrary strength, 

\begin{widetext}
\begin{align}
    C_\delta &=  \alpha \left[ 4 \zeta^{(1,0)}\left(-1,{1 \over 2b} \right) - {1 \over 4b^2} + {1 \over 3} \ln(2b) + {1 \over 2b} \ln \left( {\pi \over b}\right) - {1 \over b} \ln \left( \Gamma\left( {1 \over 2b} \right) \right) - {1 \over 6} \right], \\
    C_\epsilon &=  \alpha \left[ 4 \zeta^{(1,0)}\left(-1,{1 \over 2b} \right) - {1 \over 4b^2} + {b \over 3} - {1 \over 3} \psi\left( 1 + {1 \over 2 b} \right) + {1 \over 2b} \ln \left( {\pi \over b}\right) - {1 \over b} \ln \left( \Gamma\left( {1 \over 2b} \right) \right) - {1 \over 6} \right], \\
    C_\mu &=  \alpha \left[ {1 \over 2b} \psi\left( 1 + {1 \over 2 b} \right) - {1 \over 2 b^2} - {1 \over 2b} + {1 \over 2b} \ln(4\pi b)- {1 \over b} \ln \left( \Gamma\left( {1 \over 2b} \right) \right) + {1 \over 3}  \right],
\end{align}
\end{widetext}
where $\zeta^{(1,0)}(z, x) = \partial_z \zeta(z, x)$, $\zeta(z,x)$ is the Hurwitz zeta-function, $\Gamma(x)$ is the Gamma function, and $\psi(x) = \Gamma'(x)/\Gamma(x)$ is the polygamma function. 

The plasma susceptibility for a general anisotropic plasma can be derived by substituting a plane wave solution, $\propto e^{i \vec{k} \cdot \vec{x} - i \omega t}$, into Eqs. (\ref{eqn:gauss}--\ref{eqn:klein-gordon}). We consider a coordinate system in which $\vec{k} = k \hat{z}$, and the magnetic field lies in the $yz$ plane at an angle $\theta$ to the $z$-axis. The resulting plasma susceptibility is ~\cite{Gurevich1993, millar2021axionphotonUPDATED} 

\begin{align} \label{eqn:chiplasma}
    \chi_{ij}^{\rm plasma} = \begin{pmatrix} 0 & 0 & 0\\
    0 & - \left\langle \omega_p^2 \over \gamma^3 \tilde{\omega}^2 \right\rangle \sin^2\theta & \left\langle \omega_p^2 \over \gamma^3 \tilde{\omega}^2 \right\rangle \sin\theta \cos \theta \\ 0 & \left\langle \omega_p^2 \over \gamma^3 \tilde{\omega}^2 \right\rangle \sin\theta \cos \theta & - \left\langle \omega_p^2 \over \gamma^3 \tilde{\omega}^2 \right\rangle \cos^2 \theta\end{pmatrix},
\end{align}
where $\omega_p = \sum_s \sqrt{q_s^2 n_s/m_s}$ is the plasma frequency, summed over all species (e.g., electrons, positrons, ions, etc.), and $q_s$, $n_s$, $m_s$ are the charge, number density, and mass of species $s$, respectively. In the case of plasma surrounding magnetars, the relevant species are electrons and positrons, so we drop the species label going forward. The plasma is assumed to move along the magnetic field with speed $v_\parallel$, with corresponding boost factor $\gamma = (1 - v_\parallel^2)^{-1/2}$ and $\tilde{\omega} = \omega - k_\parallel v_\parallel$ (where $k_\parallel = \vec{k} \cdot \hat{B}$), in the NS frame\footnote{The rest-frame of a NS is technically a frame co-rotating with the star (and is thus non-inertial). Since the timescale over which axion-photon mixing takes place is much smaller than the rotation period, for the purposes of this computation, we assume the NS frame to be static.}. The one-dimensional assumption is an excellent approximation in magnetar-strength fields, as particles are generally confined to their ground Landau levels and move along magnetic fields as ``beads on a wire''. The angled brackets $\langle ... \rangle$ represents an ensemble average,
\begin{align}
    \left\langle \mathcal{O} \right\rangle &= \displaystyle\sum_s \displaystyle\int_{-\infty}^\infty \mathcal{O} f_{s, \parallel} (p_\parallel) \ dp_\parallel,
\end{align}
where $f_{s,\parallel}$ is the distribution function of species $s$. Putting together (\ref{eqn:chivac}) and (\ref{eqn:chiplasma}), the electric permittivity is
\begin{widetext}
\begin{gather}
    \varepsilon_{ij} = \begin{pmatrix} 1 - C_\delta & 0 & 0 \\
    0 & 1 - C_\delta + \left(C_\epsilon  - \left\langle \omega_p^2 \over \gamma^3 \tilde{\omega}^2 \right\rangle \right) \sin^2\theta & \left(C_\epsilon  + \left\langle \omega_p^2 \over \gamma^3 \tilde{\omega}^2 \right\rangle \right) \sin \theta \cos \theta \\
    0 & \left(C_\epsilon  + \left\langle \omega_p^2 \over \gamma^3 \tilde{\omega}^2 \right\rangle \right) \sin \theta \cos \theta & 1 - C_\delta + \left(C_\epsilon  - \left\langle \omega_p^2 \over \gamma^3 \tilde{\omega}^2 \right\rangle \right) \cos^2\theta \end{pmatrix}; \\(\mu^{-1})_{ij} = \begin{pmatrix}1 - C_\delta & 0 & 0 \\
    0 & 1 - C_\delta - C_\mu  \sin^2\theta & - C_\mu  \sin \theta \cos\theta \\
    0 & - C_\mu  \sin\theta \cos\theta & 1 - C_\delta -C_\mu  \cos^2\theta\end{pmatrix}.
\end{gather}
\end{widetext}

We have assumed above that the magnetic permeability receives no contribution from the plasma. In the limit of large magnetic field, 

\begin{align}
    C_\epsilon &= {\alpha b \over 3}, \quad C_\mu = - {\alpha \over 6}, \nonumber\\
     C_\delta &= {\alpha }\left({1 \over 3} \ln(2b) + 4 \zeta^{(1,0)}(-1,0)- {1 \over 6} \right), 
\end{align}
where $4\zeta^{(1,0)}(-1,0)-1/6 \approx -0.83$. Magnetic fields of astrophysical relevance have $b\lesssim 10^3$, for which $C_\mu, C_\epsilon \ll 1$ and can be safely neglected. The fastest-growing term, $C_\epsilon$, becomes order unity at $b = 3/\alpha \approx 400$, corresponding to a magnetic field of $B \approx 1.8 \times 10^{16}$ G,

Magnetic fields of such strength have been inferred from very long period magnetars, making this contribution significant. Keeping just terms containing $C_\epsilon$, the magnetic permeability tensor becomes trivial and the electric permittivity tensor takes the simpler form

\begin{align}
    &\varepsilon_{ij} = \nonumber \\
    &\begin{pmatrix} 1 & 0 & 0 \\
    0 & 1 + \left(C_\epsilon  - \left\langle \omega_p^2 \over \gamma^3 \tilde{\omega}^2 \right\rangle \right) \sin^2\theta & \left(C_\epsilon  + \left\langle \omega_p^2 \over \gamma^3 \tilde{\omega}^2 \right\rangle \right) \sin \theta \cos \theta \\
    0 & \left(C_\epsilon  + \left\langle \omega_p^2 \over \gamma^3 \tilde{\omega}^2 \right\rangle \right) \sin \theta \cos \theta & 1 + \left(C_\epsilon  - \left\langle \omega_p^2 \over \gamma^3 \tilde{\omega}^2 \right\rangle \right) \cos^2\theta \end{pmatrix}.
\end{align}

\section{Computationally Ray-Traced Diagrams}
\label{sec:raytracingappendix}

In this section, we show the resonant conversion surfaces (specifically, we plot point samples from the MC sampling procedure, with the color coding reflecting the relative weight of the sample itself) for all the sheared model geometries in Fig.~\ref{fig:raytraced-conversion-surfaces}, including the collisional quarter, half, and slice models. We also display Mollweide projections for the collisional quarter model in Fig.~\ref{fig:quarter_mollweide_projections}, showing both the radiated power and the relative width of the line across the sky. Here, one can see a mild azimuthal asymmetry in the flux, reflective of the `quarter' feature itself. Note that rotationally averaging implies that this translates into a time dependence of the radio line, which in turn could be used to further differentiate a potential signal from background. In the right panel of  Fig.~\ref{fig:quarter_mollweide_projections}, one can see that the characteristic width is typically $\lesssim 10^{-4}$, justifying the choice of bandwidth adopted in the previous section.

\begin{figure*}
    \centering
    \includegraphics[width=0.32\linewidth]{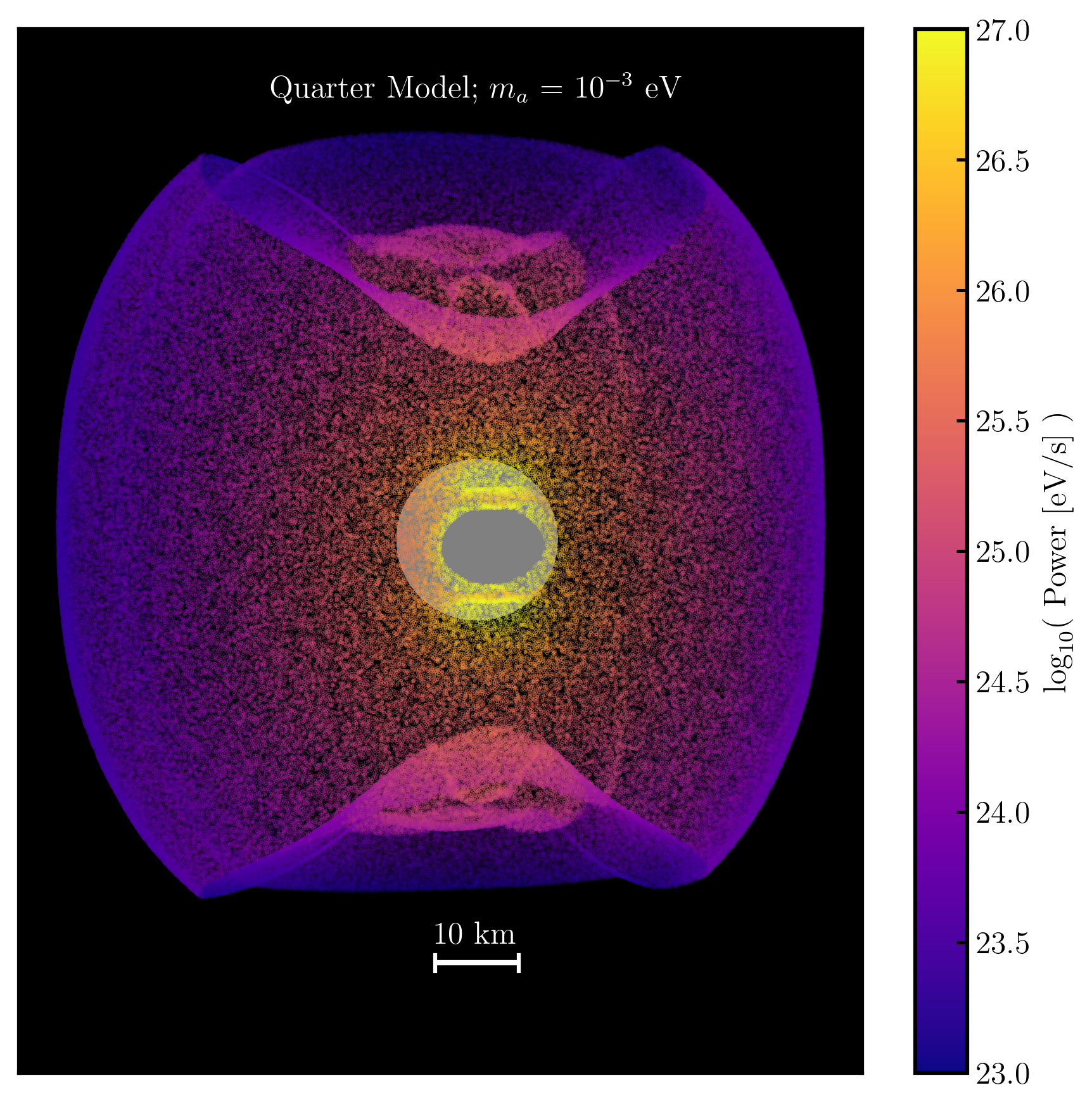}
    \includegraphics[width=0.32\linewidth]{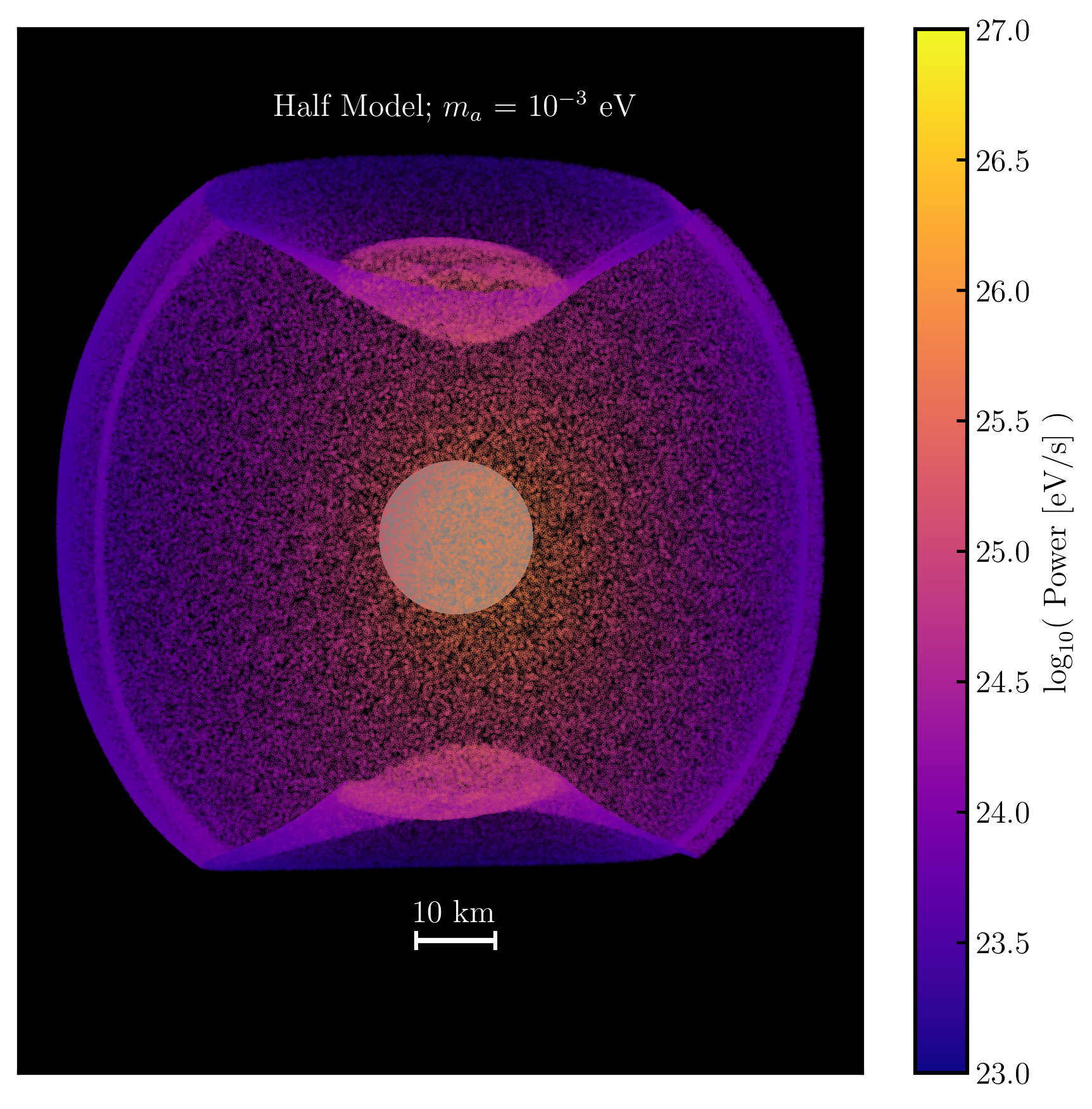}
    \includegraphics[width=0.32\linewidth]{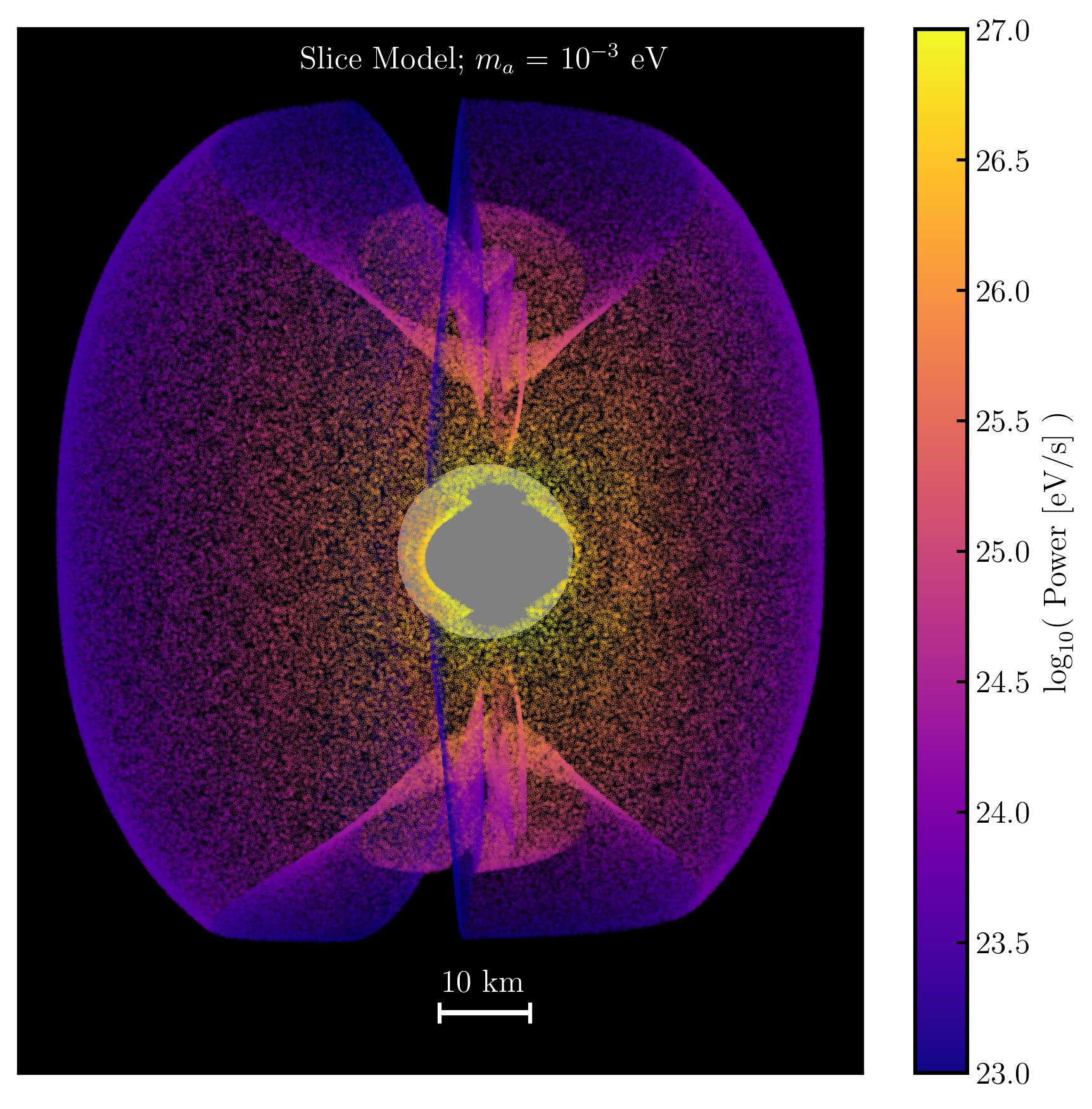}

    \caption{
    Ray-traced resonant conversion surface for the quarter, half, and slice arcade geometries {\color{black} in the collisional TK20 model, as viewed from the side of the magnetar
    opposite to the gamma-ray emitting arcade. We assume} $m_a = 10^{-3}\,$eV, $g_{a\gamma\gamma}=10^{-12}\GeV^{-1}$, $B_{\rm pole} = 10 B_Q$, and a local dark matter density of $0.4\GeV\cm^{-3}$.
    {\color{black} Color scale indicates the} resonant conversion photon power on the unit sphere. The gray sphere represents the NS surface and photons originating in front of the NS surface are masked. The highest-power {\color{black} emission zones are those behind the
    magnetar,} close to the surface where 
    the magnetic field is strongest.
    }
    \label{fig:raytraced-conversion-surfaces}
\end{figure*}

\begin{figure*}
    \centering
    \includegraphics[width=0.49\linewidth]{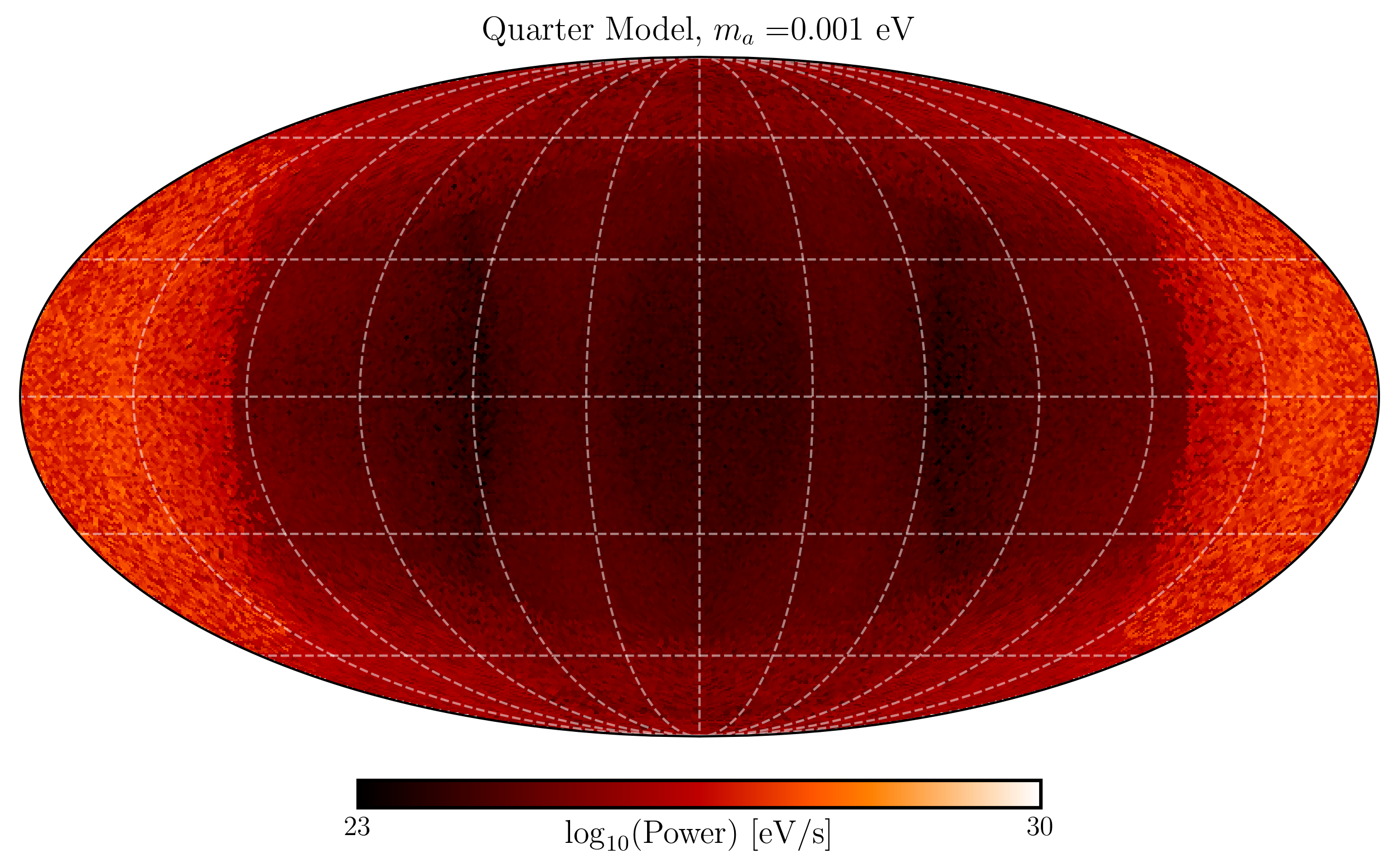}
    \includegraphics[width=0.49\linewidth]{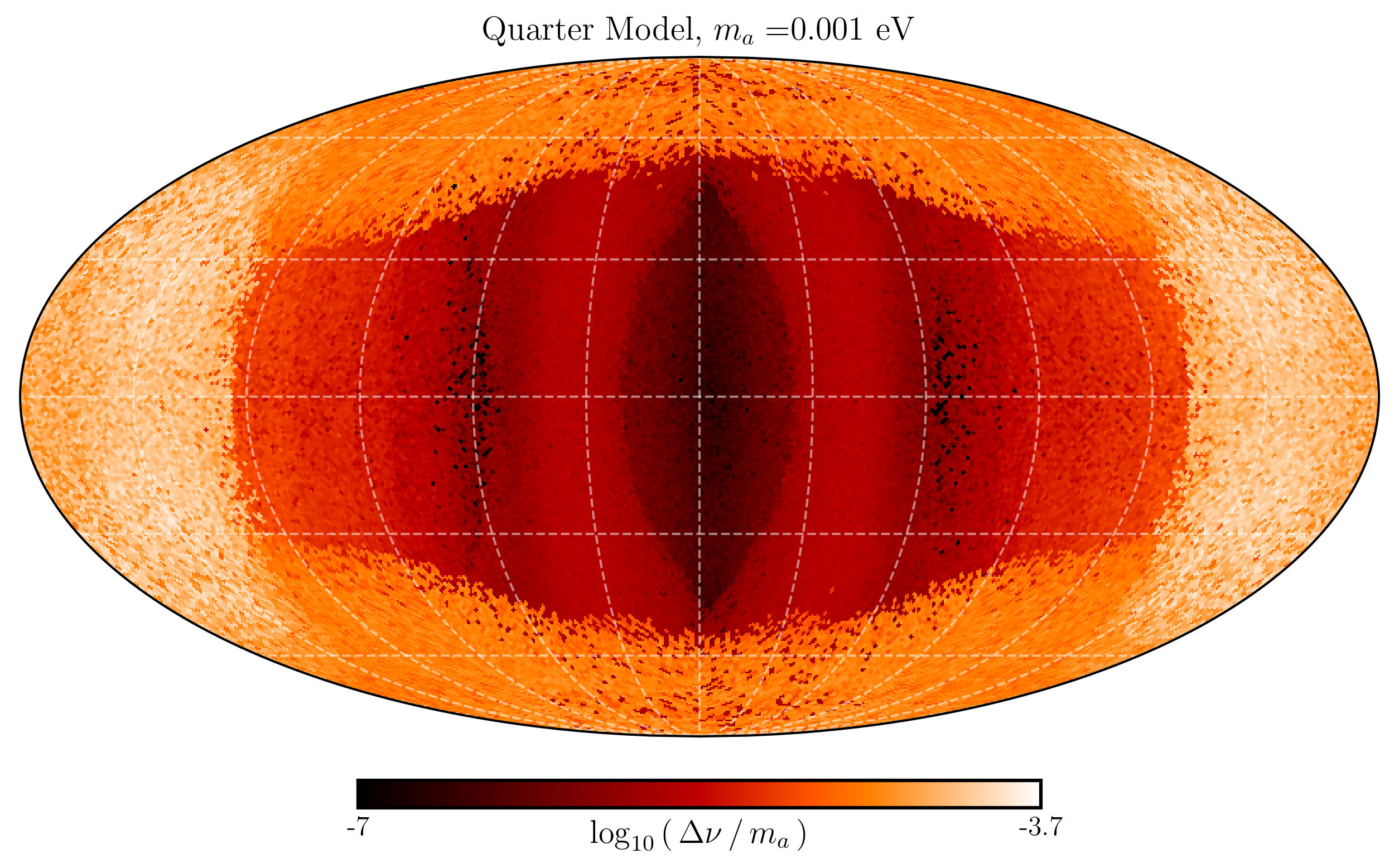}

    \caption{
    Mollweide projections {\color{black} of the escaping line power (left panel) and line width (right panel) for the collisional TK20 model with quarter arcade geometry, here} assuming $m_a = 10^{-3}\,$eV, $g_{a\gamma\gamma}=10^{-12}\GeV^{-1}$, and a local dark matter density of $0.4\GeV\cm^{-3}$. The central regions of these Mollweide projections correspond to the azimuthal angles directly in front of the gamma-ray emitting arcades. Both line power and line width
    peak on the side of the magnetar opposite to the current-carrying arcade. 
    }
    \label{fig:quarter_mollweide_projections}
\end{figure*}

\section{Telescope Sensitivities}
\label{sec:telescope_appendices}

\begin{figure}
    \centering
    \includegraphics[width=1.0\linewidth]{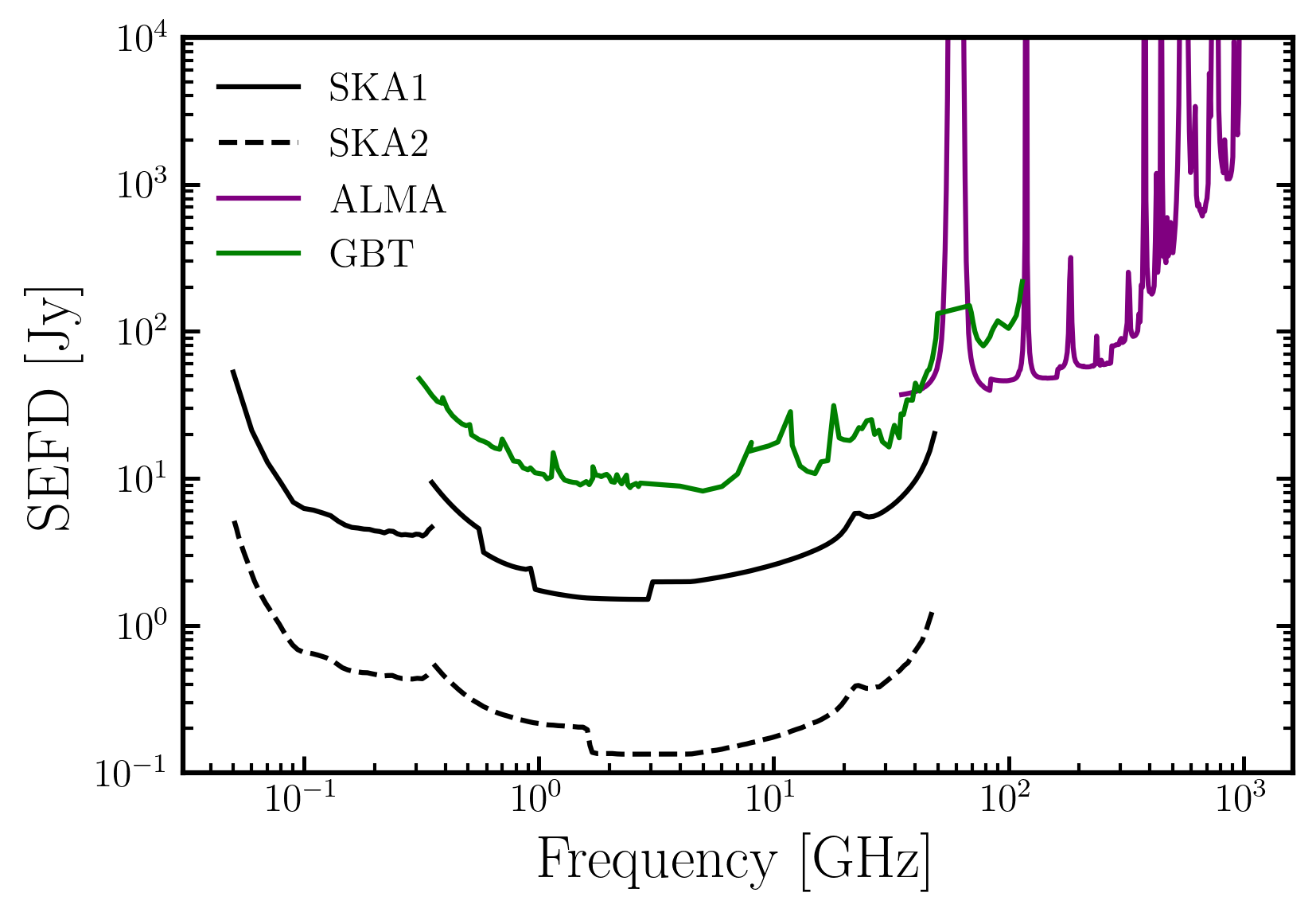}

    \vspace{-1em}
    \caption{Telescope system equivalent flux densities (SEFDs) for the four telescope used to project sensitivities in this investigation: the Green Bank Telescope (GBT), the Atacama Large Millimeter Array (ALMA) and the Square Kilometer Array's Phase 1 and Phase 2 (SKA1 and SKA2).  
    }
    \label{fig:telescope_SEFDs}
    \vspace{-2em}
\end{figure}

\begin{figure*}
    \centering
    \includegraphics[width=0.49\linewidth]{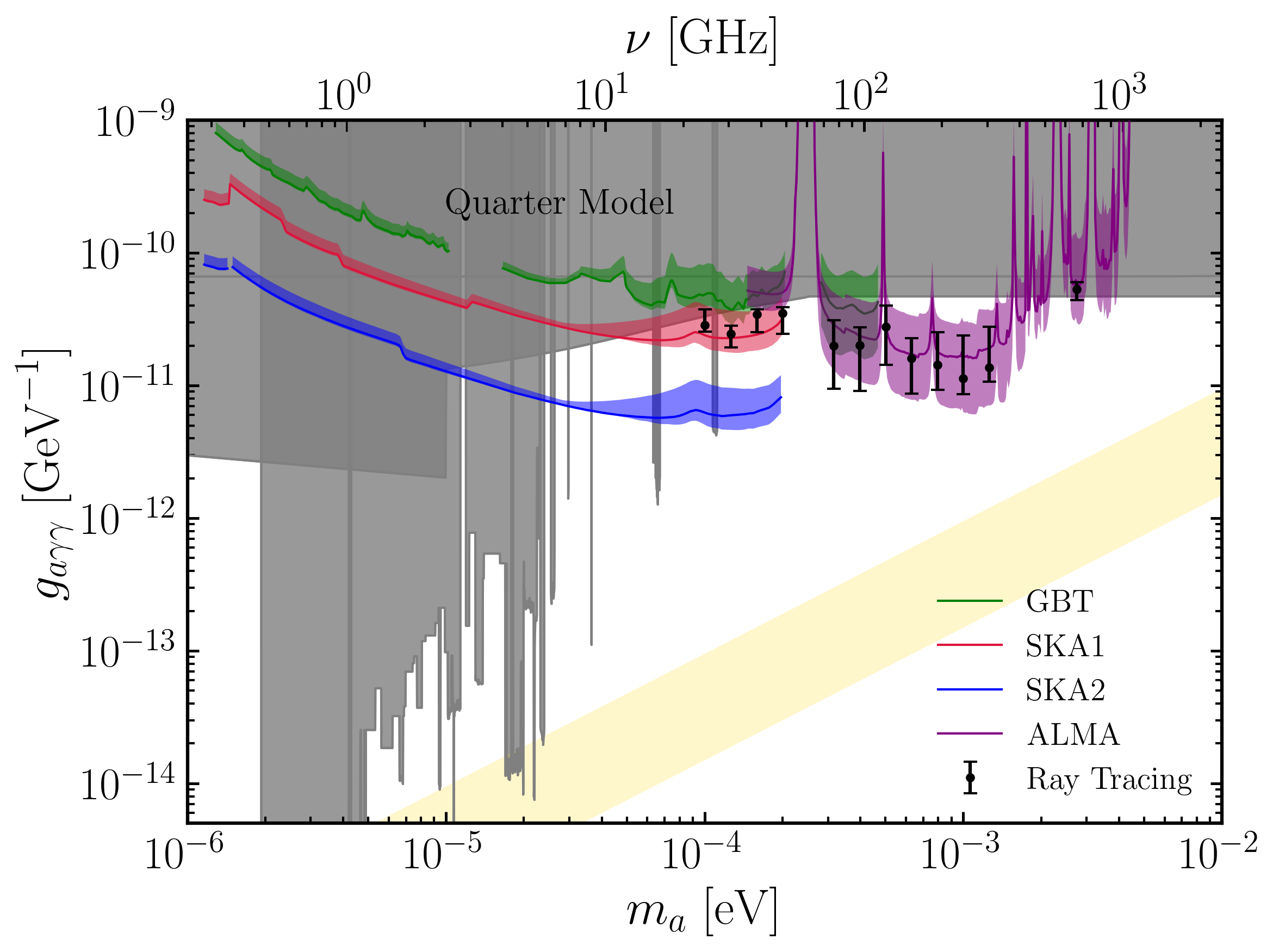}
    \includegraphics[width=0.49\linewidth]{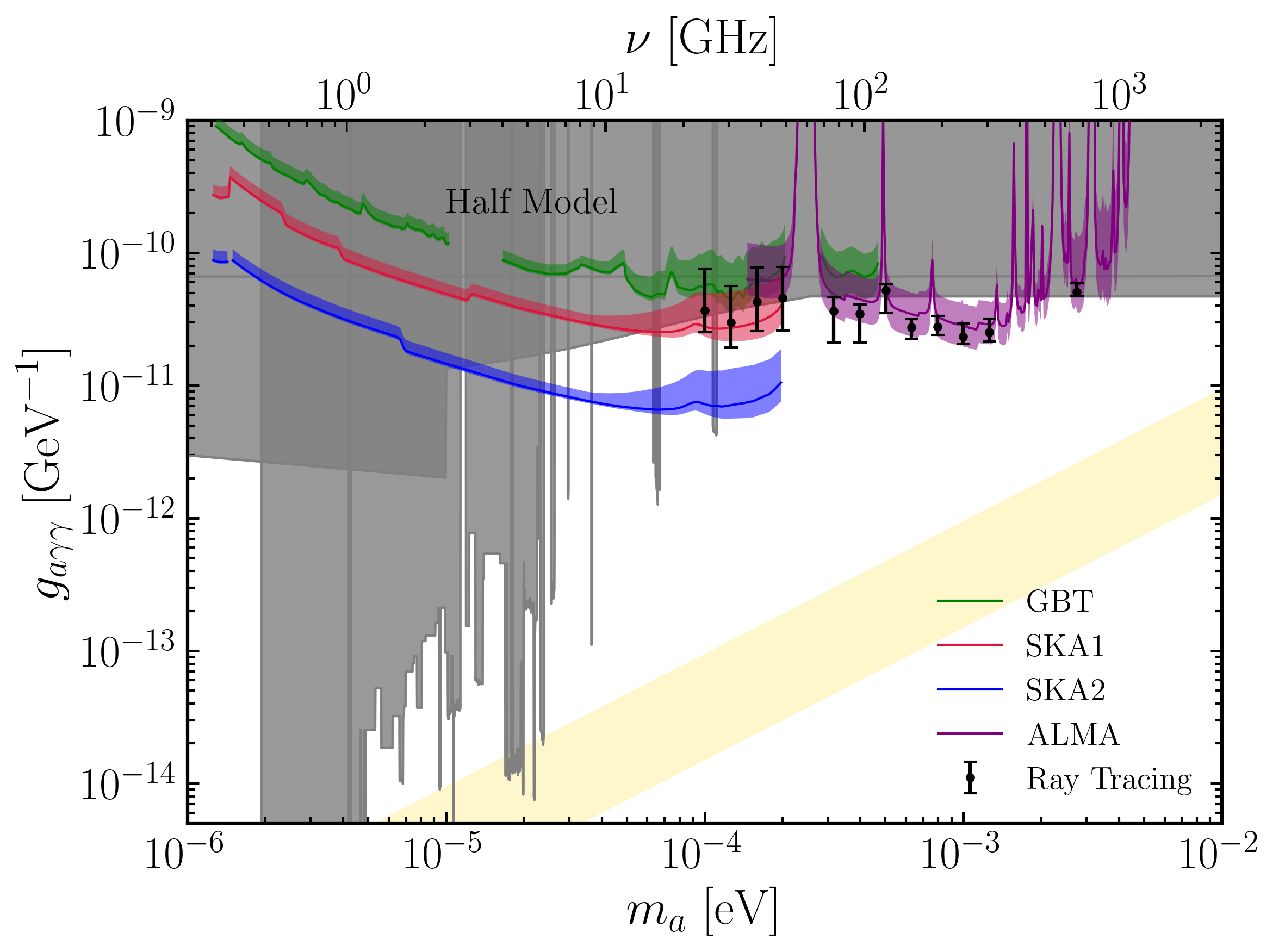}
    \includegraphics[width=0.49\linewidth]{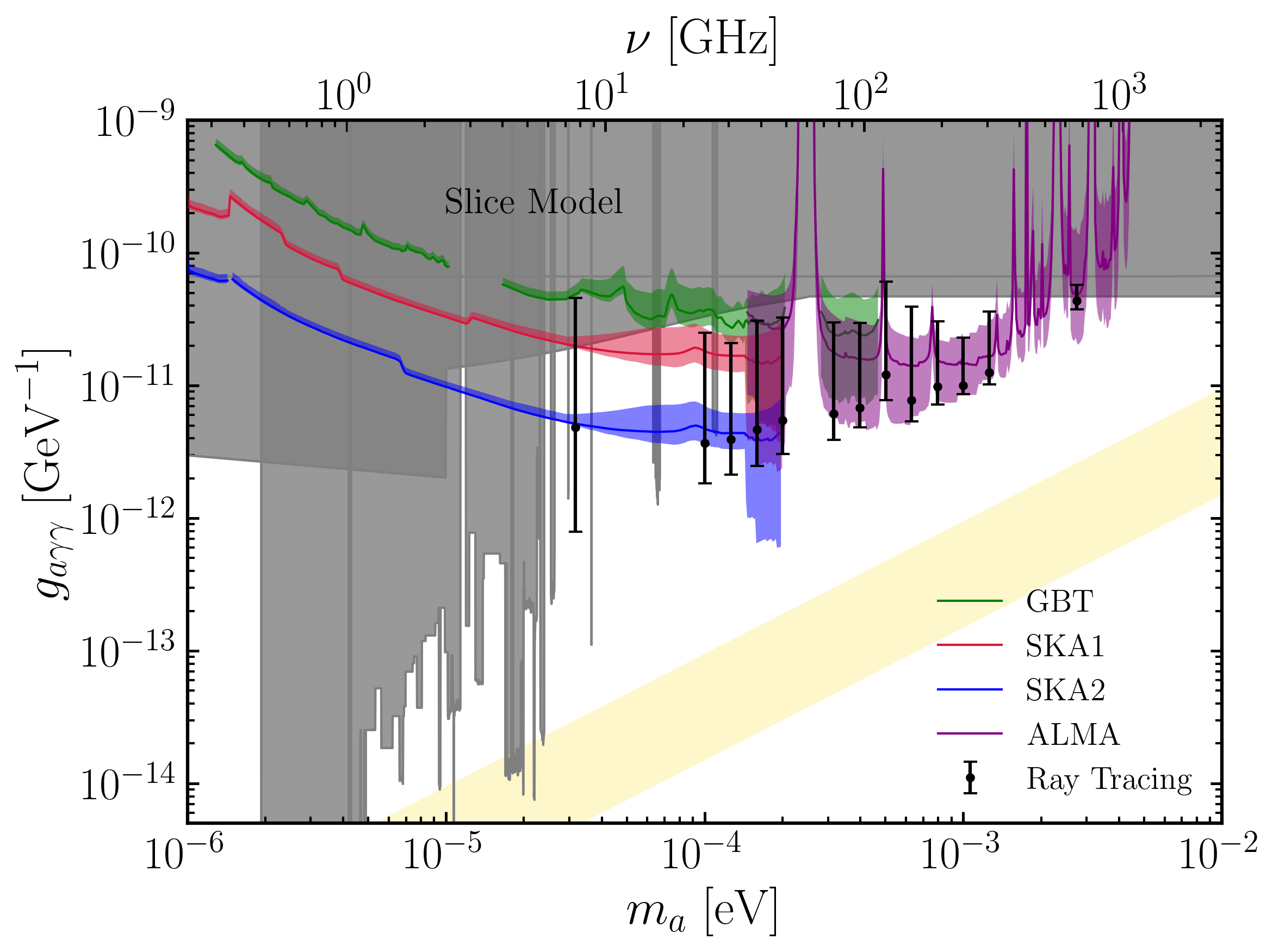}
    \includegraphics[width=0.49\linewidth]{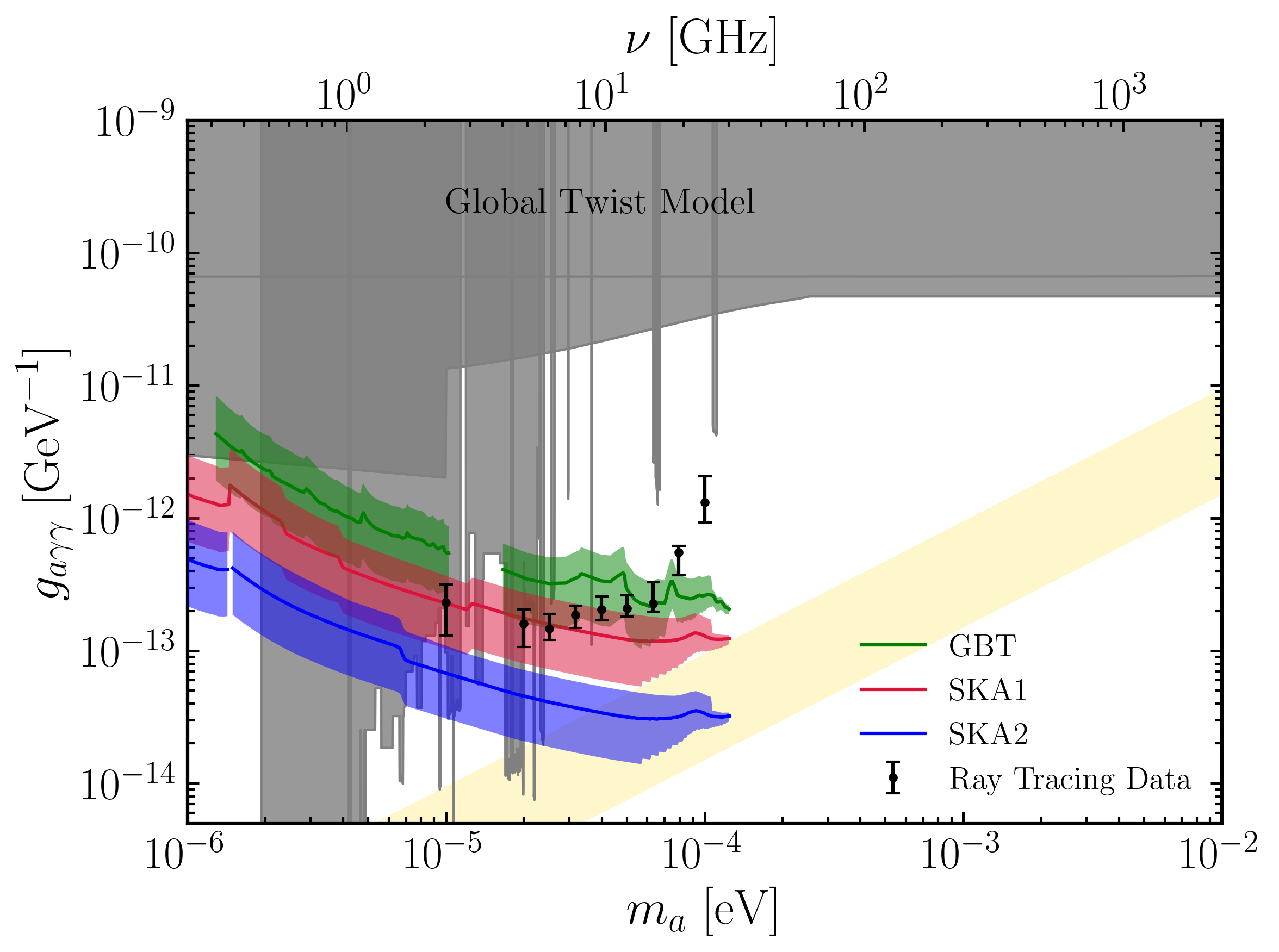}

    \vspace{-1em}
    \caption{\emph{Upper Left Panel}: Axion-photon conversion sensitivities for the GC magnetar \gcmag\, assuming $B_{\rm pole} = 10B_Q$ and the {\color{black} collisional TK20 model with quarter arcade geometry. Curves correspond to the GBT, SKA1 and SKA2, and ALMA facilities.} 
    {\color{black} As in Fig. \ref{fig:sensitivities}, the data points represent
    the 1-sigma range obtained from MC ray tracing simulations, with the line sensitivity adjusted
    to the GBT or ALMA value (whichever is greater).}
    Colored bands show sensitivities in the {\color{black} one-dimensional approximation of the
    resonant surface.} All sensitivities assume an integration time of ten hours and a bandwidth of $10^{-5}\, m_a$. 
    \emph{Upper Right Panel}: Same as upper left panel but using the half arcade model. \emph{Lower Left Panel}: Slice arcade model. \emph{Lower Right Panel}: Same as the other panels but assuming the global 
    j-bundle model.  Note that {\color{black} sensitivities computed in the one-dimensional approximation} generally agree to within $2\sigma$ with the ray tracing
    MC results, but there are notable exceptions. 
    Below $m_a \sim 3\times 10^{-4}\eV$ in the collisional model, the one-dimensional approximation predicts weaker sensitivities. 
    }
    \label{fig:sensitivities_one_dimensional_approx.}
    \vspace{-1em}
\end{figure*}

In order to project sensitivities to axion-photon conversion in the mass ranges of $10^{-6}\eV\lesssim m_a \lesssim 10^{-3}\eV$, we compute the sensitivity of current radio and millimeter telescopes, the Green Bank Telescope (GBT)~\cite{Jewell2004,White2022,GBTTelescope2025} and the Atacama
Large Millimeter Array (ALMA)~\cite{Wootten2009,almatechnicalhandbook}, and the projected sensitivities of the upcoming Square Kilometer Array (SKA)~\cite{Dewdney2009,Braun2019}. The individual SEFDs for each telescope are shown in Fig.~\ref{fig:telescope_SEFDs}. Note that in the sensitivities shown in the main text, we ignore the frequency range $2.6\GHz \lesssim \nu \lesssim 3.96\GHz$ for GBT since the official proposer's guide does not report a receiver sensitivity in that frequency range~\cite{GBTTelescope2025}.


\section{One-dimensional Approximation} \label{sec:1D_approx}

In this appendix, we apply a simplified one-dimensional model of axion-photon
conversion to the same magnetar plasma profiles, as a check of the more precise, but largely
numerical, methodology adopted in Secs. \ref{sec:mixing} and \ref{sec:sense}. 
This method has been employed in a number of early studies of axion-photon conversion in NS magnetospheres (see e.g.~\cite{Hook:2018iia}). 

The underlying assumptions of this one-dimensional model are that all in-falling axion trajectories are 
radial, that all photons propagate along straight lines from their point of production, that the conversion surface is exactly two-dimensional, and that, more generally, the effects of spacetime curvature are ignorable.
The first of these assumptions is 
perhaps the most inaccurate, as the conversion probability can depend sensitively on the local phase space; nonetheless this approach provides a reasonable estimate 
in various regimes. 

Here, an in-falling axion experiences a plasma profile $\omega_p(r, \theta, \phi)$, where $\theta$ and $\phi$ are fixed. Resonant axion-photon conversion takes place at critical radius, $r_c$, defined using the resonance condition (see, e.g., ~\cite{millar2021axionphotonUPDATED})
\begin{align} \label{eqn:resonance_condition}
    \left\langle {\omega_p^2 \over \gamma^3 (\omega - k_\parallel v_\parallel)^2} \right\rangle = {m_a^2 \over m_a^2 \cos^2\theta_B + \omega^2 \sin^2 \theta_B},
\end{align}
where $\omega$ is the frequency of produced radiation, and $\theta_B$ is the angle between the magnetic field and the photon/axion momentum. 
An important complication arising in the magnetar magnetosphere modeled considered in Sec.~\ref{sec:jbundle} and Sec.~\ref{sec:collisionalModel} is that the plasma profile along a given trajectory need not increase monotonically toward the star.\footnote{Care must be taken when computing the axion-photon conversion probability near an extremum of the plasma distribution. There, phase interference effects can lead to orders-of-magnitude deviations from standard expectations~\cite{Brahma:2023zcw}.} In the BT07 model, there is a steep plasma frequency gradient at the boundary between the equatorial cavity and the j-bundle (see the rightmost panel in Fig.~\ref{fig:plasma_profiles_global}). The steepness of the transition between the cavity and j-bundle is an artifact of assuming a threshold for pair creation~\cite{Beloborodov_2009}. In a more realistic model, the transition is expected to be smooth. A non-monotonic plasma profile implies multiple solutions to (\ref{eqn:resonance_condition}). The solution at smaller radius (usually within the cavity) necessarily leads to enhanced conversion to photons; however, these photons will be reflected at the boundary of the cavity upon their attempted exit from the magnetosphere, leading to a possible build-up of radio photons in the cavity (see main text for a discussion). For simplicity, we do not consider this scenario and instead set the conversion radius to be the largest solution to (\ref{eqn:resonance_condition}). At the conversion radius, the conversion probability is approximated by a simplified version of (\ref{eqn:conversion_probability_full}),
\begin{align}
    P_{a\gamma} = {\pi \over 3} {\gagg^2 B(r_c)^2 r_c \over m_a v_a(r_c)},
\end{align}
where $v_a(r_c)$ is the axion velocity at the conversion radius, and we have assumed the plasma profile gradient at the conversion radius to be $|\omega_p'(r_c)| = 3 m_a / 2 r_c$, consistent with the requirement that the plasma number density $n \propto B \propto r^{-3}$.

As pointed out in~\cite{Hook:2018iia}, for conversion radii $r_c \ll G M_{\rm NS}/v_0^2$, where $v_0$ is the axion velocity far from the NS, the DM density at the conversion radius is 

\begin{align}
    \rho_{\rm DM}(r_c) = \rho_{\rm DM, 0} \sqrt{{2 G M_{\rm NS} \over v_0 ^2 r_c}},
\end{align}
where $\rho_{\rm DM, 0}$ is the DM density at distances large compared to the NS radius, but small compared to the DM halo scale radius. Applying the simplifying approximation that axions and photons execute 1D motion along radial trajectories, the outgoing photon power is

\begin{align}
    {dP \over d\Omega} = 2 P_{a\to \gamma} \ \rho_{\rm DM}(r_c) \ v_c  r_c^2,
\end{align}
where $v_c = \sqrt{v_0^2 + 2 G M_{\rm NS}/r_c}$ is the axion velocity at the conversion radius. The flux density of this photon signal can then be computed as before. As in the main text, we adopt a plasma broadened bandwidth of  $\Delta \nu = 10^{-4}m_a$.

In Fig.~\ref{fig:sensitivities_one_dimensional_approx.}, we show the sensitivities of the telescopes GBT, ALMA, and SKA (phases 1 and 2) to axion-photon coupling, assuming the same magnetar magnetosphere models and observational parameters as in Fig.~\ref{fig:sensitivities}. The different colored bands show the minimum, maximum, and average sensitivities when calculating the one-dimensional rotation-averaged power distributions for different viewing angles (i.e. the minimum sensitivity corresponds to the viewing angle with the lowest rotation-averaged differential power). Note that caution should be taken when comparing with the ray tracing results, as there we plot the $1\sigma$ flux containment regions, which are necessarily narrower than the maximal range. We find that the results of the sensitivities predicted by the one-dimensional approximation agree with the computationally ray-traced sensitivities (shown as error bars) to within $\sim 2\sigma$, assuming that we use the bandwidth calibrated by the ray-tracing simulations (i.e. $\Delta \nu \sim 10^{-4}\eV$). However, for lower axion masses $m_a \lesssim 10^{-4}\eV$, the TK20 slice model's ray-traced median axion-photon coupling sensitivities are over an order of magnitude lower than those predicted by the one-dimensional approximation. This is because the plasma
frequency is significantly lower on the side of the magnetar opposite to the emitting arcade,
where 
axion-photon conversion occurs close to the NS surface. In the radial approximation, the plasma profile is not monotonic and does rise locally with distance (e.g. see Fig.~\ref{fig:plasma_profiles}) and thus photons are not able to follow purely radial trajectories. The ray-tracing scheme we employ accounts for non-radial trajectories and allows those photons to escape, resulting in greater differential power and thus sensitivity to lower axion-photon coupling strengths. We also find that the radial approximation does poorly for the BT07 global twist model when conversion occurs close to the neutron surface (i.e. $m_a \sim 10^{-4}\eV$). These results emphasize the importance of using the detailed ray-tracing procedure outlined in Sec.~\ref{sec:mixing} in order to calculate sensitivities to these complex magnetosphere geometries.


\section{Axion-Photon Conversion in a Turbulent Magnetic Field} \label{sec:conversion_turbulent}

Here, we estimate the effect of a turbulent component of the magnetic field on the axion-to-photon conversion probability. We consider anisotropic incompressible magnetohydrodynamic (MHD) turbulence in the background of a strong, constant magnetic field, $\vec{B} = B_0 \hat{x}_\parallel$, which represents the potential (untwisted) component of the NS's magnetic field. Here $\hat{x}_\parallel$ is the unit vector in the direction of the magnetic field. We add to this ordered magnetic field a turbulent field $\delta \vec{B}(x_\parallel, \vec{x}_\perp)$, where $\vec{x}_\perp$ is the position in the plane perpendicular to the background magnetic field, and $|\delta \vec{B}|/B_0 \ll 1$. We can expand the turbulent field in Fourier modes,

\begin{align}
    \delta B_i(x_\parallel, \vec{x}_\perp) = \displaystyle\int {d^3 k \over (2\pi)^3} \delta \tilde{B}_i(k_\parallel, \vec{k}_\perp) e^{i k_\parallel x_\parallel + i \vec{k}_\perp \cdot \vec{x}_\perp}.
\end{align}

The root-mean-square (RMS) magnetic field can be written in the form

\begin{align} \label{eqn:two_point_function}
    \left\langle \delta \tilde{B}_i(\vec k) \delta \tilde{B}_j(\vec q) \right\rangle = (2\pi)^6 \left( \delta_{ij} - {k_i k_j \over k^2}\right) P_B(k_\parallel, k_\perp) \ \delta(\vec k - \vec q),
\end{align}
where $\left \langle \dots\right\rangle$ denotes a statistical average over different field configurations. The amplitude $P_B(k_\parallel,k_\perp)$ represents the breakdown of isotropy due to the presence of the background magnetic field. The dependence on the magnitude of $\vec k _\perp$ is a result of the assumed axisymmetry about the background magnetic field. The root-mean-square (RMS) magnetic field is

\begin{align}
   {1 \over 2} \left\langle \delta B^2 \right\rangle = \displaystyle \int d^3 k  \ P_B(k_\parallel, k_\perp).
\end{align}
$P_B(k_\parallel, k_\perp)$ can be interpreted as the differential magnetic energy density carried by fields with wave vector $\vec k$. As a concrete example, we consider Goldreich-Sridhar turbulence~\cite{GoldreichSridhar1, GoldreichSridhar2}, a key feature of which is that eddies are highly elongated along magnetic field lines, with $k_\perp \gg k_\parallel$. The magnetic power spectrum can be written in the form

\begin{align} \label{eqn:GS_spectrum}
    P_B(k_\parallel, k_\perp) =  {\varepsilon B_0^2\over 6\pi} \ k_\perp^{-10/3}L^{-1/3} \ \delta\left( {k_\parallel L^{1/3} \over k_\perp^{2/3}} - C \right),
\end{align}
where $L$ is the scale on which energy is injected (taken to be $\sim \rns$), which sets the infrared cutoff of the turbulent spectrum. The delta function follows from the assumption of a scale-independent cascade rate, which imposes the condition that $k_\parallel L^{1/3}/k_\perp^{2/3}$ is a constant that we denote $C$. Eq.~\eqref{eqn:GS_spectrum} is normalized such that the average energy density in the turbulent field $\left\langle \delta B^2 \right\rangle/B_0^2 = \varepsilon$. The effect on the non-resonant axion-to-photon conversion probability, $P_{a\to \gamma} \propto B^2$, is therefore altered by $\mathcal{O}(\varepsilon)$, which is taken to be much less than unity. In resonant conversion, however, the turbulent spectrum can lead to the formation of plasma overdensities, which can lead to many new resonances. To demonstrate this, we observe the RMS current density, $\vec{j} = \nabla \times \delta \vec{B}$, is

\begin{align}
    \left\langle j^2\right\rangle &= 2 \displaystyle \int d^3 k \ (k_\parallel^2 + k_\perp^2) P_B(k_\parallel, k_\perp) \label{eqn:j_spectrum}\\
    &= {\varepsilon B_0^2 \over 8\pi L^2} \left(L \over \ell_{\rm diss}\right)^{4/3} + \mathcal{O}\left[ \left(L \over \ell_{\rm diss}\right)^{2/3}\right],
\end{align}
where $\ell_{\rm diss}$ is the turbulent dissipation scale, which sets the ultraviolet cutoff of the spectrum. Integrating over $k_\parallel$, and keeping the leading order terms in $k_\parallel/k_\perp \ll 1$, Eq.~\eqref{eqn:j_spectrum} can be rewritten as a one-dimensional power spectrum,

\begin{align}\label{eqn:j_spectrum_1D}
      \left\langle j^2\right\rangle &= \displaystyle\int dk_\perp P_J(k_\perp), \quad P_J(k_\perp) = {\varepsilon B_0^2 k_\perp^{1/3} \over 6\pi L^{2/3}}.
\end{align}
The blue tilt of $P_J(k_\perp)$ suggests that the current amplitude may be dominated by small-scale fluctuations. In the presence of such fluctuations, there may be many resonant crossings, but each will be suppressed by the large spatial gradient of the fluctuations. 

To estimate the effect of small-scale fluctuations, we first revisit the computation of resonant axion-to-photon conversion. For simplicity, we consider axions propagating perpendicular to a background magnetic field. We ignore the effect of the turbulent field on the angle between the axion momentum and the magnetic field since $\cos\theta \sim \delta B/B_0 \ll 1$. For an axion plane wave, $a(z,t) = a_0  \ e^{i k_a z - i \omega t}$, where $k_a = \sqrt{\omega^2 - m_a^2}$, and $\omega$ is the frequency of the wave, the outgoing photon field in the WKB limit is~\cite{Witte:2021arp}

\begin{align}
    &A_\parallel(z,t) = -{\omega a_0 \over 2\sqrt{k_\gamma(z)}} e^{- i \omega t + i \int_0^z k_\gamma(z')dz'} \times  \nonumber\\
    & \displaystyle\int_0^z dz' \  {\gagg B_0(z') \over \sqrt{k_\gamma(z')} }  \exp\left( i \displaystyle\int _{0}^{z'} dz'' (k_\gamma(z'') - k_a) \right),
\end{align}
where $k_\gamma(z) = \sqrt{\omega^2 - \omega_p(z)^2}$, and $\omega_p(z)$ is the effective plasma frequency. Using the stationary phase approximation,

\begin{align}
    &A_\parallel(z,t) \approx -{\omega a_0 \over 2\sqrt{k_\gamma(z)}} e^{- i \omega t + i \int_0^z k_\gamma(z')dz'} \times \nonumber \\
    & \displaystyle\sum_{n} {\gagg B_0(z_n) \over \sqrt{k_\gamma(z_n)}} e^{i \phi_n} \sqrt{2\pi \over |k_\gamma'(z_n)|},
\end{align}
where the sum is performed over all solutions $z_n$ to the equations $k_\gamma(z) = k_a$, and 
\begin{align}
    \phi_n = \displaystyle\int_0^{z_n} (k_\gamma (z')-k_a) \ d z' + {\pi \over 4} {\rm sign}(k_\gamma'(z_n)).
\end{align}

Most resonant crossings will still take place near the unperturbed resonant point, where the magnetic field is taken to be $B_c$. Resonant crossings may still take place away from the unperturbed resonant point in the presence of large-amplitude current fluctuations. This can lead to enhancements in the conversion probability if those fluctuations occur closer to the surface than the unperturbed conversion location, but we neglect this possibility. We can then write the effective conversion probability,

\begin{align} \label{eqn:conversion_probability_turbulent}
    P_{a \to \gamma} \equiv {|A_\parallel(z,t)|^2 \over |a_0|^2} =  {\pi \over 2 v_c} \gagg^2 B_c^2 \left| \displaystyle\sum_n {e^{i \phi_n} \over \sqrt{|\omega_p'(z_n)}|}\right|^2,
\end{align}
where $v_c$ is the axion speed at the unperturbed resonant crossing point, and the expression is taken in the limit $z \to \infty$. As a lower bound, let us assume the resonant crossings are dominated by the highest-$k$ modes possible, $|\omega_p'(z_n)| \sim \omega_p(z_n) k_{max} \sim m_a/\ell_{\rm diss}$. We also treat $\phi_n$ to be random phases chosen from the uniform distribution $\phi \in [0,2\pi]$. The magnitude of the sum of $N \gg 1$ random phases is $|\sum_n \exp(i\phi_n)|^2 = N$. Therefore, the conversion probability is

\begin{align}
    P_{a \to \gamma} \ge {\pi \over 2 v_c} \gagg^2 B_c^2 {\ell_{\rm diss} N_{\times} \over m_a},
\end{align}
where $N_\times$ is the number of resonant crossings. Assuming current fluctuations are dominated by the high-$k_\perp$ modes, the expected number of resonant crossings in a region of length $r_c$, taken to be the conversion radius, is $N_\times \sim \sqrt{r_c/\ell_{\rm diss}}$. Assuming the dissipation scale, $\ell_{\rm diss} \sim 1/\omega_p\approx m_a$ in the resonance region, we get an effective conversion probability of

\begin{align}
    P_{a\to\gamma} \sim {1\over v_c }\gagg^2 B_c^2 \sqrt{r_c \over m_a^3},
\end{align}
which is less than the conversion probability in a smooth plasma by a factor of $\sim (m_a r_c)^{-1/2}$. A few comments on this result are in order. Firstly, this result relies on the assumption that the dissipation length scale is larger than the plasma scale. Computation of the axion-to-photon conversion probability presented here and in the main text relies on the WKB approximation, which breaks down for modes with wavelength less than or comparable to the axion wavelength. Secondly, we have computed the probability in the one-dimensional approximation. In one dimension, photons from resonant conversion will not be able to escape if they encounter a downstream plasma overdensity. In two or three dimensions, photons can refract (or diffract for overdensities on scales comparable to, or smaller than, the wavelength) around overdensities and escape the magnetosphere. This may lead to interesting interference effects, a thorough investigation of which we leave to future work. 




\end{document}